\documentclass[twocolumn,showpacs,preprintnumbers,amsmath,amssymb,epsfig]{revtex4}

\usepackage{graphicx}
\usepackage{dcolumn}
\usepackage{bm}
\usepackage{epsfig}
\usepackage{hyperref}

\usepackage{float}
\usepackage{amsmath}
\usepackage{epsfig,floatflt}
\usepackage{subfigure}
\usepackage[usenames]{color}
\usepackage{comment} 
\usepackage{ulem}

\begin{document}


\title{Current data are consistent with flat spatial hypersurfaces in the $\Lambda$CDM cosmological model but favor more lensing than the model predicts}

\author{Javier de Cruz P\'erez${}^{1}$, Chan-Gyung Park${}^{2}$, and Bharat Ratra${}^{1}$}
\affiliation{
         ${}^{1}$Department of Physics, Kansas State University, 116 Cardwell Hall, Manhattan, KS 66506, USA \\
         ${}^{2}$Division of Science Education and Institute of Science Education, Jeonbuk National University, Jeonju 54896, Republic of Korea }
\email{decruz@phys.ksu.edu, park.chan.gyung@\-gmail.com, ratra@phys.ksu.edu}
\date{\today}

\begin{abstract}
We study the performance of three pairs of tilted, and a pair of untilited, $\Lambda$CDM cosmological models, with three of these four pairs allowing for non-flat spatial hypersurfaces, against cosmic microwave background (CMB) temperature and polarization power spectrum data (P18), measurements of the Planck 2018 lensing potential power spectrum (lensing), and a large compilation of non-CMB data (non-CMB). For the eight models, we measure cosmological parameters and study whether or not pairs of the data sets (as well as subsets of them) are mutually consistent in these models.
Half of these models allow the lensing consistency parameter $A_L$, which re-scales the gravitational potential power spectrum, to be an additional free parameter to be determined from data, while the other three have $A_L = 1$ which is the theoretically expected value. The pair of untilted non-flat $\Lambda$CDM models are incompatible with P18 data. The tilted spatially-flat models assume the usual primordial spatial inhomogeneity power spectrum that is a power law in wave number. The tilted non-flat models assume either the primordial power spectrum used in the Planck group analyses [Planck $P(q)$], that has recently been numerically shown to be a good approximation to what is quantum-mechanically generated from a particular choice of closed inflation model initial conditions, or a recently computed power spectrum [new $P(q)$] that quantum-mechanically follows from a different set of non-flat inflation model initial conditions. In the tilted non-flat models with $A_L=1$ we find differences between P18 data and non-CMB data cosmological parameter constraints, which are large enough to rule out the Planck $P(q)$ model at 3$\sigma$ but not the new $P(q)$ model. No significant differences are found when cosmological parameter constraints obtained with two different data sets are compared within the standard tilted flat $\Lambda$CDM model. While both P18 data and non-CMB data separately favor a closed geometry, with spatial curvature density parameter $\Omega_k<0$, when P18+non-CMB data are jointly analyzed the evidence in favor of non-flat hypersurfaces subsides. Differences between P18 data and non-CMB data cosmological constraints subside when $A_L$ is allowed to vary. From the most restrictive P18+lensing+non-CMB data combination we get almost model-independent constraints on the cosmological parameters and find that the $A_L>1$ option is preferred over the $\Omega_k<0$ one, with the $A_L$ parameter, for all models, being larger than unity by $\sim 2.5\sigma$. According to the deviance information criterion, in the P18+lensing+non-CMB analysis, the varying $A_L$ option is on the verge of being {\it strongly} favored over the $A_L=1$ one, which could indicate a problem for the standard tilted flat $\Lambda$CDM model. These data are consistent with flat spatial hypersurfaces but more and better data could improve the constraints on $\Omega_k$ and might alter this conclusion. Error bars on some cosmological parameters are significantly reduced when non-CMB data are used jointly with P18+lensing data. For example, in the tilted flat $\Lambda$CDM model for P18+lensing+non-CMB data the Hubble constant $H_0=68.09\pm 0.38$ km s$^{-1}$ Mpc$^{-1}$, which is consistent with that from a median statistics analysis of a large compilation of $H_0$  measurements as well as with a number of local measurements of the cosmological expansion rate. This $H_0$ error bar is 31\% smaller than that from P18+lensing data alone.
  \end{abstract}

\maketitle

\section{Introduction}
\label{sec:intro}

General relativity is the current best description of gravity on cosmological scales. In general relativity gravity is responsible for the observed expansion of the Universe and can be sourced by non-relativistic (cold dark and baryonic) matter, relativistic radiation/matter, a cosmological constant (or a dynamical dark energy density), and the curvature of space. In an influential 1932 paper Einstein and de Sitter, \citep{EinsteindeSitter1932}, noted that available data then were unable to measure spatial curvature and so decided to study whether a spatially-flat cosmological model was observationally consistent. They acknowledged that the cosmological model had to be dynamical, and so Einstein's original argument for a cosmological constant --- to make the Universe static --- was no longer valid and so the cosmological constant did not have to be included in this Einstein-de Sitter model. They ignored relativistic radiation/matter in this model (which was not under discussion then, and is known to be negligible at late times when the model was meant to be applicable). This Einstein-de Sitter model only included non-relativistic (and then only baryonic) matter.     

A little over half a century later, motivated by observations indicating a lower than critical non-relativistic matter energy density and the first inflation model, an improved standard model, the spatially-flat $\Lambda$CDM model, was proposed, \citep{Peebles:1984ge}. In this model the cosmological constant $\Lambda$, which has a time- and space-independent energy density, is the dominant contributor to the current cosmological energy budget, followed by non-relativistic non-baryonic cold dark matter (CDM), and then non-relativistic baryonic matter. Like the Einstein-de Sitter model, the standard spatially-flat $\Lambda$CDM model assumes vanishing spatial curvature, motivated by early models of spatially-flat inflation, \citep{Guth1981, Sato1981a, Sato1981b, Kazanas1980}. Soon after, spatially-non-flat, open and closed, inflation models were developed, \citep{Gott1982, Hawking1984, Ratra1985}.       

A decade and a half later, the observed currently accelerated cosmological expansion, discovered from type Ia supernova (SNIa) measurements \cite{Riess:1998cb, Perlmutter:1998np}, greatly strengthened support for a cosmological constant or a dynamical dark energy density that slowly varied in time and space \citep{PeeblesRatra1988, RatraPeebles1988} --- if general relativity is an accurate model for gravity on cosmological length scales --- and for the spatially-flat $\Lambda$CDM model or a model close to it. For reviews of the current situation see Refs.\ \citep{DiValentino:2021izs, Perivolaropoulos:2021jda, Abdalla:2022yfr}.

A half-decade prior to the first SNIa observations indicating currently accelerated cosmological expansion, evidence for a lower than critical value of non-relativistic matter density, along with the development of an open inflation model, \citep{Gott1982}, led to some discussion of an open CDM model, \citep{RatraPeebles1994, RatraPeebles1995, Bucheretal1995, Yamamotoetal1995, Kamionkowskietal1994, Gorskietal1998, Ratraetal1999}, but with cosmic microwave background (CMB) observations indicating that space curvature had to be a subdominant contributor to the current cosmological energy budget, \citep{WMAP:2012nax, Planck:2018vyg}, and with SNIa observations favoring a significant contribution to the energy budget from a cosmological constant, interest in open CDM models soon faded.    

More recently, especially because of results from Planck CMB anisotropy data, \citep{Planck:2018vyg}, there has been renewed interest in non-flat models. In these models the current cosmological energy budget is dominated by $\Lambda$, to be consistent with the observed currently accelerated cosmological expansion, but they now have very mildly closed spatial hypersurfaces instead of open ones. This is because from an analysis of the final Planck 2018 TT,TE,EE+lowE (hereafter P18) data, that makes use of a specific primordial power spectrum (see below for a fuller discussion of these data and the power spectrum they use in this analysis), they find a spatial curvature energy density parameter value $\Omega_k = -0.044^{+0.018}_{-0.015}$ that is closed and 2.7$\sigma$ away from flat, \citep{Planck:2018vyg}, when $\Omega_k$ is included as an additional free parameter in the analysis. We note that from a combination of Atacama Cosmology Telescope (ACT) and Wilkinson Microwave Anisotropy Probe CMB anisotropy data Ref.\ \citep{ACT:2020gnv} find $\Omega_k = -0.001^{+0.014}_{-0.010}$, which is 2.1$\sigma$ from the P18 value and consistent with flat spatial hypersurfaces, while the South Pole Telescope (SPT) CMB anisotropy data results in $\Omega_k = 0.001^{+0.018}_{-0.019}$, \citep{SPT-3G:2021wgf}, which is 1.7$\sigma$ from the P18 value and also consistent with flat spatial hypersurfaces. Both these analyses use the primordial power spectrum used in the P18 analysis.

The above result led to the study of the so-called lensing anomaly. The trajectory of CMB photons are bent by the gravitational effect of inhomogeneities present in the mass distribution along their way to us. This statistical phenomenon, predicted by general relativity, is known as weak gravitational lensing of the CMB. When computing the predicted CMB temperature and polarization spectra in a cosmological model that are to be compared to the observed spectra, it is important to account for this effect and compute what are known as lensed CMB spectra. If we use the tilted flat $\Lambda$CDM model to measure cosmological parameter values from Planck CMB data, we can use this model, with these parameter values, to compute the expected amount of CMB weak gravitational lensing, \cite{Lewis:2006fu}. Incorrectly predicting the amount of weak lensing present in the CMB power spectra would indicate an inconsistency in the standard model when it is used to fit Planck CMB temperature and polarization anisotropy data. It turns out that this is actually the case, since an excess of CMB weak lensing is observed in the CMB power spectra, compared to what is expected in the standard model with parameter values determined from CMB data  \cite{Calabreseetal2008, Planck:2018vyg}. This is known as the lensing anomaly, since the effect is not yet thought to be statistically significant enough to reject the standard spatially-flat $\Lambda$CDM model. 

A number of solutions have been proposed, with two being more widely debated. The first of these is related to the aforementioned non-zero value for $\Omega_k$ in the P18 data analysis, which favors closed spatial hypersurfaces, when $\Omega_k$ is taken to be an additional free parameter, e.g.\ \citep{Planck:2018vyg, Handley:2019tkm, DiValentino:2019qzk, DiValentino:2022rdg,Yang:2022kho}. Due to the excess of CMB weak lensing found, it is desirable to have a higher value of the non-relativistic matter energy density parameter $\Omega_m$ in order to increase the amount of gravitational lensing of CMB photons. Because of the tight constraints imposed by CMB data on this parameter there is no room, within the tilted flat $\Lambda$CDM model, to do this. By allowing non-flat spatial hypersurfaces, a closed model with $\Omega_k<0$ can resolve this problem, since the CMB power spectra are affected by the combination $(\Omega_m + \Omega_k)h^2$, where $h$ is the Hubble constant $H_0$ in units of $100~\textrm{km}~\textrm{s}^{-1}~\textrm{Mpc}^{-1}$, which can be held constant by making $\Omega_k$ slightly more negative while slightly increasing $\Omega_m$ to give more CMB weak lensing, and also slightly adjusting $h$. Cosmological distances also depend on spatial curvature, therefore in a non-flat cosmological model the positions of the acoustic peaks are shifted relative to the flat model case. This would not be a welcome change, since the constraints from the observed CMB power spectra are tight. This can be resolved by reducing the value of $h$ which shifts the acoustic peaks in the opposite direction. The fact that almost the same temperature and polarization power spectra can be produced with different combinations of the cosmological parameter values points to a geometrical degeneracy between these three parameters,  $H_0$-$\Omega_m$-$\Omega_k$. 

While the first of the more widely debated resolutions is based  on a change of more-conventional cosmological parameters, the second one is more phenomenological, e.g.\ \citep{Planck:2018vyg, SPT:2017jdf, SPT:2019fqo, DiValentino:2019qzk, DiValentino:2022rdg}. Reference \cite{Calabreseetal2008} introduces the lensing consistency parameter $A_L$ which re-scales the gravitational potential power spectrum in such a way that when $A_L=1$ we recover the theoretically predicted amount of weak lensing. If $A_L$ is allowed to vary in the analysis, to be determined from data, when $A_L>1$ the predicted amount of lensing is greater than the case when $A_L=1$. In Ref.\ \cite{Planck:2018vyg} when P18 data are used to analyze the tilted flat $\Lambda$CDM+$A_L$ model, the result is $A_L = 1.180\pm 0.065$ which represents a 2.8$\sigma$ deviation from the theoretically expected value $A_L=1$. We emphasize however that the measured Planck lensing likelihood is consistent with $A_L = 1$, see Fig.\ 3 of Ref.\ \citep{Planck:2018vyg} and Ref.\ \citep{Planck:2018lbu}. We also note that from ACT CMB anisotropy data $A_L = 1.01 \pm 0.11$, \citep{ACT:2020gnv}, consistent with $A_L = 1$ and 1.3$\sigma$ smaller than the P18 value, while from SPT CMB anisotropy data  $A_L = 0.81 \pm 0.14$, \citep{SPT:2017jdf}, 1.4$\sigma$ smaller than $A_L = 1$ and 2.4$\sigma$ smaller than the P18 value.

To analyze CMB anisotropy data one must assume a form for the primordial power spectrum of spatial inhomogeneities as a function of wavenumber. In the inflation scenario zero-point quantum-mechanical fluctuations during inflation generate the spatial inhomogeneities, \citep{Hawking:1982cz, Starobinsky:1982ee, Guth:1982ec, Bardeen:1983qw, Fischler:1985ky}. In spatially-flat inflation models, if the inflaton field slowly rolls down an almost flat potential energy density the scale factor increases exponentially with time and the primordial power spectrum is almost scale-invariant with hardly any tilt, \citep{Harrison:1969fb, Peebles:1970ag, Zeldovich:1972zz}. A steeper inflaton potential energy density makes the inflaton evolve more rapidly, can cause the scale factor to grow only as a power of time, and will increase the power spectral tilt \citep{Lucchin:1984yf, Ratra:1989uv, Ratra:1989uz}.

There has been much less study of the quantum-mechanical generation of spatial inhomgeneities in non-flat inflation models. Power spectra have been derived in spatially open and closed inflation models,  \citep{Gott1982, Hawking1984, Ratra1985}, with a slowly-rolling inflation potential energy density, \citep{RatraPeebles1995, Ratra:2017ezv}, but these are untilted power spectra. The power spectrum assumed in the non-flat analyses of Refs.\ \citep{Planck:2018vyg, Handley:2019tkm, DiValentino:2019qzk} is tilted but was not derived from an inflation model computation. Very recently, a numerical study in closed inflation models that computes primordial power spectra generated for a few different, initially slow-roll, inflation initial conditions finds that it is possible to generate, in the closed case, a tilted power spectrum very close to that used in Refs.\ \citep{Planck:2018vyg, Handley:2019tkm, DiValentino:2019qzk}, \cite{Guth:2022xyz}. Also recently, a different set of initial conditions in closed and open inflation models were used to compute a different tilted power spectrum, \citep{Ratra:2022ksb}.

In this paper we consider cosmological models with four different power spectra. In the tilted flat $\Lambda$CDM model we use the usual spatially-flat inflation model tilted power spectrum. In the untilted non-flat $\Lambda$CDM model, we use the untilted non-flat inflation model power spectrum, \citep{RatraPeebles1995, Ratra:2017ezv}. In the two different tilted non-flat $\Lambda$CDM models, we use the power spectrum assumed in Ref.\ \citep{Planck:2018vyg, Handley:2019tkm, DiValentino:2019qzk} --- which we call the Planck $P(q)$ --- as well as the power spectrum computed in Ref.\ \citep{Ratra:2022ksb}, which we call the new $P(q)$. See Sec.\ \ref{sec:method} below for a fuller description of the four power spectra we use.

We emphasize that we use only non-flat inflation model power spectra that can be derived using a straightforward extension of the spatially-flat  inflation model initial conditions to the non-flat inflation case. The issue of non-flat inflation model initial conditions is more complex than the flat inflation case,  see discussion in Ref.\ \citep{Ratra:2022ksb}, so we focus on the simplest physically-consistent options, which also makes the analysis tractable. We note that a number of other power spectra have been considered in closed cosmological models, see Refs.\ \citep{Lasenby:2003ur, Masso:2006gv, Asgari:2015spa, Bonga:2016iuf, Handley:2019wlz, Thavanesan:2020lov, Kiefer:2021iko, Hergt:2022fxk}.

A desire to measure the spatial curvature energy density parameter $\Omega_k$ provides part of the motivation for our work. The CMB anisotropy data are currently the most restrictive cosmological data, but to use these to measure $\Omega_k$ requires assumption of a primordial power spectrum for spatial inhomgeneities. Other, less-restrictive, data that do not require assuming a power spectrum can also be used to measure $\Omega_k$. These include better-established lower redshift data (that reach to $z \sim 2.3$), such as SNIa, Hubble parameter as a function of redshift [$H(z)$], and (non-growth-rate) baryon acoustic oscillation (BAO) measurements, \citep{Scolnic:2017caz, Yu:2017iju, eBOSS:2020yzd}, as well as emerging probes that reach to higher $z$, such as H \textsc{ii} starburst galaxy apparent magnitude observations as a function of $z$ that reach to $z \sim 2.5$, \citep{Gonzalez-Moran:2019uij,Cao:2020jgu,Cao:2020evz, Johnson:2021wou, Mehrabi:2021feg}; quasar angular size measurements that reach to $z \sim 2.7$, \citep{Cao:2017ivt,Ryan:2019uor, Lian:2021tca, Cao:2021cix}; Mg \textsc{ii} and C \textsc{iv} reverberation measured quasar data that reach to $z \sim 3.4$, \citep{OzDES:2021byt, Khadka:2021ukv, Khadka:2021sxe, Khadka:2022ooh, Cao:2022pdv, OzDES:2022ysr, Czerny:2022xfj}; possibly quasar flux measurements that reach to $z \sim 7.5$, \citep{Risaliti:2018reu, Khadka:2020whe, Khadka:2020vlh, Lusso:2020pdb, Khadka:2020tlm, Khadka:2021xcc, Rezaei:2021qwd, Dainotti:2022rfz, Petrosian:2022tlp, Khadka:2022aeg}; and gamma-ray burst data that reach to $z \sim 8.2$, \citep{Dirirsa:2019fcs, Khadka:2020hvb, Khadka:2021vqa, Wang:2021hcx, Hu:2021ycz, Cao:2021irf, Luongo:2021pjs, Cao:2022wlg, Liu:2022srx, Dainotti:2022wli, Cao:2022yvi}. Individually these low- and intermediate-redshift data sets are only able to provide relatively weaker constraints on cosmological parameters in general, and specifically on $\Omega_k$, compared to those from CMB data. However, when many (or all) low- and intermediate-redshift data are analyzed jointly they provide useful constraints on $\Omega_k$ --- currently still not nearly as restrictive as the CMB ones --- favoring flat spatial hypersurfaces but still allowing a small amount of spatial curvature energy density, \citep{Park:2018tgj, Cao:2021ldv, Cao:2022ugh}. For other recent discussions of constraints on spatial curvature, see Refs.\ \citep{Vagnozzi:2020dfn, Arjona:2021hmg, Dhawan:2021mel, Gonzalez:2021ojp, Geng:2021hqc, Wei:2022plg, Mukherjee:2022ujw, Wu:2022fmr} and references therein, and see Refs.\ \citep{Baumgartner:2022jdz, Anselmi:2022uvj, Jimenez:2022asc} and references therein for recent, more general, discussions of non-flat cosmological models.

While the standard spatially-flat $\Lambda$CDM cosmological model is attractive because of its simplicity --- the model only has 6 free cosmological parameters --- it is not straightforward to understand how to consistently generalize the current quantum-mechanical standard model of particle physics to one that accommodates the cosmological constant that is part of the standard $\Lambda$CDM model. Nonetheless, the standard cosmological model is consistent with a wide variety of measurements, including CMB anisotropy measurements \cite{Planck:2018vyg}, SNIa apparent magnitude observations \citep{Scolnic:2017caz}, BAO data \citep{eBOSS:2020yzd}, $H(z)$ observations \citep{Yu:2017iju}, and measurements of the growth of structure as a function of redshift ($f\sigma_8$). It is important to bear in mind that these data do not rule out mild evolution of the dark energy density \cite{Gomez-Valent:2018nib, Ooba:2018dzf, Ryan:2018aif, SolaPeracaula:2018wwm, Singh:2018izf, Park:2019emi, Gomez-Valent:2020mqn, Moreno-Pulido:2020anb, Sinha:2020vob, Cao:2020jgu, Urena-Lopez:2020npg, Cao:2021ldv, SolaPeracaula:2021gxi, Khadka:2021vqa, Cao:2021cix, Xu:2021xbt, Cao:2021irf, Jesus:2021bxq, Cao:2022wlg, Moreno-Pulido:2022phq, Cao:2022pdv, Adil:2022hkj} or, as discussed in detail above, mildly curved spatial hypersurfaces. These extensions, among others, might alleviate some of the issues affecting the standard spatially-flat $\Lambda$CDM model, such as differences in $H_0$ and $\sigma_8$ values determined using different techniques, \citep{DiValentino:2021izs, Perivolaropoulos:2021jda, Abdalla:2022yfr}. In this paper however we focus our efforts on the study of the lensing anomaly and on the measurement of $\Omega_k$.

In this paper we study eight cosmological models (six of them are tilted models and two untilted), namely, the tilted flat $\Lambda$CDM (+$A_L$) models, the untilted non-flat $\Lambda$CDM (+$A_L$) models, the tilted non-flat $\Lambda$CDM (+$A_L$) Planck $P(q)$ models, and the tilted non-flat $\Lambda$CDM(+$A_L$) new $P(q)$ models. Six of these are non-flat models, characterized by three different primordial power spectra (see Sec.\ \ref{sec:method} for the form of the power spectra). By using a number of cosmological models with compilations of observational data to test how well the models fit these data, and to constrain the cosmological parameters of the models, we can measure, among other things, $\Omega_k$ and also determine whether the cosmological parameter constraints set by different data are model-dependent or not. The data sets we employ in this work are P18 data, Planck 2018 CMB weak lensing data, non-growth-factor BAO (BAO$^{\prime}$) data, BAO (including growth-factor) data, and non-CMB data [composed of BAO, $f\sigma_8$, $H(z)$, and SNIa data]. These data are described in more detail in Sec.\ \ref{sec:data}. 

A brief summary of the more significant results we find follows. These assume that the data sets we use are correct and do not have unaccounted for systematic errors. The untilted non-flat $\Lambda$CDM model with and without a varying $A_L$ parameter is not able to properly fit the P18 CMB anisotropy power spectra, due to the lack of the tilt ($n_s$) degree of freedom. Consequently its performance in comparison with the tilted models turns out to be very poor. Significant evidence in favor of a closed Universe is found when P18 data are considered alone and the tilted non-flat models better fit these data than does the standard tilted flat $\Lambda$CDM model. There are disagreements between P18 data cosmological constraints and non-CMB data cosmological constraints in the context of the tilted non-flat models with $A_L=1$, with the tilted non-flat $\Lambda$CDM Planck $P(q)$ model ruled out at 3$\sigma$. These tensions completely fade when the $A_L$ parameter is allowed to vary. On the other hand no significant tension is found when the cosmological parameter constraints obtained with two different data sets are compared within the standard tilted flat $\Lambda$CDM model. The most-restrictive P18+lensing+non-CMB data set clearly favors the varying $A_L$ option (with $A_L>1$) over the $A_L=1$ one --- which could be a problem for the standard tilted flat $\Lambda$CDM model --- and when this data set is utilized we get almost model-independent cosmological parameter constraints. These data are consistent with flat spatial hypersurfaces --- so we conclude that current data do not favor curved geometry --- but more and better data could improve the constraints on $\Omega_k$ and might alter this conclusion.  We note that even though both P18 data and non-CMB data favor closed geometry, the larger $H_0$ and smaller $\Omega_m$ values favored by non-CMB data (compared to those favored by P18 data) result in P18+lensing+non-CMB data favoring flat spatial hypersurfaces.  The Hubble constant value measured using these data in the tilted flat $\Lambda$CDM model is $H_0=68.09\pm 0.38$ km s$^{-1}$ Mpc$^{-1}$, which is consistent with that from a median statistics analysis of a large compilation of Hubble constant measurements as well as with a number of local measurements of the cosmological expansion rate. This $H_0$ error bar is 31\% smaller than that from P18+lensing data alone; similarly augmenting the P18+lensing data with our non-CMB data compilation reduces the $\Omega_m$ error bar by 33\% and also reduces error bars on all the other cosmological parameters by smaller amounts. 

The layout of our paper is as follows. In Sec.\ \ref{sec:data} we detail the observational data sets we employ to test the different cosmological models. In Sec.\ \ref{sec:method} we describe the cosmological models and primordial power spectra we study and summarize the methods we use in our analyses. We dedicate Sec.\ \ref{sec:results} to discuss in detail the results obtained by testing the different cosmological models against the several data sets we consider. In this section we also utilize different statistical estimators to compare the performance of the models in fitting data and to study possible tensions between different data sets in a given model. In Sec.\ \ref{sec:discussion} we summarize the more significant results of the previous (long) section. Finally in Sec.\ \ref{sec:conclusion} we deliver our conclusions.

\section{Data}
\label{sec:data} 
 
We use CMB anisotropy data and non-CMB data to constrain cosmological parameters,  to determine how well the cosmological models we study fit these data, and to study how mutually consistent these data sets are in each of the cosmological models. We now list the data sets we use. 

{\bf P18}. Planck 2018 CMB temperature anisotropy data together with polarization data and their corresponding cross-spectra (TT,TE,EE+lowE), \cite{Planck:2018vyg}, which contain: TT power spectra at low-$\ell$ ($2\leq \ell \leq 29$) and high-$\ell$ ($30\leq \ell\leq 2508$) --- where $\ell$ is multipole number, TE data at high-$\ell$ ($30\leq \ell \leq 1996$), and EE data at low-$\ell$ ($2\leq \ell \leq 29$) and high-$\ell$ ($30\leq \ell\leq 1996$). We use the Planck 2018 baseline \texttt{Plik} $\ell \geq 30$ likelihood, which is described in Sec.\ 2.2.1 of Ref.\ \cite{Planck:2018vyg}. 

{\bf (P18) lensing}. Planck 2018 lensing potential power spectrum, see Sec.\ 2.3 of Ref.\ \cite{Planck:2018vyg} or Sec.\ 2 of Ref.\ \cite{Planck:2018lbu} for more details. 

{\bf BAO$^\prime$}. Twelve BAO data points from both anisotropic and isotropic BAO estimators that probe the redshift range $0.122 \leq z \leq 2.334$ \cite{Gil-Marin:2020bct,Bautista:2020ahg,Hou:2020rse,Neveux:2020voa,Carter:2018vce,DES:2017rfo,duMasdesBourboux:2020pck}. These are BAO data with growth rates excluded from the original papers, and are listed, along with the appropriate covariance matrices, in Sec.\ 3 of Ref.\ \cite{Cao:2022ugh}.

\begin{table}
\caption{BAO measurements.}
\begin{ruledtabular}
\begin{tabular}{ccc}
 $z_\textrm{eff}$                     &  Measurement                                          &   Reference    \\[+0mm]
 \hline \\[-2mm]
 $0.122$    & $D_V\left(r_{d,\textrm{fid}}/r_d\right)$ [Mpc]                                              $= 539\pm 17$ [Mpc]  &  \cite{Carter:2018vce} \\[+1mm]
  \hline \\[-2mm]
 $0.38$     & $D_M/r_d$                   $= 10.274 \pm 0.151$   &  \cite{Gil-Marin:2020bct} \\[+1mm]
 $0.38$     & $D_H/r_d$   $= 24.888\pm 0.582$   &  \cite{Gil-Marin:2020bct}  \\[+1mm]
 $0.51$     & $D_M/r_d$                    $= 13.381 \pm 0.179$  &  \cite{Gil-Marin:2020bct}  \\[+1mm]
 $0.51$     & $D_H/r_d$  $= 22.429   \pm 0.482$   &  \cite{Gil-Marin:2020bct}  \\[+1mm]
 $0.38$     & $f \sigma_8                                               =0.49729\pm0.04508$       &  \cite{Gil-Marin:2020bct}  \\[+1mm]
 $0.51$     & $f \sigma_8                                               =0.45902\pm 0.03784$       &  \cite{Gil-Marin:2020bct}  \\[+1mm]
 \hline \\[-2mm]
 $0.698$    & $D_M/ r_d$                                               $= 17.646\pm0.302$      & \cite{Gil-Marin:2020bct,Bautista:2020ahg} \\[+1mm]
 $0.698$    & $D_H / r_d$                                               $= 19.770\pm0.469$     & \cite{Gil-Marin:2020bct,Bautista:2020ahg} \\[+1mm]
 $0.698$    & $f\sigma_8$                                               $= 0.47300\pm 0.04429$     & \cite{Gil-Marin:2020bct,Bautista:2020ahg} \\[+1mm]
 \hline \\[-2mm]
 $0.81$     & $D_A/r_d$                   $= 10.75\pm 0.43$       & \cite{DES:2017rfo}  \\[+1mm]
 \hline \\[-2mm]
 $1.48$    & $D_M/ r_d$                                               $= 30.21\pm 0.79$      & \cite{Hou:2020rse,Neveux:2020voa} \\[+1mm]
 $1.48$    & $D_H / r_d$                                               $= 13.23\pm 0.47$     & \cite{Hou:2020rse,Neveux:2020voa} \\[+1mm]
 $1.48$    & $f\sigma_8$                                               $= 0.462\pm 0.045$     & \cite{Hou:2020rse,Neveux:2020voa} \\[+1mm]
 
 \hline \\[-2mm]
 $2.334$    & $D_M / r_d$                                               $= 37.5^{+1.2}_{-1.1}$      & \cite{duMasdesBourboux:2020pck} \\[+1mm]
 $2.334$    & $D_H / r_d$                                               $= 8.99^{+0.20}_{-0.19}$     & \cite{duMasdesBourboux:2020pck} \\[+0mm]
\end{tabular}
\\[+1mm]
\begin{flushleft}
Note: For the data point at $z = 0.122$ the sound horizon size (at the drag epoch) of the fiducial model is $r_{d,\textrm{fid}}=147.5~\textrm{Mpc}$ \cite{Carter:2018vce}.
\end{flushleft}
\end{ruledtabular}
\label{tab:bao}
\end{table}

{\bf BAO}. An extension of the BAO$^\prime$ data described above, that also probe the redshift range $0.122 \leq z \leq 2.334$, but now include the correlated growth rate ($f\sigma_8$) data points provided in Refs.\ \cite{Gil-Marin:2020bct,Bautista:2020ahg,Hou:2020rse,Neveux:2020voa}. Table \ref{tab:bao} lists these BAO data points. 

The quantities listed in Table \ref{tab:bao} include transverse comoving distance at redshift $z$

\begin{equation}
D_M(z) = (1+z)D_A(z), 
\end{equation}
where $D_A(z)$ is the angular size distance at $z$,
\begin{equation}
D_H(z) = \frac{c}{H(z)},     
\end{equation}
where $H(z)$ is the Hubble parameter and $c$ the speed of light, and the angle-averaged distance
\begin{equation}
D_V(z) = \left[czD^2_M(z)/H(z)\right]^{1/3}.  
\end{equation}
The measurements are provided as relative distances with respect to the radius of the sound horizon at the drag epoch redshift $z_d$
\begin{equation}
r_d = \int^{\infty}_{z_d}\frac{c_s(z)dz}{H(z)},    
\end{equation}
where $c_s(z)$ is the speed of sound in the photon-baryon fluid.

For BAO data from Ref.\ \cite{Gil-Marin:2020bct} the appropriate covariance matrix is now
\begin{widetext}
\begin{equation}
\label{eq:cov_BOSS}
\begin{pmatrix}
0.022897 & -0.02007 & 0.0026481 & 0.013487 & -0.0081402 & 0.0010292 \\
-0.02007 & 0.33849 & -0.0085213 & -0.016024 & 0.13652 & -0.0038002 \\
0.0026481 & -0.0085213 & 0.0020319 & 0.001325 & -0.0023012 & 0.000814158 \\
0.013487 & -0.016024 & 0.001325  & 0.032158 & -0.020091 & 0.0026409 \\ 
-0.0081402 & 0.13652 & -0.0023012 & -0.020091 & 0.23192 & -0.0055377 \\ 
0.0010292 & -0.0038002 & 0.000814158 & 0.0026409 & -0.0055377 & 0.0014322
\end{pmatrix},
\end{equation}
\end{widetext}
while the covariance matrix for BAO data from Refs.\ \cite{Gil-Marin:2020bct,Bautista:2020ahg} is
\begin{small}
\begin{equation}
\label{eq:cov_LRG}
\begin{pmatrix}
 0.09114 & -0.033789 & 0.0024686 \\
 -0.033789 & 0.22009 & -0.0036088 \\
 0.0024686 & -0.0036088 & 0.0019616
\end{pmatrix},
\end{equation}
\end{small} 
and that for BAO data from Refs.\ \cite{Hou:2020rse,Neveux:2020voa} is
\begin{small}
\begin{equation}
\label{eq:cov_Quasar}
\begin{pmatrix}
 0.6227 & 0.01424 & 0.02257 \\
 0.01424 & 0.2195 & -0.007315 \\
 0.02257 & -0.007315 & 0.002020
\end{pmatrix}.
\end{equation}
\end{small}

{${\boldsymbol f\bm{\sigma}_8}$}. $f\sigma_8$ data points, in addition to those correlated with BAO data that are listed in Table \ref{tab:bao}. These independent $f\sigma_8$ measurements are obtained either from peculiar velocity data \cite{Turnbull:2011ty,Hudson:2012gt,Said:2020epb} or from redshift space distortion (RSD) analyses \cite{Shi:2017qpr,Simpson:2015yfa,Blake:2013nif,Mohammad:2018mdy,Okumura:2015lvp}. These are listed in Table \ref{tab:fs8}. 

The combination $f(z)\sigma_8(z)$ is used to quantify the growth rate of the matter density perturbation. Here, the growth rate 
\begin{equation}
f(z) = -(1+z)\frac{d\ln{D(z)}}{dz}    
\end{equation}
where $D(z)$ is the growth function. The other function involved, $\sigma_8(z)$, is the root mean square of matter fluctuations smoothed over spheres of radius $R_8 = 8h^{-1}\textrm{Mpc}$ at a given value of the redshift. It is computed as
\begin{equation}
\sigma^2_8(z) = \int\frac{d^{3}k}{(2\pi)^3}P_m(z,\vec{k})W^2(k{R_8}),    
\end{equation}
where $P_m(z,\vec{k})$ is the matter power spectrum and $W(k{R_8})$ is the window function. 
\begin{table}
\caption{$f\sigma_8$ measurements.}
\begin{ruledtabular}
\begin{tabular}{ccc}
 $z_\textrm{eff}$                     &  $f\sigma_8$                                          &   Reference    \\[+0mm]
  \hline \\[-2mm]
 $0.02$     &                   $ 0.398\pm 0.065$      & \cite{Turnbull:2011ty,Hudson:2012gt}  \\[+1mm]
 \hline \\[-2mm]
 $0.035$     &              $0.338\pm 0.027$      & \cite{Said:2020epb}  \\[+1mm]
  \hline \\[-2mm]
 $0.1$     &                  $0.376\pm 0.038$      & \cite{Shi:2017qpr}  \\[+1mm]
 \hline \\[-2mm]
 $0.18$     &                 $ 0.29\pm 0.10$      & \cite{Simpson:2015yfa}  \\[+1mm]
 $0.38$     &                   $0.44\pm 0.06$      & \cite{Blake:2013nif}  \\[+1mm]
  \hline \\[-2mm]
 $0.6$     &                   $0.49\pm 0.12$      & \cite{Mohammad:2018mdy}\\[+1mm]
 $0.86$     &                    $0.46\pm 0.09$      & \cite{Mohammad:2018mdy}  \\[+1mm]
 \hline \\[-2mm]
 $1.36$     &                    $0.482\pm 0.116$       & \cite{Okumura:2015lvp}  \\[+0mm]
 \end{tabular}
\end{ruledtabular}
\label{tab:fs8}
\end{table}

{\bf SNIa}. Apparent magnitude as a function of redshift measurements for 1048 Pantheon SNIa \cite{Scolnic:2017caz}, probing the redshift range $0.01 < z < 2.3$, and 20 compressed data points, spanning the redshift range $0.015 \leq z \leq 0.7026$, representing 207 DES 3yr SNIa \cite{DES:2018paw}. The Pantheon and DES 3yr data are independent of each other, but the data points within each sample are correlated and we account for the corresponding covariance matrices in our analyses.  

{${\bm{H(z)}}$}. Hubble parameter measurements over the redshift range $0.070 \leq z \leq 1.965$ obtained using the differential age technique. The 31 data points employed are listed in Table 2 of Ref.\ \cite{Park:2017xbl}.

Hereafter we denote the combination of BAO, $f\sigma_8$, SNIa, and $H(z)$ data sets as the non-CMB data set.

\section{Methods}
\label{sec:method}

We apply the Markov chain Monte Carlo (MCMC) method, implemented in the \texttt{CAMB}/\texttt{COSMOMC} package (version of Oct.\ 2018), \cite{Challinor:1998xk,Lewis:1999bs,Lewis:2002ah},
to explore the parameter space of the different models under study. The \texttt{CAMB} program computes the matter and CMB power spectra based on the evolution of density perturbations of the matter and radiation components and the \texttt{COSMOMC} program uses the MCMC method to estimate the parameter constraints that are favored by the given observational data sets. We have performed cross-checks using the \texttt{CLASS}/\texttt{MontePython} package, \cite{Blas:2011rf,Audren:2012wb}. In general a good agreement between the results is obtained unless significant degeneracies between some of the fitting parameters are present. When this happens, differences in the central values are found, but the two sets of results remain compatible at 1$\sigma$ due to large error bars. The inclusion of more data breaks the aforementioned degeneracies and the two sets of results then agree really well.

In this paper we consider eight cosmological models: the tilted flat, (two) tilted non-flat, and the untilted non-flat $\Lambda$CDM models, as well as their extensions through the inclusion of the $A_L$ parameter, for a total of eight cases. $A_L$ is a phenomenological parameter that scales the theoretical prediction of the gravitational potential power spectrum, with its theoretical expected value being $A_L=1$, see Ref.\ \citep{Calabreseetal2008}. $A_L>1$ causes the smoothing of acoustic peaks in the CMB angular power spectrum, and Planck CMB data tend to prefer $A_L>1$ \cite{Planck:2018vyg}.  

The tilted flat $\Lambda$CDM model is characterized by six cosmological parameters ($\Omega_b h^2$, $\Omega_c h^2$, $\theta_\textrm{MC}$, $\tau$, $A_s$, $n_s$), where $\Omega_b$ and $\Omega_c$ are the current values of non-relativistic baryonic and cold dark matter density parameters, $\theta_\textrm{MC}$ is the angular size of the sound horizon at recombination defined in the \texttt{CAMB}/\texttt{COSMOMC} program, $\tau$ is the reionization optical depth, and $A_s$ and $n_s$ are the amplitude and the spectral index of the primordial scalar-type energy density perturbation power spectrum
\begin{equation}\label{eq:tilted_flat_PS}
    P_\delta(k) = A_s \left(\frac{k}{k_0} \right)^{n_s},
\end{equation}
where $k$ is wavenumber and the pivot scale for $A_s$ is $k_0=0.05~\textrm{Mpc}^{-1}$. This power spectrum is generated by quantum mechanical fluctuations during an early epoch of power-law inflation in an exponential potential energy density scalar field cosmological model with flat spatial hypersurfaces, \cite{Lucchin:1984yf, Ratra:1989uv, Ratra:1989uz}.  

In the non-flat very-slow-roll (so untilted) inflation $\Lambda$CDM model, the presence of non-zero spatial curvature determines a new length scale, and the power-law part of the primordial power spectrum is not relevant. Thus, this model still has six cosmological parameters, with the spectral index $n_s$ being replaced by the current value of the spatial curvature density parameter $\Omega_k$. For very-slow-roll inflation in this non-flat inflation mode, the primordial power spectrum is, \cite{RatraPeebles1995, Ratra:2017ezv},
\begin{equation}\label{eq:untilted_nonflat_PS}
P_\delta(q) \propto \frac{(q^2-4K)^2}{q(q^2-K)}
\end{equation}
where $q=\sqrt{k^2+K}$ is the wavenumber in a model with non-zero spatial curvature $K=-(H_0^2 / c^2)\Omega_k$, and $A_s$ is defined to be the amplitude of the power spectrum at the pivot-scale $k_0$. This power spectrum form holds in both the open ($\Omega_k > 0$) and closed ($\Omega_k < 0$) cases, with $q|K|^{-1/2} \geq 0$ and continuous in the open case and $q|K|^{-1/2} = 3, 4, 5\dots$ in the closed case. It is the power spectrum used in the non-flat model analyses in Refs.\ \cite{Ooba:2017ukj, Ooba:2017npx, Ooba:2017lng, Park:2017xbl,Park:2018bwy, Park:2018fxx, Park:2019emi}.  

For the tilted non-flat $\Lambda$CDM model, there are seven cosmological parameters, with $\Omega_k$ added to the six of the tilted flat $\Lambda$CDM model. In this model it has been usual to assume, e.g.\ \cite{Planck:2018vyg}, a primordial power spectrum of the form
\begin{equation}\label{eq:tilted_nonflat_Planck_PS}
    P_\delta(q) \propto \frac{(q^2-4K)^2}{q(q^2-K)} \left( \frac{k}{k_0} \right)^{n_s -1}, 
\end{equation}
where $q$ (and $A_s$) is defined in the previous paragraph. The above expression, which we refer to as the Planck $P(q)$, is a phenomenologically modified version of the non-flat very-slow-roll untilted primordial density perturbation, given in Eq.\ \eqref{eq:untilted_nonflat_PS}, to now also allow for tilt, \cite{Lesgourgues:2013bra}. It assumes that tilt in a non-flat space works in a way similar to how it does in flat space. This expression was known to be physically reasonable in the cases when $K = 0$ or $n_s=1$, since Eqs.\ \eqref{eq:tilted_flat_PS} and \eqref{eq:untilted_nonflat_PS} are recovered, respectively, and these two expressions hold in physically-consistent inflation models. Very recently, a numerical study in closed inflation models that computes primordial power spectra generated for a few different, initially slow-roll, inflation initial conditions finds that it is possible to generate, in the closed case, a power spectrum very close to that given in Eq.\ \eqref{eq:tilted_nonflat_Planck_PS}, \cite{Guth:2022xyz}.  

In this paper we also study another not-necessarily very-slowly-rolling non-flat (closed and open) inflation model, \cite{Ratra:2022ksb}. These tilted non-flat inflation models result in a primordial power spectrum that differs from that of eq.\ \eqref{eq:tilted_nonflat_Planck_PS} and assumes a different inflation initial condition than those studied in Ref.\ \cite{Guth:2022xyz}. For the closed and open inflation models, the resulting power spectrum
\begin{equation}
{ P_\delta(q) \propto (q^2 -4K)^2|P_{\zeta}(A)|, }
\label{eq:tilted_nonflat_new_PS}
\end{equation}
where $P_\zeta(A)$ is different in the closed and open cases. For the closed inflation model 
\begin{widetext}
\begin{equation}
\sqrt{|P_{\zeta}(A)|} = \left(\frac{16\pi}{m^2_p}\right)^{1/2}\!\!\!\!Q^{1/p}\frac{(2+q_s)p}{\sqrt{\pi q_s}}
\Bigg|-1 + \frac{W(A)}{p}\Bigg| \,\,
\frac{2^{-(6-4q_s+2A -W(A))/p}}{\sqrt{A}(A-1)(A+3)} \,\,
\Bigg|\frac{\Gamma\left(1 + W(A)/p\right)\Gamma\left((2+q_s)/(2p)\right)}{\Gamma\left((2+W(A))/p\right)}\Bigg|,
\end{equation}
\end{widetext}
with
\begin{equation}
W(A) = \sqrt{-8-4q_s + q_s^2 + 4A(A+2)},  
\end{equation}
and 
\begin{equation}
A =  \frac{q}{\sqrt{|K|}} -1.    
\end{equation}
While for the open inflation model
\begin{widetext}
\begin{equation}
\sqrt{|P_{\zeta}(A)|} = \left(\frac{16\pi}{m^2_p}\right)^{1/2}\!\!\!\!Q^{1/p}\frac{(2+q_s)p}{\sqrt{\pi q_s}}\Bigg|-1 + \frac{W(A)}{p}\Bigg|\,\, \frac{2^{-(6-4q_s)/p +\textrm{Re}(W(A)/p)}}{\sqrt{A}(A^2 + 4)}\,\,  \Bigg|\frac{\Gamma\left(1 + W(A)/p\right)\Gamma\left((2+q_s)/(2p)\right)}{\Gamma\left((2+W(A))/p\right)}\Bigg|,
\end{equation}
\end{widetext}
with
\begin{equation}
W(A) = \sqrt{-12 -4q_s + q_s^2 -4A^2},     
\end{equation}
and 
\begin{equation}
A = \frac{q}{\sqrt{|K|}}.    
\end{equation}
In these equations, $\Gamma(x)$ is the Gamma function, $m_p$ is the Planck mass, $Q$ is a normalization constant, $q_s = (2 -2n_s)/(3-n_s)$, and finally $p = 2-q_s$. In both the closed and open inflation models $0 < q_s < 2$, so $-\infty < n_s < 1$. In these equations the large vertical bars indicate that we take the absolute value of the enclosed functions. In this paper we refer to the power spectrum in this tilted non-flat $\Lambda$CDM as the new $P(q)$, which is shown in Eq.\ \eqref{eq:tilted_nonflat_new_PS}, and 
following the procedure applied to the other power spectra, $A_s$ gives the amplitude of the new $P(q)$ at the pivot-scale $k_0$.

\begin{figure*}[htbp]
\centering
\mbox{\includegraphics[width=78mm,trim=0cm 5cm 0cm 5cm]{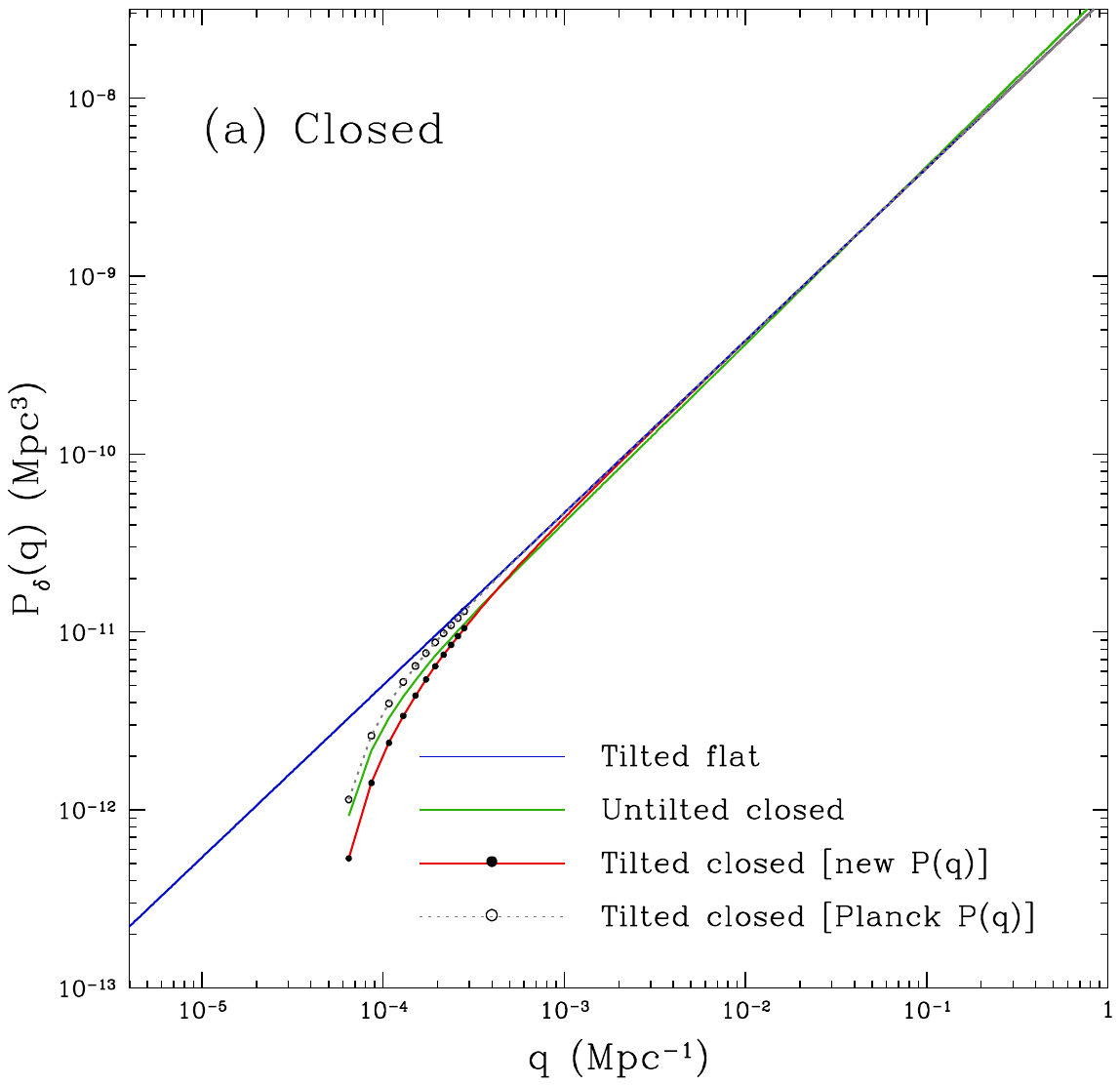}}
\mbox{\includegraphics[width=78mm,trim=0cm 5cm 0cm 5cm]{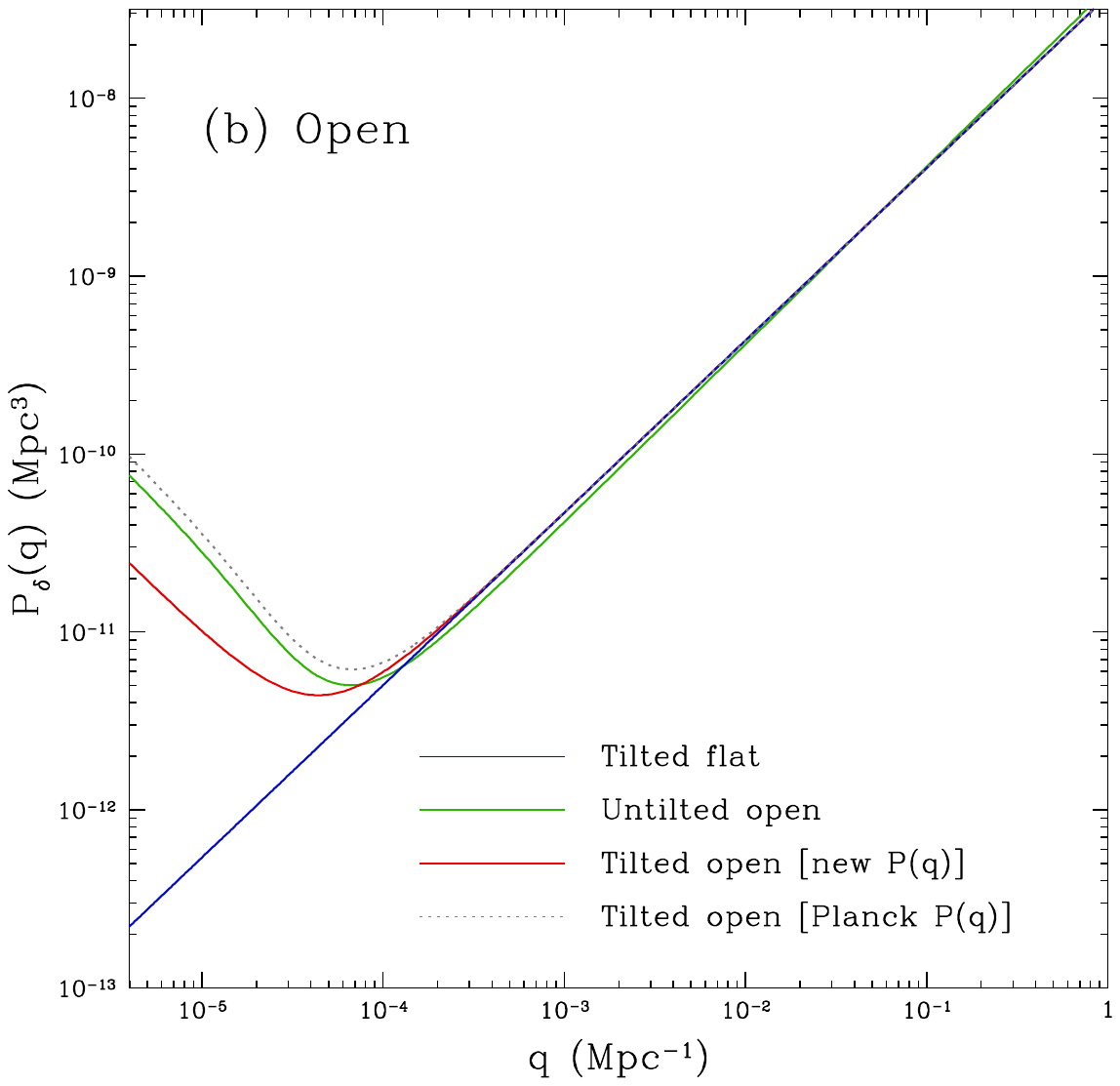}}
\caption{Initial scalar-type perturbation power spectra of the tilted flat, untilted non-flat, and two tilted non-flat $\Lambda$CDM models. For the tilted non-flat closed models, the cosmological parameters of the tilted non-flat $\Lambda$CDM model with Planck $P(q)$ constrained by using P18+lensing data (including $\Omega_k=-0.0103$) are used (see Table \ref{tab:para_NL_ns_nonCMB}). For closed models, the same value of $A_s$ was assumed for all models and the same value of $n_s$ was assumed for all tilted models. The powers at the first 11 large-scale wavenumbers are indicated by the filled (open) circles for the tilted closed model with the new (Planck) $P(q)$. For open non-flat models, $\Omega_k=+0.0103$ was assumed. For the tilted flat model, the generalized wavenumber $q$ is equivalent to $k$. 
}
\label{fig:pinit}
\end{figure*}

Figure \ref{fig:pinit} compares the initial scalar-type perturbation spectra of the tilted flat, untilted non-flat, and two tilted non-flat models with the Planck $P(q)$ and the new $P(q)$. In this figure we set the values of the cosmological parameters, for all the models, to the mean values of the tilted non-flat $\Lambda$CDM model with Planck $P(q)$ constrained by the P18+lensing data (see Table \ref{tab:para_NL_ns_nonCMB} for the parameters), except in panel (b) for the open models where we change the sign of $\Omega_k$. 

In the cases where we include the $A_L$ parameter in the analysis, this increases by one  the number of cosmological model parameters to be determined from data, so depending on model we then have either seven or eight cosmological model parameters in these cases.

At the background level, the evolution of the scale factor $a$ in all models we study is described by the Hubble function
\begin{equation}
\begin{split}
\label{eq:Hubble_function}
H^2(a) = H^2_0[\Omega_\gamma{a^{-4}} &+ (\Omega_b + \Omega_c){a^{-3}} \\
       &+ \Omega_k{a^{-2}} + \Omega_\nu(a)+ \Omega_\Lambda].
\end{split}
\end{equation}
Here $a=1/(1+z)$ is the cosmic scale factor normalized to unity at present, $\Omega_\Lambda$ represents the cosmological constant dark energy density parameter, $\Omega_\gamma$ is the current value of the CMB photon energy density parameter, and $\Omega_\nu(a)$ represents the contribution of the massless and massive neutrinos, for which it is not possible to get an analytical expression. In all cases we study, we determine the contribution of photons, and massless and massive neutrinos by assuming a present CMB temperature $T_0=2.7255$ K, the effective number of neutrino species $N_\textrm{eff}=3.046$, and a single massive neutrino species with neutrino mass $0.06$ eV. During parameter exploration using the MCMC method, we set wide flat priors on parameters in order that they not affect the parameter estimation; these priors are listed in Table \ref{tab:Priors}. 

\begin{table}
\caption{Flat priors of the fitting parameters.}
\begin{ruledtabular}
\begin{tabular}{cccc}
 $\textrm{Parameters}$      &  $\textrm{Our}$ &  $\textrm{Handley}$ &  $\textrm{Handley}$+$\Omega_k$    \\[+0mm]
\hline \\[-2mm]
 $\Omega_b h^2$     &       [0.005,0.1] &      [0.019,0.025]&       [0.019,0.025] \\[+1mm]
\hline \\[-2mm]
 $\Omega_c h^2$     &               [0.001,0.99] & [0.095,0.145] & [0.095,0.145] \\[+1mm]
  \hline \\[-2mm]
 100$\theta_\textrm{MC}$     &         [0.5,10] & [1.03,1.05]& [1.03,1.05]\\[+1mm]
 \hline \\[-2mm]
 $\tau$     &                 [0.01,0.8] & [0.01,0.4]& [0.01,0.4]\\[+1mm]
  \hline \\[-2mm]
$\Omega_k$     &                 [-0.5,0.5] & [-0.1,0.05]& [-0.3,0.15]\\[+1mm]
  \hline \\[-2mm]  
 $n_s$     &                   [0.8,1.2] & [0.885,1.04]& [0.885,1.04]\\[+1mm]
 \hline \\[-2mm]
 $\ln\left(10^{10}A_s\right)$     &     [1.61,3.91] & [2.5,3.7]& [2.5,3.7]\\[+1mm]
 \hline \\[-2mm]
 $A_L$     &     [0,10] & -& -\\[+1mm]
 \end{tabular}
\\[+1mm]
\begin{flushleft}
Note: In almost all the computations reported in this paper we use the priors listed in the Our column in this table. A general exception is that in almost all the computations in the tilted non-flat $\Lambda$CDM model with the new $P(q)$ we use a more restrictive prior range for the spectral index, $0.8\le n_s < 1$. In addition to these choices, in all cases, for the derived parameter $H_0$ we restrict its range of variation to $0.2 \le h \le 1$. In Table \ref{tab:para_sigmap} when only lensing data is used, in order to test the impact of different choices of priors, we also provide results for the narrower priors employed in Ref.\ \cite{Handley:2019tkm} (listed in the Handley column above). The Handley+$\Omega_k$ column priors above differ from Handley priors by allowing for a broader prior for the $\Omega_k$ parameter.
\end{flushleft} 
\end{ruledtabular}
\label{tab:Priors}
\end{table}

Due to the lack of constraining power of some of the data sets, when they are considered alone, we have to fix the values of some of the cosmological parameters in the analyses of these data sets. In BAO$^\prime$, BAO, (P18) lensing, or non-CMB data alone analyses we set the values of $\tau$ and $n_s$ to those obtained in the P18 data alone analysis for each model. Additionally, in BAO$^\prime$ data alone analyses we also fix the value of $\ln\left(10^{10}A_s\right)$, again, to the corresponding P18 data analysis value. Finally, in Sec. \ref{sec:P18+lensing_vs_non-CMB}, when we compare the constraints obtained from P18+lensing data and non-CMB data, in the non-CMB data analyses the values of $\tau$ and $n_s$ are fixed to the ones we get in the P18+lensing data analysis for each model.
 
We use the converged MCMC chains to compute mean values, their confidence limits, and the posterior distributions of the model parameters with the \texttt{GetDist} code \cite{Lewis:2019xzd}. The MCMC chains are considered to converge when the Gelman and Rubin $R$ statistic provided by \texttt{COSMOMC} becomes $R-1<0.01$.

In addition to using the various combinations of data sets (see Sec.\ \ref{sec:data}) for constraining cosmological parameters in the models we study, we want to also determine which of these models better fit the data sets we study. For a fair comparison between competing cosmological models with different numbers of free parameters it is necessary to be able to conveniently penalize for extra degrees of freedom. In this work we employ two different statistical criteria, that differently penalize for extra degrees of freedom, to compare the performance of the models. 

The first one we use is the Akaike information criterion (AIC) \cite{Akaike} which is defined as
\begin{equation}
\label{eq:AIC}    
\textrm{AIC} = \chi^2_{\textrm{min}} + 2n.
\end{equation}
Here $n$ is the number of independent cosmological parameters $\theta$ and $\chi^2_{\textrm{min}}\equiv \chi^2(\hat{\theta}) = -2\ln\mathcal{L}(\hat{\theta})$ is the minimum value of $\chi^2(\theta) = -2\ln\mathcal{L}(\theta)$ evaluated at the best-fit cosmological parameter values $\hat{\theta}$ where $\mathcal{L}(\theta)$ is the likelihood function.

The expression in eq.\ \eqref{eq:AIC} is valid only for a large number of data points. According to Ref.\ \cite{Burnham2002}, when the number of data points $N$ obeys $N/n<40$, the expression in eq.\ \eqref{eq:AIC} should be replaced by  
\begin{equation}\label{eq:AIC_modified}
\textrm{AIC}_{c} = \textrm{AIC} + \frac{2n(n+1)}{N-n-1} = \chi^2_{\textrm{min}} + \frac{2nN}{N-n-1}.
\end{equation}
Note that when $N$ is large compared to $n$ we have $N/(N-n-1)\simeq 1$ and then $\textrm{AIC}_c\simeq\textrm{AIC}$. This is the case for P18 data and non-CMB data but not for the BAO, BAO$^\prime$, and lensing data sets. In particular for BAO data $N = 16$, for BAO$^\prime$ data $N = 12$, for the lensing data set $N = 9$, and in all three cases $N/n<40$ so $\textrm{AIC}_c\neq \textrm{AIC}$.

The second one we use is the deviance information criterion (DIC) \cite{DIC} which is defined as  
\begin{equation}
\label{eq:DIC} 
\textrm{DIC} = \chi^2(\hat{\theta}) + 2p_D
\end{equation}
where $p_D= \overline{\chi^2} -  \chi^2(\hat{\theta}) $ is the penalization for those models with more degrees of freedom. Here an overbar represents the mean value of the corresponding quantity. Unlike the AIC, the DIC is both computed from Monte Carlo posterior samples and also uses the effective number of constrained parameters by taking into account whether or not a parameter is unconstrained by data, see Refs.\ \cite{DIC, Liddle:2007fy}. Therefore, we may say that the DIC is more reliable than the AIC.

We mostly use the differences in the AIC$_c$ and DIC values that are defined as 
\begin{eqnarray}
\label{eq:diff_AIC_BIC}
&\Delta\textrm{AIC}_c \equiv &\textrm{AIC}_{c,\textrm{X}} - \textrm{AIC}_{c,\textrm{Y}}\\
&\Delta\textrm{DIC} \equiv &\textrm{DIC}_{\textrm{X}} - \textrm{DIC}_{\textrm{Y}}.
\end{eqnarray}
Here Y represents the tilted flat $\Lambda$CDM model and X represents the model under study. When $-2 \leq \Delta\textrm{AIC}_c,\Delta\textrm{DIC}<0$ there is {\it weak} evidence in favor of the model under study relative to the tilted flat $\Lambda$CDM model. If $-6 \leq \Delta\textrm{AIC}_c,\Delta\textrm{DIC} < -2$ there is {\it positive} evidence, whereas if $-10 \leq \Delta\textrm{AIC}_c,\Delta\textrm{DIC} < -6$ there is {\it strong} evidence for the model under study. Finally if $\Delta\textrm{AIC}_c,\Delta\textrm{DIC} < -10$ there is {\it very strong} evidence in favor of the model under study relative to the tilted flat $\Lambda$CDM model. This scale also holds when $\Delta\textrm{AIC}_c$ and $\Delta\textrm{DIC}$ are positive, and then favors the tilted flat $\Lambda$CDM model over the model under study.   

We also want to determine whether some of the data sets we consider are mutually consistent (or inconsistent) in a specified cosmological model, and 
also whether or not the data set consistency (inconsistency) is model dependent. We utilize two different statistical estimators for this purpose. The first one makes use of DIC values and is presented in  Sec.\ 2.1.7 of Ref.\ \cite{Joudaki:2016mvz}). This estimator is based on
\begin{equation}
\label{eq:Tension_estimator_1}
\mathcal{I}(D_1,D_2) \equiv \textrm{exp}\left(-\frac{\mathcal{G}(D_1,D_2)}{2}\right),
\end{equation}
where
\begin{equation}
\mathcal{G}(D_1,D_2) = \textrm{DIC}(D_1\cup D_2) - \textrm{DIC}(D_1) - \textrm{DIC}(D_2). 
\end{equation}
Here $D_1$ and $D_2$ represent the two data sets under comparison, $\textrm{DIC}(D_1)$ and $\textrm{DIC}(D_2)$ are the DIC values that result when data set $D_1$ and $D_2$, respectively, are individually used to constrain cosmological parameters of the specified cosmological model, and $\textrm{DIC}(D_1\cup D_2)$ is the DIC value that results when data sets $D_1$ and $D_2$ are jointly used to constrain cosmological parameters of the specified model. The intuitive idea behind this estimator is that if two data sets are mutually consistent in a given cosmological model, which means that the cosmological parameter best-fit values determined from each data set are approximately similar, we would have $\chi^2_{\textrm{min}}(D_1\cup D_2)\simeq \chi^2_{\textrm{min}}(D_1) + \chi^2_{\textrm{min}}(D_2)$. This would lead to negative values of $\mathcal{G}(D_1,D_2)$, see eq.\ \eqref{eq:DIC}, which in turn would lead to $\mathcal{I}(D_1,D_2)>1$. However if $\chi^2_{\textrm{min}}(D_1\cup D_2) > \chi^2_{\textrm{min}}(D_1) + \chi^2_{\textrm{min}}(D_2)$, and is large enough, then we would find  $\mathcal{I}(D_1,D_2)<1$. Therefore $\log_{10}\mathcal{I}>0$ when the two data sets are mutually consistent and when $\log_{10}\mathcal{I}<0$ the two data sets are mutually inconsistent, in the cosmological model under study. Applying Jeffreys' scale, the level of consistency or inconsistency between the two data sets is {\it substantial} if $\lvert \log_{10}\mathcal{I} \rvert >0.5$, is {\it strong} if $\lvert \log_{10}\mathcal{I} \rvert >1$, and  is {\it decisive} if $\lvert \log_{10}\mathcal{I} \rvert >2$, \cite{Joudaki:2016mvz}. 

We now summarize the second statistical estimator we utilize to determine whether two data sets are mutually consistent (or inconsistent) in a specified cosmological model. This is described in Refs.\ 
\cite{Handley:2019pqx,Handley:2019wlz,Handley:2019tkm}, also see references therein. 
Given a data set $D$ and a given model $M$, we can express the posterior distribution for the independent model parameters $\theta$ through Bayes' theorem
\begin{eqnarray}\label{eq:BayesTheorem}
p(\theta|D,M)=\frac{p(D|\theta,M) p(\theta | M)}{p(D|M)}\,.
\end{eqnarray}
In the above expression $\mathcal{L}_D(\theta)\equiv p(D|\theta,M)$ is the likelihood function, $\pi(\theta) \equiv p(\theta | M) $ are the priors for the model parameters $\theta$, $\mathcal{Z}_D\equiv p(D|M)$ represents the evidence, and $\mathcal{P}_D(\theta)\equiv p(\theta|D,M)$ is the posterior distribution. Taking advantage of the fact that $\mathcal{P}_D(\theta)$ is a probability distribution function in $\theta$, which means that $\int \mathcal{P}_D(\theta)d\theta = 1$, we can express the evidence as  
\begin{equation}
\label{eq:Evidence}
\mathcal{Z}_D = \int \mathcal{L}_D(\theta)\pi(\theta)d\theta .
\end{equation}
For numerical computation of the Bayesian evidence we apply the method described in Ref.\ \cite{Heavens:2017hkr}. We are interested in quantifying the tension between two independent data sets $D_1$ and $D_2$. The total likelihood from a joint analysis of both these data sets is the product of the likelihoods for each data set, $\mathcal{L}_{12}$ = $\mathcal{L}_1\mathcal{L}_2$. Consequently, $\mathcal{Z}_{12}=\int \mathcal{L}_1(\theta)\mathcal{L}_2(\theta)\pi(\theta)d\theta$. Here and in what follows we index quantities with ``1" or ``2" when they have been computed using data set $D_1$ or $D_2$ respectively, and we use index ``12" when the two data sets are jointly used. We define the Bayes ratio as
\begin{equation}
\label{eq:Bayes_ratio}
R_D\equiv  \frac{\mathcal{Z}_{12}}{\mathcal{Z}_1\mathcal{Z}_2}.  
\end{equation}
This statistic is constructed in such a way that when $R_D\gg 1$ we can say that data sets $D_1$ and $D_2$ are consistent in the context of the particular model, while if $R_D\ll 1$ the two data sets are inconsistent. However, $R_D$ is strongly prior-dependent and to avoid this problem we instead use the suspiciousness $S_D$, \cite{Handley:2019pqx,Handley:2019wlz,Handley:2019tkm}, which we define in the following.

To define $S_D$ we will need the Shannon information \cite{Shannon:1948zz}
\begin{equation}
\mathcal{I}_{S,D}(\theta) = \ln\frac{\mathcal{P}_D(\theta)}{\pi(\theta)},    
\end{equation}
which is a measure of the amount of information, about the parameters $\theta$, that has been gained when moving from the priors to the posterior. The average value over the posterior of the Shannon information
\begin{equation}
\label{eq:KL_divergence}
\mathcal{D}_D = \int \mathcal{P}_D(\theta)\mathcal{I}_{S,D}(\theta)d\theta \equiv \langle \mathcal{I}_{S,D}\rangle_{\mathcal{P}_D},
\end{equation}
is known as the Kullback-Leibler divergence and measures how data compresses from prior to posterior. The suspiciousness $S_D$ is defined in terms of the Bayes ratio $R_D$ and the information ratio $I_D$
\begin{equation}
S_D = \frac{R_D}{I_D},
\end{equation}
where
\begin{equation}
\ln(I_D) = \mathcal{D}_1 + \mathcal{D}_2 - \mathcal{D}_{12}.     
\end{equation}
By considering a Gaussian analogy we can turn $\ln(S_D)$ into the tension probability $p$ of two data sets being inconsistent, \cite{Handley:2019pqx,Handley:2019wlz,Handley:2019tkm},
\begin{equation}
\label{eq:Tension_estimator_2}
p = \int^{\infty}_{d-2\ln(S_D)}\!\!\!\!\!\!\!\!\chi^2_d(x)dx = \int^{\infty}_{d-2\ln(S_D)}\!\!\frac{x^{d/2 -1}e^{-x/2}}{2^{d/2}\Gamma(d/2)}dx,  
\end{equation}
where $d$ is the Bayesian model dimensionality
\begin{equation}
d = \Tilde{d}_1 + \Tilde{d}_2 - \Tilde{d}_{12}, \qquad \Tilde{d}/2 = \langle \mathcal{I}_{S,D}^2\rangle_{\mathcal{P}_D} -  \langle \mathcal{I}_{S,D}\rangle^2_{\mathcal{P}_D} .
\end{equation}
If $p\lesssim 0.05$ the data sets are in moderate tension whereas if $p\lesssim 0.003$ they are in strong tension. The value of $p$ can be converted into a ``sigma value" using the Gaussian formula
\begin{equation}
\label{eq:Tension_estimator_2_sigma}
\sigma = \sqrt{2}\textrm{Erfc}^{-1}(p),    
\end{equation}
where $\textrm{Erfc}^{-1}$ is the inverse complementary error function. In particular $p\lesssim 0.05$ and $p\lesssim 0.003$ correspond to 2$\sigma$ and 3$\sigma$ Gaussian standard deviation, respectively. 

In Sec.\ \ref{subsec:data_set_tensions} we use both these statistical estimators to examine the consistency of five pairs of data, namely: P18 and lensing, P18 and BAO, P18 and BAO$^\prime$, P18 and non-CMB, and P18+lensing and non-CMB, in the context of different cosmological models. We shall see in Sec.\ \ref{subsec:cosmological_parameters} that when $A_L$ is allowed to vary error bars and two-dimensional cosmological constraint contours determined from each data set broaden (compared to the $A_L = 1$ case) and so are mutually consistent between different data sets (even if they are not mutually consistent when $A_L = 1$). We find, in Sec.\ \ref{subsec:data_set_tensions}, a similar improvement in consistency when $A_L$ is allowed to vary (compared to the $A_L = 1$ case).

\section{Results}
\label{sec:results}

\subsection{Cosmological parameters}
\label{subsec:cosmological_parameters}

\begin{figure*}[htbp]
\centering
\mbox{\includegraphics[width=170mm]{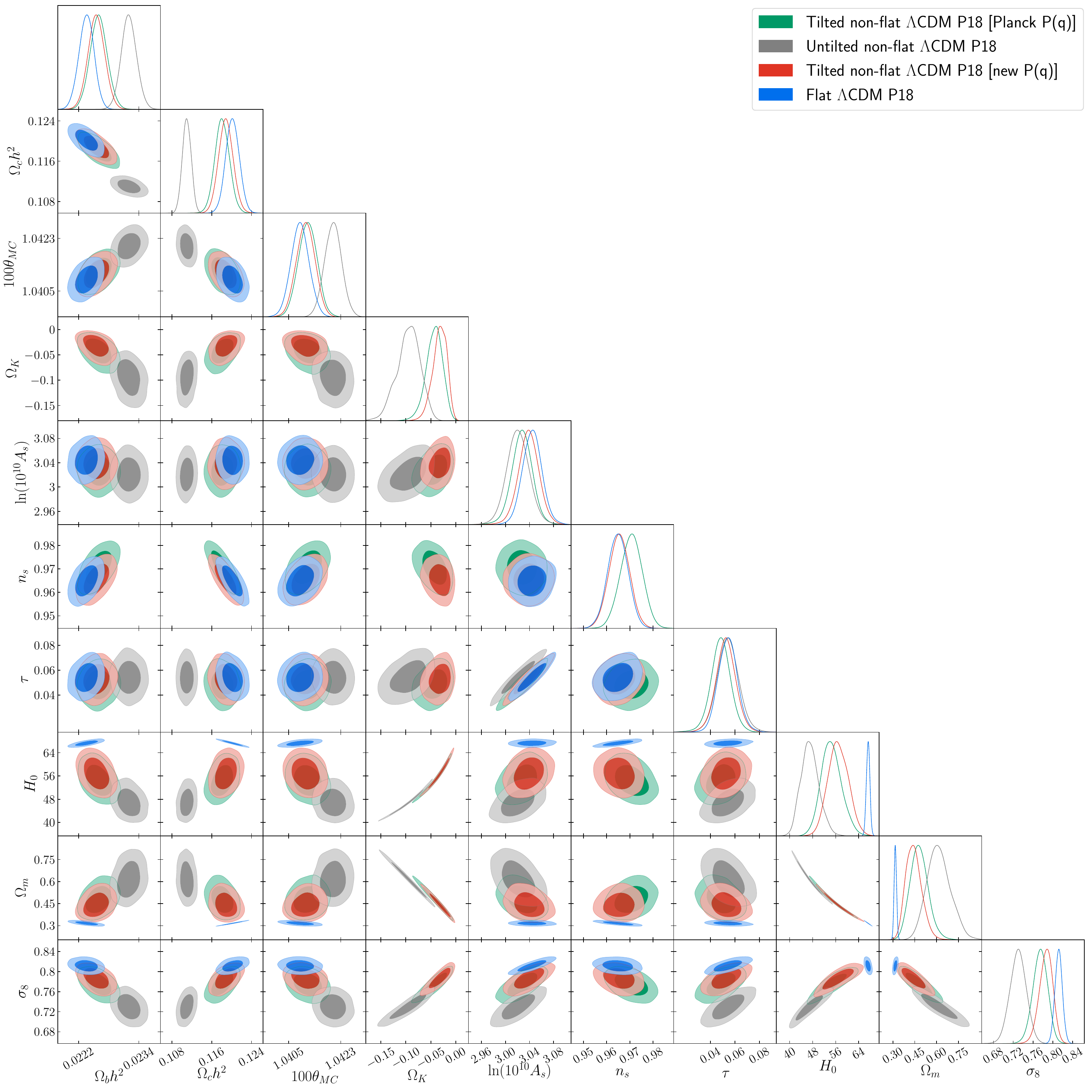}}
\caption{Planck 2018 TT,TE,EE+lowE (P18) data likelihood distributions of parameters of the tilted non-flat $\Lambda$CDM model with the new initial power spectrum [new $P(q)$] (red contours), of the tilted non-flat $\Lambda$CDM model with the Planck team's initial spectrum [Planck $P(q)$] (green), of the untitled non-flat $\Lambda$CDM model (grey), and of the tilted flat $\Lambda$CDM model (blue contours). 
}
\label{fig:like_P18}
\end{figure*}

\begin{figure*}[htbp]
\centering
\mbox{\includegraphics[width=170mm]{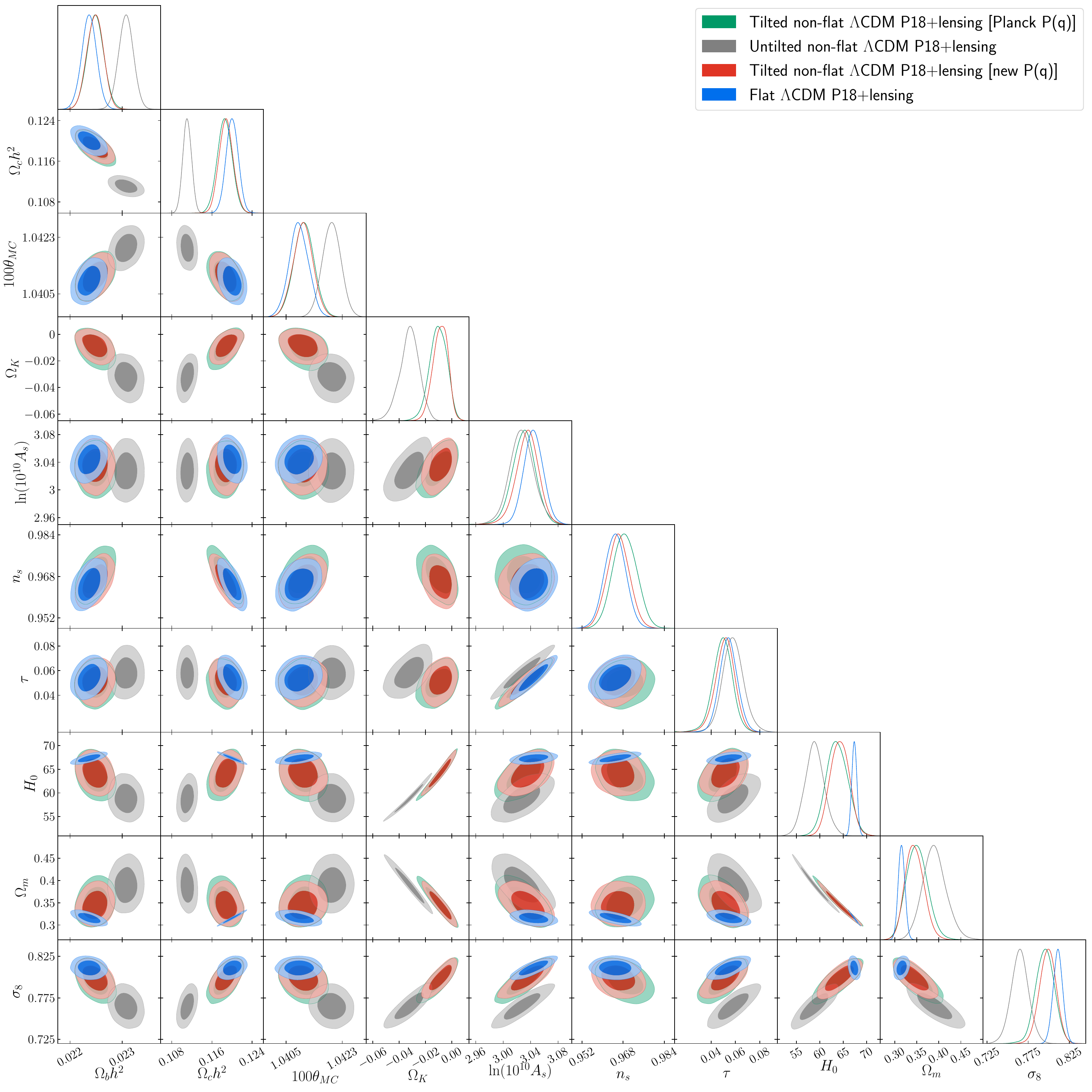}}
\caption{P18+lensing data likelihood distributions of parameters of the tilted non-flat $\Lambda$CDM model with the new initial power spectrum [new $P(q)$] (red contours), of the tilted non-flat $\Lambda$CDM model with the Planck team's initial spectrum [Planck $P(q)$] (green), of the untitled non-flat $\Lambda$CDM model (grey), and of the tilted flat $\Lambda$CDM model (blue contours). 
}
\label{fig:like_P18_lensing}
\end{figure*}

\begin{figure*}[htbp]
\centering
\mbox{\includegraphics[width=170mm]{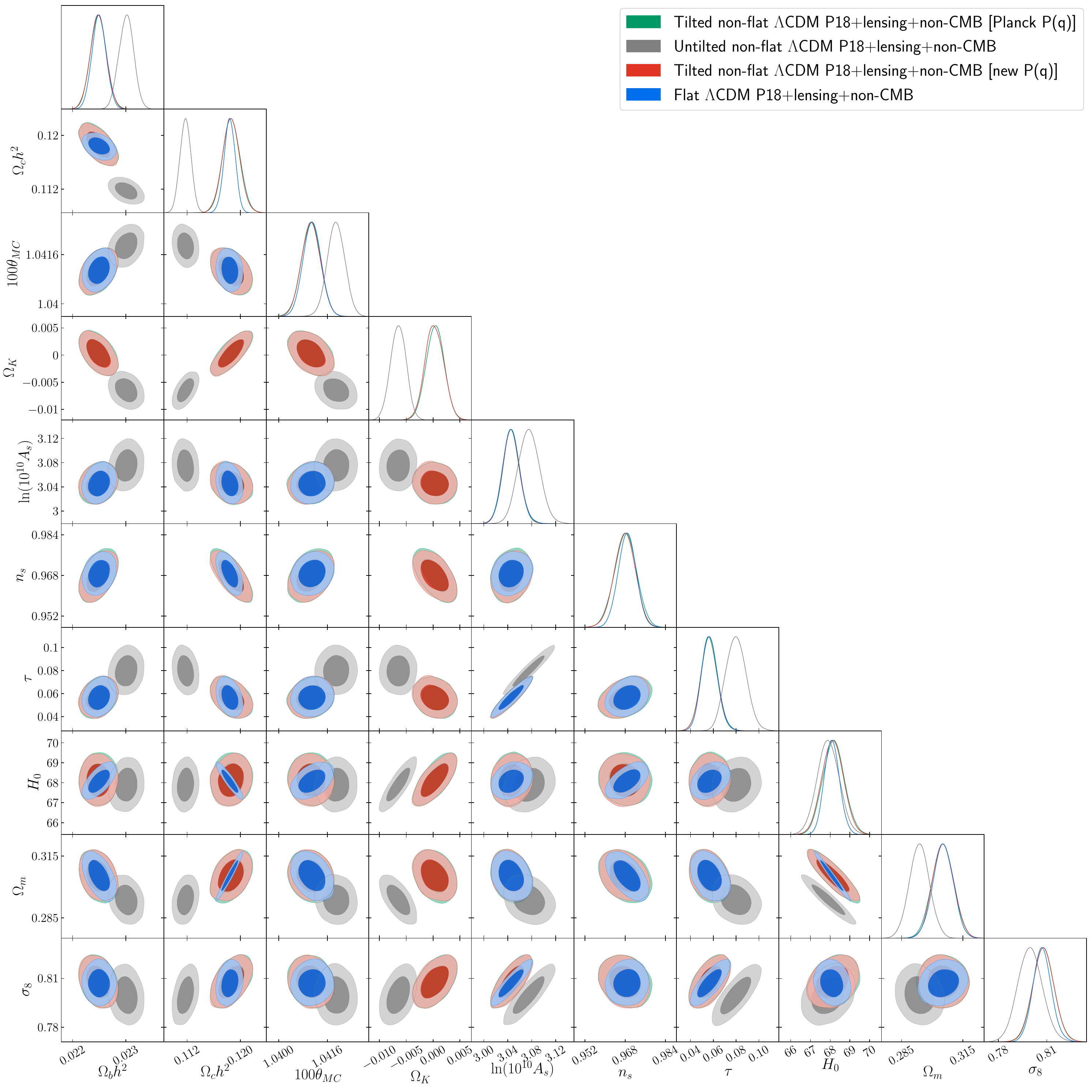}}
\caption{P18+lensing+non-CMB data likelihood distributions of parameters of the tilted non-flat $\Lambda$CDM model with the new initial power spectrum [new $P(q)$] (red contours), of the tilted non-flat $\Lambda$CDM model with the Planck team's initial spectrum [Planck $P(q)$] (green), of the untitled non-flat $\Lambda$CDM model (grey), and of the tilted flat $\Lambda$CDM model (blue contours). 
}
\label{fig:like_P18_lensing_nonCMB}
\end{figure*}

\begin{figure*}[htbp]
\centering
\mbox{\includegraphics[width=170mm]{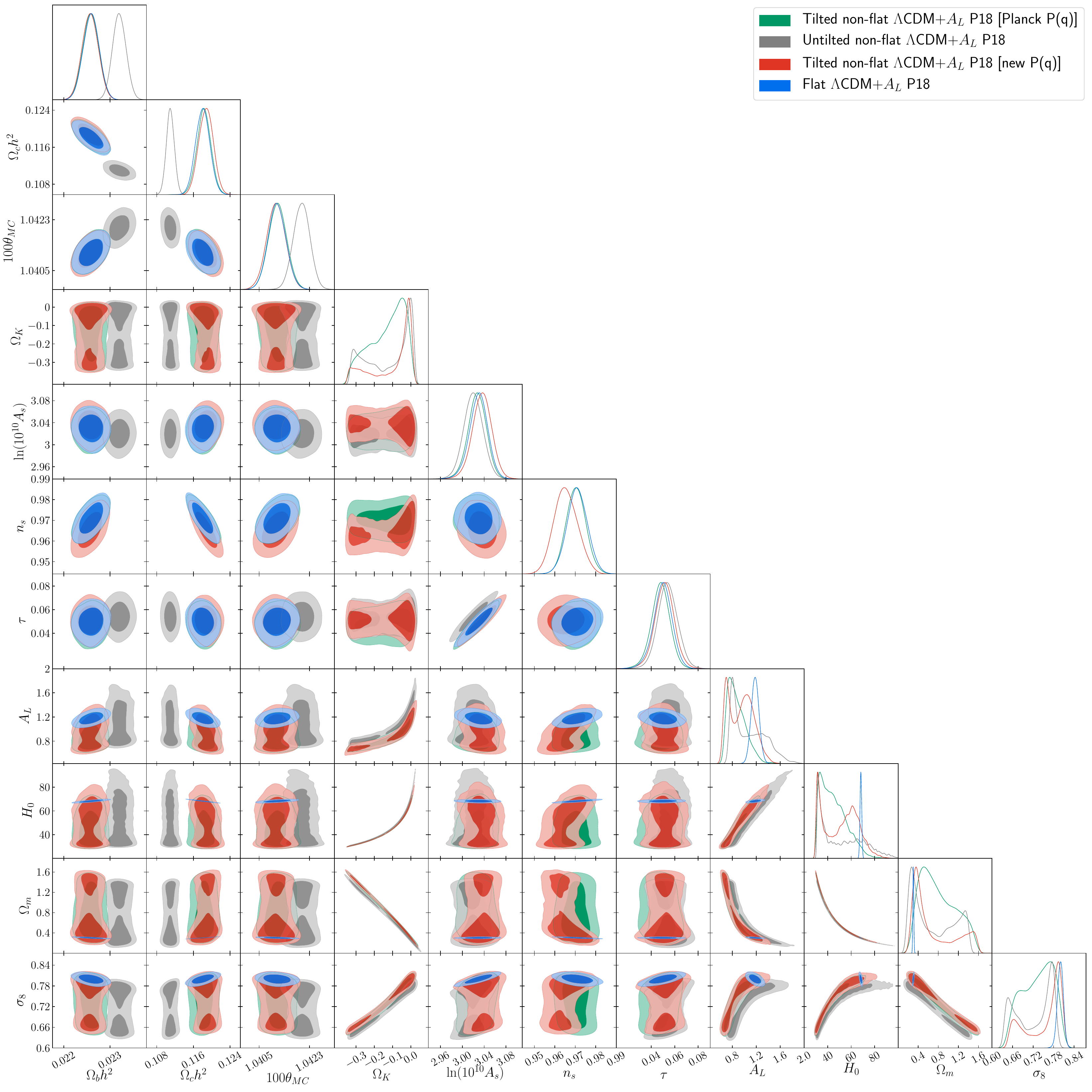}}
\caption{Planck 2018 TT,TE,EE+lowE (P18) data likelihood distributions of parameters of the tilted non-flat $\Lambda$CDM+$A_L$ model with the new initial power spectrum [new $P(q)$] (red contours), of the tilted non-flat $\Lambda$CDM+$A_L$ model with the Planck team's initial spectrum [Planck $P(q)$] (green), of the untitled non-flat $\Lambda$CDM+$A_L$ model (grey), and of the tilted flat $\Lambda$CDM+$A_L$ model (blue contours). 
}
\label{fig:like_Alens_P18}
\end{figure*}

\begin{figure*}[htbp]
\centering
\mbox{\includegraphics[width=170mm]{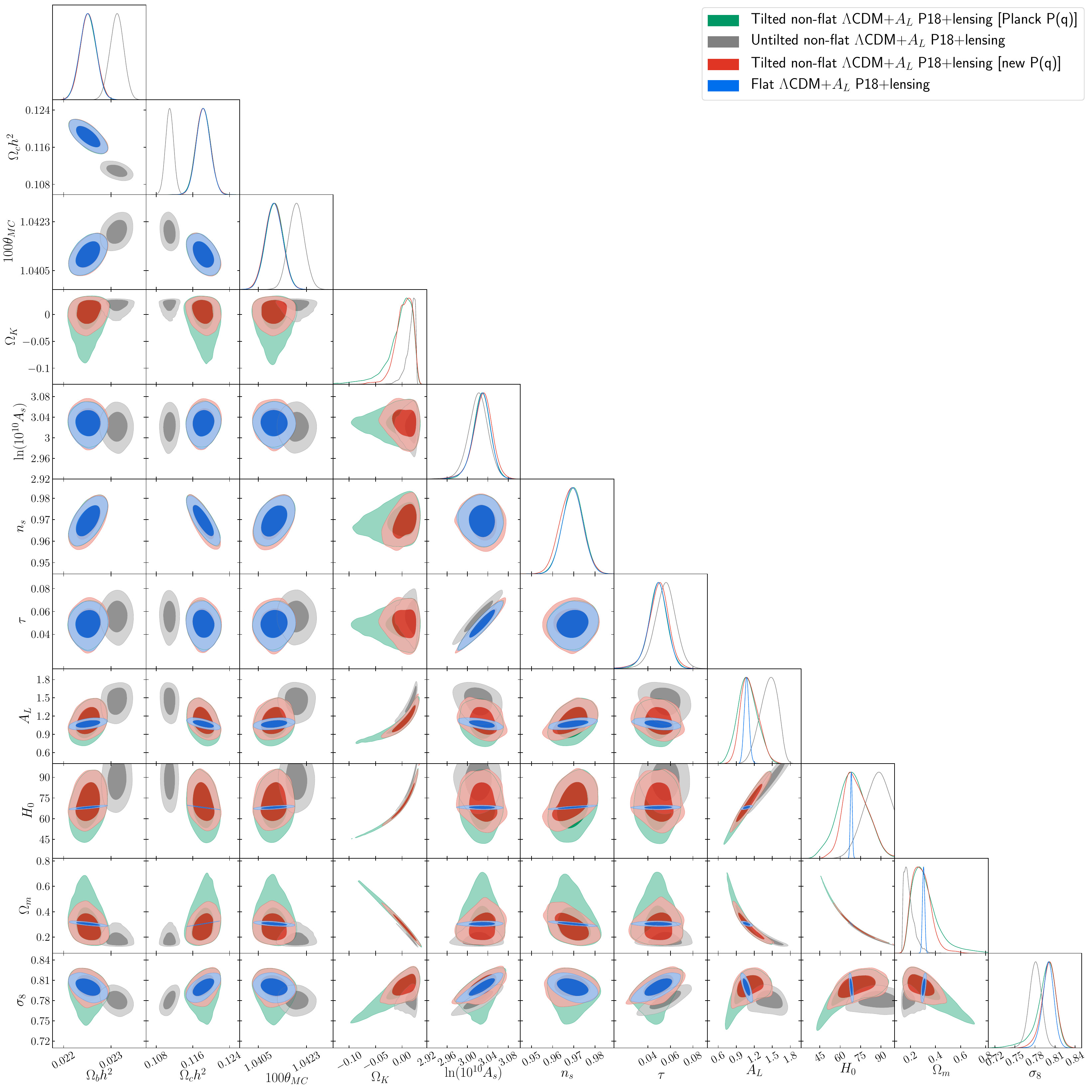}}
\caption{P18+lensing data likelihood distributions of parameters of the tilted non-flat $\Lambda$CDM+$A_L$ model with the new initial power spectrum [new $P(q)$] (red contours), of the tilted non-flat $\Lambda$CDM+$A_L$ model with the Planck team's initial spectrum [Planck $P(q)$] (green), of the untitled non-flat $\Lambda$CDM+$A_L$ model (grey), and of the tilted flat $\Lambda$CDM+$A_L$ model (blue contours).  
}
\label{fig:like_Alens_P18_lensing}
\end{figure*}

\begin{figure*}[htbp]
\centering
\mbox{\includegraphics[width=170mm]{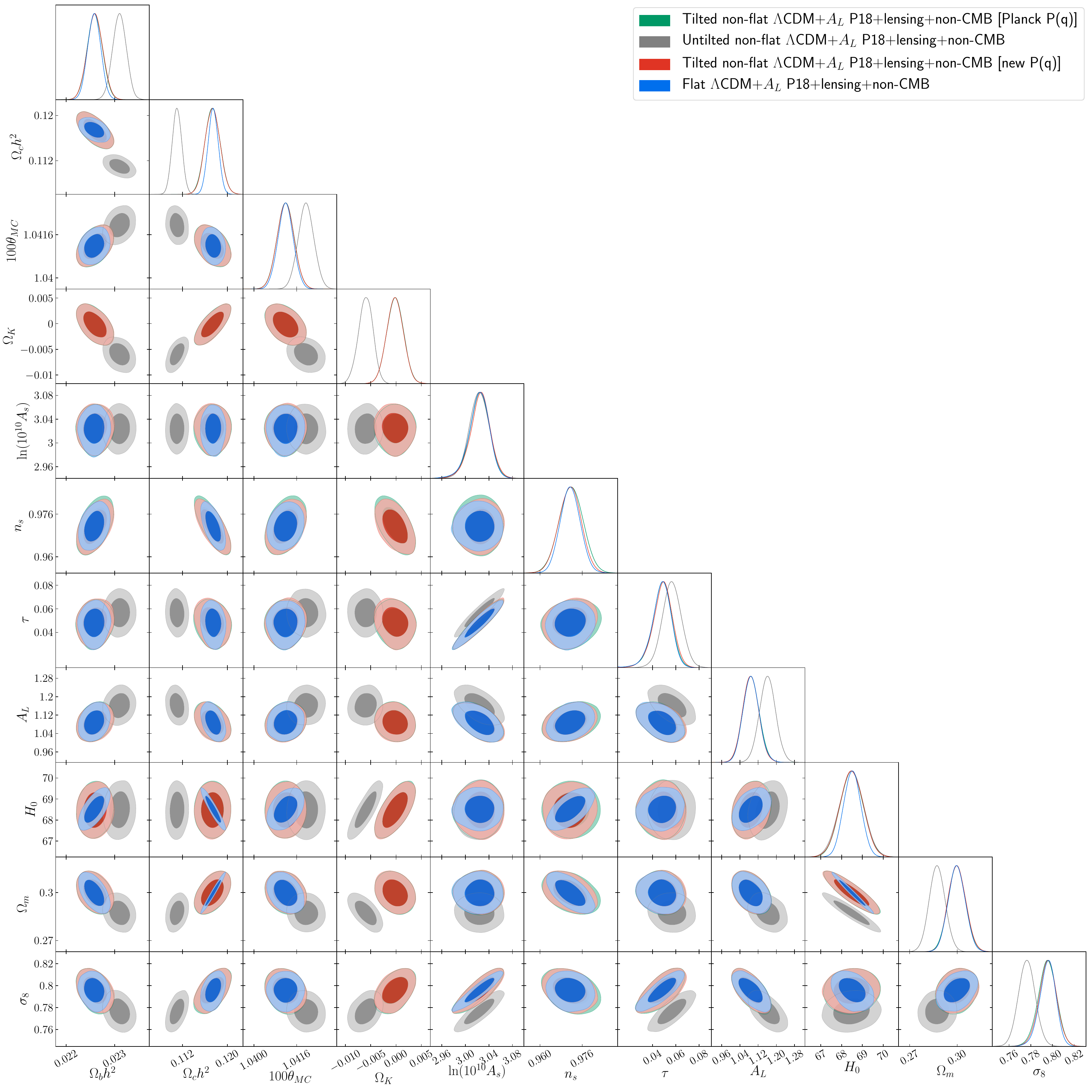}}
\caption{P18+lensing+non-CMB data likelihood distributions of parameters of the tilted non-flat $\Lambda$CDM+$A_L$ model with the new initial power spectrum [new $P(q)$] (red contours), of the tilted non-flat $\Lambda$CDM+$A_L$ model with the Planck team's initial spectrum [Planck $P(q)$] (green), of the untitled non-flat $\Lambda$CDM+$A_L$ model (grey), and of the tilted flat $\Lambda$CDM+$A_L$ model (blue contours).
}
\label{fig:like_Alens_P18_lensing_nonCMB}
\end{figure*}

\begin{figure*}[htbp]
\centering
\mbox{\includegraphics[width=170mm]{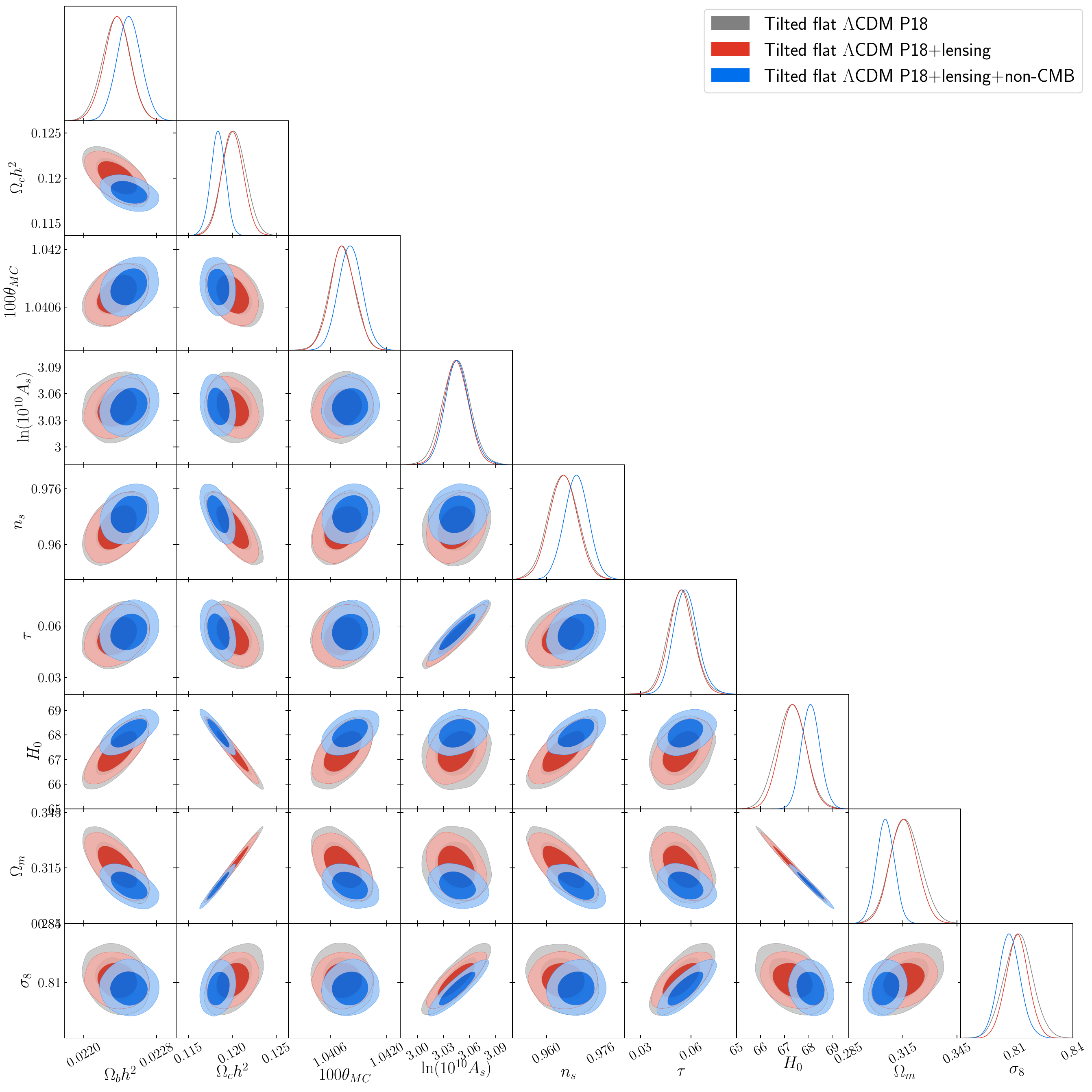}}
\caption{Likelihood distributions of tilted flat $\Lambda$CDM model parameters constrained by P18 (gray contours), P18+lensing (red contours), P18+lensing+non-CMB (blue contours) data sets.
}
\label{fig:like_FL_compar}
\end{figure*}

\begin{figure*}[htbp]
\centering
\mbox{\includegraphics[width=170mm]{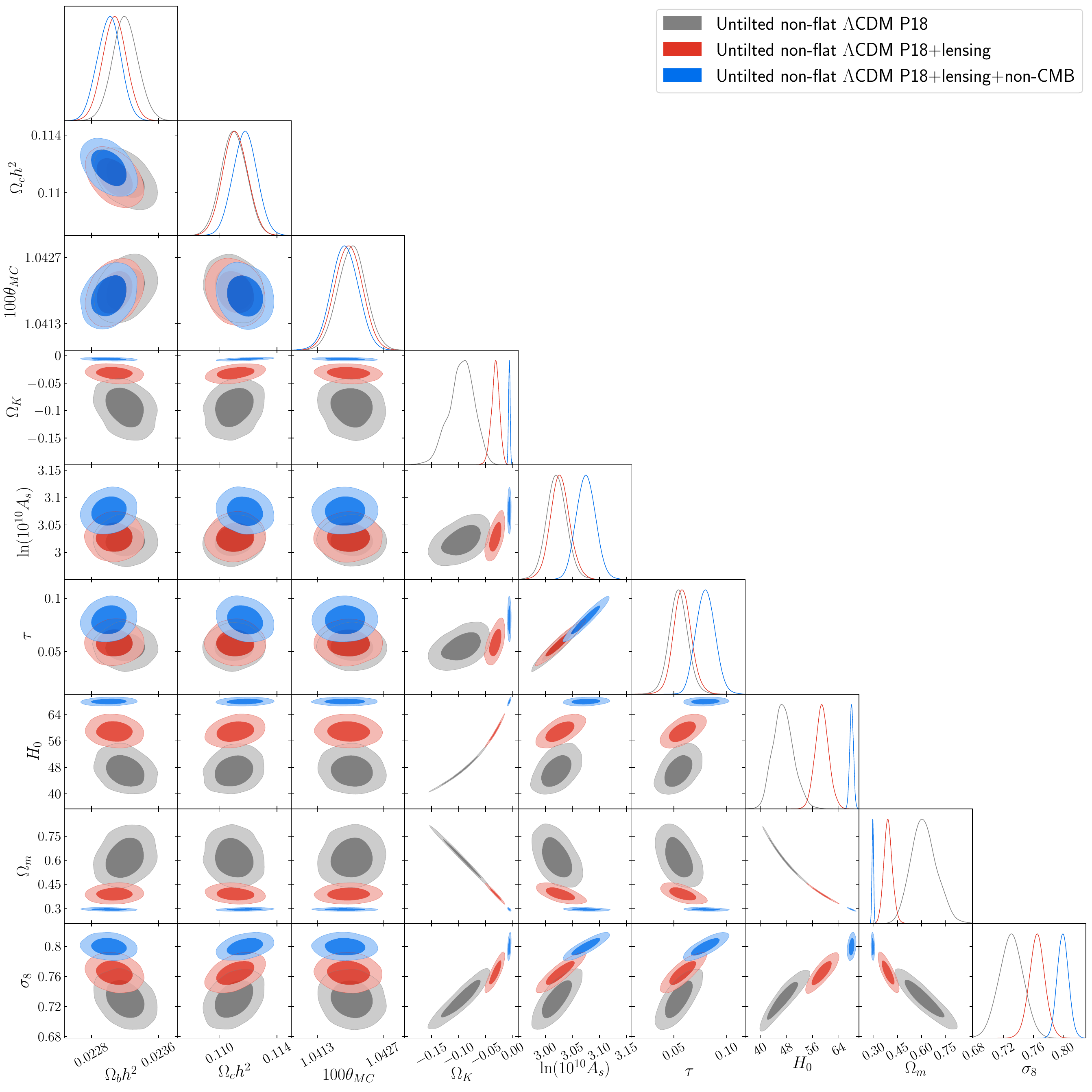}}
\caption{Likelihood distributions of untilted non-flat $\Lambda$CDM model parameters constrained by P18 (gray contours), P18+lensing (red contours), P18+lensing+non-CMB (blue contours) data sets.
}
\label{fig:like_NL_compar}
\end{figure*}

\begin{figure*}[htbp]
\centering
\mbox{\includegraphics[width=170mm]{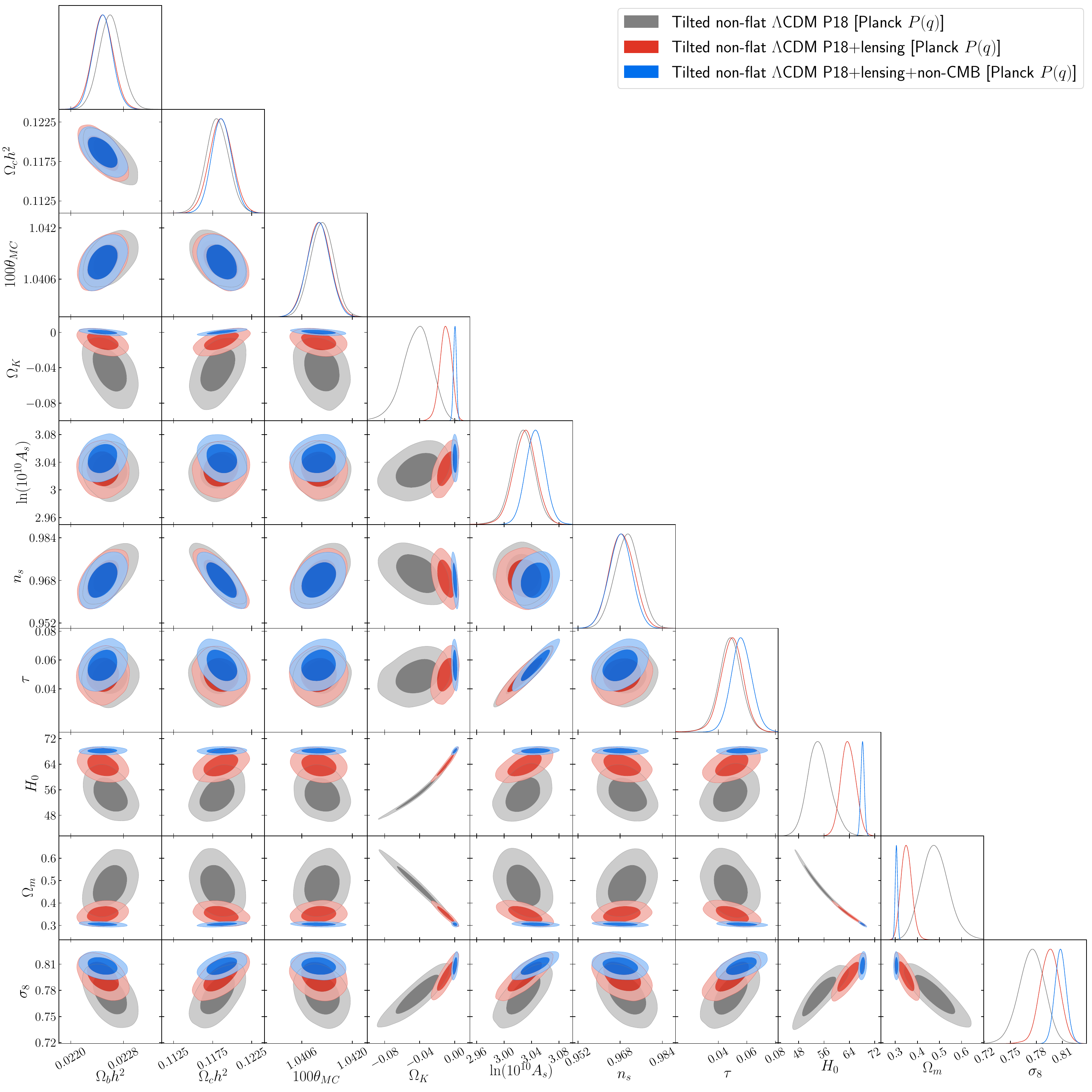}}
\caption{Likelihood distributions of tilted non-flat $\Lambda$CDM model parameters with the Planck team's initial power spectrum [Planck $P(q)$] constrained by P18 (gray contours), P18+lensing (red contours), P18+lensing+non-CMB (blue contours) data sets.
}
\label{fig:like_NL_ns_compar}
\end{figure*}

\begin{figure*}[htbp]
\centering
\mbox{\includegraphics[width=170mm]{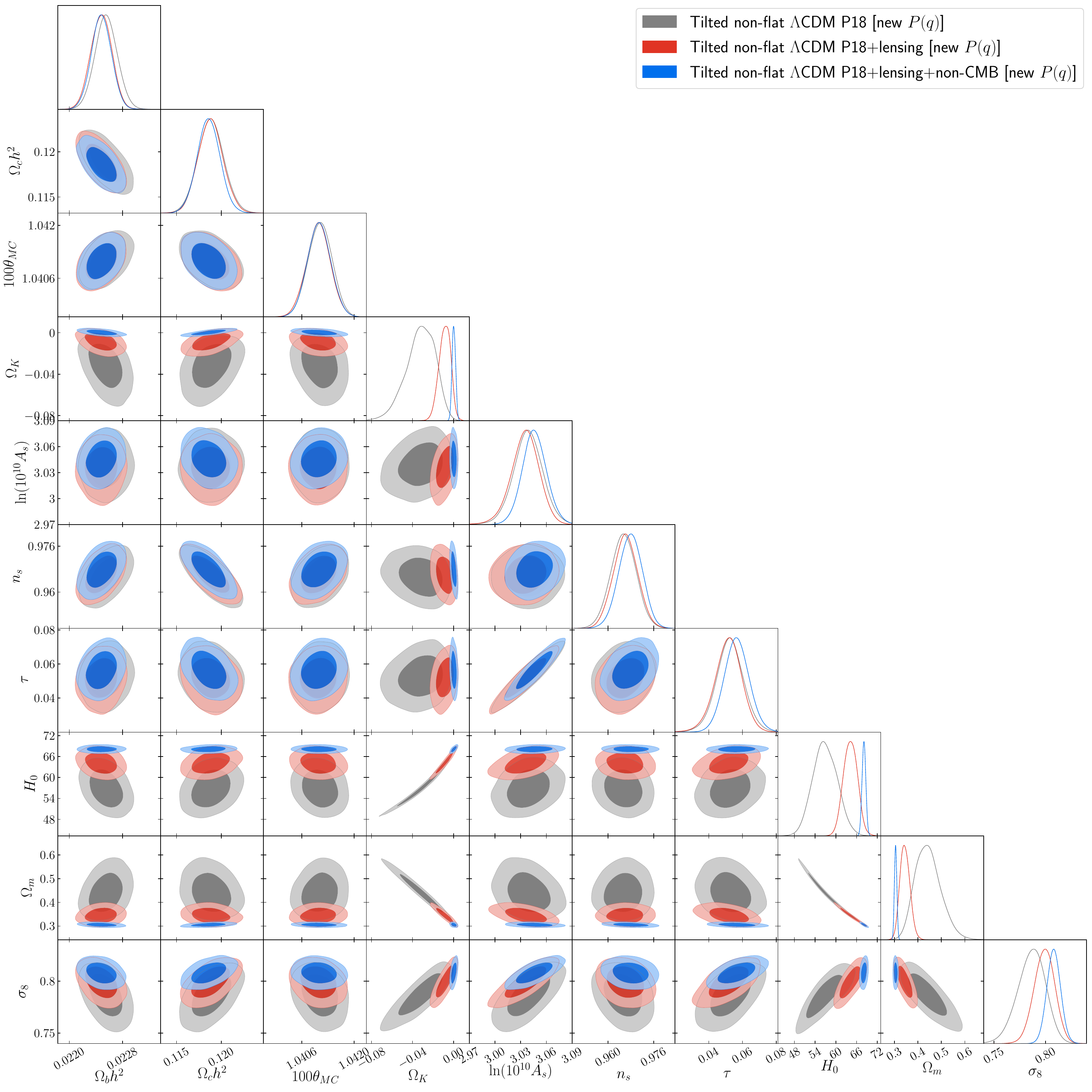}}
\caption{Likelihood distributions of tilted non-flat $\Lambda$CDM model parameters with the new initial power spectrum [new $P(q)$] constrained by P18 (gray contours), P18+lensing (red contours), P18+lensing+non-CMB (blue contours) data sets.}
\label{fig:like_TNL_compar}
\end{figure*}

\begin{figure*}[htbp]
\centering
\mbox{\includegraphics[width=170mm]{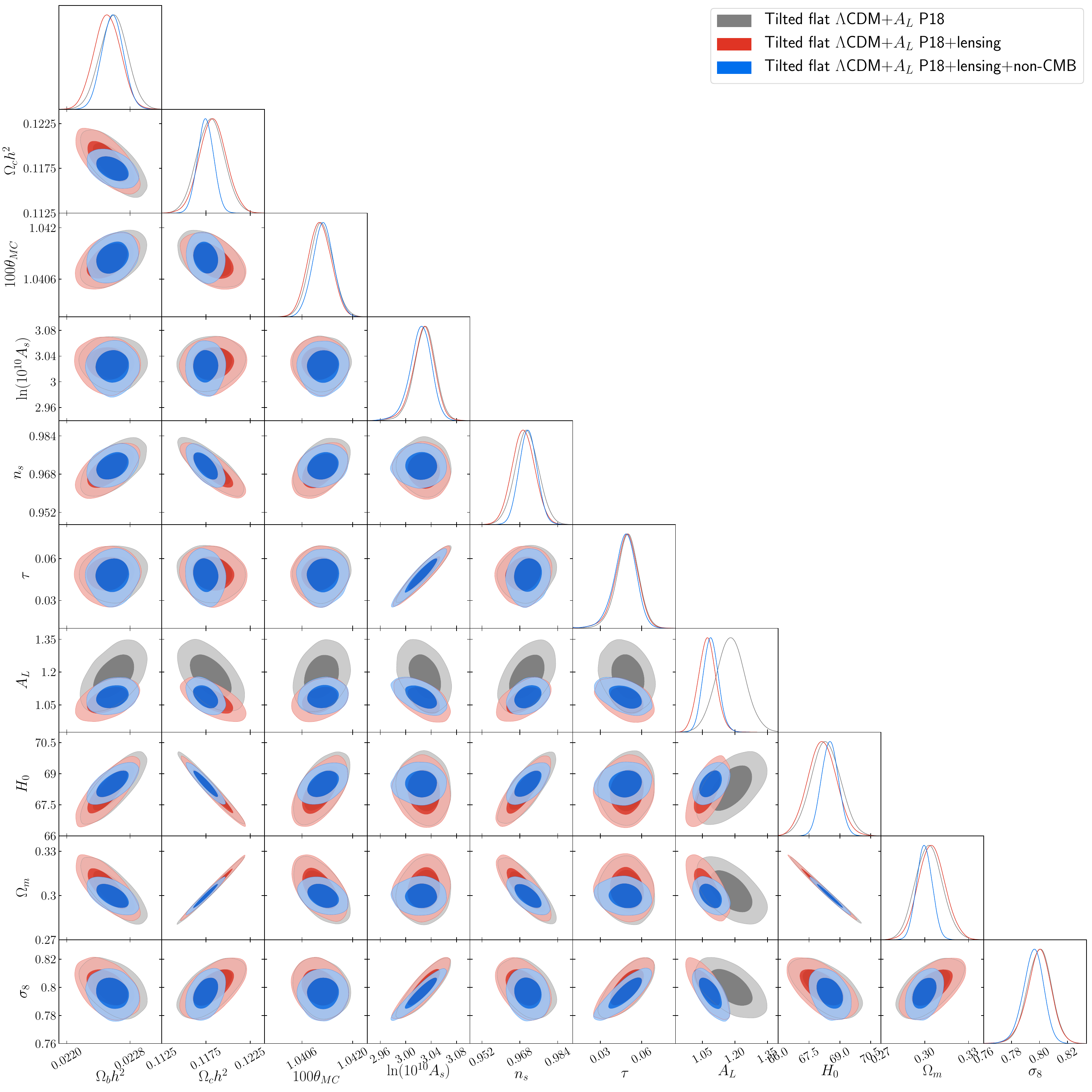}}
\caption{Likelihood distributions of tilted flat $\Lambda$CDM+$A_L$ model parameters constrained by P18 (gray contours), P18+lensing (red contours), P18+lensing+non-CMB (blue contours) data sets.
}
\label{fig:like_FL_Alens_compar}
\end{figure*}

\begin{figure*}[htbp]
\centering
\mbox{\includegraphics[width=170mm]{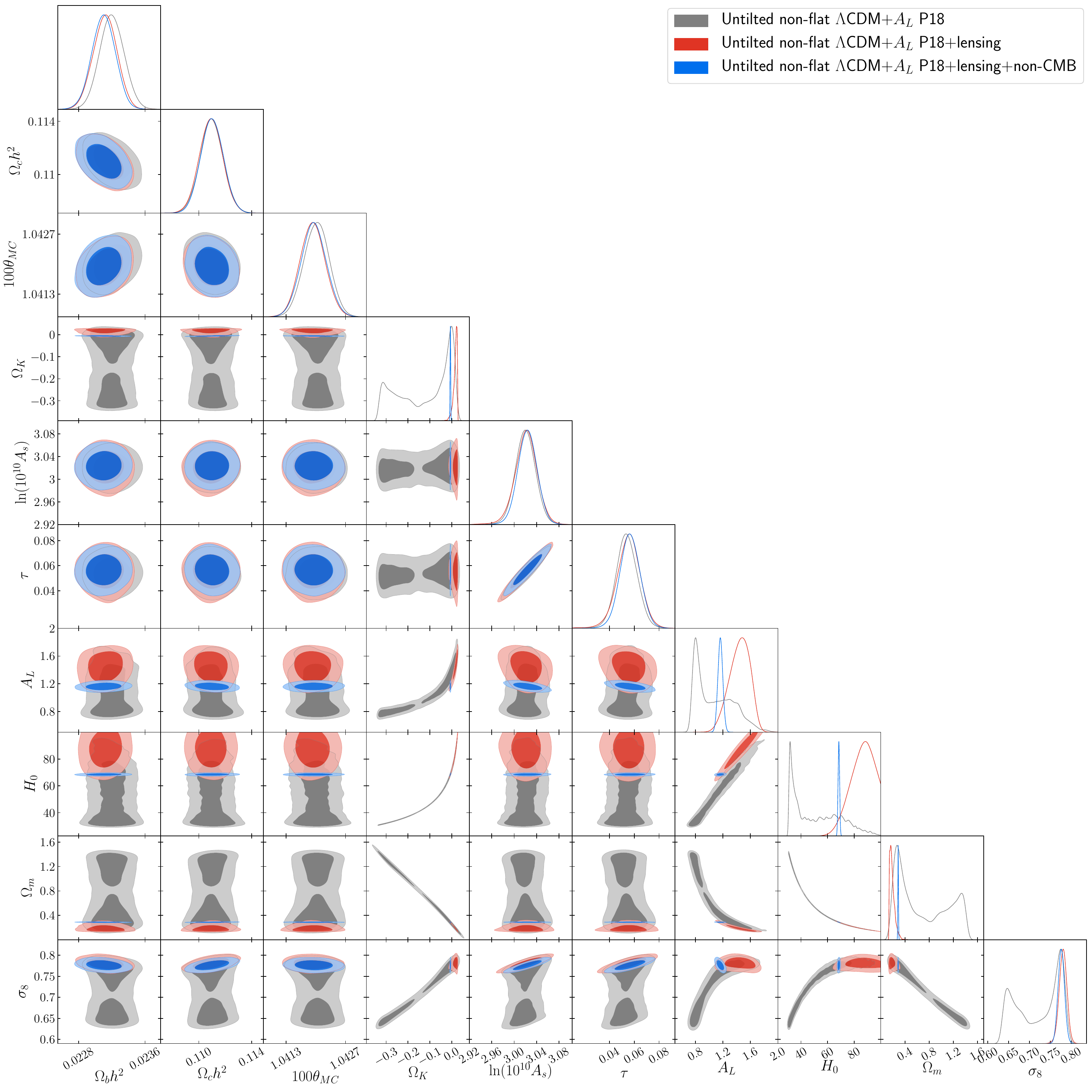}}
\caption{Likelihood distributions of untilted non-flat $\Lambda$CDM+$A_L$ model parameters constrained by P18 (gray contours), P18+lensing (red contours), P18+lensing+non-CMB (blue contours) data sets.
}
\label{fig:like_NL_Alens_compar}
\end{figure*}

\begin{figure*}[htbp]
\centering
\mbox{\includegraphics[width=170mm]{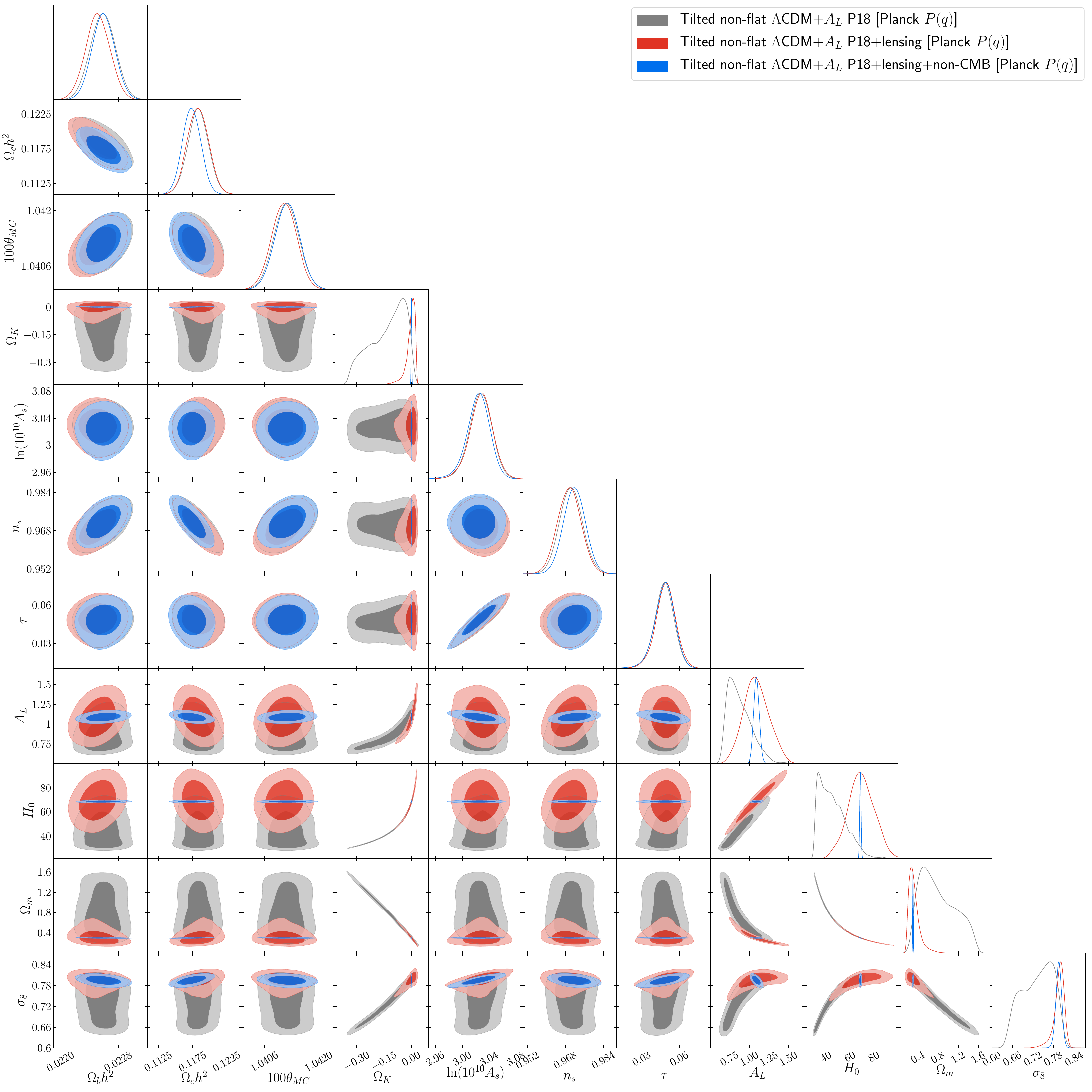}}
\caption{Likelihood distributions of tilted non-flat $\Lambda$CDM+$A_L$ model parameters with the Planck team's initial power spectrum [Planck $P(q)$] constrained by P18 (gray contours), P18+lensing (red contours), P18+lensing+non-CMB (blue contours) data sets.
}
\label{fig:like_NL_ns_Alens_compar}
\end{figure*}

\begin{figure*}[htbp]
\centering
\mbox{\includegraphics[width=170mm]{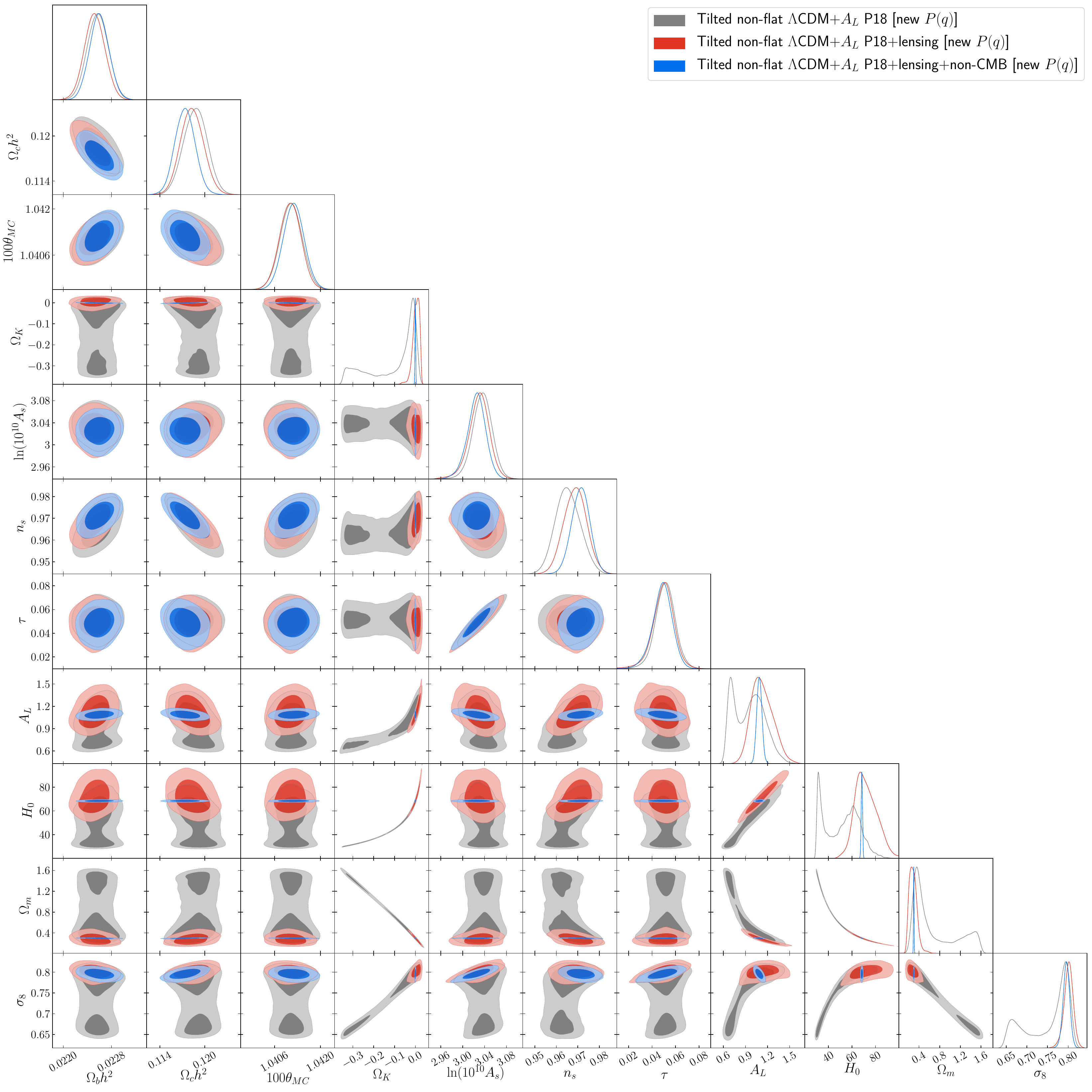}}
\caption{Likelihood distributions of tilted non-flat $\Lambda$CDM+$A_L$ model parameters with the new initial power spectrum [new $P(q)$] constrained by P18 (gray contours), P18+lensing (red contours), P18+lensing+non-CMB (blue contours) data sets.
}
\label{fig:like_TNL_Alens_compar}
\end{figure*}

The cosmological parameter mean values and error bars favored by the P18, P18+lensing, and P18+lensing+non-CMB data sets are summarized in Tables \ref{tab:para_FL_nonCMB}-\ref{tab:para_TNL_nonCMB} for the tilted flat $\Lambda$CDM (+$A_L$) models, the untilted non-flat $\Lambda$CDM (+$A_L$) models, the tilted non-flat $\Lambda$CDM (+$A_L$) models with the Planck $P(q)$, and the tilted non-flat $\Lambda$CDM ($+A_L$) models with the new $P(q)$, respectively. Likelihood distributions of cosmological parameters of the four models with $A_L=1$ are shown in Figs.\ \ref{fig:like_P18}, \ref{fig:like_P18_lensing}, and \ref{fig:like_P18_lensing_nonCMB} for the P18, P18+lensing, and P18+lensing+non-CMB data sets, respectively. The likelihood results for these four models, but now with  $A_L$ allowed to vary, are shown in Figs.\ \ref{fig:like_Alens_P18}, \ref{fig:like_Alens_P18_lensing}, and \ref{fig:like_Alens_P18_lensing_nonCMB}. Figures \ref{fig:like_FL_compar}---\ref{fig:like_TNL_Alens_compar} show, in each of the eight cosmological models we study, the cosmological parameter constraints for P18, P18+lensing, and P18+lensing+non-CMB data, to illustrate how the cosmological parameter constraints change as we include more data. These results are discussed in Secs.\ \ref{subsubsec:P18_data_constraints}-\ref{subsubsec:contour_plots}. In the third paragraph of Sec.\ \ref{subsec:data_set_tensions} we briefly discuss some cosmological parameter constraints from (P18) lensing only data and in Sec.\ \ref{sec:discussion} we discuss whether P18, P18+lensing, P18+non-CMB and P18+lensing+non-CMB data cosmological parameter constraints are model-independent or not.

Our results may indicate tensions between some of the the CMB data sets and some non-CMB low-redshift data in the context of the non-flat models. Tension between P18 data and BAO$^\prime$/BAO data in the tilted non-flat $\Lambda$CDM Planck $P(q)$ model has been noted in Refs.\  \cite{Handley:2019tkm, DiValentino:2019qzk, DiValentino:2020hov} (our updated BAO$^\prime$/BAO data differ from those used in these references, see Sec.\ \ref{sec:data}). Here we want to check whether this tension is observed for our updated BAO$^\prime$/BAO data, whether it is observed in the context of the other models we study, and how this tension is affected when we allow the $A_L$ parameter to vary. In addition to the P18 vs.\ BAO$^\prime$/BAO comparison, we also compare P18 data and non-CMB data as well as P18+lensing and non-CMB data. These comparisons are discussed in Secs.\  \ref{sec:P18_vs_BAO}-\ref{sec:P18+lensing_vs_non-CMB}.
%
\begin{table*}
\caption{Mean and 68.3\% confidence limits of tilted flat $\Lambda\textrm{CDM}$ (+$A_L$) model parameters constrained by TT,TE,EE+lowE (P18), P18+lensing, and P18+lensing+non-CMB data sets. $H_0$ has units of km s$^{-1}$ Mpc$^{-1}$.
}
\begin{ruledtabular}
\begin{tabular}{lccc}
\\[-1mm]                         & \multicolumn{3}{c}{Tilted flat $\Lambda$CDM model}        \\[+1mm]
\cline{2-4}\\[-1mm]
  Parameter                      & P18  &   P18+lensing  &  P18+lensing+non-CMB     \\[+1mm]
 \hline \\[-1mm]
  $\Omega_b h^2$                 & $0.02236 \pm 0.00015$ & $0.02237 \pm 0.00014$ &  $0.02250 \pm 0.00013$    \\[+1mm]
  $\Omega_c h^2$                 & $0.1202 \pm 0.0014$   & $0.1200 \pm 0.0012$   &  $0.11838 \pm 0.00083$    \\[+1mm]
  $100\theta_\textrm{MC}$        & $1.04090 \pm 0.00031$ & $1.04091 \pm 0.00031$ &  $1.04110 \pm 0.00029$    \\[+1mm]
  $\tau$                         & $0.0542 \pm 0.0079$   & $0.0543 \pm 0.0073$   &  $0.0569 \pm 0.0071$      \\[+1mm]
  $n_s$                          & $0.9649 \pm 0.0043$   & $0.9649 \pm 0.0041$   &  $0.9688 \pm 0.0036$      \\[+1mm]
  $\ln(10^{10} A_s)$             & $3.044 \pm 0.016$     & $3.044 \pm 0.014$     &  $3.046 \pm 0.014$        \\[+1mm]
 \hline \\[-1mm]
  $H_0$                          & $67.28 \pm 0.61$      & $67.34 \pm 0.55$      &  $68.09 \pm 0.38$         \\[+1mm]
  $\Omega_m$                     & $0.3165 \pm 0.0084$   & $0.3155 \pm 0.0075$   &  $0.3053 \pm 0.0050$      \\[+1mm]
  $\sigma_8$                     & $0.8118 \pm 0.0074$   & $0.8112 \pm 0.0059$   &  $0.8072 \pm 0.0058$      \\[+1mm]
 \hline \hline \\[-1mm]
                                 & \multicolumn{3}{c}{Tilted flat $\Lambda$CDM+$A_L$ model}              \\[+1mm]
\cline{2-4}\\[-1mm]
  Parameter                      & P18  & P18+lensing    &  P18+lensing+non-CMB     \\[+1mm]
 \hline \\[-1mm]
  $\Omega_b h^2$                 & $0.02259 \pm 0.00017$ & $0.02251 \pm 0.00017$ & $0.02258 \pm 0.00014$     \\[+1mm]
  $\Omega_c h^2$                 & $0.1180 \pm 0.0015$   & $0.1183 \pm 0.0015$   & $0.11747 \pm 0.00091$     \\[+1mm]
  $100\theta_\textrm{MC}$        & $1.04114 \pm 0.00032$ & $1.04109 \pm 0.00032$ & $1.04118 \pm 0.00029$     \\[+1mm]
  $\tau$                         & $0.0496 \pm 0.0082$   & $0.0487 \pm 0.0087$   & $0.0476 \pm 0.0085$       \\[+1mm]
  $n_s$                          & $0.9710 \pm 0.0050$   & $0.9695 \pm 0.0048$   & $0.9715 \pm 0.0038$       \\[+1mm]
  $\ln(10^{10} A_s)$             & $3.030 \pm 0.017$     & $3.028 \pm 0.018$     & $3.023 \pm 0.018$         \\[+1mm]
  $A_{L}$                        & $1.181 \pm 0.067$     & $1.073 \pm 0.041$     & $1.089 \pm 0.035$         \\[+1mm]
 \hline \\[-1mm]
  $H_0$                          & $68.31 \pm 0.71$      & $68.14 \pm 0.69$      & $68.52 \pm 0.42$          \\[+1mm]
  $\Omega_m$                     & $0.3029 \pm 0.0093$   & $0.3048 \pm 0.0091$   & $0.2998 \pm 0.0053$       \\[+1mm]
  $\sigma_8$                     & $0.7997 \pm 0.0088$   & $0.7996 \pm 0.0089$   & $0.7955 \pm 0.0075$       \\[+1mm]
\end{tabular}
\\[+1mm]
\end{ruledtabular}
\label{tab:para_FL_nonCMB}
\end{table*}

\begin{table*}
\caption{Mean and 68.3\% confidence limits of untilted non-flat $\Lambda\textrm{CDM}$ (+$A_L$) model parameters constrained by TT,TE,EE+lowE (P18), P18+lensing, and P18+lensing+non-CMB data sets. $H_0$ has units of km s$^{-1}$ Mpc$^{-1}$.
}
\begin{ruledtabular}
\begin{tabular}{lccc}
\\[-1mm]                         & \multicolumn{3}{c}{Untilted non-flat $\Lambda$CDM model}        \\[+1mm]
\cline{2-4}\\[-1mm]
  Parameter                      & P18  &   P18+lensing   &  P18+lensing+non-CMB     \\[+1mm]
 \hline \\[-1mm]
  $\Omega_b h^2$                 & $0.02320 \pm 0.00015$  & $0.02307 \pm 0.00014$  &  $0.02301 \pm 0.00014$   \\[+1mm]
  $\Omega_c h^2$                 & $0.11098 \pm 0.00088$  & $0.11108 \pm 0.00086$  &  $0.11176 \pm 0.00083$   \\[+1mm]
  $100\theta_\textrm{MC}$        & $1.04204 \pm 0.00030$  & $1.04196 \pm 0.00029$  &  $1.04189 \pm 0.00029$   \\[+1mm]
  $\tau$                         & $0.0543 \pm 0.0091$    & $0.0580 \pm 0.0087$    &  $0.0799 \pm 0.0089$     \\[+1mm]
  $\Omega_k$                     & $-0.095 \pm 0.024$     & $-0.0322 \pm 0.0075$   &  $-0.0065 \pm 0.0014$    \\[+1mm]
  $\ln(10^{10} A_s)$             & $3.021 \pm 0.019$      & $3.027 \pm 0.018$      &  $3.075 \pm 0.018$       \\[+1mm]
 \hline \\[-1mm]
  $H_0$                          & $47.1 \pm 3.2$         & $58.9 \pm 2.1$         &  $67.90 \pm 0.56$        \\[+1mm]
  $\Omega_m$                     & $0.617 \pm 0.082$      & $0.390 \pm 0.027$      &  $0.2938 \pm 0.0049$     \\[+1mm]
  $\sigma_8$                     & $0.730 \pm 0.017$      & $0.765 \pm 0.011$      &  $0.7997 \pm 0.0076$     \\[+1mm]
 \hline \hline \\[-1mm]
                                 & \multicolumn{3}{c}{Untilted non-flat $\Lambda$CDM+$A_L$ model}              \\[+1mm]
\cline{2-4}\\[-1mm]
  Parameter                      & P18  & P18+lensing     &  P18+lensing+non-CMB     \\[+1mm]
 \hline \\[-1mm]
  $\Omega_b h^2$                 & $0.02320 \pm 0.00015$  & $0.02312 \pm 0.00014$  & $0.02310 \pm 0.00014$    \\[+1mm]
  $\Omega_c h^2$                 & $0.11097 \pm 0.00087$  & $0.11092 \pm 0.00087$  & $0.11100 \pm 0.00085$    \\[+1mm]
  $100\theta_\textrm{MC}$        & $1.04202 \pm 0.00030$  & $1.04193 \pm 0.00029$  & $1.04195 \pm 0.00030$    \\[+1mm]
  $\tau$                         & $0.0540 \pm 0.0087$    & $0.0554 \pm 0.0097$    & $0.0566 \pm 0.0083$      \\[+1mm]
  $\Omega_k$                     & $-0.12 \pm 0.12$       & $0.0161 \pm 0.0094$    & $-0.0060 \pm 0.0014$     \\[+1mm]
  $\ln(10^{10} A_s)$             & $3.020 \pm 0.018$      & $3.021 \pm 0.020$      & $3.024 \pm 0.017$        \\[+1mm]
  $A_{L}$                        & $1.08 \pm 0.27$        & $1.44 \pm 0.15$        & $1.162 \pm 0.036$        \\[+1mm]
 \hline \\[-1mm]
  $H_0$                          & $52 \pm 18$            & $85.7 \pm 8.5$         & $68.48 \pm 0.58$         \\[+1mm]
  $\Omega_m$                     & $0.70 \pm 0.42$        & $0.190 \pm 0.043$      & $0.2874 \pm 0.0050$      \\[+1mm]
  $\sigma_8$                     & $0.721 \pm 0.053$      & $0.7805 \pm 0.0094$    & $0.7764 \pm 0.0078$      \\[+1mm]
\end{tabular}
\\[+1mm]
\end{ruledtabular}
\label{tab:para_NL_nonCMB}
\end{table*}

\begin{table*}
\caption{Mean and 68.3\% confidence limits of Planck-$P(q)$-based tilted non-flat $\Lambda\textrm{CDM}$ ($+A_L$) model parameters constrained by TT,TE,EE+lowE (P18), P18+lensing, and P18+lensing+non-CMB data sets. $H_0$ has units of km s$^{-1}$ Mpc$^{-1}$.
}
\begin{ruledtabular}
\begin{tabular}{lccc}
\\[-1mm]                         & \multicolumn{3}{c}{Tilted non-flat $\Lambda$CDM model [Planck $P(q)$]}      \\[+1mm]
\cline{2-4}\\[-1mm]
  Parameter                      & P18  &   P18+lensing  &  P18+lensing+non-CMB     \\[+1mm]
 \hline \\[-1mm]
  $\Omega_b h^2$                 & $0.02260 \pm 0.00017$ & $0.02249 \pm 0.00016$  &  $0.02249 \pm 0.00015$   \\[+1mm]
  $\Omega_c h^2$                 & $0.1181 \pm 0.0015$   & $0.1186 \pm 0.0015$    &  $0.1187 \pm 0.0013$     \\[+1mm]
  $100\theta_\textrm{MC}$        & $1.04116 \pm 0.00032$ & $1.04107 \pm 0.00032$  &  $1.04106 \pm 0.00031$   \\[+1mm]
  $\tau$                         & $0.0483 \pm 0.0083$   & $0.0495 \pm 0.0082$    &  $0.0563 \pm 0.0073$     \\[+1mm]
  $\Omega_k$                     & $-0.043 \pm 0.017$    & $-0.0103 \pm 0.0066$   &  $0.0004 \pm 0.0017$     \\[+1mm]
  $n_s$                          & $0.9706 \pm 0.0047$   & $0.9687 \pm 0.0046$    &  $0.9681 \pm 0.0044$     \\[+1mm]
  $\ln(10^{10} A_s)$             & $3.027 \pm 0.017$     & $3.030 \pm 0.017$      &  $3.046 \pm 0.014$       \\[+1mm]
 \hline \\[-1mm]
  $H_0$                          & $54.5 \pm 3.6$        & $63.7 \pm 2.3$         &  $68.17 \pm 0.55$        \\[+1mm]
  $\Omega_m$                     & $0.481 \pm 0.062$     & $0.351 \pm 0.024$      &  $0.3051 \pm 0.0053$     \\[+1mm]
  $\sigma_8$                     & $0.775 \pm 0.015$     & $0.796 \pm 0.011$      &  $0.8080 \pm 0.0066$     \\[+1mm]
 \hline \hline \\[-1mm]
                                 & \multicolumn{3}{c}{Tilted non-flat $\Lambda$CDM+$A_L$ model [Planck $P(q)$]}     \\[+1mm]
\cline{2-4}\\[-1mm]
  Parameter                      & P18  & P18+lensing    &  P18+lensing+non-CMB     \\[+1mm]
 \hline \\[-1mm]
  $\Omega_b h^2$                 & $0.02258 \pm 0.00017$ & $0.02251 \pm 0.00017$  & $0.02259 \pm 0.00016$    \\[+1mm]
  $\Omega_c h^2$                 & $0.1183 \pm 0.0015$   & $0.1183 \pm 0.0015$    & $0.1173 \pm 0.0014$      \\[+1mm]
  $100\theta_\textrm{MC}$        & $1.04116 \pm 0.00033$ & $1.04110 \pm 0.00032$  & $1.04118 \pm 0.00032$    \\[+1mm]
  $\tau$                         & $0.0478 \pm 0.0081$   & $0.0489 \pm 0.0085$    & $0.0479 \pm 0.0085$      \\[+1mm]
  $\Omega_k$                     & $-0.130 \pm 0.095$    & $-0.005 \pm 0.027$     & $-0.0002 \pm 0.0017$     \\[+1mm]
  $n_s$                          & $0.9704 \pm 0.0048$   & $0.9696 \pm 0.0049$    & $0.9718 \pm 0.0045$      \\[+1mm]
  $\ln(10^{10} A_s)$             & $3.027 \pm 0.017$     & $3.028 \pm 0.018$      & $3.024 \pm 0.017$        \\[+1mm]
  $A_{L}$                        & $0.88 \pm 0.15$       & $1.09 \pm 0.16$       & $1.090 \pm 0.036$        \\[+1mm]
 \hline \\[-1mm]
  $H_0$                          & $45 \pm 11$           & $69 \pm 11$            & $68.49 \pm 0.56$         \\[+1mm]
  $\Omega_m$                     & $0.80 \pm 0.35$       & $0.32 \pm 0.11$        & $0.2998 \pm 0.0055$      \\[+1mm]
  $\sigma_8$                     & $0.733 \pm 0.045$     & $0.796 \pm 0.016$      & $0.7952 \pm 0.0085$        \\[+1mm]
\end{tabular}
\\[+1mm]
\end{ruledtabular}
\label{tab:para_NL_ns_nonCMB}
\end{table*}

\begin{table*}
\caption{Mean and 68.3\% confidence limits of new-$P(q)$-based tilted non-flat $\Lambda\textrm{CDM}$ ($+A_L$) model parameters constrained by TT,TE,EE+lowE (P18), P18+lensing, and P18+lensing+non-CMB data sets. $H_0$ has units of km s$^{-1}$ Mpc$^{-1}$.
}
\begin{ruledtabular}
\begin{tabular}{lccc}
\\[-1mm]                         & \multicolumn{3}{c}{Tilted non-flat $\Lambda$CDM model [new $P(q)$]}   \\[+1mm]
\cline{2-4}\\[-1mm]
  Parameter                      & P18  &   P18+lensing   &  P18+lensing+non-CMB     \\[+1mm]
 \hline \\[-1mm]
  $\Omega_b h^2$                 & $0.02255 \pm 0.00017$  & $0.02248 \pm 0.00016$  &  $0.02248 \pm 0.00015$   \\[+1mm]
  $\Omega_c h^2$                 & $0.1188 \pm 0.0015$    & $0.1188 \pm 0.0014$    &  $0.1186 \pm 0.0013$     \\[+1mm]
  $100\theta_\textrm{MC}$        & $1.04109 \pm 0.00032$  & $1.04104 \pm 0.00032$  &  $1.04106 \pm 0.00031$   \\[+1mm]
  $\tau$                         & $0.0525 \pm 0.0083$    & $0.0515 \pm 0.0081$    &  $0.0566 \pm 0.0074$     \\[+1mm]
  $\Omega_k$                     & $-0.033 \pm 0.014$     & $-0.0086 \pm 0.0057$   &  $0.0003 \pm 0.0017$    \\[+1mm]
  $n_s$                          & $0.9654 \pm 0.0045$    & $0.9661 \pm 0.0043$    &  $0.9679 \pm 0.0042$     \\[+1mm]
  $\ln(10^{10} A_s)$             & $3.039 \pm 0.017$      & $3.035 \pm 0.016$      &  $3.046 \pm 0.014$   \\[+1mm]
 \hline \\[-1mm]
  $H_0$                          & $56.9 \pm 3.6$         & $64.2 \pm 2.0$         &  $68.13 \pm 0.54$    \\[+1mm]
  $\Omega_m$                     & $0.444 \pm 0.055$      & $0.345 \pm 0.021$      &  $0.3054 \pm 0.0051$ \\[+1mm]
  $\sigma_8$                     & $0.786 \pm 0.014$      & $0.799 \pm 0.010$      &  $0.8079 \pm 0.0067$  \\[+1mm]
 \hline \hline \\[-1mm]
 $              $                  & \multicolumn{3}{c}{Tilted non-flat $\Lambda$CDM+$A_L$ model [new $P(q)$]}  \\[+1mm]
\cline{2-4}\\[-1mm]
  Parameter                      & P18  & P18+lensing     &  P18+lensing+non-CMB     \\[+1mm]
 \hline \\[-1mm]
  $\Omega_b h^2$                 & $0.02257 \pm 0.00017$  & $0.02252 \pm 0.00017$  & $0.02260 \pm 0.00016$    \\[+1mm]
  $\Omega_c h^2$                 & $0.1187 \pm 0.0016$    & $0.1183 \pm 0.0015$    & $0.1174 \pm 0.0013$    \\[+1mm]
  $100\theta_\textrm{MC}$        & $1.04111 \pm 0.00033$  & $1.04108 \pm 0.00032$  & $1.04118 \pm 0.00032$    \\[+1mm]
  $\tau$                         & $0.0512 \pm 0.0086$    & $0.0495 \pm 0.0093$    & $0.0486 \pm 0.0086$    \\[+1mm]
  $\Omega_k$                     & $-0.10 \pm 0.11$       & $0.003 \pm 0.016$      & $-0.0002 \pm 0.0017$    \\[+1mm]
  $n_s$                          & $0.9654 \pm 0.0057$    & $0.9688 \pm 0.0053$    & $0.9713 \pm 0.0042$    \\[+1mm]
  $\ln(10^{10} A_s)$             & $3.036 \pm 0.018$      & $3.030 \pm 0.019$      & $3.025 \pm 0.017$    \\[+1mm]
  $A_{L}$                        & $0.94 \pm 0.20$        & $1.13 \pm 0.15$        & $1.088 \pm 0.035$    \\[+1mm]
 \hline \\[-1mm]
  $H_0$                          & $51 \pm 14$            & $72.0 \pm 9.2$         & $68.48 \pm 0.56$    \\[+1mm]
  $\Omega_m$                     & $0.70 \pm 0.43$        & $0.287 \pm 0.076$      & $0.2999 \pm 0.0055$    \\[+1mm]
  $\sigma_8$                     & $0.752 \pm 0.052$      & $0.801 \pm 0.011$      & $0.7956 \pm 0.0082$    \\[+1mm]
\end{tabular}
\\[+1mm]
\end{ruledtabular}
\label{tab:para_TNL_nonCMB}
\end{table*}

We now discuss the results obtained from the different data sets we consider.

\subsubsection{P18 data cosmological constraints}
\label{subsubsec:P18_data_constraints}

In the case of the tilted flat $\Lambda$CDM model, with just six primary (not derived) cosmological parameters, and with $\Omega_k = 0$, from P18 data alone (see Table \ref{tab:para_FL_nonCMB} and Figs.\ \ref{fig:like_P18} and \ref{fig:like_Alens_P18}) the derived parameters $\Omega_m = 0.3165\pm 0.0084$ and $H_0 = 67.28\pm 0.61$ km s$^{-1}$ Mpc$^{-1}$, which are consistent with many other measurements of these quantities and which differ from the low-redshift data measurements of Ref.\ \cite{Cao:2022ugh}, $\Omega_m = 0.295\pm 0.017$ and $H_0 = 69.7\pm 1.2$ km s$^{-1}$ Mpc$^{-1}$, by $1.1\sigma$ and $1.8\sigma$. 

The improvement in the fit to P18 data when the $A_L$ parameter is allowed to vary, in the tilted flat $\Lambda$CDM$+A_L$ model, is positive, according to the DIC statistical criterion described in Sec. \ref{sec:method} (see the results in Sec. \ref{subsec:model_selection} ). This fact is reflected in the measured (from P18 data) value of this phenomenological parameter, $A_L=1.181\pm 0.067$, which differs from the theoretically expected $A_L = 1$ by $2.7\sigma$. The inclusion of the $A_L$ parameter, introduced to deal with the lensing anomaly, does not significantly affect the values of the other six primary parameters, leaving them close to the values found for the six parameter ($\Omega_k = 0$) tilted flat $\Lambda$CDM model with $A_L = 1$ (the largest difference is for $\Omega_c h^2$, where it is 1.1$\sigma$ of the quadrature sum of the two error bars); it does however increase the error bars somewhat, with the largest increase being 16\% for $n_s$. In addition, in the case when $A_L$ is allowed to vary, the derived parameters $\Omega_m$ and $H_0$ (as well as $\sigma_8$) errors bars mildly increase, by 11\% and 16\%, resulting in (for P18 data) $\Omega_m = 0.3029\pm 0.0093$ and $H_0 = 68.31\pm 0.71$ km s$^{-1}$ Mpc$^{-1}$, that are consistent with many other measurements of these quantities, now differing only by $0.41\sigma$ and $1.0\sigma$, respectively, from those of Ref.\ \cite{Cao:2022ugh}. These derived $\Omega_m$ and $H_0$ parameter values in the $A_L$-varying case also differ from those in the $A_L = 1$ case by at most 1.1$\sigma$. 

The addition of the $\Omega_k$ parameter to the six primary cosmological parameters of the tilted flat $\Lambda$CDM model introduces a strong degeneracy between increasing $\Omega_m$ and decreasing $H_0$. The non-flat models also show some degeneracy between $\Omega_m$ and $\Omega_k$ as well as between $H_0$ and $\Omega_k$. These degeneracies can be seen in the corresponding panels in Fig.\ \ref{fig:like_P18}. In the tilted non-flat $\Lambda$CDM models (see Tables \ref{tab:para_NL_ns_nonCMB} and \ref{tab:para_TNL_nonCMB} and Fig.\ \ref{fig:like_P18}) we see that P18 data alone is unable to break the strong geometrical degeneracy between $\Omega_m$-$H_0$-$\Omega_k$. For the Planck $P(q)$ and the new $P(q)$, the measured values (from P18 data) $\Omega_m = 0.481\pm 0.062$ and $0.444\pm 0.055$, as well as $H_0 = 54.5\pm 3.6$ and $56.9\pm 3.6$ km s$^{-1}$ Mpc$^{-1}$, respectively, are in conflict with most other measurements of these parameters; for example, see the low-redshift data measurements of Ref.\ \cite{Cao:2022ugh} in the paragraph before last. Note that even though the values of, and error bars on, the six primary cosmological parameters in common between the two tilted non-flat models and the tilted flat model are very similar (the largest difference is 1.1$\sigma$ for $\Omega_b h^2$ between the tilted flat and the tilted non-flat Planck $P(q)$ models, and the biggest increase, 13\%, in the error bars is also for $\Omega_b h^2$, in both tilted non-flat models relative to the tilted flat model), the additional primary cosmological parameter $\Omega_k$ in the two tilted non-flat models is relatively poorly constrained, and the derived cosmological parameters $\Omega_m$ and $H_0$ error bars in the two tilted non-flat $\Lambda$CDM models are approximately factors of 7 and 6 larger than those in the tilted flat $\Lambda$CDM model (and $\Omega_m$ and $H_0$ in these tilted non-flat models differ by between 2.3$\sigma$ and 3.5$\sigma$ from the tilted flat model values). The evidence in favor of $\Omega_k < 0$ is significant in both of the tilted non-flat $\Lambda$CDM models. For the Planck $P(q)$ case we find $\Omega_k=-0.043\pm 0.017$ while for the new $P(q)$ case $\Omega_k= -0.033\pm 0.014$, being 2.5$\sigma$ and 2.4$\sigma$ away from flat respectively from flat spatial hypersurfaces. In both cases there is clear preference for a closed over an open spatial geometry. And we shall see in Sec. \ref{subsec:model_selection} that the P18 data DIC statistical criterion strongly favors both tilted non-flat models over the tilted flat $\Lambda$CDM model. 

Allowing the $A_L$ parameter to vary in the non-flat models introduces an additional strong degeneracy between $\Omega_k$, $\Omega_m$, $H_0$, and $A_L$; compare the corresponding panels in Figs.\ \ref{fig:like_P18} and \ref{fig:like_Alens_P18}. In the tilted non-flat $\Lambda$CDM+$A_L$ models with the Planck $P(q)$ and with the new $P(q)$, where the $A_L$ parameter is allowed to vary (see Tables \ref{tab:para_NL_ns_nonCMB} and \ref{tab:para_TNL_nonCMB} and Fig.\ \ref{fig:like_Alens_P18}) P18 data alone is unable to break the strong geometrical degeneracy between $\Omega_m$-$H_0$-$\Omega_k$-$A_L$. (In the tilted non-flat $\Lambda$CDM+$A_L$ new $P(q)$ model some parameters have a somewhat bimodal distribution for P18 data, see the one-dimensional posterior distributions in Fig.\ \ref{fig:like_Alens_P18}. This is not the case for the tilted non-flat $\Lambda$CDM+$A_L$ Planck $P(q)$ model.) Like in the tilted flat $\Lambda$CDM case discussed in the paragraph before last, the extra $A_L$ parameter does not significantly affect any of the (primary, not derived) cosmological parameter constraints, compared to the $A_L = 1$ case, except, because of the additional $\Omega_k$-$A_L$ degeneracy, the $\Omega_k$ constraints which are now $\Omega_k=-0.130\pm 0.095$ for the Planck $P(q)$ case and $\Omega_k= -0.10\pm 0.11$ for the new $P(q)$, being only 1.4$\sigma$ and 0.91$\sigma$, respectively, away from flat spatial hypersurfaces, with the $\Omega_k$ error bars now being factors of 6 and 8, respectively, larger than those in the $A_L = 1$ case. Also, unlike the tilted flat $\Lambda$CDM case of the paragraph before last, we measure, from the P18 data, $A_L=0.88\pm 0.15$ and $0.94\pm0.20$, which differ from the theoretically expected $A_L = 1$ by only 0.80$\sigma$ and 0.30$\sigma$. We will see in Sec. \ref{subsec:model_selection} that in both these models the fit to P18 data is weakly or positively better when $A_L = 1$ compared to the case when the $A_L$ parameter is allowed to vary. However, when $A_L$ varies the DIC statistical criterion weakly favors [positively disfavors] the tilted non-flat Planck $P(q)$ [new $P(q)$] model over the tilted flat $\Lambda$CDM$+A_L$ model. In addition, in both these cases when $A_L$ is allowed to vary, the $\Omega_m$ and $H_0$ (as well as $\sigma_8$) errors bars significantly increase, resulting in (for P18 data) $\Omega_m = 0.80\pm 0.35$ and $0.70\pm 0.43$, as well as $H_0 = 45\pm 11$ and $51\pm 14$ km s$^{-1}$ Mpc$^{-1}$, that are consistent with many other measurements of these quantities. Again, the error bars on $\Omega_b h^2$, $\Omega_c h^2$, $\theta_{\rm MC}$, $\tau$, $n_s$, and $A_s$ are similar in the two tilted non-flat $\Lambda$CDM$+A_L$ models and the tilted flat $\Lambda$CDM$+A_L$ model, however the $A_L$ error bars are approximately a factor of 2.5 larger in the tilted non-flat models, with the introduction of the seventh primary cosmological parameter $\Omega_k$ (that is poorly constrained) also resulting in the $\Omega_m$ error bars being a factor $\sim$42 larger and the $H_0$ error bars being a factor $\sim$18 larger in the tilted non-flat $\Lambda$CDM$+A_L$ models compared to the tilted flat $\Lambda$CDM$+A_L$ model.   

The restriction that $n_s=1$ in the untilted non-flat $\Lambda$CDM (+$A_L$) models is an unwelcome feature when fitting the P18 CMB anisotropy spectra, according to the statistical criteria outlined in Sec. \ref{subsec:model_selection}. Because of this we will focus less attention on the untilted non-flat $\Lambda$CDM model compared to the two tilted non-flat models. Despite the poor performance of the untilted non-flat $\Lambda$CDM model in this case (which also affects what happens when additional data are jointly analyzed with P18 data), the model shares some features with the two tilted non-flat $\Lambda$CDM models (see Table \ref{tab:para_NL_nonCMB} and Fig.\ \ref{fig:like_P18}), namely, the evidence in favor of closed spatial geometry, now with $\Omega_k= -0.095\pm 0.024$ (4.0$\sigma$), and the presence of the aforementioned geometrical degeneracy between $\Omega_m$-$H_0$-$\Omega_k$. Also, as in the two tilted non-flat models, in the untilted non-flat case the measured values (from P18 data) $\Omega_m = 0.617\pm 0.082$ and $H_0 = 47.1\pm 3.2$ km s$^{-1}$ Mpc$^{-1}$ are in conflict with most other measurements of these parameters. 

In the untilted non-flat $\Lambda$CDM+$A_L$ model where the $A_L$ parameter is allowed to vary (see Table \ref{tab:para_NL_nonCMB} and Fig.\ \ref{fig:like_Alens_P18}) P18 data alone is again unable to break the larger geometrical degeneracy between $\Omega_m$-$H_0$-$\Omega_k$-$A_L$. Like in the tilted flat and non-flat $\Lambda$CDM cases discussed earlier, the extra $A_L$ parameter does not significantly affect any of the (primary, not derived) cosmological parameter constraints in the untilted non-flat model, compared to the $A_L = 1$ case, except, because of the additional $\Omega_k$-$A_L$ degeneracy, for the $\Omega_k$ constraint which is now $\Omega_k=-0.12\pm 0.12$ and only 1.0$\sigma$ away from flat spatial hypersurfaces. Also, unlike the tilted flat $\Lambda$CDM case, but like the tilted non-flat cases of the paragraph before last, we measure, from the P18 data, $A_L=1.08\pm 0.27$ which differs from the theoretically expected value of $A_L = 1$ by only 0.30$\sigma$. We will see in Sec.\ \ref{subsec:model_selection} that in this model the fit to P18 data is slightly better when $A_L = 1$ compared to the case when the $A_L$ parameter is allowed to vary. Similar to the two tilted non-flat models of the paragraph before last, when $A_L$ is allowed to vary in the untilted non-flat model, the $\Omega_m$ and $H_0$ (as well as $\sigma_8$) errors bars significantly increase, resulting in (for P18 data) $\Omega_m = 0.70\pm 0.42$ and $H_0 = 52\pm 18$ km s$^{-1}$ Mpc$^{-1}$, that are consistent with most other measurements of these quantities.

In Fig.\ \ref{fig:like_P18} we provide the 2$\sigma$ contour plots for all four of the $A_L = 1$ models. The contours for the untilted non-flat $\Lambda$CDM model overlap minimally or even do not overlap at all with those corresponding to the other models. This is likely due to the lack of the degree of freedom encapsulated in $n_s$ in the untilted non-flat model, which greatly hinders the fit of the CMB anisotropy power spectra and causes the other parameters to shift from the ranges preferred in the context of the other three models. As for the other three cosmological models, there is a significant overlap of contours, except when $\Omega_m$ or $H_0$ (or less so $\sigma_8$) is involved, which can even lead to the contours not overlapping at all. This is presumably related with the geometrical degeneracy previously mentioned. 

The corresponding plots for the four models now including allowing $A_L$ to vary are in Fig.\ \ref{fig:like_Alens_P18}. Allowing $A_L$ to vary broadens the contours, and for some parameters there are two disconnected 1$\sigma$ regions. While the untilted non-flat model contours still do not overlap in many cases with those of the other three models, in the other three models the contours overlap even when $\Omega_m$ or $H_0$ or $\sigma_8$ is involved.

\subsubsection{ P18+lensing data cosmological constraints} 
\label{subsubsec:P18_lensing_data_constraints}

Constraints on primary parameters derived from joint analyses of P18 and lensing data are quite similar to those derived from P18 data alone (see Tables \ref{tab:para_FL_nonCMB}-\ref{tab:para_TNL_nonCMB} and Figs. \ref{fig:like_P18_lensing} and \ref{fig:like_Alens_P18_lensing}), except for $\Omega_k$ and $A_L$ constraints. On the other hand, constraints on derived parameters $\Omega_m$ and $H_0$ are, in most non-flat cases, greatly affected by lensing data. In this subsubsection we discuss parameter constraints from jointly analyzed P18 and lensing data and compare these to the P18 data alone constraints. 

Ideally one would like to establish that cosmological parameter constraints derived from P18 data and from lensing data are mutually consistent, prior to using P18+lensing data in joint analyses. While it is not straightforward to derive (P18) lensing data alone cosmological parameter constraints for the wide flat Our priors of Table \ref{tab:Priors}, we shall see, in Sec.\ \ref{subsec:data_set_tensions} (where we do briefly discuss some of these cosmological constraints), that P18 data and lensing data are not significantly mutually inconsistent. This is also consistent with the results we discuss in this subsubsection.   

Comparing the six-parameter tilted flat $\Lambda$CDM primary cosmological parameter constraints for P18 data and P18+lensing data, shown in the upper half of Table \ref{tab:para_FL_nonCMB}, we see that there are no significant changes in parameter values (the largest change is that $\Omega_c h^2$ is 0.11$\sigma$ smaller in the P18+lensing case) with all but the $\theta_{\rm MC}$ error bars being smaller in the P18+lensing case (the $\theta_{\rm MC}$ error bar is unchanged and the largest decrease is 14\%  for the $\Omega_c h^2$ error bar). For the derived parameters, the largest change is the 0.089$\sigma$ decrease in $\Omega_m$ relative to the P18 data value, and the 20\% smaller $\sigma_8$ error bar. For P18+lensing data we find $\Omega_m = 0.3155\pm 0.0075$ and $H_0 = 67.34\pm 0.55$ km s$^{-1}$ Mpc$^{-1}$ which are consistent with many other measurements of these quantities and 1.1$\sigma$ larger and 1.8$\sigma$ lower than the low-redshift data measurements of Ref.\ \cite{Cao:2022ugh}.

Comparing the seven-parameter tilted flat $\Lambda$CDM$+A_L$ primary cosmological parameter constraints for P18 data and P18+lensing data, shown in the lower half of Table \ref{tab:para_FL_nonCMB}, we see more significant changes in the parameter values (the largest change is that $A_L$ is 1.4$\sigma$ smaller in the P18+lensing case, with the next largest being $\Omega_b h^2$ which is 0.33$\sigma$ smaller) with some of the error bars being larger in the P18+lensing case (the largest increase is 6\% for the $\tau$ and $\ln(10^{10}A_s)$ error bars) and some of the error bars being smaller (the largest decrease is 39\% for $A_L$). The reason the error bars of $\tau$ and $\ln (10^{10} A_s)$ increase, contrary to the common expectation that the error bars of the parameters will become smaller as more data is added, appears to be that the degeneracy between parameters is only partially broken by the lensing data. Interestingly, these characteristics are common to all other $A_L$-varying models (see Tables \ref{tab:para_NL_nonCMB}-\ref{tab:para_TNL_nonCMB}). For the derived parameters, the largest change is the 0.17$\sigma$ decrease in $H_0$ relative to the P18 data value, and the 3\% smaller $H_0$ error bar. For P18+lensing data in the varying $A_L$ case we measure $\Omega_m = 0.3048\pm 0.0091$ and $H_0 = 68.14\pm 0.69$ km s$^{-1}$ Mpc$^{-1}$ which are consistent with many other measurements of these quantities and 0.51$\sigma$ larger and 1.1$\sigma$ lower than the low-redshift data measurements of Ref.\ \cite{Cao:2022ugh}.

The improvement in the fit to P18+lensing data when the $A_L$ parameter is allowed to vary, in the tilted flat $\Lambda$CDM$+A_L$ model, is only weak, as discussed in Sec.\ \ref{subsec:model_selection}. We now find $A_L = 1.073\pm 0.041$ which is 1.8$\sigma$ away from the theoretically expected $A_L = 1$. While there is still a deviation from the predicted value, the tendency of the lensing data is to push $A_L$ closer to 1, resulting in a smaller deviation than the 2.7$\sigma$ one found for $A_L = 1.181\pm 0.067$ from P18 data in the tilted flat $\Lambda$CDM$+A_L$ model. The inclusion of the $A_L$ parameter does not significantly affect the values of the other six primary parameters, leaving them close to the values found for the six parameter tilted flat $\Lambda$CDM model with $A_L = 1$ (the largest difference is for $\Omega_c h^2$, where it is 0.88$\sigma$ of the quadrature sum of the two error bars); it does however increase the error bars, more than what happens in the P18 alone data case discussed in Sec. \ref{subsubsec:P18_data_constraints}, with largest increase being 29\% for $\ln(10^{10}A_s)$. In addition, in the case when $A_L$ is allowed to vary, the derived parameters change somewhat and their error bars increase, with the largest changes associated with $\sigma_8$, where it is now 1.1$\sigma$ smaller with a 51\% larger error bar.

In the six-parameter untilted non-flat $\Lambda$CDM model, including lensing data in the mix results in a reduction in the size of the cosmological parameter error bars relative to those from P18 data (see Table \ref{tab:para_NL_nonCMB} and Fig. \ref{fig:like_P18_lensing}). The most affected parameters are the primary parameter $\Omega_k$, whose error bars decrease by 69\%, and the derived parameters $H_0$, $\Omega_m$ and $\sigma_8$, for which we observe a shrinkage of the error bars by 34\%, 67\%, and 35\%, respectively. As happens in the tilted flat $\Lambda$CDM model, here also there are no significant changes in the values of the primary parameters, with the exception of the curvature parameter $\Omega_k$. This is not true for two of the derived parameters, $H_0$ and $\Omega_m$, which together with the curvature parameter are involved in the $\Omega_m$-$H_0$-$\Omega_k$ geometrical degeneracy. From P18+lensing data we find $\Omega_k=-0.095\pm 0.024$, $H_0=47.1\pm 3.2$ km s$^{-1}$ Mpc$^{-1}$, and $\Omega_m=0.390\pm 0.027 $. These values differ by 2.5$\sigma$, 3.1$\sigma$, and 2.6$\sigma$, respectively, from the corresponding values obtained in the P18 data alone analysis. 

From the results obtained for the untilted non-flat $\Lambda$CDM+$A_L$ model (see Table \ref{tab:para_NL_nonCMB} and Fig. \ref{fig:like_Alens_P18_lensing}), we observe significant changes in the values of the primary parameters $\Omega_k$ and $A_L$, as well as in the derived parameters $H_0$ and $\Omega_m$. For the P18+lensing data set we get $\Omega_k = 0.0161\pm 0.0094$ (1.7$\sigma$ away from flat) and $A_L = 1.44\pm 0.15$ (2.9$\sigma$ away from $A_L = 1$). These values differ by 1.1$\sigma$ and 1.2$\sigma$, respectively, from the corresponding values obtained in the P18 data alone analysis. For the derived parameters, from P18+lensing data we find $H_0=85.7\pm 8.5$ km s$^{-1}$ Mpc$^{-1}$ and $\Omega_m=0.190\pm 0.043$, which differ by 1.7$\sigma$ and 1.2$\sigma$ from the corresponding P18 data alone values. 

Joint analyses of the P18 and lensing data in the tilted non-flat models result in constraints that differ more from those derived using just P18 data compared to what happens in the tilted flat model case (see Tables \ref{tab:para_NL_ns_nonCMB} and \ref{tab:para_TNL_nonCMB} and Fig.\ \ref{fig:like_P18_lensing}). This is because lensing data partially breaks the $\Omega_m$-$H_0$-$\Omega_k$ geometrical P18 alone data degeneracy of the tilted non-flat models (compare the corresponding panels in Figs.\ \ref{fig:like_P18} and \ref{fig:like_P18_lensing}). 

Comparing the seven-parameter tilted non-flat $\Lambda$CDM Planck $P(q)$ and new $P(q)$ primary cosmological parameter constraints for P18 data and P18+lensing data, shown in the upper halves of Tables \ref{tab:para_NL_ns_nonCMB} and \ref{tab:para_TNL_nonCMB}, we see that aside from $\Omega_k$ (discussed next) there are no significant changes in parameter values [the largest change is that $\Omega_b h^2$ is 0.47$\sigma$ (0.30$\sigma$) smaller in the P18+lensing case, for the Planck (new) $P(q)$] with some of the error bars being smaller in the P18+lensing case [leaving aside $\Omega_k$ (discussed next) the largest decrease is 6\% (7\%) for the $\Omega_b h^2$ ($\Omega_c h^2$) error bar, for the Planck (new) $P(q)$]. On the other hand, $\Omega_k$ changes significantly when lensing data are added to the mix, becoming 1.8$\sigma$ (1.6$\sigma$) larger, and closer to flat for the Planck (new) $P(q)$, with 61\% (59\%) smaller error bars, still favoring closed geometry over flat but only by 1.6$\sigma$ (1.5$\sigma$), respectively. For the derived parameters, the largest change is the 2.2$\sigma$ (1.8$\sigma$) increase in $H_0$ relative to the P18 data value for the Planck (new) $P(q)$, with 61\% (62\%) smaller error bars for $\Omega_m$. For P18+lensing data we find $\Omega_m = 0.351\pm 0.024$ ($0.345\pm 0.021$) and $H_0 = 63.7\pm 2.3$ ($64.2\pm 2.0$) km s$^{-1}$ Mpc$^{-1}$ for the Planck (new) $P(q)$, which are consistent with many other measurements of these quantities and 1.9$\sigma$ (1.9$\sigma$) larger and 2.3$\sigma$ (2.4$\sigma$) lower, respectively, than the low-redshift data measurements of Ref.\ \cite{Cao:2022ugh}.

Comparing the eight-parameter tilted non-flat $\Lambda$CDM$+A_L$ Planck (new) $P(q)$ primary cosmological parameter constraints for P18 data and P18+lensing data, shown in the lower half of Table \ref{tab:para_NL_ns_nonCMB} (\ref{tab:para_TNL_nonCMB}), we see that there are smaller differences compared to the tilted flat $\Lambda$CDM$+A_L$ case. For the Planck $P(q)$ case we mostly find less significant changes (the largest changes are that $\Omega_k$ and $A_L$ are 1.3$\sigma$ and 0.95$\sigma$ larger in the P18+lensing case, with the next largest being $\Omega_b h^2$ which is 0.29$\sigma$ smaller) with some of the error bars being larger in the P18+lensing case (the largest increase is 7\% for the $A_L$ error bar, and this is the only model where the $A_L$ error bar is larger for P18+lensing data than for P18 data) and some of the error bars being smaller (the largest decrease is 72\% for $\Omega_k$). In the new $P(q)$ case we find roughly half the parameters change more significantly (the largest changes again are that $\Omega_k$ and $A_L$ are 0.93$\sigma$ and 0.76$\sigma$ larger in the P18+lensing case, with the next largest being $n_s$ which is 0.44$\sigma$ larger) with some of the error bars being larger in the P18+lensing case (the largest increase is 8\% for the $\tau$ error bar) and some of the error bars being smaller (the largest decrease is 85\% for $\Omega_k$). From the P18+lensing analyses, we measure $\Omega_k=-0.005\pm 0.027$ for the Planck $P(q)$ case and $\Omega_k= 0.003\pm 0.016$ for the new $P(q)$, both being only 0.19$\sigma$ away from flat spatial hypersurfaces, very different from the P18 data alone results. For the derived parameters, the largest change is the 1.5$\sigma$ (1.3$\sigma$) increase in $H_0$ relative to the P18 data value for the Planck (new) $P(q)$, with 69\% (82\%) smaller error bars for $\Omega_m$. For P18+lensing data we find $\Omega_m = 0.32\pm 0.11$ ($0.287\pm 0.076$) and $H_0 = 69\pm 11$ ($72.0\pm 9.2$) km s$^{-1}$ Mpc$^{-1}$ for the Planck (new) $P(q)$, which are consistent with many other measurements of these quantities and 0.22$\sigma$ larger (0.10$\sigma$ smaller) and 0.063$\sigma$ lower (0.25$\sigma$ higher), respectively, than the low-redshift data measurements of Ref.\ \cite{Cao:2022ugh}. (Note that the P18+lensing data Planck $P(q)$ $H_0$ error bar is unchanged, $\pm 11$ km s$^{-1}$, from the P18 data value, and this is the only model where this happens.)

We will see in Sec.\ \ref{subsec:model_selection} that in both tilted non-flat $\Lambda$CDM$+A_L$ models the fit to P18+lensing data is weakly better when $A_L = 1$ compared to the case when the $A_L$ parameter is allowed to vary; this differs from what happens in the tilted flat $\Lambda$CDM$+A_L$ model. Also, unlike the tilted flat $\Lambda$CDM$+A_L$ P18+lensing case, we measure, from P18+lensing data, $A_L=1.089\pm 0.16$ and $1.13\pm0.15$, for the Planck $P(q)$ and the new $P(q)$, respectively, which differ from the theoretically expected $A_L = 1$ by only 0.56$\sigma$ and 0.87$\sigma$. The inclusion of the $A_L$ parameter does not significantly affect the values of the other seven primary parameters, leaving them close to the values found for the seven parameter tilted non-flat $\Lambda$CDM models with $A_L = 1$ [the largest difference is for $\Omega_k$, where it is 0.19$\sigma$ (0.68$\sigma$) for the Planck (new) $P(q)$]; it does however increase the error bars, but less than what happens in the P18 alone data case discussed in Sec.\ \ref{subsubsec:P18_data_constraints}, with largest factor being 4 (3) for $\Omega_k$ for the Planck (new) $P(q)$. In addition, in the case when $A_L$ is allowed to vary, the derived parameters change somewhat and their error bars increase, with the largest changes associated with $H_0$, where it is now 0.47$\sigma$ (0.83$\sigma$) larger for the Planck (new) $P(q)$ with a factor of 5 (5) larger error bar.

From the discussion above in this subsubsection, the fact that the cosmological constraint contours displayed in Fig.\ \ref{fig:like_P18_lensing} for the three tilted models overlap should not come as a surprise. Unlike in the previous P18 data alone case, the P18+lensing data contours that involve $\Omega_m$, $H_0$, or $\Omega_k$ now overlap for the tilted models, indicating that the geometrical degeneracy is, at least, partially broken. Figure \ref{fig:like_Alens_P18_lensing} shows the results when the $A_L$ parameter is included in the analysis. While the overlap already found in the P18 data alone analysis (see Fig.\ \ref{fig:like_Alens_P18}) remains, the bimodal 1$\sigma$ regions of that plot have now disappeared.

\subsubsection{P18+lensing+non-CMB data cosmological constraints} 
\label{subsubsec:P18_lensing_nonCMB_data_constraints}

In this subsubsection we comment on the results obtained from a joint analysis of the P18+lensing+non-CMB data set and how the cosmological constraints change when compared to those obtained using P18+lensing data. As outlined in Sec.\ \ref{sec:data} non-CMB data we use here is comprised of BAO, $f\sigma_8$, SNIa, and $H(z)$ data, all of which provide useful information on the late-time Universe. 

Ideally one would like to establish that cosmological parameter constraints derived from P18+lensing data and from non-CMB data are mutually consistent, prior to using P18+lensing+non-CMB data in joint analyses. Given that P18 data dominate the P18+lensing data compilation, it is instructive to also study whether P18 data cosmological constraints are consistent with those from non-CMB data. We shall see in Sec.\ \ref{sec:P18_vs_BAO} that, in some of the models we study here, cosmological constraints from BAO$^\prime$ and BAO data, the dominant part of the non-CMB data compilation, are somewhat inconsistent with those derived using P18 data. This is also consistent with the results we discuss in this subsubsection, as well as, with the results presented in Sec.\ \ref{sec:P18_vs_non-CMB}, where we compare the cosmological parameter values obtained using P18 data and using non-CMB data. In Sec.\ \ref{sec:P18+lensing_vs_non-CMB} we compare P18+lensing data cosmological constraints and non-CMB data cosmological constraints, and find similar tensions. In addition, in Sec.\ \ref{subsec:data_set_tensions}, we study tensions between some of the CMB data sets and some of the low-redshift data sets, including the case of P18+lensing data vs.\ non-CMB data, by using the two statistical estimators presented in Sec. \ref{sec:method}.

Comparing the six-parameter tilted flat $\Lambda$CDM primary cosmological parameter constraints for P18+lensing data and P18+lensing+non-CMB data, shown in the upper half of Table \ref{tab:para_FL_nonCMB}, we see that there are no significant changes in parameter values (the largest change is that $\Omega_c h^2$ is 1.1$\sigma$ smaller in the P18+lensing+non-CMB case) with all but the ln($10^{10}A_s)$ error bars being smaller in the P18+lensing+non-CMB case (the ln($10^{10}A_s)$ error bar is unchanged and the largest decrease is 31\% for the $\Omega_c h^2$ error bar). For the derived parameters, the largest changes are the 1.1$\sigma$ decrease in $\Omega_m$ and the 1.1$\sigma$ increase in $H_0$ relative to the P18+lensing data values, and the 33\% (31\%) smaller $\Omega_m$ ($H_0$) error bar. For P18+lensing+non-CMB data we find $\Omega_m = 0.3053\pm 0.0050$ and $H_0 = 68.09\pm 0.38$ km s$^{-1}$ Mpc$^{-1}$ which are consistent with many other measurements of these quantities and 0.58$\sigma$ larger and 1.3$\sigma$ lower than the low-redshift data measurements of Ref.\ \cite{Cao:2022ugh}.

Comparing the seven-parameter tilted flat $\Lambda$CDM$+A_L$ primary cosmological parameter constraints for P18+lensing data and P18+lensing+non-CMB data, shown in the lower half of Table \ref{tab:para_FL_nonCMB}, we see smaller changes in the parameter values (the largest change is that $\Omega_c h^2$ is 0.47$\sigma$ smaller in the P18+lensing+non-CMB case, with the next largest being $n_s$ which is 0.33$\sigma$ larger) with all but the ln($10^{10}A_s)$ error bars being smaller in the P18+lensing+non-CMB case (the ln$(10^{10}A_s)$ error bar is unchanged and the largest decrease is 39\% for the $\Omega_c h^2$ error bars).  For the derived parameters, the largest changes are the 0.47$\sigma$ increase in $H_0$ and the 0.47$\sigma$ decrease in $\Omega_m$ relative to the P18+lensing data values, and the 42\% smaller $\Omega_m$ error bar. For P18+lensing+non-CMB data in the varying $A_L$ case we measure $\Omega_m = 0.2998\pm 0.0053$ and $H_0 = 68.52\pm 0.42$ km s$^{-1}$ Mpc$^{-1}$ which are consistent with many other measurements of these quantities and 0.27$\sigma$ larger and 0.93$\sigma$ lower than the low-redshift data measurements of Ref.\ \cite{Cao:2022ugh}.

The improvement in the fit to P18+lensing+non-CMB data when the $A_L$ parameter is allowed to vary, in the tilted flat $\Lambda$CDM$+A_L$ model, is positive, as discussed in Sec.\ \ref{subsec:model_selection}. We now find $A_L = 1.089\pm 0.035$ which is now 2.5$\sigma$ away from the theoretically expected $A_L = 1$, larger than the 1.8$\sigma$ deviation for the P18+lensing case of Sec.\ \ref{subsubsec:P18_lensing_data_constraints}; the tendency of the non-CMB data is to push $A_L$ farther away from 1. The inclusion of the $A_L$ parameter does not significantly affect the values of the other six primary parameters, leaving them close to the values found for the six parameter tilted flat $\Lambda$CDM model with $A_L = 1$ (the largest difference is for ln$(10^{10}A_s)$, where it is 1.0$\sigma$ lower); it does however increase the error bars, comparable to what happens in the P18+lensing data case discussed in Sec.\ \ref{subsubsec:P18_lensing_data_constraints}, with largest increase being 29\% for $\ln(10^{10}A_s)$. In addition, in the case when $A_L$ is allowed to vary, the derived parameters change somewhat and their error bars increase, with the largest changes associated with $\sigma_8$, where it is now 1.2$\sigma$ smaller with a 29\% larger error bar.

Adding non-CMB data to P18+lensing data strongly suppresses P18+lensing data support for non-zero spatial curvature, (see Tables \ref{tab:para_NL_ns_nonCMB} and \ref{tab:para_TNL_nonCMB}), except in the case of the untilted non-flat $\Lambda$CDM model, for which $\Omega_k= -0.0065\pm 0.0014$ (4.6$\sigma$ away from flat) and also for the untilted non-flat $\Lambda$CDM$+A_L$ model where $\Omega_k = -0.0060\pm 0.0014$ (4.3$\sigma$ away from flat) (see Table \ref{tab:para_NL_nonCMB}).

Comparing the seven-parameter tilted non-flat $\Lambda$CDM Planck $P(q)$ and new $P(q)$ primary cosmological parameter constraints for P18+lensing data and P18+lensing+non-CMB data, shown in the upper halves of Tables \ref{tab:para_NL_ns_nonCMB} and \ref{tab:para_TNL_nonCMB}, we see that aside from $\Omega_k$ (discussed next) there are no significant changes in parameter values [the largest change is that ln($10^{10}A_s$) is 0.73$\sigma$ (0.52$\sigma$) larger in the P18+lensing+non-CMB case, for the Planck (new) $P(q)$] with all of the error bars being smaller in the P18+lensing+non-CMB case [leaving aside $\Omega_k$ (discussed next) the largest decrease is 18\% (13\%) for the log($10^{10}A_s$) error bar, for the Planck (new) $P(q)$]. On the other hand, $\Omega_k$ changes significantly when non-CMB data are added to the mix, becoming 1.6$\sigma$ (1.5$\sigma$) larger, and closer to flat for the Planck (new) $P(q)$, with 74\% (70\%) smaller error bars, now favoring open geometry over flat but only by 0.24$\sigma$ (0.18$\sigma$), respectively. For the derived parameters, the largest changes are the 1.9$\sigma$ (1.9$\sigma$) increase in $H_0$ and the 1.9$\sigma$ (1.8$\sigma$) decrease in $\Omega_m$ relative to the P18+lensing data values for the Planck (new) $P(q)$, with 78\% (76\%) smaller error bars for $\Omega_m$ and 76\% (73\%) smaller error bars for $H_0$. For P18+lensing+non-CMB data we find $\Omega_m = 0.3051\pm 0.0053$ ($0.3054\pm 0.0051$) and $H_0 = 68.17\pm 0.55$ ($68.13\pm 0.54$) km s$^{-1}$ Mpc$^{-1}$ for the Planck (new) $P(q)$, which are consistent with many other measurements of these quantities and 0.57$\sigma$ (0.59$\sigma$) larger and 1.2$\sigma$ (1.2$\sigma$) lower, respectively, than the low-redshift data measurements of Ref.\ \cite{Cao:2022ugh}.

Comparing the eight-parameter tilted non-flat $\Lambda$CDM$+A_L$ Planck (new) $P(q)$ primary cosmological parameter constraints for P18+lensing data and P18+lensing+non-CMB data, shown in the lower half of Table \ref{tab:para_NL_ns_nonCMB} (\ref{tab:para_TNL_nonCMB}), we see that there are approximately comparable differences to the tilted flat $\Lambda$CDM$+A_L$ case. For the Planck (new) $P(q)$ case the largest change is that $\Omega_c h^2$ is 0.49$\sigma$ (0.45$\sigma$) smaller in the P18+lensing+non-CMB case, with the next largest being $\Omega_b h^2$ ($n_s$) which is 0.34$\sigma$ (0.37$\sigma$) smaller, with most of the error bars being smaller in the P18+lensing+non-CMB case (the largest decreases are 94\% (89\%) for $\Omega_k$ and 77\% (77\%) for $A_L$). From the P18+lensing+non-CMB analyses, we measure $\Omega_k=-0.0002\pm 0.0017$ for both $P(q)$ cases, both being only 0.12$\sigma$ away from flat spatial hypersurfaces, very different from the P18 data alone results. For the derived parameters, the largest change is the 0.18$\sigma$ (0.39$\sigma$) decrease in $\Omega_m$ ($\sigma_8$) relative to the P18+lensing data value for the Planck (new) $P(q)$, with 95\% (93\%) smaller error bars for $\Omega_m$ and 95\% (94\%) smaller error bars for $H_0$. For P18+lensing+non-CMB data we find $\Omega_m = 0.2998\pm 0.0055$ ($0.2999\pm 0.0055$) and $H_0 = 68.49\pm 0.56$ ($68.48\pm 0.56$) km s$^{-1}$ Mpc$^{-1}$ for the Planck (new) $P(q)$, which are consistent with many other measurements of these quantities and 0.27$\sigma$ (0.27$\sigma$) larger and 0.91$\sigma$ (0.92$\sigma$) lower, respectively, than the low-redshift data measurements of Ref.\ \cite{Cao:2022ugh}. 

We will see in Sec.\ \ref{subsec:model_selection} that in both tilted non-flat $\Lambda$CDM$+A_L$ models the fit to P18+lensing+non-CMB data is positively better when the $A_L$ parameter is allowed to vary compared to the $A_L = 1$ case; this is similar to what happens in the tilted flat $\Lambda$CDM$+A_L$ model. Also, like the tilted flat $\Lambda$CDM$+A_L$ P18+lensing+non-CMB case, we measure, from P18+lensing+non-CMB data, $A_L=1.090\pm 0.036$ and $1.088\pm0.035$, for the Planck $P(q)$ and the new $P(q)$, respectively, which both differ from the theoretically expected $A_L = 1$ by 2.5$\sigma$. The inclusion of the $A_L$ parameter does not significantly affect the values of the other seven primary parameters, leaving them close to the values found for the seven-parameter tilted non-flat $\Lambda$CDM models with $A_L = 1$ [the largest difference is for ln($10^{10}A_s$), where it is 1.0$\sigma$ (0.95$\sigma$) smaller for the Planck (new) $P(q)$]; it does however increase the error bars, but less than what happens in the P18 alone and P18+lensing data cases discussed in Secs.\ \ref{subsubsec:P18_data_constraints} and \ref{subsubsec:P18_lensing_data_constraints}, with largest increase being 21\% for ln($10^{10}A_s$) for both $P(q)$ cases. In addition, in the case when $A_L$ is allowed to vary, the derived parameters change somewhat and their error bars increase, with the largest changes associated with $\sigma_8$, where it is now 1.3$\sigma$ (1.2$\sigma$) smaller for the Planck (new) $P(q)$ with a 21\% (22\%) larger error bar.

When non-CMB data (that include $f \sigma_8$ data) are added to the mix and the $A_L$ parameter is allowed to vary, $A_L > 1$ is favored and there is a decrease in the value of $\sigma_8$ compared to the $A_L = 1$ case, which helps to alleviate the corresponding tension. Since $A_L>1$ helps to resolve the lensing anomaly, there is less or no need to increase the value of $\Omega_m$ to predict more lensing. A lower value of $\Omega_m$ means less structure formation in the past, consequently slightly alleviating the $\sigma_8$ tension. While $\Omega_k$ plays a role at both early and late times, the $A_L$ parameter only has an impact on CMB data. Since, as we shall see in Sec. \ref{sec:P18_vs_non-CMB}, non-CMB data prefer a flatter geometry than do P18 data, it is possible to understand why the evidence in favor of $\Omega_k\neq 0$ subsides, while the evidence for $A_L>1$ does not, when non-CMB data is added to the mix. A fairly large negative value of $\Omega_k$ is required to resolve the P18 data lensing anomaly, thus improving upon the performance shown by the tilted flat $\Lambda$CDM model, however, such a large value of the curvature parameter is not supported by lensing data or by non-CMB data. This fact raises the issue of whether it is consistent to jointly use P18, lensing, and non-CMB data sets in the context of the non-flat models. We try to answer this question, through the use of different statistical criteria, in Sec.\ \ref{subsec:data_set_tensions}. Note that Figs.\ \ref{fig:like_P18_lensing_nonCMB} and \ref{fig:like_Alens_P18_lensing_nonCMB} show that when P18+lensing+non-CMB data is used it is not necessary to consider $A_L\neq 1$ in order to make the three sets of tilted model contours overlap.

\subsubsection{Comparing P18, P18+lensing, and P18+lensing+non-CMB data cosmological constraints for each model}
\label{subsubsec:contour_plots}

Cosmological parameter contour plots allow us to easily see the degree of correlation between the two variables. If the two variables are more correlated then the corresponding constraint contours are more line-like, on the other hand, if they are less correlated the contours are broader and enclose 2-dimensional areas. In this subsubsection we comment on how the constraint contours, for each cosmological model, change depending on whether we consider P18, P18+lensing, or P18+lensing+non-CMB data. Figures \ref{fig:like_FL_compar}-\ref{fig:like_TNL_Alens_compar} show, for each of the eight cosmological models we study, the cosmological parameter constraints for P18, P18+lensing, and P18+lensing+non-CMB data. The constraint contours shrink as more data is included in the analysis used to determine them.

From Fig.\ \ref{fig:like_FL_compar} for the six-parameter tilted flat $\Lambda$CDM we see that there are significant overlaps between the contours obtained in the three data sets considered. Along with the results discussed in Secs.\ \ref{subsubsec:P18_lensing_data_constraints} and \ref{subsubsec:P18_lensing_nonCMB_data_constraints} this is an indication that there is not significant tension between P18, P18+lensing, and P18+lensing+non-CMB data when these data are analyzed in the tilted flat $\Lambda$CDM model. The $\Omega_m$-$H_0$ panel contours indicate that these two parameters are strongly correlated. The inclusion of lensing data and/or non-CMB data, which provide information about the late-time Universe, partially breaks this correlation and induces a shift in the one-dimensional posterior distributions of not only these two parameters but also other parameters. Non-CMB data cause a larger shift. 

For the six-parameter untilted non-flat $\Lambda$CDM model (see Fig.\ \ref{fig:like_NL_compar}) constraint contours determined from the three different data sets overlap only for some parameters. In particular, for constraint contours in panels that involve $\Omega_k$, $\Omega_m$, or $H_0$ there is no overlap between those determined using P18 data and those determined using P18+lensing+non-CMB data (there are larger than 2$\sigma$ differences between these contours when one of these three parameters are involved and the differences are larger when two of these three parameters are involved), and there is only a slight amount of overlap between the P18 data contours and the P18+lensing data contours. The $\Omega_m$-$H_0$-$\Omega_k$ geometrical degeneracy is prominent for P18 data and is clearly seen in the $\Omega_k$-$\Omega_m$, $\Omega_k$-$H_0$, and $\Omega_m$-$H_0$ panels, as these three parameters are strongly correlated. Including lensing data and/or non-CMB data partially breaks this degeneracy causing significant shifts in the one-dimensional posterior distributions of not only these three parameters but also other parameters. The shifts are bigger here than in the tilted flat $\Lambda$CDM model and indicate significant tension between the data sets, especially between the P18 and P18+lensing+non-CMB data sets, when they are analyzed in the untilted non-flat $\Lambda$CDM model. Non-CMB data again appear to cause the larger shift. As discussed in more detail in Sec.\ \ref{subsec:data_set_tensions}, these shifts mean that P18 and non-CMB data are mutually inconsistent in the untilted non-flat $\Lambda$CDM model and so cannot be jointly used to derive cosmological parameter constraints in this model. 

Similar, but quantitatively less discrepant, results are obtained for the seven-parameter tilted non-flat $\Lambda$CDM Planck $P(q)$ and the tilted non-flat $\Lambda$CDM new $P(q)$ models (see Figs.\ \ref{fig:like_NL_ns_compar} and \ref{fig:like_TNL_compar}). The differences between the untilted non-flat and tilted non-flat results is likely a consequence of the additional $n_s$ parameter in the tilted non-flat models. In the tilted non-flat models, the more-discrepant P18 and P18+lensing+non-CMB data constraint contours overlap in all panels for pairs of the six primary cosmological parameters excluding the $\Omega_k$ parameter as well as the derived $\Omega_m$ and $H_0$ parameters. The differences are larger in the Planck $P(q)$ case than in the new $P(q)$ case, largest for $H_0$, smallest for $\Omega_m$, with $\Omega_k$ being in-between. In the new $P(q)$ case, the 2$\sigma$ contours overlap for $\Omega_m$ and almost overlap for $\Omega_k$. These results may be an indication of the tension found, in the context of the tilted non-flat models,  between P18 data and the BAO data set. We shall study this tension in more detail in Sec.\ \ref{subsec:data_set_tensions}. As in the untilted non-flat $\Lambda$CDM model, the geometrical degeneracy between $\Omega_m$-$H_0$-$\Omega_k$ affects the tilted non-flat models. Again, including lensing data and/or non-CMB data partially breaks this degeneracy causing significant shifts in the one-dimensional posterior distributions of not only these three parameters but also other parameters. The shifts are bigger here than in the tilted flat $\Lambda$CDM model and but smaller than in the untilted non-flat $\Lambda$CDM model, but still indicate some tension between the data sets, especially between the P18 and P18+lensing+non-CMB data sets, especially when they are analyzed in the tilted non-flat $\Lambda$CDM Planck $P(q)$ model. Non-CMB data again appear to cause the larger shift. 

When the $A_L$ parameter is allowed to vary, the three different sets of constraint contours overlap in all four models (see Figs.\ \ref{fig:like_FL_Alens_compar}--\ref{fig:like_TNL_Alens_compar}). In the non-flat models there now is a bigger degeneracy between the cosmological parameters $\Omega_m$-$H_0$-$\Omega_k$-$A_L$ which causes the constraint contours to expand relative to the $A_L = 1$ case, especially for P18 data. For some parameters in the untilted non-flat $\Lambda$CDM model and the tilted non-flat $\Lambda$CDM new $P(q)$ model we observe a bimodal distribution when only P18 data is used, and the same parameters in the tilted non-flat $\Lambda$CDM Planck $P(q)$ model have an almost bimodal distribution for P18 data. These bimodalities are likely a consequences of the above-mentioned geometrical degeneracy.

%
\begin{table*}
\caption{Mean and 68.3\% confidence limits of tilted flat $\Lambda\textrm{CDM}$ (+$A_L$) model parameters
        constrained by TT,TE,EE+lowE (P18), BAO, and BAO$^\prime$ data.
        $H_0$ has units of km s$^{-1}$ Mpc$^{-1}$. 
}
\begin{ruledtabular}
\begin{tabular}{lccccc}
\\[-1mm]                         & \multicolumn{5}{c}{Tilted flat $\Lambda$CDM model}        \\[+1mm]
\cline{2-6}\\[-1mm]
  Parameter                      & P18                   & P18+BAO                &  BAO                  &  P18+BAO$^\prime$      & BAO$^\prime$  \\[+1mm]
 \hline \\[-1mm]
  $\Omega_b h^2$                 & $0.02236 \pm 0.00015$ & $0.02243 \pm 0.00013$  &  $0.043 \pm 0.016$    & $0.02241 \pm 0.00014$  & $0.043 \pm 0.016$   \\[+1mm]
  $\Omega_c h^2$                 & $0.1202 \pm 0.0014$   & $0.11926 \pm 0.00097$  &  $0.163 \pm 0.042$    & $0.11946 \pm 0.00098$  & $0.168 \pm 0.044$   \\[+1mm]
  $100\theta_\textrm{MC}$        & $1.04090 \pm 0.00031$ & $1.04102 \pm 0.00029$  &  $1.054 \pm 0.026$    & $1.04099 \pm 0.00029$  & $1.059 \pm 0.025$   \\[+1mm]
  $\tau$                         & $0.0542 \pm 0.0079$   & $0.0581 \pm 0.0081$    &  $0.0542$             & $0.0555 \pm 0.0077$    & $0.0542$           \\[+1mm]
  $n_s$                          & $0.9649 \pm 0.0043$   & $0.9673 \pm 0.0037$    &  $0.9649$             & $0.9665 \pm 0.0038$    & $0.9649$           \\[+1mm]
  $\ln(10^{10} A_s)$             & $3.044 \pm 0.016$     & $3.051 \pm 0.017$      &  $3.01 \pm 0.27$      & $3.045 \pm 0.016$      & $3.044$           \\[+1mm]
 \hline \\[-1mm]
  $H_0$                          & $67.28 \pm 0.61$      & $67.70 \pm 0.43$       &  $83 \pm 12$          & $67.60 \pm 0.44$       & $83 \pm 12$         \\[+1mm]
  $\Omega_m$                     & $0.3165 \pm 0.0084$   & $0.3106 \pm 0.0058$    &  $0.294 \pm 0.015$    & $0.3119 \pm 0.0059$    & $0.300 \pm 0.016$   \\[+1mm]
  $\sigma_8$                     & $0.8118 \pm 0.0074$   & $0.8119 \pm 0.0073$    &  $0.874 \pm 0.037$    & $0.8102 \pm 0.0070$    & $0.92 \pm 0.12$     \\[+1mm]
 \hline \\[-1mm]
  $\chi_{\textrm{min}}^2$ (Total)& $2765.80$             & $2786.66$              &  $15.92$              & $2777.75$              & $10.98$     \\[+1mm]
  $\chi_{\textrm{min}}^2$ (BAO/BAO$^\prime$) & $\cdots$  & $20.22$                &  $15.92$              & $11.61$                & $10.98$     \\[+1mm]
  $\chi_{\textrm{BAO/BAO}^\prime}^2$ (at P18 B-F) & $\cdots$   & $22.24$          &  $22.24$              & $12.58$                & $12.58$     \\[+1mm]
  $\textrm{DIC}$                 & $2817.93$             & $2839.25$              &  $21.93$              & $2829.61$              & $14.93$     \\[+1mm]
  $\textrm{AIC}_c$               & $2819.80$             & $2840.66$              &  $27.56$              & $2831.75$              & $19.98$     \\[+1mm]
 \hline \hline \\[-1mm]
                                 & \multicolumn{5}{c}{Tilted flat $\Lambda$CDM+$A_L$ model}              \\[+1mm]
\cline{2-6}\\[-1mm]
  Parameter                      & P18                   & P18+BAO                & BAO                  &  P18+BAO$^\prime$       & BAO$^\prime$  \\[+1mm]
 \hline \\[-1mm]
  $\Omega_b h^2$                 & $0.02259 \pm 0.00017$ & $0.02258 \pm 0.00015$  & $0.043 \pm 0.015$    & $0.02256 \pm 0.00014$   & $0.045 \pm 0.013$  \\[+1mm]
  $\Omega_c h^2$                 & $0.1180 \pm 0.0015$   & $0.1183 \pm 0.0010$    & $0.163 \pm 0.042$    & $0.1185 \pm 0.0010$     & $0.177 \pm 0.042$  \\[+1mm]
  $100\theta_\textrm{MC}$        & $1.04114 \pm 0.00032$ & $1.04113 \pm 0.00029$  & $1.055 \pm 0.024$    & $1.04109 \pm 0.00030$   & $1.065 \pm 0.018$  \\[+1mm]
  $\tau$                         & $0.0496 \pm 0.0082$   & $0.0522 \pm 0.0080$    & $0.0496$             & $0.0492 \pm 0.0084$     & $0.0496$           \\[+1mm]
  $n_s$                          & $0.9710 \pm 0.0050$   & $0.9705 \pm 0.0038$    & $0.9710$             & $0.9698 \pm 0.0039$     & $0.9710$           \\[+1mm]
  $\ln(10^{10} A_s)$             & $3.030 \pm 0.017$     & $3.036 \pm 0.017$      & $3.00 \pm 0.27$      & $3.030 \pm 0.018$       & $3.030$           \\[+1mm]
  $A_{L}$                        & $1.181 \pm 0.067$     & $1.170 \pm 0.060$      & $1.181$              & $1.174 \pm 0.061$       & $1.181$           \\[+1mm]
 \hline \\[-1mm]
  $H_0$                          & $68.31 \pm 0.71$      & $68.21 \pm 0.46$       & $83 \pm 12$          & $68.11 \pm 0.47$        & $85 \pm 10$        \\[+1mm]
  $\Omega_m$                     & $0.3029 \pm 0.0093$   & $0.3042 \pm 0.0060$    & $0.294 \pm 0.015$    & $0.3055 \pm 0.0061$     & $0.302 \pm 0.017$  \\[+1mm]
  $\sigma_8$                     & $0.7997 \pm 0.0088$   & $0.8031 \pm 0.0077$    & $0.875 \pm 0.037$    & $0.8011 \pm 0.0079$     & $0.93 \pm 0.11$    \\[+1mm]
 \hline \\[-1mm]
  $\chi_{\textrm{min}}^2$ (Total)& $2756.12$             & $2776.71$              & $15.91$              & $2767.77$                & $10.98$             \\[+1mm]
  $\chi_{\textrm{min}}^2$ (BAO/BAO$^\prime$) & $\cdots$  & $20.47$                & $15.91$              & $11.37$                  & $10.98$     \\[+1mm]
  $\chi_{\textrm{BAO/BAO}^\prime}^2$ (at P18 B-F) & $\cdots$ & $20.78$            & $20.78$              & $11.88$                  & $11.88$     \\[+1mm]
  $\textrm{DIC}$                 & $2812.41$             & $2832.92$              & $21.83$              & $2823.77$                & $15.04$     \\[+1mm]
  $\Delta\textrm{DIC}$           & $-5.52$               & $-6.33$                & $-0.10$              & $-5.90$                  & $0.11$     \\[+1mm]
  $\textrm{AIC}_c$               & $2812.12$             & $2832.71$              & $27.55$              & $2823.77$                & $19.98$     \\[+1mm]
  $\Delta\textrm{AIC}_c$         & $-7.68$               & $-7.95$                & $-0.01$              & $-7.98$                  & $0.00$     \\[+1mm]
\end{tabular}
\\[+1mm]
\begin{flushleft}
Note: $\Delta\textrm{DIC}$ ($\Delta\textrm{AIC}_c$) indicates an excess value relative to that of the tilted flat $\Lambda$CDM model constrained with the same data. 
The number of free parameters of the tilted flat $\Lambda$CDM model is 27 for P18, P18+BAO, and P18+BAO$^\prime$ data sets (including 21 internal calibration parameters), 4 for BAO data, and 3 for BAO$^\prime$ data. 

\end{flushleft}
\end{ruledtabular}
\label{tab:para_FL_BAO}
\end{table*}

%
\begin{table*}
\caption{Mean and 68.3\% confidence limits of untilted non-flat $\Lambda\textrm{CDM}$ (+$A_L$) model parameters
        constrained by TT,TE,EE+lowE (P18), BAO, and BAO$^\prime$ data.
        $H_0$ has units of km s$^{-1}$ Mpc$^{-1}$.
}
\begin{ruledtabular}
\begin{tabular}{lccccc}
\\[-1mm]                         & \multicolumn{5}{c}{Untilted non-flat $\Lambda$CDM model}        \\[+1mm]
\cline{2-6}\\[-1mm]
  Parameter                      & P18                   & P18+BAO                &  BAO                &  P18+BAO$^\prime$      & BAO$^\prime$     \\[+1mm]
 \hline \\[-1mm]
  $\Omega_b h^2$                 & $0.02320 \pm 0.00015$ & $0.02298 \pm 0.00014$  &  $0.040 \pm 0.015$  & $0.02299 \pm 0.00014$  & $0.040 \pm 0.015$   \\[+1mm]
  $\Omega_c h^2$                 & $0.11098 \pm 0.00088$ & $0.11184 \pm 0.00089$  &  $0.175 \pm 0.046$  & $0.11171 \pm 0.00089$  & $0.175 \pm 0.047$   \\[+1mm]
  $100\theta_\textrm{MC}$        & $1.04204 \pm 0.00030$ & $1.04188 \pm 0.00029$  &  $1.16 \pm 0.13$    & $1.04189 \pm 0.00030$  & $1.13 \pm 0.12$   \\[+1mm]
  $\tau$                         & $0.0543 \pm 0.0091$   & $0.077 \pm 0.010$      &  $0.0543$           & $0.073 \pm 0.010$      & $0.0543$          \\[+1mm]
  $\Omega_k$                     & $-0.095 \pm 0.024$    & $-0.0066 \pm 0.0015$   &  $-0.047 \pm 0.059$ & $-0.0074 \pm 0.0016$   & $-0.034 \pm 0.057$      \\[+1mm]
  $\ln(10^{10} A_s)$             & $3.021 \pm 0.019$     & $3.069 \pm 0.021$      &  $2.70 \pm 0.43$    & $3.059 \pm 0.021$      & $3.021$           \\[+1mm]
 \hline \\[-1mm]
  $H_0$                          & $47.1 \pm 3.2$        & $67.77 \pm 0.60$       &  $84 \pm 12$        & $67.46 \pm 0.63$       & $83 \pm 12$        \\[+1mm]
  $\Omega_m$                     & $0.617 \pm 0.082$     & $0.2950 \pm 0.0055$    &  $0.303 \pm 0.019$  & $0.2975 \pm 0.0057$    & $0.307 \pm 0.019$   \\[+1mm]
  $\sigma_8$                     & $0.730 \pm 0.017$     & $0.7977 \pm 0.0093$    &  $0.850 \pm 0.048$  & $0.7927 \pm 0.0090$    & $1.00 \pm 0.18$     \\[+1mm]
 \hline \\[-1mm]
  $\chi_{\textrm{min}}^2$ (Total)& $2789.77$             & $2837.93$              &  $15.91$            & $2828.81$              & $10.67$ \\[+1mm]
  $\chi_{\textrm{min}}^2$ (BAO/BAO$^\prime$) & $\cdots$  & $20.34$                &  $15.91$            & $11.68$                & $10.67$     \\[+1mm]
  $\chi_{\textrm{BAO/BAO}^\prime}^2$ (at P18 B-F) & $\cdots$ & $1987.47$          &  $1987.47$          & $1765.08$              & $1765.08$     \\[+1mm]
  $\textrm{DIC}$                 & $2847.14$             & $2895.04$              &  $24.31$            & $2884.90$              & $17.55$     \\[+1mm]
  $\Delta\textrm{DIC}$           & $29.21$               & $55.79$                &  $2.38$             & $55.29$                & $2.62$     \\[+1mm]
  $\textrm{AIC}_c$               & $2843.77$             & $2891.93$              &  $31.91$            & $2882.81$              & $24.39$    \\[+1mm]
  $\Delta\textrm{AIC}_c$         & $23.97$               & $51.27$                &  $4.35$             & $51.06$                & $4.41$    \\[+1mm]
 \hline \hline \\[-1mm]
                                 & \multicolumn{5}{c}{Untilted non-flat $\Lambda$CDM+$A_L$ model}              \\[+1mm]
\cline{2-6}\\[-1mm]
Parameter                      & P18                   & P18+BAO                & BAO                  &  P18+BAO$^\prime$          & BAO$^\prime$     \\[+1mm]
 \hline \\[-1mm]
  $\Omega_b h^2$                 & $0.02320 \pm 0.00015$ & $0.02318 \pm 0.00015$  & $0.041 \pm 0.015$    & $0.02320 \pm 0.00015$   & $0.042 \pm 0.014$  \\[+1mm]
  $\Omega_c h^2$                 & $0.11097 \pm 0.00087$ & $0.11117 \pm 0.00086$  & $0.176 \pm 0.045$    & $0.11095 \pm 0.00087$   & $0.180 \pm 0.044$  \\[+1mm]
  $100\theta_\textrm{MC}$        & $1.04202 \pm 0.00030$ & $1.04198 \pm 0.00030$  & $1.16 \pm 0.13$      & $1.04199 \pm 0.00030$   & $1.14 \pm 0.12$  \\[+1mm]
  $\tau$                         & $0.0540 \pm 0.0087$   & $0.0598 \pm 0.0087$    & $0.0540$             & $0.0557 \pm 0.0089$     & $0.0540$           \\[+1mm]
  $\Omega_k$                     & $-0.12 \pm 0.12$      & $-0.0064 \pm 0.0015$   & $-0.050 \pm 0.060$   & $-0.0073 \pm 0.0015$    & $-0.035 \pm 0.058$           \\[+1mm]
  $\ln(10^{10} A_s)$             & $3.020 \pm 0.018$     & $3.033 \pm 0.018$      & $2.68 \pm 0.41$      & $3.023 \pm 0.019$       & $3.020$           \\[+1mm]
  $A_{L}$                        & $1.08 \pm 0.27$       & $1.310 \pm 0.062$      & $1.08$               & $1.319 \pm 0.063$       & $1.08$           \\[+1mm]
 \hline \\[-1mm]
  $H_0$                          & $52 \pm 18$           & $68.27 \pm 0.61$       & $84 \pm 12$          & $67.93 \pm 0.62$        & $84 \pm 11$        \\[+1mm]
  $\Omega_m$                     & $0.70 \pm 0.42$       & $0.2897 \pm 0.0054$    & $0.304 \pm 0.018$    & $0.2921 \pm 0.0055$     & $0.307 \pm 0.020$  \\[+1mm]
  $\sigma_8$                     & $0.721 \pm 0.053$     & $0.7799 \pm 0.0083$    & $0.848 \pm 0.049$    & $0.7750 \pm 0.0085$     & $1.01 \pm 0.18$    \\[+1mm]
 \hline \\[-1mm]
  $\chi_{\textrm{min}}^2$ (Total)& $2787.76$             & $2809.82$              & $15.89$              & $2799.18$               & $10.68$           \\[+1mm]
  $\chi_{\textrm{min}}^2$ (BAO/BAO$^\prime$) & $\cdots$  & $21.96$                & $15.89$              & $11.38$                 & $10.68$     \\[+1mm]
  $\chi_{\textrm{BAO/BAO}^\prime}^2$ (at P18 B-F) & $\cdots$    & $106.63$        & $106.63$             & $80.18$                 & $80.18$  \\[+1mm]
  $\textrm{DIC}$                 & $2846.45$             & $2869.28$              & $24.63$              & $2857.90$               & $17.89$    \\[+1mm]
  $\Delta\textrm{DIC}$           & $28.52$               & $30.03$                & $2.70$               & $28.29$                 & $2.96$     \\[+1mm]
  $\textrm{AIC}_c$               & $2843.76$             & $2865.82$              & $31.89$              & $2855.18$               & $24.39$    \\[+1mm]
  $\Delta\textrm{AIC}_c$         & $23.96$               & $25.16$                & $4.33$               & $23.43$                 & $4.41$     \\[+1mm]
\end{tabular}
\\[+1mm]
\begin{flushleft}
Note: $\Delta\textrm{DIC}$ ($\Delta\textrm{AIC}_c$) indicates an excess value relative to that of the tilted flat $\Lambda$CDM model constrained with the same data. 
\end{flushleft}
\end{ruledtabular}
\label{tab:para_NL_BAO}
\end{table*}

\begin{table*}
\caption{Mean and 68.3\% confidence limits of Planck-$P(q)$-based tilted non-flat $\Lambda\textrm{CDM}$ (+$A_L$) model parameters
        constrained by TT,TE,EE+lowE (P18), BAO, and BAO$^\prime$ data.
        $H_0$ has units of km s$^{-1}$ Mpc$^{-1}$.
}
\begin{ruledtabular}
\begin{tabular}{lccccc}
\\[-1mm]                         & \multicolumn{5}{c}{Tilted non-flat $\Lambda$CDM model [Planck $P(q)$]}        \\[+1mm]
\cline{2-6}\\[-1mm]
  Parameter                      & P18                   & P18+BAO                &  BAO                &  P18+BAO$^\prime$      & BAO$^\prime$     \\[+1mm]
 \hline \\[-1mm]
  $\Omega_b h^2$                 & $0.02260 \pm 0.00017$ & $0.02241 \pm 0.00015$  & $0.040 \pm 0.015$   & $0.02241 \pm 0.00015$  & $0.040 \pm 0.016$   \\[+1mm]
  $\Omega_c h^2$                 & $0.1181 \pm 0.0015$   & $0.1195 \pm 0.0014$    & $0.174 \pm 0.047$   & $0.1195 \pm 0.0014$    & $0.172 \pm 0.047$   \\[+1mm]
  $100\theta_\textrm{MC}$        & $1.04116 \pm 0.00032$ & $1.04099 \pm 0.00032$  & $1.15 \pm 0.13$     & $1.04099 \pm 0.00032$  & $1.13 \pm 0.12$   \\[+1mm]
  $\tau$                         & $0.0483 \pm 0.0083$   & $0.0578 \pm 0.0077$    & $0.0483$            & $0.0550 \pm 0.0078$    & $0.0483$          \\[+1mm]
  $\Omega_k$                     & $-0.043 \pm 0.017$    & $0.0005 \pm 0.0018$    & $-0.046 \pm 0.060$  & $-0.0001 \pm 0.0018$   & $-0.033 \pm 0.055$      \\[+1mm]
  $n_s$                          & $0.9706 \pm 0.0047$   & $0.9667 \pm 0.0045$    & $0.9706$            & $0.9666 \pm 0.0044$    & $0.9706$          \\[+1mm]
  $\ln(10^{10} A_s)$             & $3.027 \pm 0.017$     & $3.051 \pm 0.016$      & $2.74 \pm 0.43$     & $3.044 \pm 0.016$      & $3.027$           \\[+1mm]
 \hline \\[-1mm]
  $H_0$                          & $54.5 \pm 3.6$        & $67.83 \pm 0.58$       & $83 \pm 12$         & $67.58 \pm 0.62$       & $83 \pm 12$         \\[+1mm]
  $\Omega_m$                     & $0.481 \pm 0.062$     & $0.3100 \pm 0.0060$    & $0.303 \pm 0.019$   & $0.3122 \pm 0.0063$    & $0.306 \pm 0.019$   \\[+1mm]
  $\sigma_8$                     & $0.775 \pm 0.015$     & $0.8130 \pm 0.0079$    & $0.850 \pm 0.049$   & $0.8099 \pm 0.0081$    & $0.98 \pm 0.17$     \\[+1mm]
 \hline \\[-1mm]
  $\chi_{\textrm{min}}^2$ (Total)& $2754.73$             & $2786.20$              & $15.88$             & $2776.90$              & $10.68$     \\[+1mm]
  $\chi_{\textrm{min}}^2$ (BAO/BAO$^\prime$) & $\cdots$  & $20.09$                & $15.88$             & $11.71$                & $10.68$    \\[+1mm]
  $\chi_{\textrm{BAO/BAO}^\prime}^2$ (at P18 B-F) & $\cdots$    & $665.90$        & $665.90$            & $582.59$               & $582.59$     \\[+1mm]
  $\textrm{DIC}$                 & $2810.59$             & $2840.62$              & $24.34$             & $2832.28$              & $17.58$    \\[+1mm]
  $\Delta\textrm{DIC}$           & $-7.34$               & $1.37$                 & $2.41$              & $2.67$                 & $2.65$   \\[+1mm]
  $\textrm{AIC}_c$               & $2810.73$             & $2842.20$              & $31.88$             & $2832.90$              & $24.39$    \\[+1mm]
  $\Delta\textrm{AIC}_c$         & $-9.07$               & $1.54$                 & $4.32$              & $1.15$                 & $4.41$     \\[+1mm]
 \hline \hline \\[-1mm]
                                 & \multicolumn{5}{c}{Tilted non-flat $\Lambda$CDM+$A_L$ model [Planck $P(q)$]}              \\[+1mm]
\cline{2-6}\\[-1mm]
  Parameter                      & P18                   & P18+BAO                & BAO                 &  P18+BAO$^\prime$     & BAO$^\prime$      \\[+1mm]
 \hline \\[-1mm]
  $\Omega_b h^2$                 & $0.02258 \pm 0.00017$ & $0.02260 \pm 0.00017$  & $0.041 \pm 0.014$   & $0.02262 \pm 0.00017$ & $0.044 \pm 0.013$  \\[+1mm]
  $\Omega_c h^2$                 & $0.1183 \pm 0.0015$   & $0.1180 \pm 0.0015$    & $0.174 \pm 0.045$   & $0.1178 \pm 0.0015$   & $0.182 \pm 0.043$  \\[+1mm]
  $100\theta_\textrm{MC}$        & $1.04116 \pm 0.00033$ & $1.04115 \pm 0.00033$  & $1.16 \pm 0.14$     & $1.04118 \pm 0.00032$ & $1.12 \pm 0.11$  \\[+1mm]
  $\tau$                         & $0.0478 \pm 0.0081$   & $0.0522 \pm 0.0081$    & $0.0478$            & $0.0496 \pm 0.0085$   & $0.0478$           \\[+1mm]
  $\Omega_k$                     & $-0.130 \pm 0.095$    & $-0.0004 \pm 0.0018$   & $-0.045 \pm 0.063$  & $-0.0012 \pm 0.0018$  & $-0.026 \pm 0.054$           \\[+1mm]
  $n_s$                          & $0.9704 \pm 0.0048$   & $0.9712 \pm 0.0047$    & $0.9704$            & $0.9716 \pm 0.0047$   & $0.9704$           \\[+1mm]
  $\ln(10^{10} A_s)$             & $3.027 \pm 0.017$     & $3.035 \pm 0.017$      & $2.74 \pm 0.45$     & $3.029 \pm 0.018$     & $3.027$           \\[+1mm]
  $A_{L}$                        & $0.88 \pm 0.15$       & $1.170 \pm 0.061$      & $0.88$              & $1.178 \pm 0.061$     & $0.88$           \\[+1mm]
 \hline \\[-1mm]
  $H_0$                          & $45 \pm 11$           & $68.13 \pm 0.60$       & $84 \pm 11$         & $67.85 \pm 0.61$      & $85 \pm 10$        \\[+1mm]
  $\Omega_m$                     & $0.80 \pm 0.35$       & $0.3044 \pm 0.0062$    & $0.303 \pm 0.019$   & $0.3064 \pm 0.0063$   & $0.307 \pm 0.019$  \\[+1mm]
  $\sigma_8$                     & $0.733 \pm 0.045$     & $0.8020 \pm 0.0089$    & $0.851 \pm 0.048$   & $0.7983 \pm 0.0091$   & $0.99 \pm 0.16$    \\[+1mm]
 \hline \\[-1mm]
  $\chi_{\textrm{min}}^2$ (Total)& $2754.99$             & $2776.32$              & $15.91$             & $2767.04$             & $10.73$           \\[+1mm]
  $\chi_{\textrm{min}}^2$ (BAO/BAO$^\prime$) & $\cdots$  & $20.38$                & $15.91$             & $11.22$               & $10.73$  \\[+1mm]
  $\chi_{\textrm{BAO/BAO}^\prime}^2$ (at P18 B-F) & $\cdots$  & $593.77$          & $593.77$            & $518.08$              & $518.08$     \\[+1mm]
  $\textrm{DIC}$                 & $2811.63$             & $2835.10$              & $24.31$             & $2825.27$             & $17.54$   \\[+1mm]
  $\Delta\textrm{DIC}$           & $-6.30$               & $-4.15$                & $2.38$              & $-4.34$               & $2.61$     \\[+1mm]
  $\textrm{AIC}_c$               & $2812.99$             & $2834.32$              & $31.91$             & $2825.04$             & $24.45$    \\[+1mm]
  $\Delta\textrm{AIC}_c$         & $-6.81$               & $-6.34$                & $4.35$              & $-6.71$               & $4.47$    \\[+1mm]
\end{tabular}
\\[+1mm]
\begin{flushleft}
Note: $\Delta\textrm{DIC}$ ($\Delta\textrm{AIC}_c$) indicates an excess value relative to that of the tilted flat $\Lambda$CDM model constrained with the same data. 
\end{flushleft}
\end{ruledtabular}
\label{tab:para_NL_ns_BAO}
\end{table*}
%

\begin{table*}
\caption{Mean and 68.3\% confidence limits of tilted new-$P(q)$-based non-flat $\Lambda\textrm{CDM}$ (+$A_L$) model parameters with the new $P(q)$
        constrained by TT,TE,EE+lowE (P18), BAO, and BAO$^\prime$ data.
        $H_0$ has units of km s$^{-1}$ Mpc$^{-1}$.
}
\begin{ruledtabular}
\begin{tabular}{lccccc}
\\[-1mm]                         & \multicolumn{5}{c}{Tilted non-flat $\Lambda$CDM model [new $P(q)$]}        \\[+1mm]
\cline{2-6}\\[-1mm]
  Parameter                      & P18                   & P18+BAO                &  BAO                &  P18+BAO$^\prime$      & BAO$^\prime$     \\[+1mm]
 \hline \\[-1mm]
  $\Omega_b h^2$                 & $0.02255 \pm 0.00017$ & $0.02242 \pm 0.00015$  &  $0.039 \pm 0.015$  & $0.02243 \pm 0.00016$  & $0.041 \pm 0.016$   \\[+1mm]
  $\Omega_c h^2$                 & $0.1188 \pm 0.0015$   & $0.1194 \pm 0.0014$    &  $0.173 \pm 0.048$  & $0.1193 \pm 0.0014$    & $0.177 \pm 0.048$   \\[+1mm]
  $100\theta_\textrm{MC}$        & $1.04109 \pm 0.00032$ & $1.04100 \pm 0.00032$  &  $1.16 \pm 0.14$    & $1.04102 \pm 0.00032$  & $1.13 \pm 0.12$   \\[+1mm]
  $\tau$                         & $0.0525 \pm 0.0083$   & $0.0582 \pm 0.0081$    &  $0.0525$           & $0.0562 \pm 0.0080$    & $0.0525$          \\[+1mm]
  $\Omega_k$                     & $-0.033 \pm 0.014$    & $0.0003 \pm 0.0018$    &  $-0.051 \pm 0.061$ & $-0.0004 \pm 0.0018$   & $-0.032 \pm 0.059$      \\[+1mm]
  $n_s$                          & $0.9654 \pm 0.0045$   & $0.9665 \pm 0.0043$    &  $0.9654$           & $0.9665 \pm 0.0043$    & $0.9654$          \\[+1mm]
  $\ln(10^{10} A_s)$             & $3.039 \pm 0.017$     & $3.051 \pm 0.016$      &  $2.72 \pm 0.45$    & $3.046 \pm 0.016$      & $3.039$           \\[+1mm]
 \hline \\[-1mm]
  $H_0$                          & $56.9 \pm 3.6$        & $67.79 \pm 0.59$       &  $83 \pm 12$        & $67.52 \pm 0.61$       & $83 \pm 12$         \\[+1mm]
  $\Omega_m$                     & $0.444 \pm 0.055$     & $0.3102 \pm 0.0060$    &  $0.304 \pm 0.019$  & $0.3124 \pm 0.0063$    & $0.307 \pm 0.020$   \\[+1mm]
  $\sigma_8$                     & $0.786 \pm 0.014$     & $0.8128 \pm 0.0079$    &  $0.846 \pm 0.048$  & $0.8098 \pm 0.0080$    & $0.99 \pm 0.18$     \\[+1mm]
 \hline \\[-1mm]
  $\chi_{\textrm{min}}^2$ (Total)& $2757.38$             & $2786.27$              &  $15.90$            & $2777.01$              & $10.67$   \\[+1mm]
  $\chi_{\textrm{min}}^2$ (BAO/BAO$^\prime$) & $\cdots$  & $20.66$                &  $15.90$            & $11.82$                & $10.67$   \\[+1mm]
  $\chi_{\textrm{BAO/BAO}^\prime}^2$ (at P18 B-F) & $\cdots$   & $278.54$         &  $278.54$           & $236.71$               & $236.71$     \\[+1mm]
  $\textrm{DIC}$                 & $2811.54$             & $2840.16$              &  $24.57$            & $2831.65$              & $17.69$   \\[+1mm]
  $\Delta\textrm{DIC}$           & $-6.39$               & $0.91$                 &  $2.64$             & $2.04$                 & $2.76$   \\[+1mm]
  $\textrm{AIC}_c$               & $2813.38$             & $2842.27$              &  $31.90$            & $2833.01$              & $24.39$    \\[+1mm]
  $\Delta\textrm{AIC}_c$         & $-6.42$               & $1.61$                 &  $4.34$             & $1.26$                 & $4.41$     \\[+1mm]
 \hline \hline \\[-1mm]
                                 & \multicolumn{5}{c}{Tilted non-flat $\Lambda$CDM+$A_L$ model [new $P(q)$]}              \\[+1mm]
\cline{2-6}\\[-1mm]
  Parameter                      & P18                   & P18+BAO                & BAO                  &  P18+BAO$^\prime$       & BAO$^\prime$     \\[+1mm]
 \hline \\[-1mm]
  $\Omega_b h^2$                 & $0.02257 \pm 0.00017$ & $0.02260 \pm 0.00017$  & $0.039 \pm 0.015$    & $0.02261 \pm 0.00017$   & $0.042 \pm 0.015$  \\[+1mm]
  $\Omega_c h^2$                 & $0.1187 \pm 0.0016$   & $0.1180 \pm 0.0014$    & $0.174 \pm 0.047$    & $0.1178 \pm 0.0015$     & $0.177 \pm 0.046$  \\[+1mm]
  $100\theta_\textrm{MC}$        & $1.04111 \pm 0.00033$ & $1.04117 \pm 0.00033$  & $1.17 \pm 0.14$      & $1.04117 \pm 0.00032$   & $1.13 \pm 0.13$  \\[+1mm]
  $\tau$                         & $0.0512 \pm 0.0086$   & $0.0532 \pm 0.0081$    & $0.0512$             & $0.0495 \pm 0.0084$     & $0.0512$           \\[+1mm]
  $\Omega_k$                     & $-0.10 \pm 0.11$      & $-0.0005 \pm 0.0017$   & $-0.055 \pm 0.060$   & $-0.0012 \pm 0.0018$    & $-0.035 \pm 0.059$    \\[+1mm]
  $n_s$                          & $0.9654 \pm 0.0057$   & $0.9707 \pm 0.0044$    & $0.9654$             & $0.9715 \pm 0.0047$     & $0.9654$        \\[+1mm]
  $\ln(10^{10} A_s)$             & $3.036 \pm 0.018$     & $3.038 \pm 0.017$      & $2.69 \pm 0.43$      & $3.029 \pm 0.018$       & $3.036$           \\[+1mm]
  $A_{L}$                        & $0.94 \pm 0.20$       & $1.168 \pm 0.061$      & $0.94$               & $1.176 \pm 0.062$       & $0.94$           \\[+1mm]
 \hline \\[-1mm]
  $H_0$                          & $51 \pm 14$           & $68.09 \pm 0.60$       & $83 \pm 12$          & $67.85 \pm 0.63$        & $84 \pm 11$        \\[+1mm]
  $\Omega_m$                     & $0.70 \pm 0.43$       & $0.3048 \pm 0.0062$    & $0.304 \pm 0.019$    & $0.3065 \pm 0.0065$     & $0.306 \pm 0.020$  \\[+1mm]
  $\sigma_8$                     & $0.752 \pm 0.052$     & $0.8026 \pm 0.0086$    & $0.844 \pm 0.048$    & $0.7982 \pm 0.0092$     & $0.99 \pm 0.18$    \\[+1mm]
 \hline \\[-1mm]
  $\chi_{\textrm{min}}^2$ (Total)& $2756.33$             & $2776.32$              & $15.90$              & $2767.43$               & $10.68$        \\[+1mm]
  $\chi_{\textrm{min}}^2$ (BAO/BAO$^\prime$) & $\cdots$  & $20.30$                & $15.90$              & $11.21$                 & $10.68$   \\[+1mm]
  $\chi_{\textrm{BAO/BAO}^\prime}^2$ (at P18 B-F) & $\cdots$  & $194.81$          & $194.81$             & $160.72$                & $160.72$     \\[+1mm]
  $\textrm{DIC}$                 & $2814.83$             & $2834.67$              & $24.75$              & $2824.97$               & $17.76$   \\[+1mm]
  $\Delta\textrm{DIC}$           & $-3.10$               & $-4.58$                & $2.82$               & $-4.64$                 & $2.83$     \\[+1mm]
  $\textrm{AIC}_c$               & $2814.33$             & $2834.32$              & $31.90$              & $2825.43$               & $24.39$   \\[+1mm]
  $\Delta\textrm{AIC}_c$         & $-5.47$               & $-6.34$                & $4.34$               & $-6.32$                 & $4.41$    \\[+1mm]
\end{tabular}
\\[+1mm]
\begin{flushleft}
Note: $\Delta\textrm{DIC}$ ($\Delta\textrm{AIC}_c$) indicates an excess value relative to that of the tilted flat $\Lambda$CDM model constrained with the same data. 
\end{flushleft}
\end{ruledtabular}
\label{tab:para_TNL_ns_BAO}
\end{table*}


\begin{figure*}[htbp]
\centering
\mbox{\includegraphics[width=170mm]{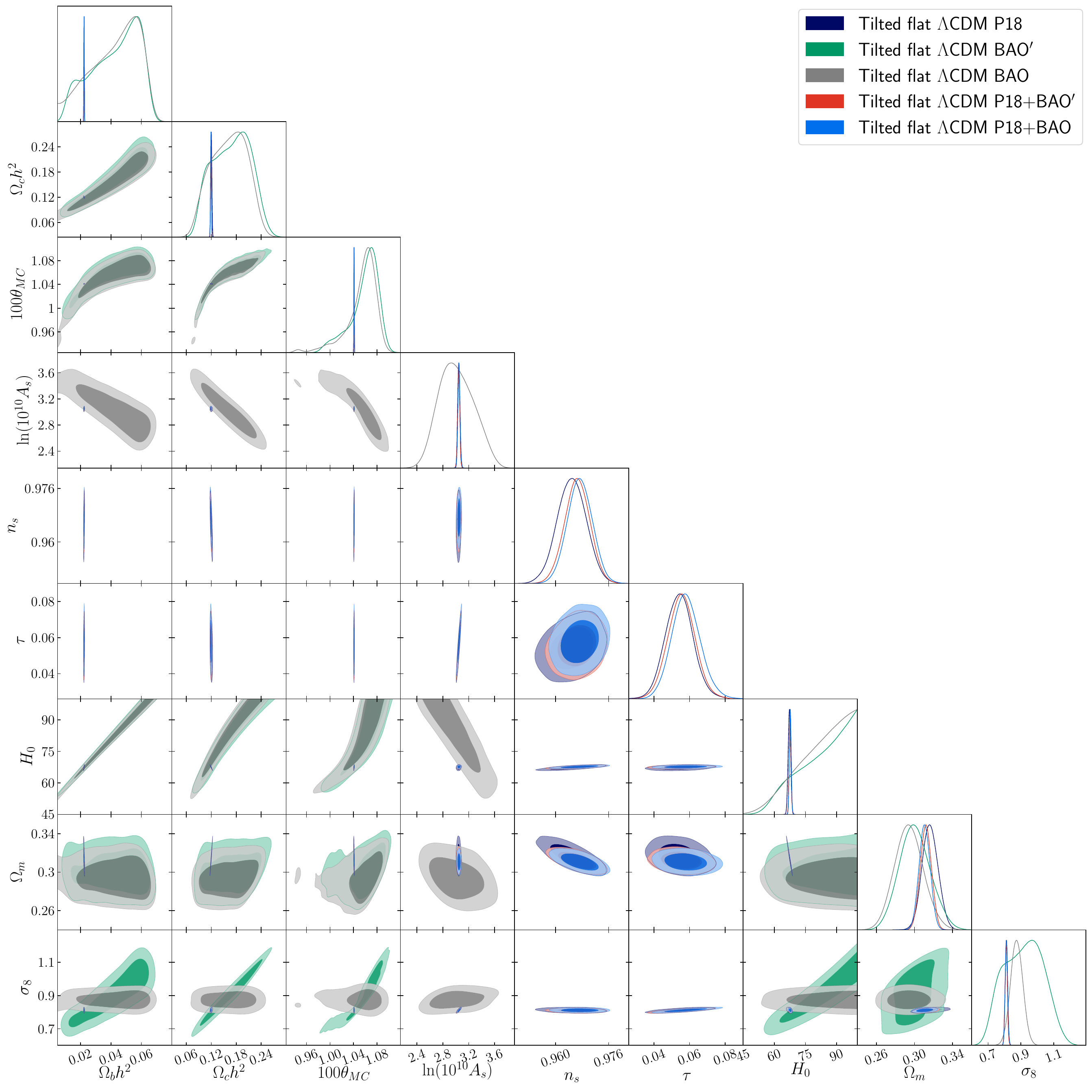}}
\caption{Likelihood distributions constrained by the Planck 2018 TT,TE,EE+lowE (P18), BAO, and BAO$^\prime$ data sets
         in the tilted flat $\Lambda$CDM model.
}
\label{fig:like_FL_BAO}
\end{figure*}
\begin{figure*}[htbp]
\centering
\mbox{\includegraphics[width=170mm]{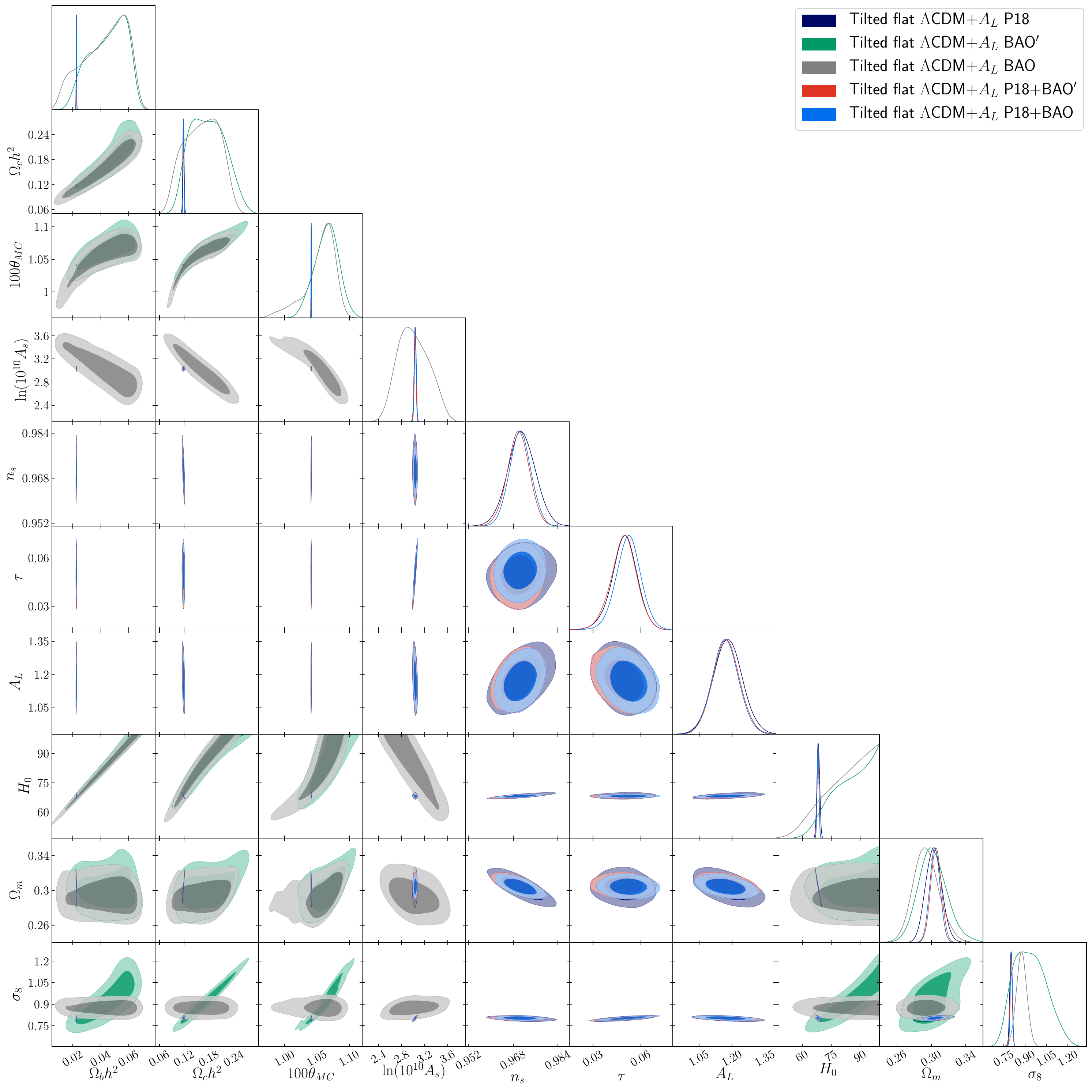}}
\caption{Likelihood distributions constrained by the Planck 2018 TT,TE,EE+lowE (P18), BAO, and BAO$^\prime$ data sets
         in the tilted flat $\Lambda$CDM$+A_L$ model.
}
\label{fig:like_FL_Alens_BAO}
\end{figure*}
\begin{figure*}[htbp]
\centering
\mbox{\includegraphics[width=170mm]{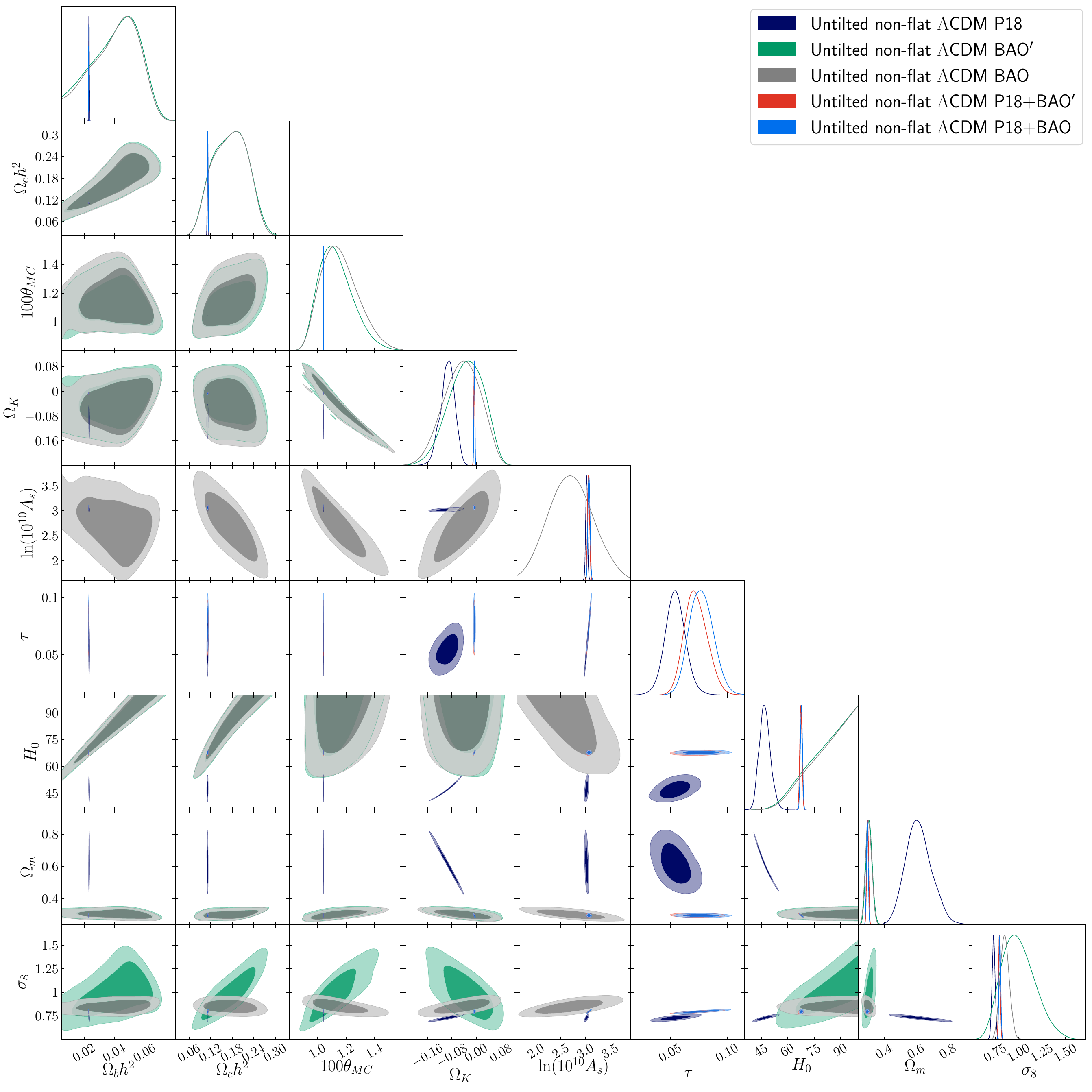}}
\caption{Likelihood distributions constrained by the Planck 2018 TT,TE,EE+lowE (P18), BAO, and BAO$^\prime$ data sets
         in the untilted non-flat $\Lambda$CDM model.
}
\label{fig:like_NL_BAO}
\end{figure*}
\begin{figure*}[htbp]
\centering
\mbox{\includegraphics[width=170mm]{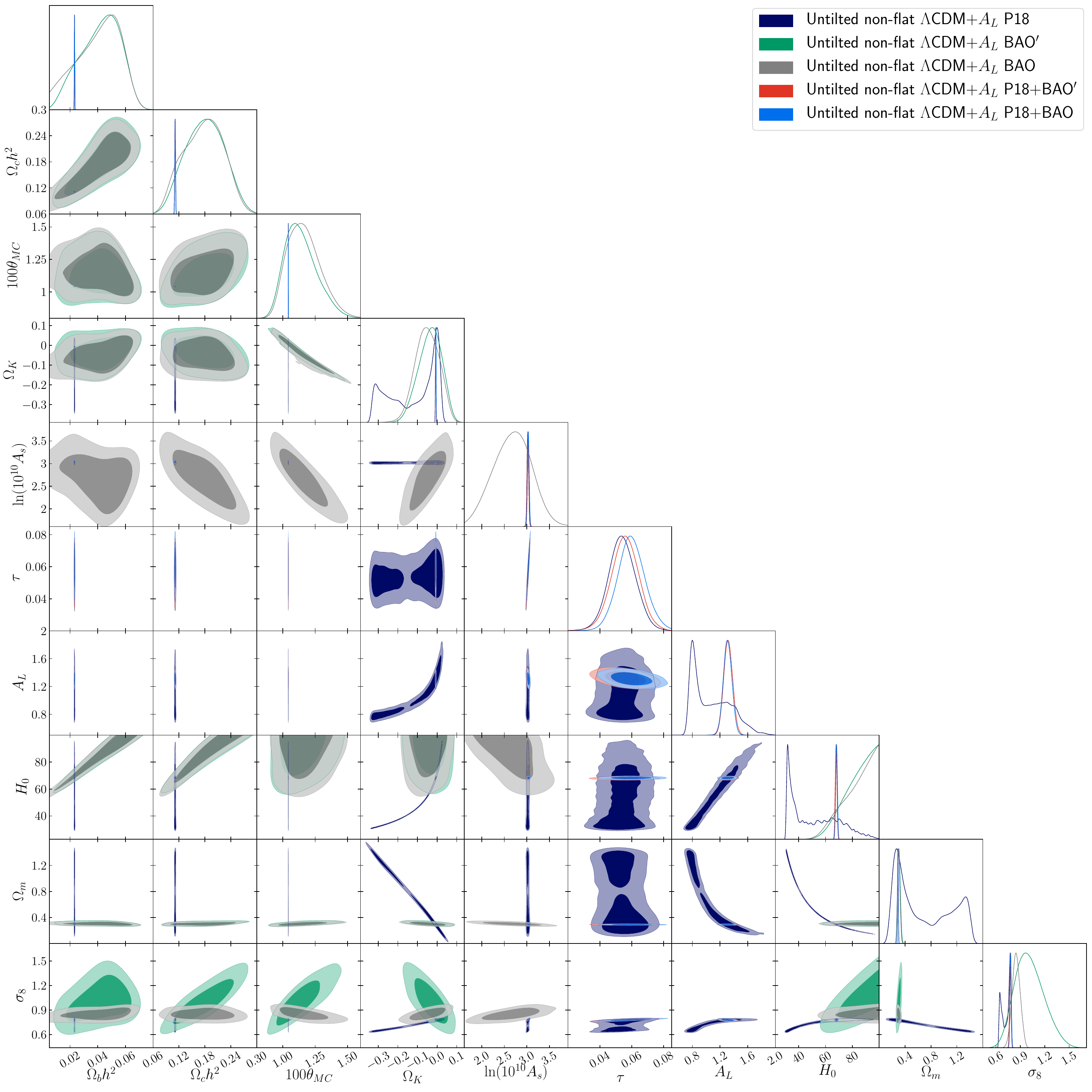}}
\caption{Likelihood distributions constrained by the Planck 2018 TT,TE,EE+lowE (P18), BAO, and BAO$^\prime$ data sets
         in the untilted non-flat $\Lambda$CDM$+A_L$ model.
}
\label{fig:like_NL_Alens_BAO}
\end{figure*} 

\begin{figure*}[htbp]
\centering
\mbox{\includegraphics[width=170mm]{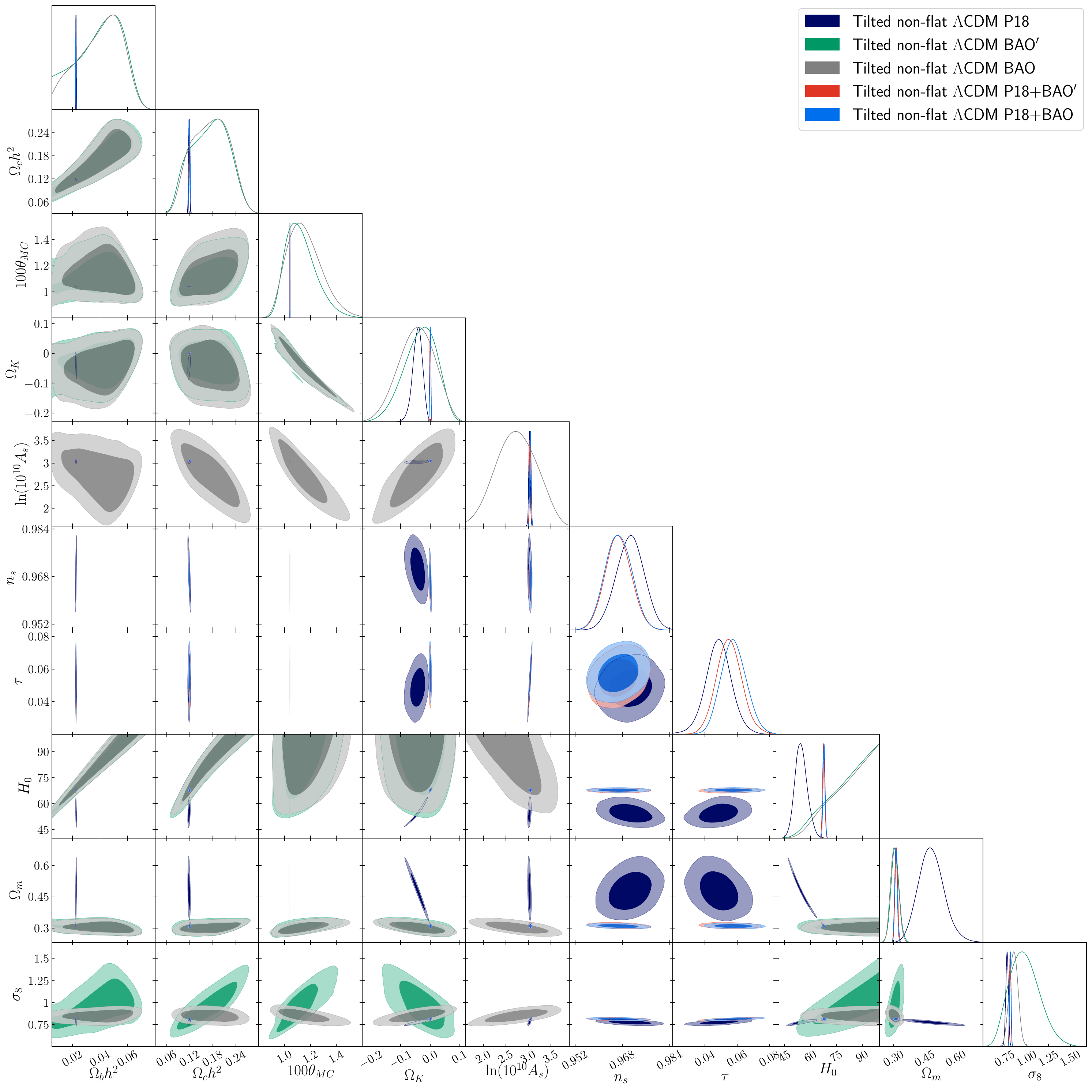}}
\caption{Likelihood distributions constrained by the Planck 2018 TT,TE,EE+lowE (P18), BAO, and BAO$^\prime$ data sets
         in the tilted non-flat $\Lambda$CDM model with the Planck $P(q)$.
}
\label{fig:like_NL_ns_BAO}
\end{figure*}
\begin{figure*}[htbp]
\centering
\mbox{\includegraphics[width=170mm]{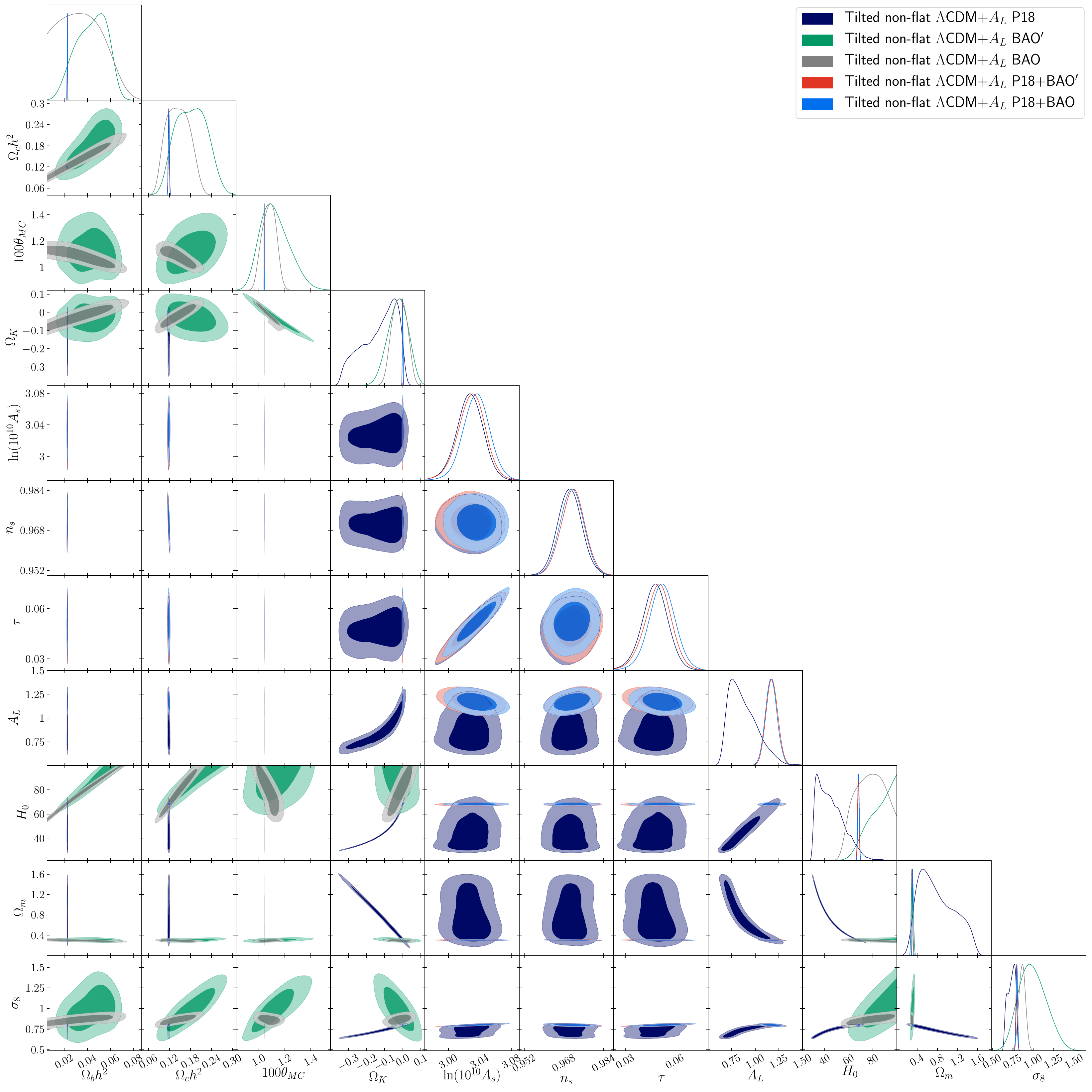}}
\caption{Likelihood distributions constrained by the Planck 2018 TT,TE,EE+lowE (P18), BAO, and BAO$^\prime$ data sets
         in the tilted non-flat $\Lambda$CDM$+A_L$ model with the Planck $P(q)$.
}
\label{fig:like_NL_Alens_ns_BAO}
\end{figure*}
\begin{figure*}[htbp]
\centering
\mbox{\includegraphics[width=170mm]{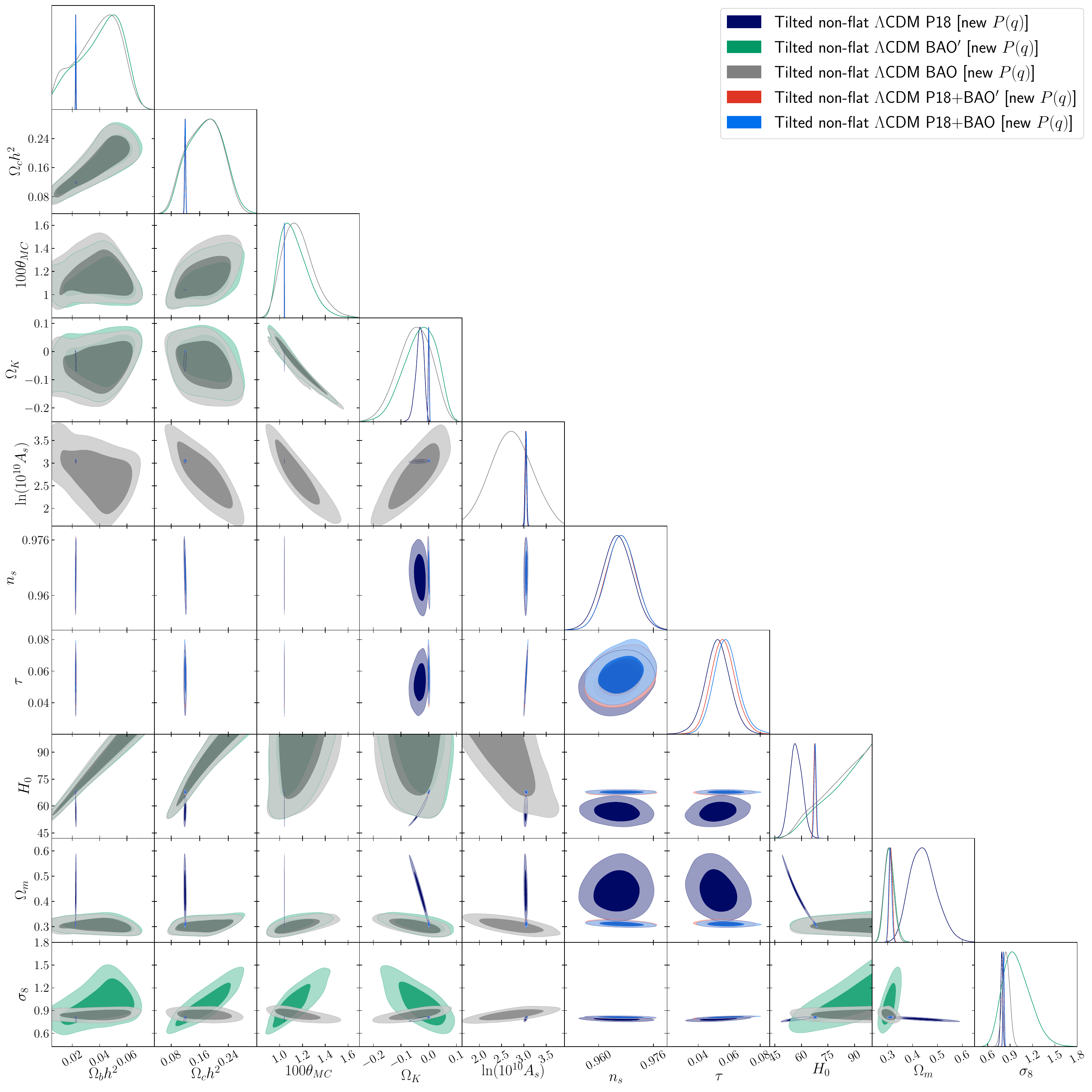}}
\caption{Likelihood distributions constrained by the Planck 2018 TT,TE,EE+lowE (P18), BAO, and BAO$^\prime$ data sets
         in the tilted non-flat $\Lambda$CDM model with the new $P(q)$.
}
\label{fig:like_TNL_ns1_BAO}
\end{figure*}
\begin{figure*}[htbp]
\centering
\mbox{\includegraphics[width=170mm]{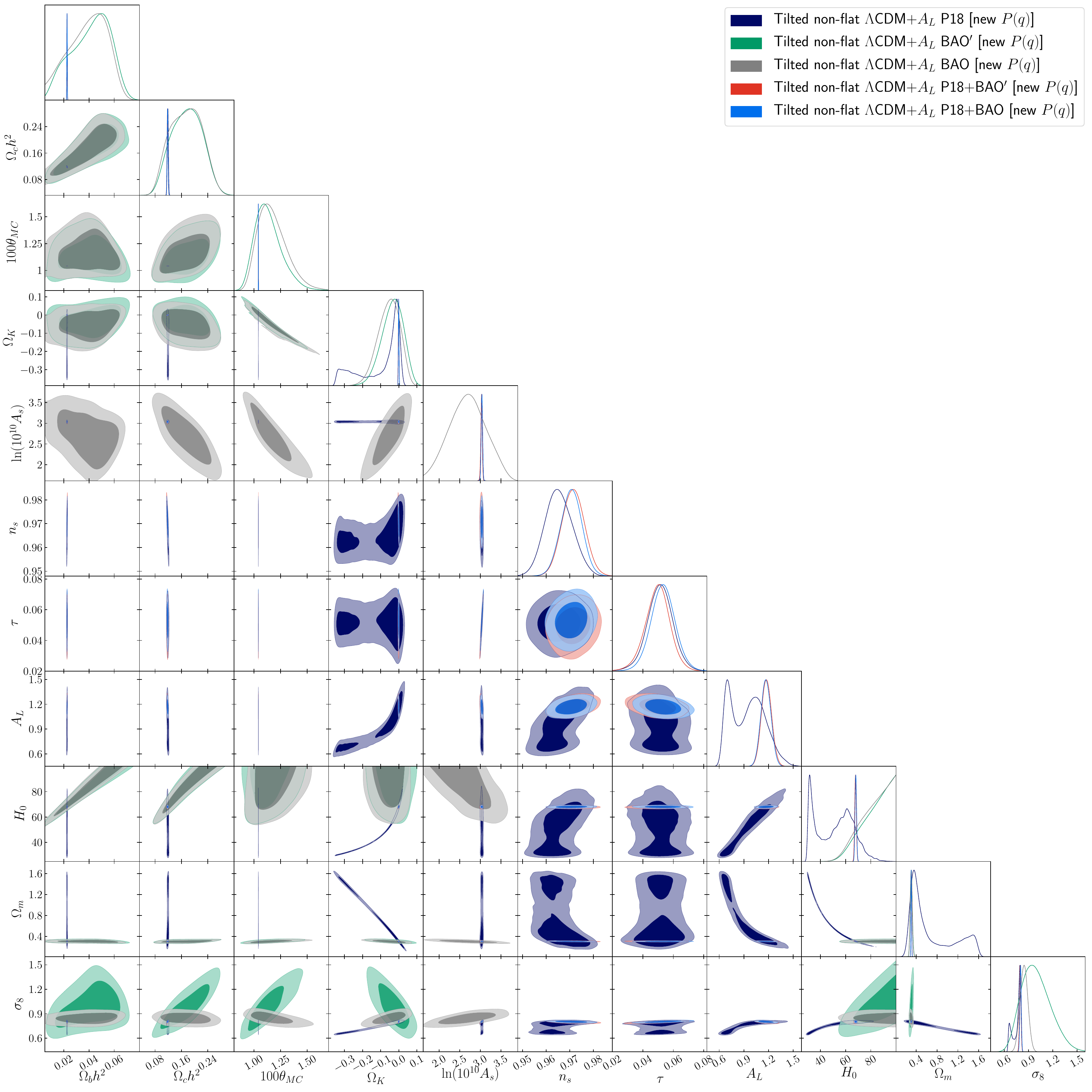}}
\caption{Likelihood distributions constrained by the Planck 2018 TT,TE,EE+lowE (P18), BAO, and BAO$^\prime$ data sets
         in the tilted non-flat $\Lambda$CDM$+A_L$ model with the new $P(q)$.
}
\label{fig:like_TNL_Alens_ns1_BAO}
\end{figure*}

\subsubsection{Comparing P18 data and BAO/BAO$^{\prime}$ data cosmological constraints}\label{sec:P18_vs_BAO}

In this subsubsection we compare BAO and BAO$^\prime$ data cosmological constraints to those obtained from P18 data. Prior to jointly analyzing P18+BAO/BAO$^\prime$ data, we need to determine whether P18 and BAO/BAO$^\prime$ data cosmological constraints are mutually consistent. In Sec.\ \ref{subsec:data_set_tensions} we use two other statistical estimators to examine whether or not P18 and BAO/BAO$^\prime$ data are in tension. 

The cosmological parameter mean values and error bars favored by the P18, BAO, BAO$^\prime$, P18+BAO, and P18+BAO$^\prime$ data sets are summarized in Tables \ref{tab:para_FL_BAO}-\ref{tab:para_TNL_ns_BAO} for the tilted flat $\Lambda$CDM (+$A_L$) models, the untilted non-flat $\Lambda$CDM (+$A_L$) models, the tilted non-flat $\Lambda$CDM (+$A_L$) models with the Planck $P(q)$, and the tilted non-flat $\Lambda$CDM ($+A_L$) models with the new $P(q)$, respectively. Likelihood distributions of cosmological parameters of the four models with $A_L=1$ and $A_L$ varying are shown in Figs.\ \ref{fig:like_FL_BAO}-\ref{fig:like_TNL_Alens_ns1_BAO} for the P18, BAO, BAO$^\prime$, P18+BAO, and P18+BAO$^\prime$ data sets.

Since neither BAO$^{\prime}$ nor BAO data have the ability to constrain $\tau$ or $n_s$ or $A_L$, we set their values to those found in the corresponding P18 data analysis. In addition, for the same reason, in the BAO$^\prime$ data analyses, we also set the value of $\ln(10^{10}A_s)$ to that found in the  corresponding P18 data analysis. We see from the upper and lower panels of Tables \ref{tab:para_FL_BAO}-\ref{tab:para_TNL_ns_BAO} that the BAO and BAO$^\prime$ data results for the $A_L=1$ and $A_L$-varying cases are similar, even though the fixed $\tau$ and $n_s$ [and $\ln(10^{10}A_s)$] values are slightly different for the $A_L = 1$ and $A_L$-varying cases.   

From Tables \ref{tab:para_FL_BAO}-\ref{tab:para_TNL_ns_BAO} we see that, in the six non-flat $\Lambda$CDM (+$A_L$) models the constraints set by  BAO$^\prime$/BAO data on $\Omega_m$ are tighter than the ones imposed by P18 data and in the three non-flat $\Lambda$CDM+$A_L$ models the constraints set by  BAO$^\prime$/BAO data on $\Omega_k$ are tighter than the ones imposed by P18 data. P18 data more restrictively constrains all other parameters in all eight cosmological models.

As we discuss below, there is a significant level of disagreement in the non-flat models between P18 data cosmological constraints and BAO$^\prime$/BAO data cosmological constraints, in most cases. From Tables \ref{tab:para_NL_BAO}-\ref{tab:para_TNL_ns_BAO} we see that all three data sets, P18, BAO$^\prime$, and BAO, favor negative values of the curvature parameter, with BAO$^\prime$ and BAO data favoring closed geometry only weakly, at 0.48$\sigma$ to 0.96$\sigma$. However, we should take into account the geometrical degeneracy between $H_0$-$\Omega_k$-$\Omega_m$ and note that both BAO$^\prime$ and BAO data favor higher values of $H_0$ and lower values of $\Omega_m$ than do P18 data and this is what causes the P18 and BAO/BAO$^\prime$ cosmological constraint differences. 

We first discuss BAO$^\prime$ data results (BAO$^\prime$ data do not include $f\sigma_8$ data points, see Sec.\ \ref{sec:data}) and then consider results from BAO data. This will allow us to test the impact of some $f\sigma_8$ data on the cosmological constraints.

Comparing the six-parameter and the three-parameter tilted flat $\Lambda$CDM model primary cosmological parameter constraints for P18 and BAO$^\prime$ data, shown in the upper half of Table \ref{tab:para_FL_BAO}, we see that the values of $\Omega_b h^2$ and $\Omega_c h^2$ are in mild disagreement, at 1.3$\sigma$ and 1.1$\sigma$ respectively. We also observe a similar 1.3$\sigma$ level of tension in the derived $H_0$ values, whereas the other two derived parameters, $\Omega_m$ and $\sigma_8$, show a better agreement, disagreeing by only 0.91$\sigma$ and 0.90$\sigma$ respectively. 

Comparing the seven-parameter and the three-parameter tilted flat $\Lambda$CDM+$A_L$ model primary cosmological parameter constraints for P18 and BAO$^\prime$ data, shown in the lower half of Table \ref{tab:para_FL_BAO}, we see that the values of $\Omega_b h^2$, $\Omega_c h^2$, and $100\theta_{\textrm{MC}}$ are in 1.7$\sigma$, 1.4$\sigma$ and 1.3$\sigma$ tension respectively.
 As for the derived parameters, we find $H_0$ and $\sigma_8$ values are in 1.7$\sigma$ and 1.2$\sigma$ disagreement while $\Omega_m$ values differ by only 0.046$\sigma$.
 This means that only for the $\Omega_m$ parameter does the inclusion of a varying $A_L$ reduce the disagreement found in the $A_L=1$ case, while increasing the disagreement for a number of other parameters.

P18 and BAO$^\prime$ data results obtained for the six-parameter and the three-parameter untilted non-flat $\Lambda$CDM model, shown in the upper half of Table \ref{tab:para_NL_BAO}, indicate more significant differences than found in the tilted flat $\Lambda$CDM model. The primary cosmological parameters $\Omega_b h^2$ and $\Omega_c h^2$ values disagree at 1.1$\sigma$ and 1.4$\sigma$, while the primary spatial curvature parameter value is $\Omega_k=-0.034\pm 0.057$ for  BAO$^\prime$ data, which is 0.60$\sigma$ away from flat and in 0.99$\sigma$ tension with the P18 value $\Omega_k=-0.095\pm 0.024$, which is 4.0$\sigma$ away from flat. Regarding the derived parameters, we find that $\Omega_m$, $H_0$, and $\sigma_8$ values are in 3.7$\sigma$, 2.9$\sigma$, and 1.5$\sigma$ disagreement. According to these results, P18 and BAO$^\prime$ data probably should not be jointly analyzed in the context of the untilted non-flat $\Lambda$CDM model. 

The results for the seven-parameter and the three-parameter untilted non-flat $\Lambda$CDM+$A_L$ model, obtained considering P18 and BAO$^\prime$ data, are in the lower half of Table \ref{tab:para_NL_BAO}. While there is a slight increase in the disagreement between the values of the primary parameters $\Omega_b h^2$  (1.3$\sigma$) and $\Omega_c h^2$ (1.6$\sigma$), there are significant decreases for the derived parameters $\Omega_m$, and $H_0$, but not for $\sigma_8$, that now disagree by  0.93$\sigma$, 1.5$\sigma$, and 1.5$\sigma$ respectively. This is caused by the increase in the size of the error bars in the $A_L$-varying P18 case with respect to the corresponding values obtained with $A_L=1$. For the BAO$^\prime$ data primary spatial curvature parameter we find $\Omega_k=-0.035\pm 0.058$, which is 0.60$\sigma$ away from flat hypersurfaces and only in 0.64$\sigma$ tension with the P18 value $\Omega_k=-0.12\pm0.12$, which is now only 1.0$\sigma$ away from flat. According to these results, unlike in the $A_L=1$ case, in the $A_L$-varying case P18 and BAO$^\prime$ data can probably be jointly analyzed in the context of the untilted non-flat $\Lambda$CDM model. Note that in this case a joint analysis of P18+BAO$^\prime$ data favors closed geometry at 4.9$\sigma$, with $\Omega_k=-0.0073\pm0.0015$, although because of the lack of the tilt ($n_s$) degree of freedom this untilted non-flat $\Lambda$CDM+$A_L$ model does not provide a good fit to smaller-angular-scale P18 data, which is reflected in the large $\Delta$DIC and $\Delta$AIC$_c$ values for the P18+BAO$^\prime$ case in the lower half of Table \ref{tab:para_NL_BAO}.

Comparing the seven-parameter and the four-parameter tilted non-flat $\Lambda$CDM Planck $P(q)$ model primary cosmological parameter constraints for P18 and BAO$^\prime$ data, we see, in the upper half of Table \ref{tab:para_NL_ns_BAO}, that the values of $\Omega_b h^2$ and $\Omega_c h^2$ are both in 1.1$\sigma$ disagreement. The BAO$^\prime$ data primary spatial curvature parameter value $\Omega_k=-0.033\pm 0.055$ is 0.6$\sigma$ away from flat and only in 0.17$\sigma$ tension with the P18 value $\Omega_k=-0.043\pm0.017$, which is 2.5$\sigma$ in favor of closed geometry. The derived parameters $\Omega_m$, $H_0$, and $\sigma_8$ are in 2.7$\sigma$, 2.3$\sigma$, and 1.2$\sigma$ tension. These results reveal that P18 and BAO$^\prime$ data cosmological constraints are somewhat inconsistent in the tilted non-flat $\Lambda$CDM Planck $P(q)$ model and these data probably should not be used jointly to constrain this model. 

Looking at the lower half of Table \ref{tab:para_NL_ns_BAO} we can compare results obtained for the eight-parameter and the four-parameter tilted non-flat $\Lambda$CDM+$A_L$ Planck $P(q)$ model from P18 and BAO$^\prime$ data respectively. We observe that the primary parameters $\Omega_b h^2$ and $\Omega_c h^2$ are in 1.6$\sigma$ and 1.5$\sigma$ tension. For the BAO$^\prime$ data primary spatial curvature parameter we find $\Omega_k= -0.026\pm 0.054$, which is only 0.48$\sigma$ away from flat and in 0.95$\sigma$ tension with the P18 value $-0.130\pm0.095$, which is 1.4$\sigma$ away from flat. Regarding the derived parameters we find that $\Omega_m$, $H_0$, and $\sigma_8$ are in 1.4$\sigma$, 2.7$\sigma$ and 1.6$\sigma$ disagreement. Compared to the $A_L = 1$ case, in the $A_L$-varying case we find a significant reduction only in the $\Omega_m$ tension, with most of the other parameter disagreements being more significant, which again suggests that P18 and BAO$^\prime$ data should not be jointly analyzed within the tilted non-flat $\Lambda$CDM+$A_L$ Planck $P(q)$ model. 

Comparing the seven-parameter and the four-parameter tilted non-flat $\Lambda$CDM new $P(q)$ model primary cosmological parameter constraints for P18 and BAO$^\prime$ data, from the upper half of Table \ref{tab:para_TNL_ns_BAO} we see that the values of $\Omega_b h^2$ and $\Omega_c h^2$ both disagree at 1.2$\sigma$. The BAO$^\prime$ data primary spatial curvature parameter value is $\Omega_k=-0.032\pm 0.059$, which is only a 0.54$\sigma$ deviation from flat and, similar to the Planck $P(q)$ model, is only in 0.016$\sigma$ disagreement with the P18 value $-0.033\pm 0.014$, which is 2.4$\sigma$ away from flat. Regarding the derived parameters $\Omega_m$, $H_0$, and $\sigma_8$, we find that their values disagree at 2.3$\sigma$, 2.1$\sigma$, and 1.1$\sigma$ respectively. While these disagreements are smaller than the ones found in the tilted non-flat $\Lambda$CDM Planck $P(q)$ model, they still are large enough to require we more carefully test whether P18 and BAO$^\prime$ data can be jointly used to constrain cosmological parameters in this cosmological model. 
 
The results for the eight-parameter and the four-parameter tilted non-flat $\Lambda$CDM+$A_L$ new $P(q)$ model are in the lower half of Table \ref{tab:para_TNL_ns_BAO}, for P18 and BAO$^\prime$ data, respectively. As happens in the Planck $P(q)$ model, when the $A_L$ parameter is allowed to vary the tensions found for the primary parameters $\Omega_b h^2$ and $\Omega_c h^2$ do not decrease (in fact they slightly increase) with respect to the $A_L=1$ case, both now being 1.3$\sigma$. For the BAO$^\prime$ data primary spatial curvature parameter we find $\Omega_k= -0.035\pm 0.059$, which is 0.59$\sigma$ away from flat hypersurfaces and only in 0.52$\sigma$ tension with the P18 value $\Omega_k=-0.10\pm 0.11$, which is 0.91$\sigma$ away from flat. As for the value of the derived parameters  $\Omega_m$, $H_0$ and $\sigma_8$ we find disagreements at 0.92$\sigma$, 1.9$\sigma$, and 1.3$\sigma$ respectively. The tensions are reduced with respect to the case with $A_L=1$, due to the increase of the error bars, but possibly still are not small enough to allow the joint use of P18+BAO$^\prime$ data for constraining tilted non-flat $\Lambda$CDM+$A_L$ new $P(q)$ model cosmological parameters. 

We now comment on the consistency between the cosmological constraints obtained using the BAO data set (which contain some $f\sigma_8$ data points) and the P18 data cosmological constraints. Here we also have to deal with the $\sigma_8$ tension, namely the discrepancy between the larger value for $\sigma_8$ obtained when P18 data are considered and the typically smaller values that one gets from low-redshift structure formation data (the $f\sigma_8$ data points we consider) or from weak lensing measurements. Note that since BAO data include some $f\sigma_8$ measurements we allow for ln($10^{10}A_s$) to vary in the BAO data only analyses (unlike the BAO$^\prime$ data only analyses where we fix the value of this parameter). We shall see that the tilted non-flat $\Lambda$CDM new $P(q)$ model is the model that best reconciles these measurements. 

Comparing the six-parameter and the four-parameter tilted flat $\Lambda$CDM primary cosmological parameter constraints for P18 and BAO data, shown in the upper half of Table \ref{tab:para_FL_BAO}, we see that the values of $\Omega_b h^2$ and $\Omega_c h^2$ are in 1.3$\sigma$ and 1.0$\sigma$ tension, respectively. A similar level of disagreement is found if we look at the values of the derived parameters. In particular for $\Omega_m$, $H_0$, and $\sigma_8$ we find 1.3$\sigma$, 1.3$\sigma$, and 1.6$\sigma$ disagreement. Here the greatest disagreement is that affecting $\sigma_8$, which has to do with the $\sigma_8$ tension mentioned above. 

Considering the results presented in the lower half of Table \ref{tab:para_FL_BAO} for the seven-parameter and the four-parameter tilted flat $\Lambda$CDM+$A_L$ model, obtained for P18 and BAO data, respectively, we find that including a varying $A_L$ parameter does not decrease the primary parameter tensions found when $A_L=1$. For $\Omega_b h^2$ and $\Omega_c h^2$ the disagreement is now 1.4$\sigma$ and 1.1$\sigma$. On the other hand for the derived $\Omega_m$, $H_0$, and $\sigma_8$ we find that their corresponding values disagree at 0.50$\sigma$, 1.2$\sigma$, and 2.0$\sigma$. Once again, allowing $A_L$ to vary reduces the $\Omega_m$ disagreement and the largest  disagreement is between the $\sigma_8$ values. 

Comparing the six-parameter and the five-parameter untilted non-flat $\Lambda$CDM model primary cosmological parameter constraints for P18 and BAO data, provided in the upper half of Table \ref{tab:para_NL_BAO}, we observe that the values of $\Omega_b h^2$ and $\Omega_c h^2$ show a disagreement of  1.1$\sigma$ and 1.4$\sigma$, respectively. The BAO data value for the primary spatial curvature parameter is $\Omega_k=-0.047\pm 0.059$, which is 0.80$\sigma$ away from flat hypersurfaces and in 0.75$\sigma$ tension with the P18 value $-0.095\pm 0.024$, which represents a 4.0$\sigma$ deviation from flat. The level of tension is worse for the derived parameters $\Omega_m$, $H_0$, and $\sigma_8$, the disagreements now being 3.7$\sigma$, 3.0$\sigma$, and 2.4$\sigma$. We may say that P18 and BAO data should not be jointly used to constrain cosmological parameters in the untilted non-flat $\Lambda$CDM model. 

Results for the seven-parameter and the five-parameter untilted non-flat $\Lambda$CDM+$A_L$ model for P18 and BAO data, respectively, can be seen in the lower half of Table \ref{tab:para_NL_BAO}. Again we do not observe a reduction in the tension for the primary parameters $\Omega_b h^2$ (1.2$\sigma$) and $\Omega_c h^2$ (1.4$\sigma$) compared with the results found for the $A_L =1$ case. On the other hand, there is an important decrease for the derived parameters $\Omega_m$, $H_0$, and $\sigma_8$, the disagreement now being 0.94$\sigma$, 1.5$\sigma$, and 1.8$\sigma$, respectively. This is probably caused by the increase in the size of the error bars in the $A_L$-varying P18 case, with respect to the corresponding values obtained with $A_L=1$. For the BAO data primary spatial curvature parameter we find $\Omega_k=-0.050\pm 0.060$, which is 0.83$\sigma$ away from flat and in 0.52$\sigma$ tension with the P18 value $\Omega_k = -0.12\pm 0.12$, which is 1.0$\sigma$ in favor of a closed geometry. 

Comparing the seven-parameter and the five-parameter tilted non-flat $\Lambda$CDM Planck $P(q)$ model primary cosmological parameter constraints for P18 and BAO data, shown in the upper half of Table \ref{tab:para_NL_ns_BAO}, we see that the values of $\Omega_b h^2$ and $\Omega_c h^2$ are both in 1.2$\sigma$ tension.
The BAO data primary spatial curvature parameter $\Omega_k=-0.046\pm 0.060$ is 0.77$\sigma$ away from flat hypersurfaces and, as in the BAO$^\prime$ case, in good agreement with, differing only by 0.048$\sigma$ from, the P18 result $-0.043\pm 0.017$, which is 2.5$\sigma$ away from flat). As for the derived parameters $\Omega_m$, $H_0$, and $\sigma_8$ we observe  disagreements of 2.7$\sigma$, 2.3$\sigma$ and 1.5$\sigma$. These results reveal an inconsistency between P18 and BAO cosmological constraints that probably mean P18 and BAO data should not be used to jointly constrain cosmological parameters in the tilted non-flat $\Lambda$CDM Planck $P(q)$ model. 

We provide results for the eight-parameter and the five-parameter tilted non-flat $\Lambda$CDM+$A_L$ Planck $P(q)$ model, from P18 and BAO data, in the lower half of Table \ref{tab:para_NL_ns_BAO}. For the primary parameters $\Omega_b h^2$ and $\Omega_c h^2$ we find a tension between the P18 and BAO values of 1.3$\sigma$ and 1.2$\sigma$, respectively. For the BAO data primary spatial curvature parameter we find $\Omega_k= -0.045\pm 0.063$, which represents a 0.71$\sigma$ evidence in favor of closed geometry and is only in 0.75$\sigma$ tension with respect to the P18 value $-0.130\pm0.095$, which represents a 1.4$\sigma$ deviation from flat. Regarding the derived $\Omega_m$, $H_0$, and $\sigma_8$ parameters, the observed disagreements are 1.4$\sigma$, 2.5$\sigma$, and 1.8$\sigma$. The tension for $\Omega_m$ has reduced significantly with respect to the $A_L=1$ case, however overall the disagreements are still large enough to not allow one to jointly analyze P18 and BAO data in this cosmological model. 

Comparing the seven-parameter and the five-parameter tilted non-flat $\Lambda$CDM new $P(q)$ model primary cosmological parameter constraints for P18 and BAO data, shown in the upper half of Table \ref{tab:para_TNL_ns_BAO}, we see that the values of $\Omega_b h^2$ and $\Omega_c h^2$ are both in 1.1$\sigma$ disagreement. The BAO data value of the primary spatial curvature parameter is $\Omega_k=-0.051\pm 0.061$, which represents a 0.84$\sigma$ deviation from a flat geometry and is only in 0.29$\sigma$ disagreement with the P18 value $\Omega_k=-0.033\pm 0.014$, which  is 2.4$\sigma$ away from flat. Regarding the derived parameters $\Omega_m$, $H_0$, and $\sigma_8$, we find 2.4$\sigma$, 2.1$\sigma$, and 1.2$\sigma$ disagreements between the corresponding values. It is necessary to further study the possible tension between P18 and BAO within this model.

Results for the eight-parameter and the five-parameter tilted non-flat $\Lambda$CDM+$A_L$ new $P(q)$ model, obtained from P18 and BAO data, can be seen in the lower half of Table \ref{tab:para_TNL_ns_BAO}. For the primary parameters $\Omega_b h^2$ and $\Omega_c h^2$ the disagreement is at 1.1$\sigma$ and 1.2$\sigma$ respectively. For the BAO data primary spatial curvature parameter we find $\Omega_k= -0.055\pm 0.060$, which represents 0.92$\sigma$ evidence in favor of closed geometry and is in only 0.36$\sigma$ disagreement with the P18 value $-0.10\pm 0.11$, which represents a 0.91$\sigma$ deviation from flat. Regarding the derived parameters, $\Omega_m$, $H_0$, and $\sigma_8$ we find 0.92$\sigma$, 1.7$\sigma$, and 1.3$\sigma$ disagreements. The tensions for $H_0$ and $\Omega_m$ have reduced with respect to the case with $A_L=1$, however they are still large enough to wonder whether we can jointly analyze P18 and BAO data in the context of this model. 

In Tables \ref{tab:para_FL_BAO}-\ref{tab:para_TNL_ns_BAO} $\chi^2_{\textrm{min}}$ (BAO/BAO$^{\prime}$) is the value of $\chi^2$ for BAO or BAO$^\prime$ data respectively, at the best-fit position for BAO or BAO$^\prime$ data, while $\chi^2_{\textrm{BAO/BAO}^\prime}$ (at P18 B-F) is the value of $\chi^2$ for BAO or BAO$^{\prime}$ data evaluated at the best-fit position for P18 data. The values of $\chi_{\textrm{min}}^2$ (BAO/BAO$^\prime$) and $\chi_{\textrm{BAO/BAO}^{\prime}}^2$ (at P18 B-F) gives a qualitative indication of the agreement or disagreement in the values of the cosmological parameters obtained by considering P18 data and by considering BAO/BAO$^\prime$ data. If the cosmological parameters agree one might expect that $\chi_{\textrm{min}}^2$ (BAO/BAO$^\prime$)$\simeq$ $\chi_{\textrm{BAO/BAO}^\prime}^2$ (at P18 B-F). We see that this is the case only for the tilted flat $\Lambda$CDM(+$A_L$) models for the BAO$^\prime$ data, but again emphasize that this is only a qualitative probe.

Figures \ref{fig:like_FL_BAO}-\ref{fig:like_TNL_Alens_ns1_BAO} show one-dimensional likelihoods and two-dimensional contours for cosmological parameters obtained using P18,  BAO$^\prime$, BAO, P18+BAO$^\prime$, and P18+BAO data. As mentioned above, BAO$^\prime$ data constraints (shown in green) and BAO data constraints (shown in grey) are comparatively less restrictive than P18 constraints (shown in dark blue), are unable to put tight constraints on the primary cosmological parameters (except for $\Omega_k$ in the three non-flat $\Lambda$CDM$+A_L$ models), in most cases overlap at 2$\sigma$ with each other, and in many cases they also overlap with the P18 data constraints. Since the BAO data set contains more measurements than the BAO$^\prime$ data set, the BAO constraints are typically more restrictive, and BAO data, which includes $f\sigma_8$ measurements, are much more effective at constraining $\sigma_8$ than are BAO$^\prime$ data.

Figures \ref{fig:like_FL_BAO} and \ref{fig:like_FL_Alens_BAO} are for tilted flat $\Lambda$CDM (+$A_L$) models. The $\sim 1 \sigma$ disagreements between the BAO$^\prime$/BAO constraints and those obtained with P18 data, discussed above, can be clearly seen in the contour plots. For the tilted flat $\Lambda$CDM model the larger disagreements are in panels for derived cosmological parameters, with the largest for $\sigma_8$. Some of these disagreements decrease when the $A_L$ parameter is allowed to vary.

Looking at the contour plots for the untilted non-flat $\Lambda$CDM (+$A_L$) models (see Figs.\ \ref{fig:like_NL_BAO} and \ref{fig:like_NL_Alens_BAO}) we observe non-overlapping contours in those panels that involve the derived parameters $\Omega_m$ and $H_0$. These disagreements are smaller when $A_L$ is allowed to vary. This may indicate that in the context of this cosmological model we may jointly analyze P18 data with BAO$^\prime$/BAO data only when $A_L$ is allowed to vary. 

Figures \ref{fig:like_NL_ns_BAO} and \ref{fig:like_NL_Alens_ns_BAO} show cosmological parameter constraints for the tilted non-flat $\Lambda$CDM (+$A_L$) Planck $P(q)$ models, while the ones for the tilted non-flat $\Lambda$CDM(+$A_L$) new $P(q)$ models are displayed in Figs.\ \ref{fig:like_TNL_ns1_BAO} and \ref{fig:like_TNL_Alens_ns1_BAO}. As expected, considering the results discussed above in this subsubsection, the contour plots for these tilted non-flat models are quite similar. We see in the panels that involve the primary cosmological parameters there is overlap at 1$\sigma$, not only when $A_L$ is allowed to vary but also when $A_L=1$. When $A_L=1$, for the Planck $P(q)$ model some P18 and BAO$^\prime$/BAO data constraint contours that involve $\Omega_m$ and $H_0$ do not overlap even at 2$\sigma$. This is not true for the new $P(q)$ model with $A_L=1$, where overlap is reached at $< 2 \sigma$. This may indicate that the new $P(q)$ model is better able to reconcile P18 and BAO$^\prime$/BAO data. 

In view of the results discussed in this subsubsection, further tests are needed to properly quantify the level of disagreement, in the context of non-flat models, between P18 data and BAO$^\prime$/BAO data cosmological constraints. We return to this issue in Sec.\ \ref{subsec:data_set_tensions}.

\begin{table*}
\caption{Mean and 68.3\% confidence limits of tilted flat $\Lambda\textrm{CDM}$ (+$A_L$) model parameters
        constrained by non-CMB, P18, and P18+non-CMB data sets.
        $H_0$ has units of km s$^{-1}$ Mpc$^{-1}$. 
}
\begin{ruledtabular}
\begin{tabular}{lccccc}
\\[-1mm]                         
                                 & \multicolumn{3}{c}{Tilted flat $\Lambda$CDM}  & \multicolumn{2}{c}{Tilted flat $\Lambda$CDM$+A_L$}      \\[+1mm]
\cline{2-4}\cline{5-6}\\[-1mm]
Parameter                        & Non-CMB  & P18  &  P18+non-CMB    & P18  & P18+non-CMB  \\[+1mm]
\hline \\[-1mm]
$\Omega_b h^2$                 & $0.0256 \pm 0.0025$ & $0.02236 \pm 0.00015$ & $0.02250 \pm 0.00012$  & $0.02259 \pm 0.00017$ & $0.02265 \pm 0.00014$          \\[+1mm]
$\Omega_c h^2$                 & $0.1129 \pm 0.0062$ & $0.1202 \pm 0.0014$   & $0.11825 \pm 0.00087$  & $0.1180 \pm 0.0015$   & $0.11736 \pm 0.00092$   \\[+1mm]
$100\theta_\textrm{MC}$        & $1.0323 \pm 0.0082$ & $1.04090 \pm 0.00031$ & $1.04112 \pm 0.00029$  & $1.04114 \pm 0.00032$ & $1.04120 \pm 0.00029$     \\[+1mm]
$\tau$                         & $0.0542$            & $0.0542 \pm 0.0079$   & $0.0548 \pm 0.0076$    & $0.0496 \pm 0.0082$   & $0.0484 \pm 0.0083$   \\[+1mm]
$n_s$                          & $0.9649$            & $0.9649 \pm 0.0043$   & $0.9692 \pm 0.0036$    & $0.9710 \pm 0.0050$   & $0.9726 \pm 0.0038$   \\[+1mm]
$\ln(10^{10} A_s)$             & $3.10 \pm 0.11$     & $3.044 \pm 0.016$     & $3.041 \pm 0.015$      & $3.030 \pm 0.017$     & $3.026 \pm 0.017$  \\[+1mm]
$A_{L}$                        & $\cdots$            & $\cdots$              & $\cdots$               & $1.181 \pm 0.067$     & $1.201 \pm 0.061$  \\[+1mm]
  \hline \\[-1mm]
$H_0$                          & $69.8 \pm 1.7$      & $67.28 \pm 0.61$      & $68.15 \pm 0.39$       & $68.31 \pm 0.71$      & $68.62 \pm 0.43$ \\[+1mm]
$\Omega_m$                     & $0.286 \pm 0.011$   & $0.3165 \pm 0.0084$   & $0.3045 \pm 0.0051$    & $0.3029 \pm 0.0093$   & $0.2988 \pm 0.0054$    \\[+1mm]
$\sigma_8$                     & $0.787 \pm 0.027$   & $0.8118 \pm 0.0074$   & $0.8048 \pm 0.0068$    & $0.7997 \pm 0.0088$   & $0.7961 \pm 0.0074$   \\[+1mm]
 \hline \\[-1mm]
$\chi_{\textrm{min}}^2$ (Total)   & $1106.54$        & $2765.80$             & $3879.35$              & $2756.12$             & $3865.90$       \\[+1mm]
$\chi_{\textrm{min}}^2$ (Non-CMB) & $1106.54$        & $\cdots$              & $1111.57$              & $\cdots$              & $1109.54$    \\[+1mm]
$\textrm{DIC}$                    & $1114.45$        & $2817.93$             & $3931.02$              & $2812.41$             & $3922.11$ \\[+1mm]
$\Delta\textrm{DIC}$              & $\cdots$         & $\cdots$              & $\cdots$               & $-5.52$               & $-8.91$          \\[+1mm]
$\textrm{AIC}_c$                  & $1114.54$        & $2819.80$             & $3933.35$              & $2812.1$              & $3921.90$   \\[+1mm]
$\Delta\textrm{AIC}_c$            & $\cdots$         & $\cdots$              & $\cdots$               & $-7.68$               & $-11.45$          \\[+1mm]
\end{tabular}
\\[+1mm]
\begin{flushleft}
Note: $\Delta\textrm{DIC}$ ($\Delta\textrm{AIC}_c$) indicates an excess value relative to that of the tilted flat $\Lambda$CDM model constrained with the same data. 
\end{flushleft}
\end{ruledtabular}
\label{tab:para_FL_P18_nonCMB}
\end{table*}

\begin{table*}
\caption{Mean and 68.3\% confidence limits of untilted non-flat $\Lambda\textrm{CDM}$ (+$A_L$) model parameters
        constrained by non-CMB, P18, and P18+non-CMB data sets.
        $H_0$ has units of km s$^{-1}$ Mpc$^{-1}$.
}
\begin{ruledtabular}
\begin{tabular}{lccccc}
\\[-1mm]                        
                                 & \multicolumn{3}{c}{Untilted non-flat $\Lambda$CDM}  & \multicolumn{2}{c}{Untilted non-flat $\Lambda$CDM$+A_L$}      \\[+1mm]
\cline{2-4}\cline{5-6}\\[-1mm]
Parameter                        & Non-CMB  & P18  &  P18+non-CMB    & P18  & P18+non-CMB  \\[+1mm]
\hline \\[-1mm]
$\Omega_b h^2$                 & $0.0243 \pm 0.0033$ & $0.02320 \pm 0.00015$   & $0.02300 \pm 0.00014$  & $0.02320 \pm 0.00015$ & $0.02320 \pm 0.00015$           \\[+1mm]
$\Omega_c h^2$                 & $0.120 \pm 0.013$   & $0.11098 \pm 0.00088$   & $0.11161 \pm 0.00086$  & $0.11097 \pm 0.00087$ & $0.11097 \pm 0.00085$ \\[+1mm]
$100\theta_\textrm{MC}$        & $1.10 \pm 0.10$     & $1.04204 \pm 0.00030$   & $1.04189 \pm 0.00029$  & $1.04202 \pm 0.00030$ & $1.04199 \pm 0.00030$       \\[+1mm]
$\tau$                         & $0.0543$            & $0.0543 \pm 0.0091$     & $0.0717 \pm 0.0095$    & $0.0540 \pm 0.0087$   & $0.0562 \pm 0.0086$   \\[+1mm]
$\Omega_k$                     & $-0.033 \pm 0.050$  & $-0.095 \pm 0.024$      & $-0.0062 \pm 0.0014$   & $-0.12 \pm 0.12$      & $-0.0062 \pm 0.0014$    \\[+1mm]
$\ln(10^{10} A_s)$             & $2.87 \pm 0.34$     & $3.021 \pm 0.019$       & $3.057 \pm 0.019$      & $3.020 \pm 0.018$     & $3.024 \pm 0.018$ \\[+1mm]
$A_{L}$                        & $\cdots$            & $\cdots$                & $\cdots$               & $1.08 \pm 0.27$       & $1.324 \pm 0.063$          \\[+1mm]
  \hline \\[-1mm]
$H_0$                          & $70.2 \pm 1.7$      & $47.1 \pm 3.2$          & $68.07 \pm 0.56$       & $52 \pm 18$           & $68.45 \pm 0.58$ \\[+1mm]
$\Omega_m$                     & $0.294 \pm 0.018$   & $0.617 \pm 0.082$       & $0.2920 \pm 0.0050$    & $0.70 \pm 0.42$       & $0.2878 \pm 0.0050$ \\[+1mm]
$\sigma_8$                     & $0.771 \pm 0.034$   & $0.730 \pm 0.017$       & $0.7921 \pm 0.0085$    & $0.721 \pm 0.053$     & $0.7759 \pm 0.0078$  \\[+1mm]
 \hline \\[-1mm]
$\chi_{\textrm{min}}^2$ (Total)&           $1106.53$  & $2789.77$                &  $3926.27$              & $2787.76$              & $3895.24$    \\[+1mm]
$\chi_{\textrm{min}}^2$ (Non-CMB) &        $1106.53$  & $\cdots$                &  $1107.71$              & $\cdots$              & $1107.45$  \\[+1mm]
$\textrm{DIC}$                 &           $1116.95$  & $2847.14$                &  $3982.38$              & $2846.45$              & $3954.21$   \\[+1mm]
$\Delta\textrm{DIC}$           &           $2.50$    & $29.21$                 &  $51.36$               & $28.52$               & $23.19$        \\[+1mm]
$\textrm{AIC}_c$                 &           $1116.53$  & $2843.77$                &  $3980.27$              & $2843.76$              & $3951.24$        \\[+1mm]
$\Delta\textrm{AIC}_c$           &           $1.99$    & $23.97$                 &  $46.92$               & $23.96$               & $17.89$         \\[+1mm]
\end{tabular}
\\[+1mm]
\begin{flushleft}
Note: $\Delta\textrm{DIC}$ ($\Delta\textrm{AIC}_c$) indicates an excess value relative to that of the tilted flat $\Lambda$CDM model constrained with the same data. 
\end{flushleft}
\end{ruledtabular}
\label{tab:para_NL_P18_nonCMB}
\end{table*}

\begin{table*}
\caption{Mean and 68.3\% confidence limits of Planck-$P(q)$-based tilted non-flat $\Lambda\textrm{CDM}$ ($+A_L$) model parameters
        constrained by non-CMB, P18, and P18+non-CMB data sets.
        $H_0$ has units of km s$^{-1}$ Mpc$^{-1}$.
}
\begin{ruledtabular}
\begin{tabular}{lccccc}
\\[-1mm]                        
                                 & \multicolumn{3}{c}{Tilted non-flat $\Lambda$CDM Planck $P(q)$}  & \multicolumn{2}{c}{Tilted non-flat $\Lambda$CDM$+A_L$ Planck $P(q)$}      \\[+1mm]
\cline{2-4}\cline{5-6}\\[-1mm]
Parameter                        & Non-CMB  & P18  &  P18+non-CMB    & P18  & P18+non-CMB  \\[+1mm]
 \hline \\[-1mm]
$\Omega_b h^2$                 & $0.0242 \pm 0.0033$ & $0.02260 \pm 0.00017$ & $0.02248 \pm 0.00015$  & $0.02258 \pm 0.00017$ & $0.02268 \pm 0.00017$       \\[+1mm]
$\Omega_c h^2$                 & $0.120 \pm 0.012$   & $0.1181 \pm 0.0015$   & $0.1185 \pm 0.0013$    & $0.1183 \pm 0.0015$   & $0.1170 \pm 0.0014$         \\[+1mm]
$100\theta_\textrm{MC}$        & $1.10 \pm 0.11$     & $1.04116 \pm 0.00032$ & $1.04107 \pm 0.00031$  & $1.04116 \pm 0.00033$ & $1.04125 \pm 0.00032$ \\[+1mm]
$\tau$                         & $0.0483$            & $0.0483 \pm 0.0083$   & $0.0543 \pm 0.0077$    & $0.0478 \pm 0.0081$   & $0.0485 \pm 0.0087$              \\[+1mm]
$\Omega_k$                     & $-0.032 \pm 0.051$  & $-0.043 \pm 0.017$    & $0.0004 \pm 0.0017$    & $-0.130 \pm 0.095$    & $-0.0006 \pm 0.0017$            \\[+1mm]
$n_s$                          & $0.9706$            & $0.9706 \pm 0.0047$   & $0.9687 \pm 0.0043$    & $0.9704 \pm 0.0048$   & $0.9735 \pm 0.0046$   \\[+1mm]
$\ln(10^{10} A_s)$             & $2.90 \pm 0.34$     & $3.027 \pm 0.017$     & $3.040 \pm 0.016$      & $3.027 \pm 0.017$     & $3.025 \pm 0.018$    \\[+1mm]
$A_{L}$                        & $\cdots$            & $\cdots$              & $\cdots$               & $0.88 \pm 0.15$       & $1.203 \pm 0.062$              \\[+1mm]
 \hline \\[-1mm]
$H_0$                          & $70.1 \pm 1.7$      & $54.5 \pm 3.6$        & $68.25 \pm 0.56$       & $45 \pm 11$           & $68.48 \pm 0.56$                          \\[+1mm]
$\Omega_m$                     & $0.294 \pm 0.018$   & $0.481 \pm 0.062$     & $0.3040 \pm 0.0055$    &  $0.80 \pm 0.35$      & $0.2994 \pm 0.0055$   \\[+1mm]
$\sigma_8$                     & $0.771 \pm 0.035$   & $0.775 \pm 0.015$     & $0.8055 \pm 0.0076$    & $0.733 \pm 0.045$     & $0.7946 \pm 0.0088$  \\[+1mm]
 \hline \\[-1mm]
$\chi_{\textrm{min}}^2$ (Total)          & $1106.53$  & $2754.73$              &  $3878.77$              & $2754.99$              & $3865.53$   \\[+1mm]
$\chi_{\textrm{min}}^2$ (Non-CMB)        & $1106.53$  & $\cdots$              &  $1111.36$              & $\cdots$              & $1109.27$  \\[+1mm]
$\textrm{DIC}$                           & $1116.92$  & $2810.59$              &  $3933.33$              & $2811.63$              & $3924.07$    \\[+1mm]
$\Delta\textrm{DIC}$                     & $2.47$    & $-7.34$                &  $2.31$                & $-6.30$                & $-6.95$       \\[+1mm]
$\textrm{AIC}_c$                           & $1116.53$  & $2810.73$              &  $3934.77$              & $2812.99$              & $3923.53$     \\[+1mm]
$\Delta\textrm{AIC}_c$                     & $1.99$  & $-9.07$                &  $1.42$                & $-6.81$                & $-9.82$        \\[+1mm]
\end{tabular}
\\[+1mm]
\begin{flushleft}
Note: $\Delta\textrm{DIC}$ ($\Delta\textrm{AIC}_c$) indicates an excess value relative to that of the tilted flat $\Lambda$CDM model constrained with the same data. 
\end{flushleft}
\end{ruledtabular}
\label{tab:para_NL_ns_P18_nonCMB}
\end{table*}

\begin{table*}
\caption{Mean and 68.3\% confidence limits of new-$P(q)$-based tilted non-flat $\Lambda\textrm{CDM}$ ($+A_L$) model parameters
        constrained by non-CMB, P18, and P18+non-CMB data sets.
        $H_0$ has units of km s$^{-1}$ Mpc$^{-1}$.
}
\begin{ruledtabular}
\begin{tabular}{lccccc}
\\[-1mm]                        
                                 & \multicolumn{3}{c}{Tilted non-flat $\Lambda$CDM new $P(q)$}  & \multicolumn{2}{c}{Tilted non-flat $\Lambda$CDM$+A_L$ new $P(q)$}      \\[+1mm]
\cline{2-4}\cline{5-6}\\[-1mm]
Parameter                        & Non-CMB  & P18  &  P18+non-CMB    & P18  & P18+non-CMB  \\[+1mm]
 \hline \\[-1mm]
$\Omega_b h^2$                 & $0.0241 \pm 0.0033$ & $0.02255 \pm 0.00017$  & $0.02249 \pm 0.00015$  & $0.02257 \pm 0.00017$ & $0.02269 \pm 0.00016$  \\[+1mm]
$\Omega_c h^2$                 & $0.120 \pm 0.013$   & $0.1188 \pm 0.0015$    & $0.1184 \pm 0.0013$    & $0.1187 \pm 0.0016$   & $0.1170 \pm 0.0013$       \\[+1mm]
$100\theta_\textrm{MC}$        & $1.11 \pm 0.11$     & $1.04109 \pm 0.00032$  & $1.04108 \pm 0.00031$  & $1.04111 \pm 0.00033$ & $1.04125 \pm 0.00032$\\[+1mm]
$\tau$                         & $0.0525$            & $0.0525 \pm 0.0083$    & $0.0549 \pm 0.0077$    & $0.0512 \pm 0.0086$   & $0.0490 \pm 0.0086$       \\[+1mm]
$\Omega_k$                     & $-0.036 \pm 0.051$  & $-0.033 \pm 0.014$     & $0.0003 \pm 0.0017$    & $-0.10 \pm 0.11$      & $-0.0006 \pm 0.0017$\\[+1mm]
$n_s$                          & $0.9654$            & $0.9654 \pm 0.0045$    & $0.9684 \pm 0.0041$    & $0.9654 \pm 0.0057$   & $0.9730 \pm 0.0043$  \\[+1mm]
$\ln(10^{10} A_s)$             & $2.88 \pm 0.34$     & $3.039 \pm 0.017$      & $3.042 \pm 0.016$      & $3.036 \pm 0.018$     & $3.026 \pm 0.018$          \\[+1mm]
$A_{L}$                        & $\cdots$            & $\cdots$               & $\cdots$               & $0.94 \pm 0.20$       & $1.204 \pm 0.061$          \\[+1mm]
 \hline \\[-1mm]
$H_0$                          & $70.1 \pm 1.8$      & $56.9 \pm 3.6$         & $68.21 \pm 0.55$       & $51 \pm 14$           & $68.47 \pm 0.56$     \\[+1mm]
$\Omega_m$                     & $0.295 \pm 0.018$   & $0.444 \pm 0.055$      & $0.3043 \pm 0.0054$    & $0.70 \pm 0.43$       & $0.2994 \pm 0.0056$  \\[+1mm]
$\sigma_8$                     & $0.770 \pm 0.035$   & $0.786 \pm 0.014$      & $0.8057 \pm 0.0074$    & $0.752 \pm 0.052$     & $0.7948 \pm 0.0083$           \\[+1mm]
 \hline \\[-1mm]
$\chi_{\textrm{min}}^2$ (Total)&           $1106.49$  & $2757.38$               &  $3878.76$              & $2756.33$              & $3865.41$                \\[+1mm]
$\chi_{\textrm{min}}^2$ (Non-CMB) &        $1106.49$  & $\cdots$               &  $1111.36$              & $\cdots$              & $1109.32$     \\[+1mm]
$\textrm{DIC}$                 &           $1117.31$  & $2811.54$               &  $3932.56$              & $2814.83$              & $3923.86$                \\[+1mm]
$\Delta\textrm{DIC}$           &           $2.86$    & $-6.39$                 &  $1.54$                & $-3.10$                & $-7.16$              \\[+1mm]
$\textrm{AIC}_c$                 &           $1116.49$  & $2813.38$               &  $3934.76$              & $2814.33$              & $3923.41$                 \\[+1mm]
$\Delta\textrm{AIC}_c$           &           $1.95$    & $-6.42$                 &  $1.41$                & $-5.47$                & $-9.94$  \\[+1mm]
\end{tabular}
\\[+1mm]
\begin{flushleft}
Note: $\Delta\textrm{DIC}$ ($\Delta\textrm{AIC}_c$) indicates an excess value relative to that of the tilted flat $\Lambda$CDM model constrained with the same data. 
\end{flushleft}
\end{ruledtabular}
\label{tab:para_TNL_P18_nonCMB}
\end{table*}

\begin{figure*}[htbp]
\centering
\mbox{\includegraphics[width=170mm]{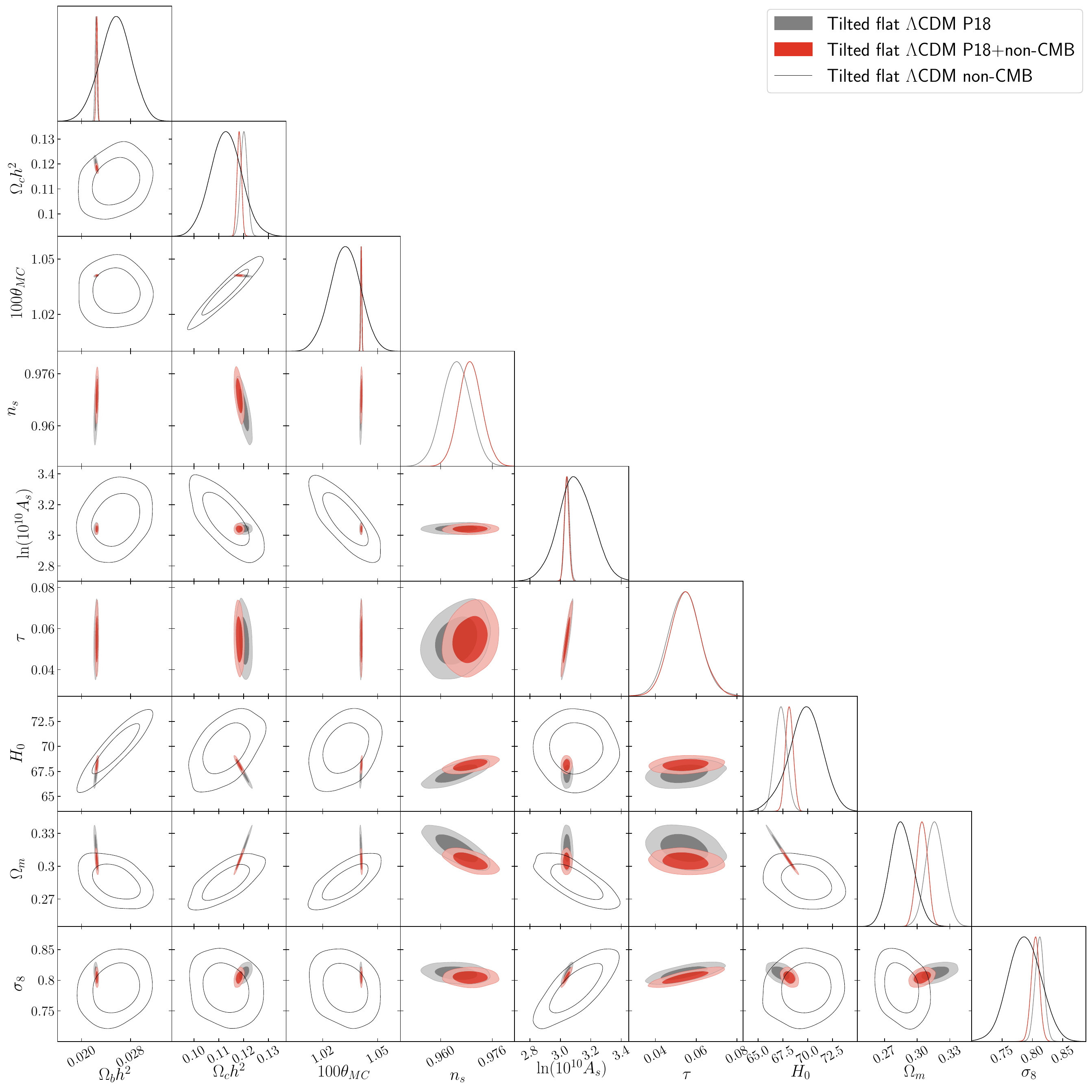}}
\caption{Likelihoods of the tilted flat $\Lambda$CDM model parameters constrained by P18, non-CMB, and P18+non-CMB data sets.
}
\label{fig:like_FL_P18_nonCMB}
\end{figure*}

\begin{figure*}[htbp]
\centering
\mbox{\includegraphics[width=170mm]{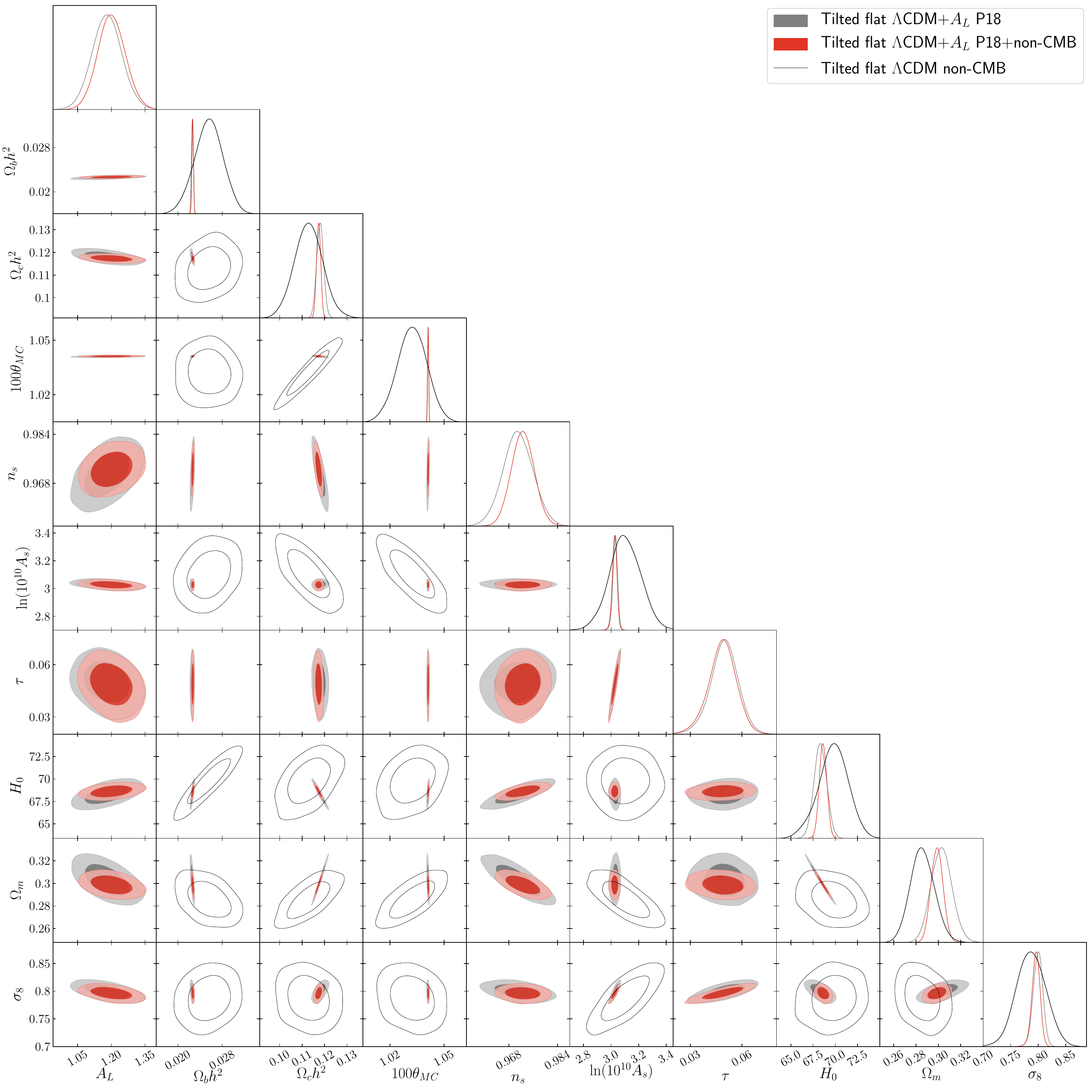}}
\caption{Likelihoods of the tilted flat $\Lambda$CDM$+A_L$ model parameters constrained by P18, non-CMB, and P18+non-CMB data sets. The likelihoods for the non-CMB data set, which do not depend on $A_L$, are the same as in Fig.\ \ref{fig:like_FL_P18_nonCMB}. 
}
\label{fig:like_FL_Alens_P18_nonCMB}
\end{figure*}

\begin{figure*}[htbp]
\centering
\mbox{\includegraphics[width=170mm]{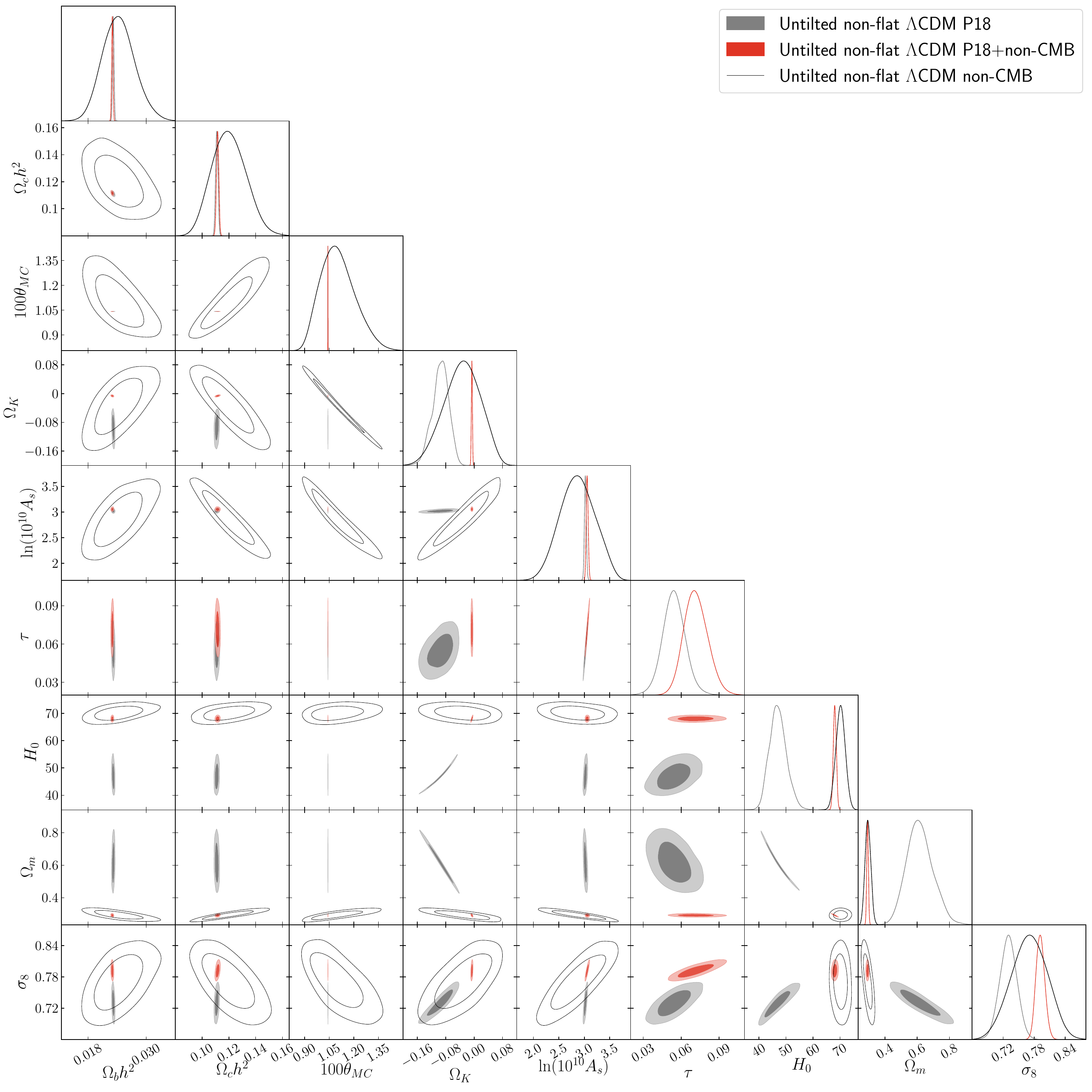}}
\caption{Likelihoods of the untilted non-flat $\Lambda$CDM model parameters constrained by P18, non-CMB, and P18+non-CMB data sets.
}
\label{fig:like_NL_P18_nonCMB}
\end{figure*}

\begin{figure*}[htbp]
\centering
\mbox{\includegraphics[width=170mm]{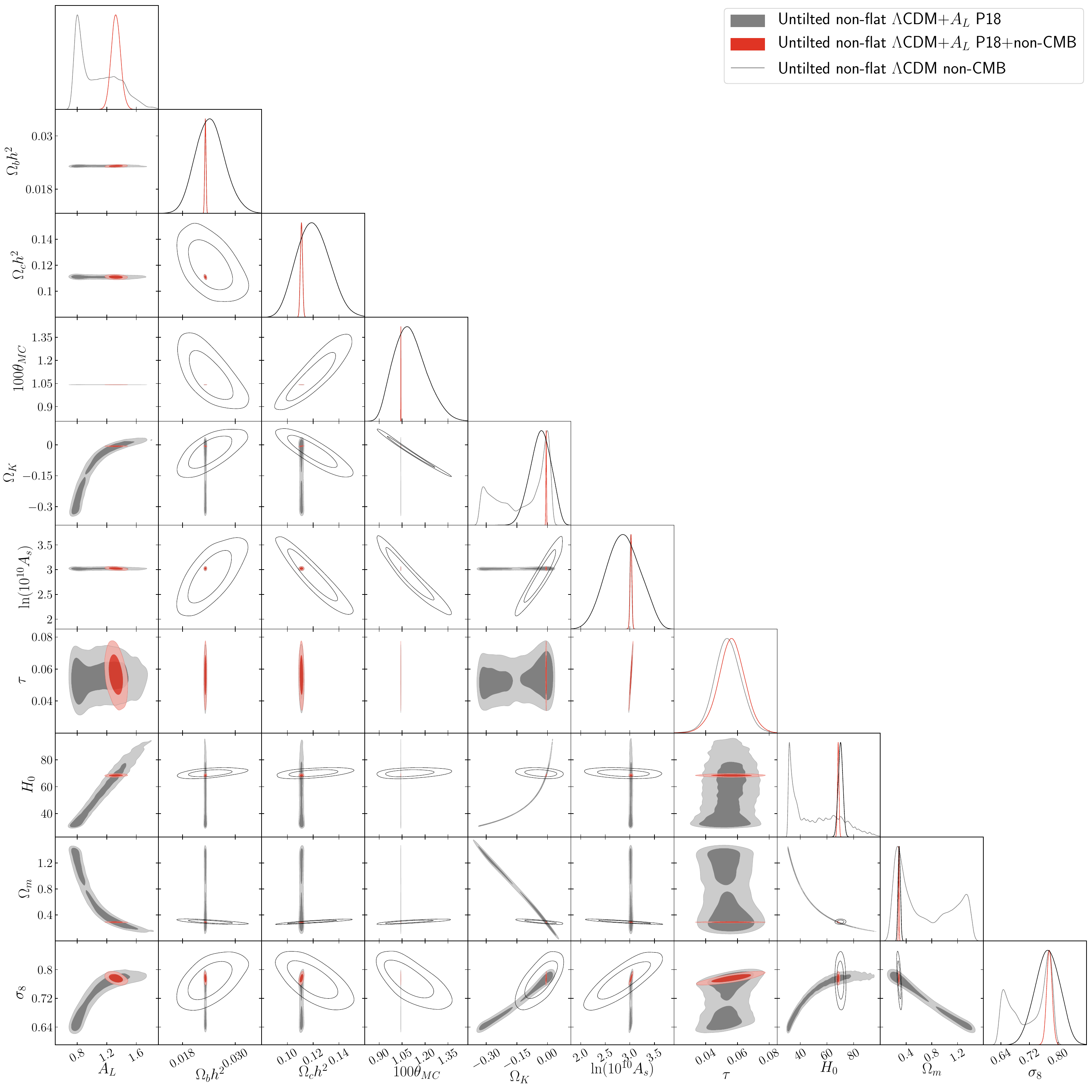}}
\caption{Likelihoods of the untilted non-flat $\Lambda$CDM$+A_L$ model parameters constrained by P18, non-CMB, and P18+non-CMB data sets. The likelihoods for the non-CMB data set, which do not depend on $A_L$, are the same as in Fig.\ \ref{fig:like_NL_P18_nonCMB}.
}
\label{fig:like_NL_Alens_P18_nonCMB}
\end{figure*}

\begin{figure*}[htbp]
\centering
\mbox{\includegraphics[width=170mm]{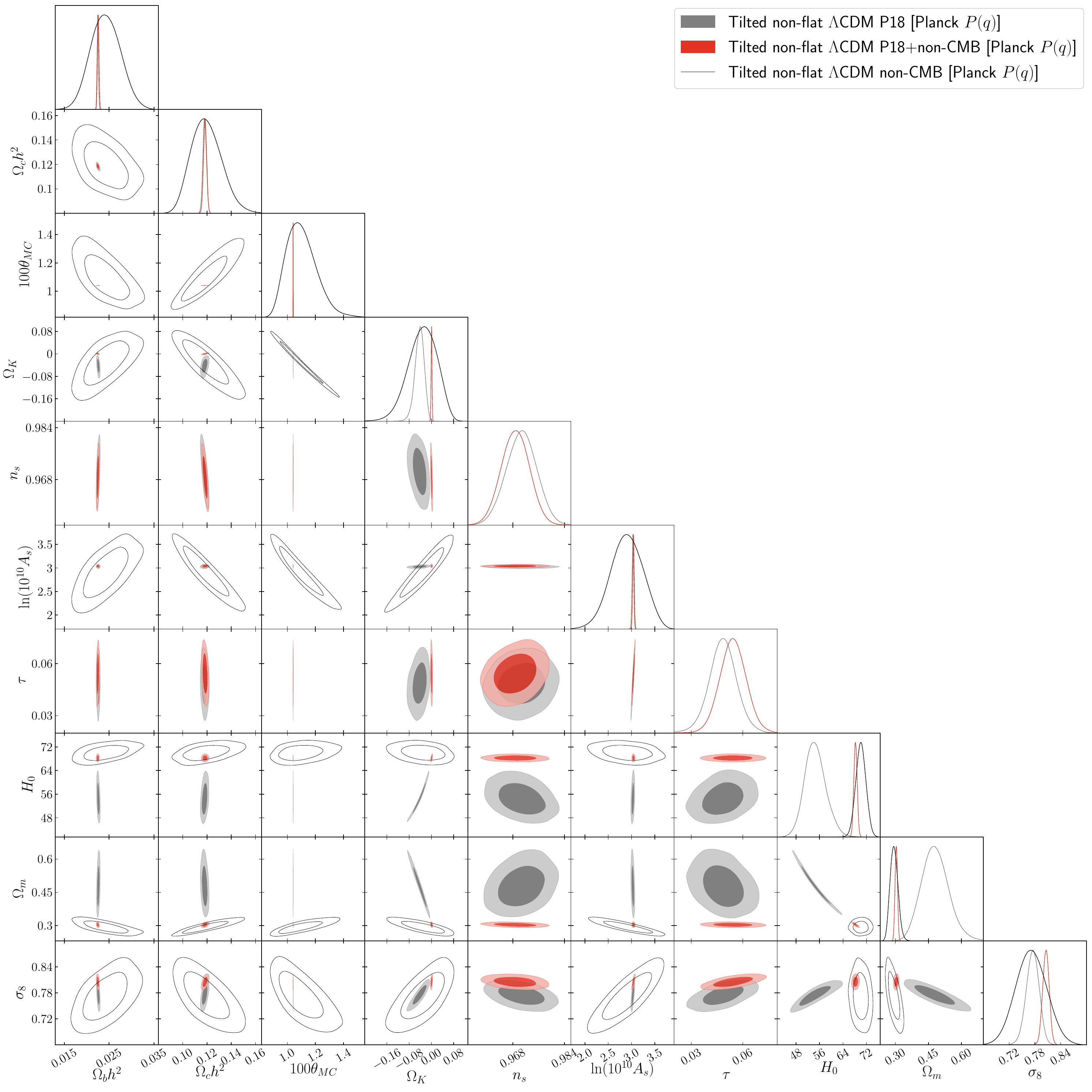}}
\caption{Likelihoods of the tilted non-flat $\Lambda$CDM model [with Planck $P(q)$] parameters constrained by P18, non-CMB, and P18+non-CMB data sets.
}
\label{fig:like_NL_ns_P18_nonCMB}
\end{figure*}

\begin{figure*}[htbp]
\centering
\mbox{\includegraphics[width=170mm]{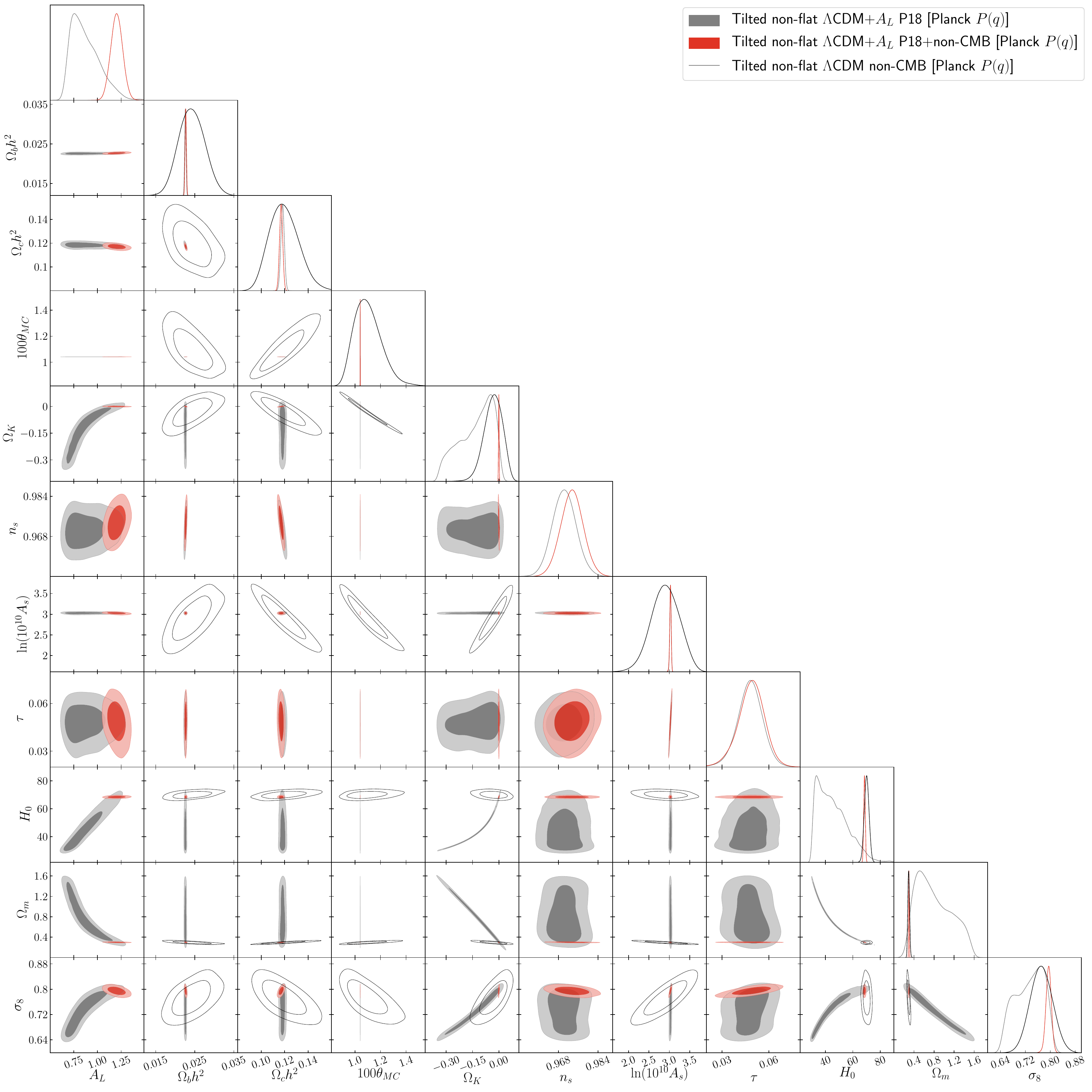}}
\caption{Likelihoods of the tilted non-flat $\Lambda$CDM$+A_L$ model [with Planck $P(q)$] parameters constrained by P18, non-CMB, and P18+non-CMB data sets. The likelihoods for the non-CMB data set, which do not depend on $A_L$, are the same as in Fig.\ \ref{fig:like_NL_ns_P18_nonCMB}.
}
\label{fig:like_NL_Alens_ns_P18_nonCMB}
\end{figure*}

\begin{figure*}[htbp]
\centering
\mbox{\includegraphics[width=170mm]{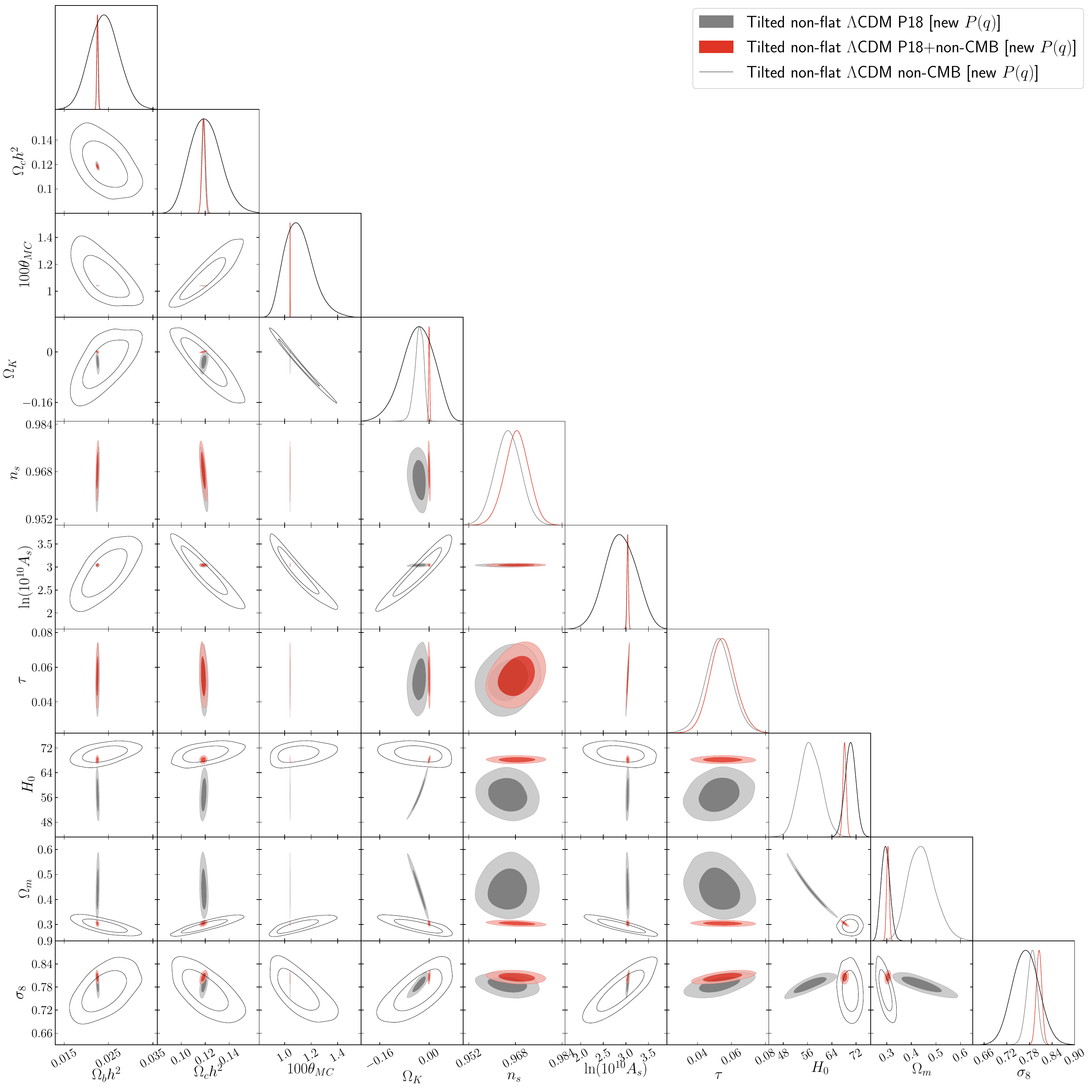}}
\caption{Likelihoods of the tilted non-flat $\Lambda$CDM model [with new $P(q)$] parameters constrained by P18, non-CMB, and P18+non-CMB data sets.
}
\label{fig:like_TNL_P18_nonCMB}
\end{figure*}

\begin{figure*}[htbp]
\centering
\mbox{\includegraphics[width=170mm]{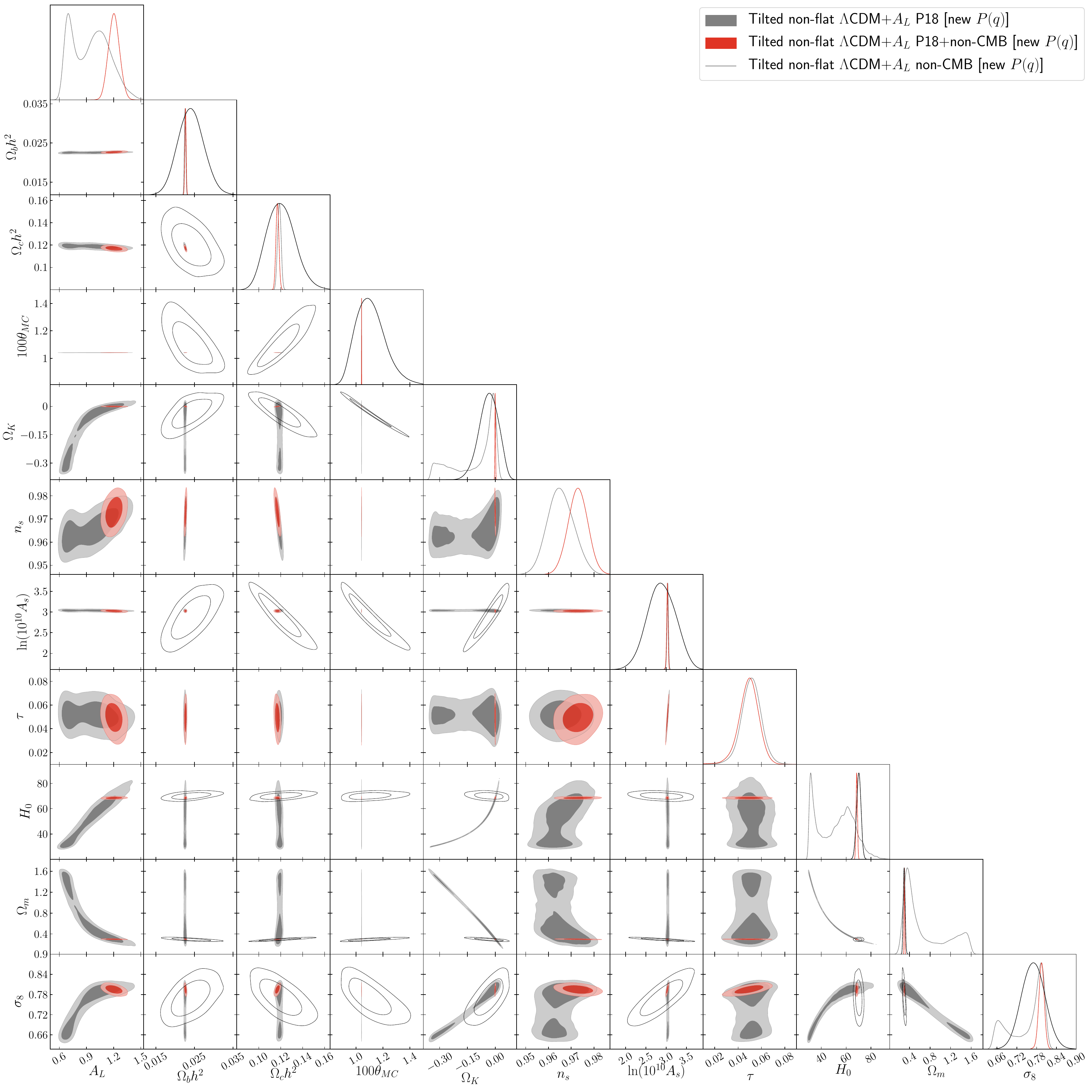}}
\caption{Likelihoods of the tilted non-flat $\Lambda$CDM$+A_L$ model [with new $P(q)$] parameters constrained by P18, non-CMB, and P18+non-CMB data sets. The likelihoods for the non-CMB data set, which do not depend on $A_L$, are the same as in Fig.\ \ref{fig:like_TNL_P18_nonCMB}.
}
\label{fig:like_TNL_Alens_P18_nonCMB}
\end{figure*}

\subsubsection{Comparing P18 data and non-CMB data cosmological constraints}\label{sec:P18_vs_non-CMB}

In the previous subsubsection we compared BAO and BAO$^\prime$ data cosmological constraints to those obtained from P18 data. In the non-flat models with $A_L =1$ there is somewhat significant disagreement between the values of the cosmological parameters (especially the derived parameters $\Omega_m$, $H_0$, and $\sigma_8$) determined using P18 data and those determined from BAO or BAO$^\prime$ data. This disagreement motivates additional tests to decide whether P18 data and BAO$^{\prime}$/BAO data can be used together to constrain parameters of the non-flat models. While both P18 data and BAO$^{\prime}$/BAO data favor negative $\Omega_k$ values, BAO$^{\prime}$/BAO data favor higher values of $H_0$ and lower values of $\Omega_m$ relative to the values obtained in the P18 analysis.  Allowing for a varying $A_L$ parameter resolves these tensions, which may indicate that we can only jointly analyze P18 data and BAO$^{\prime}$/BAO data in the non-flat models when $A_L$ is allowed to vary. 

To further examine these inconsistencies, in this subsubsection we compare non-CMB data (which include BAO as well as BAO$^\prime$ data) cosmological constraints to those obtained from P18 data. (Prior to jointly analyzing P18+non-CMB data, we need to determine whether P18 and non-CMB data cosmological constraints are mutually consistent.)  This allows us to examine how the inclusion of SNIa, $H(z)$, and $f\sigma_8$ data affects the P18 data vs.\ BAO$^\prime$/BAO data conclusions of Sec.\ \ref{sec:P18_vs_BAO} and provides a different, perhaps more expansive, test of the consistency of cosmological parameters obtained from high-redshift data and from low-redshift data.  In Sec.\ \ref{subsec:data_set_tensions} we use two other statistical estimators to examine whether or not P18 and non-CMB data are in tension. 

The cosmological parameter mean values and error bars favored by the P18, non-CMB, and P18+non-CMB data sets are summarized in Tables \ref{tab:para_FL_P18_nonCMB}-\ref{tab:para_TNL_P18_nonCMB} for the tilted flat $\Lambda$CDM (+$A_L$) models, the untilted non-flat $\Lambda$CDM (+$A_L$) models, the tilted non-flat $\Lambda$CDM (+$A_L$) models with the Planck $P(q)$, and the tilted non-flat $\Lambda$CDM ($+A_L$) models with the new $P(q)$, respectively. Likelihood distributions of cosmological parameters of the four models with $A_L=1$ and with $A_L$ varying are shown in Figs.\ 
\ref{fig:like_FL_P18_nonCMB}-\ref{fig:like_TNL_Alens_P18_nonCMB} for the P18, non-CMB, and P18+non-CMB data sets.

Since non-CMB data do not have the ability to constrain $\tau$ or $n_s$, we set their values to those found in the corresponding P18 data analysis. $A_L$ does not affect predictions for the non-CMB measurements we study so we do not include $A_L$ in the non-CMB data analyses. (We saw, in Sec.\ \ref{sec:P18_vs_BAO}, that BAO$^\prime$/BAO data constraints for $A_L = 1$ and for varying $A_L$ were very similar, see Tables \ref{tab:para_FL_BAO}-\ref{tab:para_TNL_ns_BAO}.) 

From Tables \ref{tab:para_NL_P18_nonCMB}-\ref{tab:para_TNL_P18_nonCMB} we see, in the six non-flat $\Lambda$CDM (+$A_L$) models that the constraints set by non-CMB data on $H_0$ and $\Omega_m$ are tighter than the ones imposed by P18 data, and in the three non-flat $\Lambda$CDM+$A_L$ models that the constraints set by non-CMB data on $\Omega_k$ and $\sigma_8$ are tighter than the ones imposed by P18 data. P18 data more restrictively constrain all other parameters in all eight cosmological models.

As we discuss below, there is at least one parameter in the three non-flat models with $A_L = 1$ with a more than 3$\sigma$ level of disagreement between P18 data cosmological constraints and non-CMB data cosmological constraints and one parameter in the tilted flat $\Lambda$CDM model with $A_L = 1$ and in the tilted non-flat $\Lambda$CDM$+A_L$ model with the Planck $P(q)$ with a more than 2$\sigma$ level of disagreement between P18 data cosmological constraints and non-CMB data cosmological constraints. From Tables \ref{tab:para_NL_P18_nonCMB}-\ref{tab:para_TNL_P18_nonCMB} we see that both P18 data and non-CMB data favor negative values of the curvature parameter, with non-CMB data only weakly favoring closed spatial hypersurfaces, at 0.66$\sigma$ to 0.71$\sigma$. However, we should take into account the geometrical degeneracy between $H_0$-$\Omega_k$-$\Omega_m$ and note that, like both BAO$^\prime$ and BAO data, non-CMB data favor higher values of $H_0$ and lower values of $\Omega_m$ than do P18 data and this is what causes the P18 and non-CMB cosmological constraint differences. 

The dominant component of non-CMB data is BAO$^\prime$/BAO data. This is why the cosmological constraints obtained from BAO$^\prime$/BAO data are similar to the ones obtained from the complete non-CMB low-redshift data set. However, there are some differences between these sets of constraints that are worth mentioning. As expected, the error bars obtained considering non-CMB data are smaller than the ones from BAO$^\prime$/BAO data. While similar values for $\Omega_m$ are found in both cases, the values of $H_0$ favored by non-CMB data are $\sim 1\sigma$ smaller than those favored by BAO$^\prime$/BAO data. BAO$^\prime$ data favor closed spatial hypersurfaces at 0.48$\sigma$ to 0.60$\sigma$ while BAO data favor them by 0.71$\sigma$ to 0.96$\sigma$, which are on either side of the 0.66$\sigma$ to 0.71$\sigma$ favoring of closed spatial hypersurfaces from non-CMB data.  
We also find smaller values for the $\sigma_8$ parameter when non-CMB data are considered, with BAO$^\prime$ data favoring 1.1$\sigma$-1.3$\sigma$ larger values while BAO data favor $\sim 1.3 \sigma$ larger values in the non-flat models and a 1.9$\sigma$ larger value in the tilted flat $\Lambda$CDM case. This might be because the non-CMB data set contain additional $f\sigma_8$ data points that favor lower values of $\sigma_8$ than those in the BAO data set.   

Comparing the six-parameter and the four-parameter tilted flat $\Lambda$CDM model primary cosmological parameter constraints for P18 and non-CMB data, shown in the left half of Table \ref{tab:para_FL_P18_nonCMB}, we see that the values of $\Omega_b h^2$, $\Omega_c h^2$, and $\theta_\textrm{MC}$ are in mild disagreement, at 1.3$\sigma$, 1.1$\sigma$, and 1.0$\sigma$, respectively. We also observe a more significant 2.2$\sigma$ level of tension in the derived $\Omega_m$ values,  the derived $H_0$ values differ by 1.4$\sigma$, and $\sigma_8$ values show a better agreement, disagreeing by only 0.89$\sigma$. 

Comparing the seven-parameter and the four-parameter tilted flat $\Lambda$CDM+$A_L$ model primary cosmological parameter constraints for P18 and non-CMB data, shown in Table \ref{tab:para_FL_P18_nonCMB}, we see that the values of $\Omega_b h^2$ and $\theta_{\textrm{MC}}$ are in 1.2$\sigma$ and 1.1$\sigma$ tension respectively. As for the derived parameters, we find $\Omega_m$ values differ by 1.2$\sigma$ while $H_0$ and $\sigma_8$ values are in only 0.81$\sigma$ and 0.45$\sigma$ disagreement. So unlike in the BAO data and the BAO$^\prime$ data comparisons with P18 data of Sec.\ \ref{sec:P18_vs_BAO}, the inclusion of a varying $A_L$ reduces the disagreement for all three derived parameters, but less successfully for $\Omega_m$ in the non-CMB case here compared to the BAO/BAO$^\prime$ cases there. 

P18 and non-CMB data results obtained for the six-parameter and the four-parameter untilted non-flat $\Lambda$CDM model, shown in the left half of Table \ref{tab:para_NL_P18_nonCMB}, indicate mostly less significant differences in primary parameters but more significant differences in derived parameters than found in the tilted flat $\Lambda$CDM model. The primary spatial curvature parameter value is $\Omega_k=-0.033\pm 0.050$ for non-CMB data, which is 0.66$\sigma$ away from flat and in 1.1$\sigma$ tension with the P18 value $\Omega_k=-0.095\pm 0.024$, which is 4.0$\sigma$ away from flat. Regarding the derived parameters, we find that $H_0$, $\Omega_m$, and $\sigma_8$ values are in 6.4$\sigma$, 3.8$\sigma$, and 1.1$\sigma$ disagreement. These results probably mean that P18 and non-CMB data should not be jointly analyzed in the context of the untilted non-flat $\Lambda$CDM model. 

The results for the seven-parameter and the four-parameter untilted non-flat $\Lambda$CDM+$A_L$ model, obtained considering P18 and non-CMB data, are in Table \ref{tab:para_NL_P18_nonCMB}. There is a slight increase in the disagreement between the values of the primary spatial curvature parameter $\Omega_k$  (now 0.67$\sigma$) and decreases for the derived parameters $H_0$, $\Omega_m$, and $\sigma_8$, that now disagree by  1.0$\sigma$, 0.97$\sigma$, and 0.79$\sigma$ respectively. This is caused by the larger error bars in the $A_L$-varying P18 case compared to the corresponding values obtained with $A_L=1$. According to these results, unlike in the $A_L=1$ case, in the $A_L$-varying case P18 and non-CMB data can probably be jointly analyzed in the context of the untilted non-flat $\Lambda$CDM model. Note that in this case a joint analysis of P18+non-CMB data favors closed geometry at 4.4$\sigma$, with $\Omega_k=-0.0062\pm0.0014$, although because of the lack of the tilt ($n_s$) degree of freedom this untilted non-flat $\Lambda$CDM+$A_L$ model does not provide a good fit to smaller-angular-scale P18 data, which is reflected in the large $\Delta$DIC and $\Delta$AIC$_c$ values for the P18+non-CMB case in the lower half of Table \ref{tab:para_NL_P18_nonCMB}.

Comparing the seven-parameter and the five-parameter tilted non-flat $\Lambda$CDM Planck $P(q)$ model primary cosmological parameter constraints for P18 and non-CMB data, we see, in the left half of Table \ref{tab:para_NL_ns_P18_nonCMB}, that the primary parameter values do not much disagree. The non-CMB data primary spatial curvature parameter value $\Omega_k=-0.032\pm 0.051$ is 0.63$\sigma$ away from flat and only in 0.20$\sigma$ tension with the P18 value $\Omega_k=-0.043\pm0.017$, which is 2.5$\sigma$ in favor of closed geometry. The derived parameters $H_0$, $\Omega_m$, and $\sigma_8$ are in 3.9$\sigma$, 2.9$\sigma$, and 0.11$\sigma$ tension. These results show that P18 and non-CMB data cosmological constraints are inconsistent in the tilted non-flat $\Lambda$CDM Planck $P(q)$ model and these data probably should not be used jointly to constrain this model.

Looking at Table \ref{tab:para_NL_ns_P18_nonCMB} we can compare results obtained for the eight-parameter and the five-parameter tilted non-flat $\Lambda$CDM+$A_L$ Planck $P(q)$ model from P18 and non-CMB data respectively. Aside from $\Omega_k$, the primary parameter disagreements do not change much compared to the $A_L=1$ case. For the non-CMB data primary spatial curvature parameter we have $\Omega_k= -0.032\pm 0.051$, which is 0.63$\sigma$ away from flat and in 0.91$\sigma$ tension with the P18 value $-0.130\pm0.095$, which is 1.4$\sigma$ away from flat. Regarding the derived parameters we find that $H_0$, $\Omega_m$, and $\sigma_8$ are in 2.3$\sigma$, 1.4$\sigma$ and 0.67$\sigma$ disagreement. Compared to the $A_L = 1$ case, in the $A_L$-varying case we find significant reductions in the $H_0$ and $\Omega_m$ tensions, with both disagreements still being significant, which suggest that P18 and non-CMB data should not be jointly analyzed within the tilted non-flat $\Lambda$CDM+$A_L$ Planck $P(q)$ model.

Comparing the seven-parameter and the five-parameter tilted non-flat $\Lambda$CDM new $P(q)$ model primary cosmological parameter constraints for P18 and non-CMB data, from the left half of Table \ref{tab:para_TNL_P18_nonCMB} we see that the primary parameter values do not much disagree. The non-CMB data primary spatial curvature parameter value is $\Omega_k=-0.036\pm 0.051$, which is only a 0.71$\sigma$ deviation from flat and, similar to the Planck $P(q)$ model, is only in 0.057$\sigma$ disagreement with the P18 value $-0.033\pm 0.014$, which is 2.4$\sigma$ away from flat. Regarding the derived parameters $H_0$, $\Omega_m$, and $\sigma_8$, we find that their values disagree at 3.3$\sigma$, 2.6$\sigma$, and 0.42$\sigma$ respectively. While the $H_0$ and $\Omega_m$ disagreements are a little smaller than the ones found in the tilted non-flat $\Lambda$CDM Planck $P(q)$ model, they still are large enough to require we more carefully test whether P18 and non-CMB data can be jointly used to constrain cosmological parameters in this model. 

The results for the eight-parameter and the five-parameter tilted non-flat $\Lambda$CDM+$A_L$ new $P(q)$ model are in Table \ref{tab:para_TNL_P18_nonCMB}, for P18 and non-CMB data, respectively. As happens in the Planck $P(q)$ model, when the $A_L$ parameter is allowed to vary the mild tensions found for the primary parameters, except for $\Omega_k$, do not change much compared to the $A_L=1$ case. For the non-CMB data primary spatial curvature parameter we have $\Omega_k= -0.036\pm 0.051$, which is 0.71$\sigma$ away from flat hypersurfaces and now in 0.53$\sigma$ tension with the P18 value $\Omega_k=-0.10\pm 0.11$, which is 0.91$\sigma$ away from flat. As for the value of the derived parameters $H_0$, $\Omega_m$, and $\sigma_8$ we find disagreements at 1.4$\sigma$, 0.94$\sigma$, and 0.29$\sigma$ respectively. The tensions are reduced with respect to the case with $A_L=1$, due to the increase of the error bars, but possibly the $H_0$ tension is still not small enough to allow the joint use of P18+non-CMB data for constraining tilted non-flat $\Lambda$CDM+$A_L$ new $P(q)$ model cosmological parameters. 

Figures \ref{fig:like_FL_P18_nonCMB}-\ref{fig:like_TNL_Alens_P18_nonCMB} show one-dimensional likelihoods and two-dimensional contours for cosmological parameters obtained using P18,  non-CMB, and P18+non-CMB data. As mentioned above, non-CMB data constraints (shown with unfilled black lines) are comparatively less restrictive than P18 constraints (shown in grey), are unable to put tight constraints on the primary cosmological parameters (except on $\Omega_k$ in the three non-flat $\Lambda$CDM$+A_L$ models), and in many cases they at least partially overlap with the P18 data constraints. 

Figures \ref{fig:like_FL_P18_nonCMB} and \ref{fig:like_FL_Alens_P18_nonCMB} are for tilted flat $\Lambda$CDM (+$A_L$) models. The $\sim 1 \sigma$ disagreements between the non-CMB constraints and those obtained with P18 data, discussed above, can be clearly seen in the contour plots. For the tilted flat $\Lambda$CDM model the larger disagreements are in panels for derived cosmological parameters, with the largest for $\Omega_m$ and the next largest for $H_0$. These disagreements decrease when the $A_L$ parameter is allowed to vary.

Looking at the contour plots for the untilted non-flat $\Lambda$CDM (+$A_L$) models (see Figs.\ \ref{fig:like_NL_P18_nonCMB} and \ref{fig:like_NL_Alens_P18_nonCMB}) we observe non-overlapping contours in those panels that involve the derived parameters $H_0$ and $\Omega_m$ or the primary parameter $\Omega_k$, especially in the $\Omega_k$-$\theta_{\rm MC}$ subpanel of Fig.\ \ref{fig:like_NL_P18_nonCMB}. These disagreements largely disappear when $A_L$ is allowed to vary, except perhaps for the $H_0$ one. This may indicate that in the context of this cosmological model we may jointly analyze P18 data with non-CMB data only when $A_L$ is allowed to vary. 

Figures \ref{fig:like_NL_ns_P18_nonCMB} and \ref{fig:like_NL_Alens_ns_P18_nonCMB} show cosmological parameter constraints for the tilted non-flat $\Lambda$CDM (+$A_L$) Planck $P(q)$ models, while the ones for the tilted non-flat $\Lambda$CDM(+$A_L$) new $P(q)$ models are displayed in Figs.\ \ref{fig:like_TNL_P18_nonCMB} and \ref{fig:like_TNL_Alens_P18_nonCMB}. As expected, considering the results discussed above in this subsubsection, the contour plots for these tilted non-flat models are quite similar. We see in the panels that involve the primary cosmological parameters that there is overlap at 1$\sigma$, not only when $A_L$ is allowed to vary but also when $A_L=1$. When $A_L=1$, for the Planck $P(q)$ model, P18 and non-CMB data constraint contours that involve $H_0$ and $\Omega_m$ do not overlap even at 2$\sigma$. These disagreements are less severe for the new $P(q)$ model with $A_L=1$, where overlap is reached in most cases at a little over $2 \sigma$. 

In view of the results discussed in this subsubsection, further tests are needed to properly quantify the level of disagreement, in the context of non-flat models, between P18 data and non-CMB data cosmological constraints. We return to this issue in Sec.\ \ref{subsec:data_set_tensions}.

\begin{table*}
\caption{Mean and 68.3\% confidence limits of tilted flat $\Lambda\textrm{CDM}$ (+$A_L$) model parameters
        constrained by non-CMB, P18+lensing, and P18+lensing+non-CMB data sets.
        The Hubble constant $H_0$ has a unit of km s$^{-1}$ Mpc$^{-1}$.
}
\begin{ruledtabular}
\begin{tabular}{lccccc}
\\[-1mm]                        
                                 & \multicolumn{3}{c}{Tilted flat $\Lambda$CDM}  & \multicolumn{2}{c}{Tilted flat $\Lambda$CDM$+A_L$}      \\[+1mm]
\cline{2-4}\cline{5-6}\\[-1mm]
Parameter                      & Non-CMB   & P18+lensing  &  P18+lensing+non-CMB    & P18+lensing & P18+lensing+non-CMB  \\[+1mm]
\hline \\[-1mm]
$\Omega_b h^2$                 & $0.0257 \pm 0.0026$ & $0.02237 \pm 0.00014$ & $0.02250 \pm 0.00013$  & $0.02251 \pm 0.00017$ & $0.02258 \pm 0.00014$          \\[+1mm]
$\Omega_c h^2$                 & $0.1128 \pm 0.0061$ & $0.1200 \pm 0.0012$   & $0.11838 \pm 0.00083$  & $0.1183 \pm 0.0015$   & $0.11747 \pm 0.00091$   \\[+1mm]
$100\theta_\textrm{MC}$        & $1.0321 \pm 0.0080$ & $1.04091 \pm 0.00031$ & $1.04110 \pm 0.00029$  & $1.04109 \pm 0.00032$ & $1.04118 \pm 0.00029$     \\[+1mm]
$\tau$                         & $0.0543$            & $0.0543 \pm 0.0073$   & $0.0569 \pm 0.0071$    & $0.0487 \pm 0.0087$   & $0.0476 \pm 0.0085$   \\[+1mm]
$n_s$                          & $0.9649$            & $0.9649 \pm 0.0041$   & $0.9688 \pm 0.0036$    & $0.9695 \pm 0.0048$   & $0.9715 \pm 0.0038$   \\[+1mm]
$\ln(10^{10} A_s)$             & $3.10 \pm 0.11$     & $3.044 \pm 0.014$     & $3.046 \pm 0.014$      & $3.028 \pm 0.018$     & $3.023 \pm 0.018$  \\[+1mm]
$A_{L}$                        & $\cdots$            & $\cdots$              & $\cdots$               & $1.073 \pm 0.041$     & $1.089 \pm 0.035$  \\[+1mm]
  \hline \\[-1mm]
$H_0$                          & $69.9 \pm 1.7$      & $67.34 \pm 0.55$      & $68.09 \pm 0.38$       & $68.14 \pm 0.69$      & $68.52 \pm 0.42$ \\[+1mm]
$\Omega_m$                     & $0.285 \pm 0.011$   & $0.3155 \pm 0.0075$   & $0.3053 \pm 0.0050$    & $0.3048 \pm 0.0091$   & $0.2998 \pm 0.0053$    \\[+1mm]
$\sigma_8$                     & $0.787 \pm 0.026$   & $0.8112 \pm 0.0059$   & $0.8072 \pm 0.0058$    & $0.7996 \pm 0.0089$   & $0.7955 \pm 0.0075$   \\[+1mm]
 \hline \\[-1mm]
$\chi_{\textrm{min}}^2$ (Total)   & $1106.55$        & $2774.71$             & $3888.41$              & $2771.24$             & $3881.55$       \\[+1mm]
$\chi_{\textrm{min}}^2$ (Non-CMB) & $1106.55$        & $\cdots$              & $1112.05$              & $\cdots$              & $1109.64$    \\[+1mm]
$\textrm{DIC}$                    & $1114.38$        & $2826.45$             & $3940.70$              & $2825.53$             & $3935.15$ \\[+1mm]
$\Delta\textrm{DIC}$              & $\cdots$         & $\cdots$              & $\cdots$               & $-0.92$               & $-5.55$          \\[+1mm]
$\textrm{AIC}_c$                  & $1114.55$        & $2828.71$             & $3942.41$              & $2827.24$             & $3937.55$   \\[+1mm]
$\Delta\textrm{AIC}_c$            & $\cdots$         & $\cdots$              & $\cdots$               & $-1.47$               & $-4.86$          \\[+1mm]
\end{tabular}
\\[+1mm]
\begin{flushleft}
Note: $\Delta\textrm{DIC}$ ($\Delta\textrm{AIC}_c$) indicates an excess value relative to that of the tilted flat $\Lambda$CDM model constrained with the same data. 
\end{flushleft}
\end{ruledtabular}
\label{tab:para_FL_P18_lensing_nonCMB}
\end{table*}

\begin{table*}
\caption{Mean and 68.3\% confidence limits of untilted non-flat $\Lambda\textrm{CDM}$ (+$A_L$) model parameters
        constrained by non-CMB, P18+lensing, and P18+lensing+non-CMB data sets.
        The Hubble constant $H_0$ has a unit of km s$^{-1}$ Mpc$^{-1}$.
}
\begin{ruledtabular}
\begin{tabular}{lccccc}
\\[-1mm]                        
                                 & \multicolumn{3}{c}{Untilted non-flat $\Lambda$CDM}  & \multicolumn{2}{c}{Untilted non-flat $\Lambda$CDM$+A_L$}      \\[+1mm]
\cline{2-4}\cline{5-6}\\[-1mm]
Parameter                        & Non-CMB   & P18+lensing  &  P18+lensing+non-CMB    & P18+lensing  & P18+lensing+non-CMB  \\[+1mm]
\hline \\[-1mm]
$\Omega_b h^2$                 & $0.0241 \pm 0.0033$ & $0.02307 \pm 0.00014$   & $0.02301 \pm 0.00014$  & $0.02312 \pm 0.00014$ & $0.02310 \pm 0.00014$           \\[+1mm]
$\Omega_c h^2$                 & $0.121 \pm 0.013$   & $0.11108 \pm 0.00086$   & $0.11176 \pm 0.00083$  & $0.11092 \pm 0.00087$ & $0.11100 \pm 0.00085$ \\[+1mm]
$100\theta_\textrm{MC}$        & $1.11 \pm 0.11$     & $1.04196 \pm 0.00029$   & $1.04189 \pm 0.00029$  & $1.04193 \pm 0.00029$ & $1.04195 \pm 0.00030$       \\[+1mm]
$\tau$                         & $0.0580$            & $0.0580 \pm 0.0087$     & $0.0799 \pm 0.0089$    & $0.0554 \pm 0.0097$   & $0.0566 \pm 0.0083$   \\[+1mm]
$\Omega_k$                     & $-0.037 \pm 0.050$  & $-0.0322 \pm 0.0075$    & $-0.0065 \pm 0.0014$   & $0.0161 \pm 0.0094$   & $-0.0060 \pm 0.0014$    \\[+1mm]
$\ln(10^{10} A_s)$             & $2.84 \pm 0.34$     & $3.027 \pm 0.018$       & $3.075 \pm 0.018$      & $3.021 \pm 0.020$     & $3.024 \pm 0.017$ \\[+1mm]
$A_{L}$                        & $\cdots$            & $\cdots$                & $\cdots$               & $1.44 \pm 0.15$       & $1.162 \pm 0.036$          \\[+1mm]
  \hline \\[-1mm]
$H_0$                          & $70.2 \pm 1.7$      & $58.9 \pm 2.1$          & $67.90 \pm 0.56$       & $85.7 \pm 8.5$        & $68.48 \pm 0.58$ \\[+1mm]
$\Omega_m$                     & $0.295 \pm 0.018$   & $0.390 \pm 0.027$       & $0.2938 \pm 0.0049$    & $0.190 \pm 0.043$     & $0.2874 \pm 0.0050$ \\[+1mm]
$\sigma_8$                     & $0.769 \pm 0.035$   & $0.765 \pm 0.011$       & $0.7997 \pm 0.0076$    & $0.7805 \pm 0.0094$   & $0.7764 \pm 0.0078$  \\[+1mm]
 \hline \\[-1mm]
$\chi_{\textrm{min}}^2$ (Total)&           $1106.51$  & $2813.13$                &  $3938.22$              & $2807.91$              & $3915.05$    \\[+1mm]
$\chi_{\textrm{min}}^2$ (Non-CMB) &        $1106.51$  & $\cdots$                &  $1108.60$              & $\cdots$              & $1107.39$  \\[+1mm]
$\textrm{DIC}$                 &           $1117.24$  & $2869.06$                &  $3992.71$              & $2856.10$              & $3973.55$   \\[+1mm]
$\Delta\textrm{DIC}$           &           $2.86$    & $42.61$                 &  $52.01$               & $29.65$               & $32.85$        \\[+1mm]
$\textrm{AIC}_c$                 &           $1116.51$  & $2867.13$                &  $3992.22$              & $2863.91$              & $3971.05$        \\[+1mm]
$\Delta\textrm{AIC}_c$           &           $1.96$    & $38.42$                 &  $49.81$               & $35.20$               & $28.64$        \\[+1mm]
\end{tabular}
\\[+1mm]
\begin{flushleft}
Note: $\Delta\textrm{DIC}$ ($\Delta\textrm{AIC}_c$) indicates an excess value relative to that of the tilted flat $\Lambda$CDM model constrained with the same data. 
\end{flushleft}
\end{ruledtabular}
\label{tab:para_NL_P18_lensing_nonCMB}
\end{table*}

\begin{table*}
\caption{Mean and 68.3\% confidence limits of Planck-$P(q)$-based tilted nonflat $\Lambda\textrm{CDM}$ ($+A_L$) model parameters
        constrained by non-CMB, P18+lensing, and P18+lensing+non-CMB data sets.
        The Hubble constant $H_0$ has a unit of km s$^{-1}$ Mpc$^{-1}$.
}
\begin{ruledtabular}
\begin{tabular}{lccccc}
\\[-1mm]                         
                                 & \multicolumn{3}{c}{Tilted non-flat $\Lambda$CDM Planck $P(q)$}  & \multicolumn{2}{c}{Tilted non-flat $\Lambda$CDM$+A_L$ Planck $P(q)$}      \\[+1mm]
\cline{2-4}\cline{5-6}\\[-1mm]
Parameter                        & Non-CMB  & P18+lensing  &  P18+lensing+non-CMB       & P18+lensing     & P18+lensing+non-CMB  \\[+1mm]
 \hline \\[-1mm]
$\Omega_b h^2$                 & $0.0242 \pm 0.0033$ & $0.02249 \pm 0.00016$ & $0.02249 \pm 0.00015$  & $0.02251 \pm 0.00017$ & $0.02259 \pm 0.00016$       \\[+1mm]
$\Omega_c h^2$                 & $0.120 \pm 0.013$   & $0.1186 \pm 0.0015$   & $0.1187 \pm 0.0013$    & $0.1183 \pm 0.0015$   & $0.1173 \pm 0.0014$         \\[+1mm]
$100\theta_\textrm{MC}$        & $1.10 \pm 0.10$     & $1.04107 \pm 0.00032$ & $1.04106 \pm 0.00031$  & $1.04110 \pm 0.00032$ & $1.04118 \pm 0.00032$ \\[+1mm]
$\tau$                         & $0.0495$            & $0.0495 \pm 0.0082$   & $0.0563 \pm 0.0073$    & $0.0489 \pm 0.0085$   & $0.0479 \pm 0.0085$              \\[+1mm]
$\Omega_k$                     & $-0.033 \pm 0.050$  & $-0.0103 \pm 0.0066$  & $0.0004 \pm 0.0017$    & $-0.005 \pm 0.027$    & $-0.0002 \pm 0.0017$            \\[+1mm]
$n_s$                          & $0.9687$            & $0.9687 \pm 0.0046$   & $0.9681 \pm 0.0044$    & $0.9696 \pm 0.0049$   & $0.9718 \pm 0.0045$   \\[+1mm]
$\ln(10^{10} A_s)$             & $2.90 \pm 0.34$     & $3.030 \pm 0.017$     & $3.046 \pm 0.014$      & $3.028 \pm 0.018$     & $3.024 \pm 0.017$    \\[+1mm]
$A_{L}$                        & $\cdots$            & $\cdots$              & $\cdots$               & $1.09 \pm 0.16$       & $1.090 \pm 0.036$              \\[+1mm]
 \hline \\[-1mm]
$H_0$                          & $70.1 \pm 1.8$      & $63.7 \pm 2.3$        & $68.17 \pm 0.55$       & $69 \pm 11$           & $68.49 \pm 0.56$                          \\[+1mm]
$\Omega_m$                     & $0.294 \pm 0.018$   & $0.351 \pm 0.024$     & $0.3051 \pm 0.0053$    & $0.32 \pm 0.11$       & $0.2998 \pm 0.0055$   \\[+1mm]
$\sigma_8$                     & $0.771 \pm 0.036$   & $0.796 \pm 0.011$     & $0.8080 \pm 0.0066$    & $0.796 \pm 0.016$     & $0.7952 \pm 0.0085$  \\[+1mm]
 \hline \\[-1mm]
$\chi_{\textrm{min}}^2$ (Total)     & $1106.51$      & $2771.53$             &  $3887.99$             & $2771.14$             & $3881.37$ \\[+1mm]
$\chi_{\textrm{min}}^2$ (Non-CMB)   & $1106.51$      & $\cdots$              &  $1111.94$             & $\cdots$              & $1110.31$ \\[+1mm]
$\textrm{DIC}$                      & $1117.27$      & $2826.17$             &  $3942.07$             & $2827.14$             & $3936.85$ \\[+1mm]
$\Delta\textrm{DIC}$                & $2.89$         & $-0.28$               &  $1.37$                & $0.69$                & $-3.85$       \\[+1mm]
$\textrm{AIC}_c$                    & $1116.51$      & $2827.53$             &  $3943.99$             & $2829.14$             & $3939.37$ \\[+1mm]
$\Delta\textrm{AIC}_c$              & $1.96$         & $-1.18$               &  $1.58$                & $0.43$                & $-3.04$       \\[+1mm]
\end{tabular}
\\[+1mm]
\begin{flushleft}
Note: $\Delta\textrm{DIC}$ ($\Delta\textrm{AIC}_c$) indicates an excess value relative to that of the tilted flat $\Lambda$CDM model constrained with the same data. 
\end{flushleft}
\end{ruledtabular}
\label{tab:para_NL_ns_P18_lensing_nonCMB}
\end{table*}

\begin{table*}
\caption{Mean and 68.3\% confidence limits of new-$P(q)$-based tilted nonflat $\Lambda\textrm{CDM}$ ($+A_L$) model parameters
        constrained by non-CMB, P18+lensing, and P18+lensing+non-CMB data sets.
        The Hubble constant $H_0$ has a unit of km s$^{-1}$ Mpc$^{-1}$.
}
\begin{ruledtabular}
\begin{tabular}{lccccc}

\\[-1mm]                                 & \multicolumn{3}{c}{Tilted non-flat $\Lambda$CDM new $P(q)$}  & \multicolumn{2}{c}{Tilted non-flat $\Lambda$CDM$+A_L$ new $P(q)$}      \\[+1mm]
\cline{2-4}\cline{5-6}\\[-1mm]
Parameter                        & Non-CMB  & P18+lensing  &  P18+lensing+non-CMB    & P18+lensing  & P18+lensing+non-CMB  \\[+1mm]
 \hline \\[-1mm]
$\Omega_b h^2$                 & $0.0242 \pm 0.0033$ & $0.02248 \pm 0.00016$  & $0.02248 \pm 0.00015$  & $0.02252 \pm 0.00017$ & $0.02260 \pm 0.00016$  \\[+1mm]
$\Omega_c h^2$                 & $0.120 \pm 0.013$   & $0.1188 \pm 0.0014$    & $0.1186 \pm 0.0013$    & $0.1183 \pm 0.0015$   & $0.1174 \pm 0.0013$       \\[+1mm]
$100\theta_\textrm{MC}$        & $1.10 \pm 0.10$     & $1.04104 \pm 0.00032$  & $1.04106 \pm 0.00031$  & $1.04108 \pm 0.00032$ & $1.04118 \pm 0.00032$\\[+1mm]
$\tau$                         & $0.0515$            & $0.0515 \pm 0.0081$    & $0.0566 \pm 0.0074$    & $0.0495 \pm 0.0093$   & $0.0486 \pm 0.0086$       \\[+1mm]
$\Omega_k$                     & $-0.033 \pm 0.050$  & $-0.0086 \pm 0.0057$   & $0.0003 \pm 0.0017$    & $0.003 \pm 0.016$     & $-0.0002 \pm 0.0017$\\[+1mm]
$n_s$                          & $0.9654$            & $0.9661 \pm 0.0043$    & $0.9679 \pm 0.0042$    & $0.9688 \pm 0.0053$   & $0.9713 \pm 0.0042$  \\[+1mm]
$\ln(10^{10} A_s)$             & $2.89 \pm 0.34$     & $3.035 \pm 0.016$      & $3.046 \pm 0.014$      & $3.030 \pm 0.019$     & $3.025 \pm 0.017$          \\[+1mm]
$A_{L}$                        & $\cdots$            & $\cdots$               & $\cdots$               & $1.13 \pm 0.15$       & $1.088 \pm 0.035$          \\[+1mm]
 \hline \\[-1mm]
$H_0$                          & $70.1 \pm 1.8$      & $64.2 \pm 2.0$         & $68.13 \pm 0.54$       & $72.0 \pm 9.2$        & $68.48 \pm 0.56$     \\[+1mm]
$\Omega_m$                     & $0.295 \pm 0.017$   & $0.345 \pm 0.021$      & $0.3054 \pm 0.0051$    & $0.287 \pm 0.076$     & $0.2999 \pm 0.0055$  \\[+1mm]
$\sigma_8$                     & $0.771 \pm 0.036$   & $0.799 \pm 0.010$      & $0.8079 \pm 0.0067$    & $0.801 \pm 0.011$     & $0.7956 \pm 0.0082$           \\[+1mm]
 \hline \\[-1mm]
$\chi_{\textrm{min}}^2$ (Total)&           $1106.49$  & $2771.75$               &  $3887.55$              & $2770.45$              & $3880.69$                \\[+1mm]
$\chi_{\textrm{min}}^2$ (Non-CMB) &        $1106.49$  & $\cdots$               &  $1111.65$              & $\cdots$              & $1109.43$     \\[+1mm]
$\textrm{DIC}$                 &           $1117.14$  & $2825.74$               &  $3942.22$              & $2827.29$              & $3937.52$                \\[+1mm]
$\Delta\textrm{DIC}$           &           $2.76$    & $-0.71$                 &  $1.52$                & $0.84$                & $-3.18$              \\[+1mm]
$\textrm{AIC}_c$                 &           $1116.49$  & $2827.75$               &  $3943.55$              & $2828.45$              & $3938.69$          \\[+1mm]
$\Delta\textrm{AIC}_c$           &           $1.94$    & $-0.96$                 &  $1.14$                & $-0.26$                & $-3.72$  \\[+1mm]
\end{tabular}
\\[+1mm]
\begin{flushleft}
Note: $\Delta\textrm{DIC}$ ($\Delta\textrm{AIC}_c$) indicates an excess value relative to that of the tilted flat $\Lambda$CDM model constrained with the same data. 
\end{flushleft}
\end{ruledtabular}
\label{tab:para_TNL_P18_lensing_nonCMB}
\end{table*}

%
\begin{figure*}[htbp]
\centering
\mbox{\includegraphics[width=170mm]{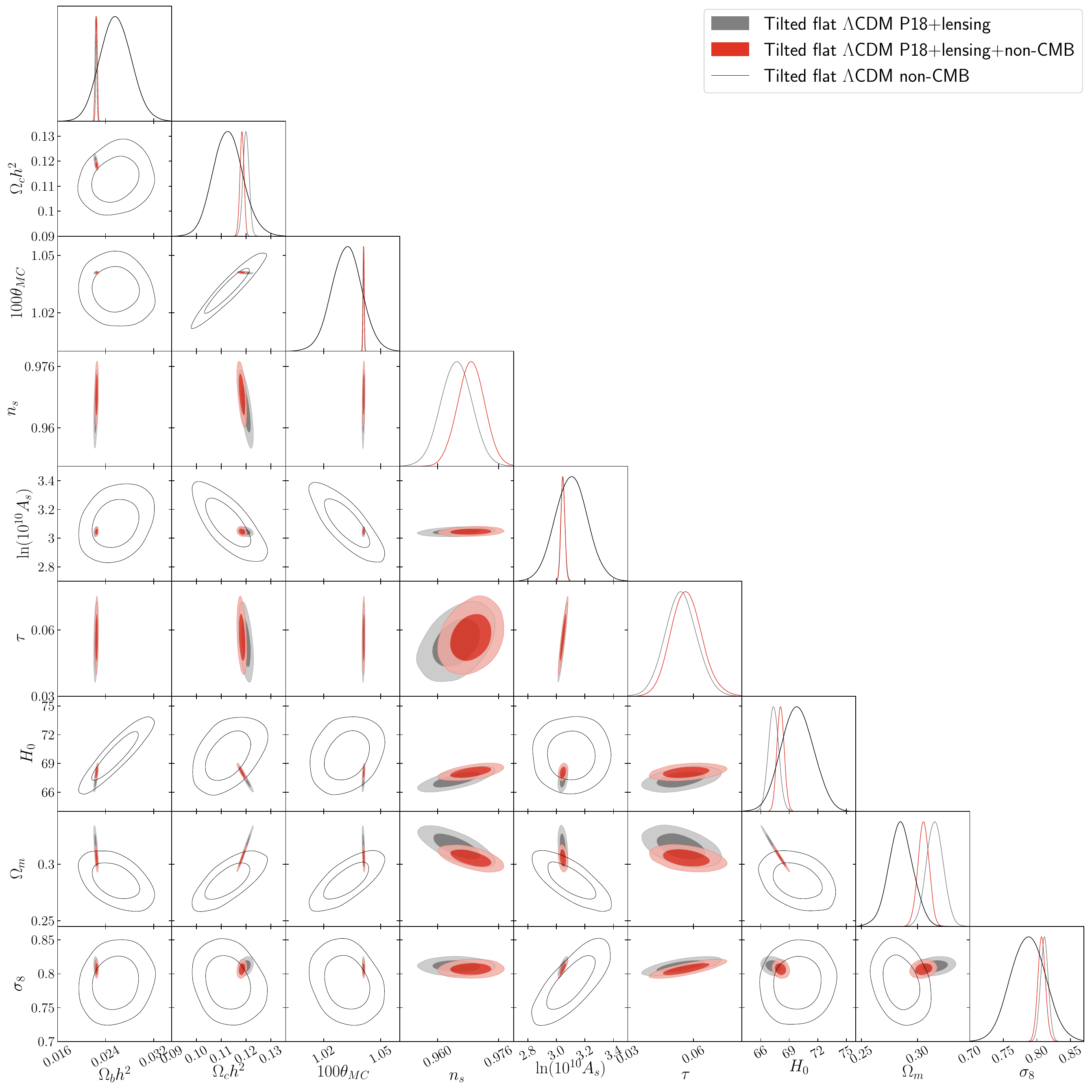}}
\caption{Likelihoods of the tilted flat $\Lambda$CDM model parameters constrained by P18+lensing, non-CMB, and P18+lensing+non-CMB data sets.
}
\label{fig:like_FL_P18_lensing_nonCMB}
\end{figure*}
\begin{figure*}[htbp]
\centering
\mbox{\includegraphics[width=170mm]{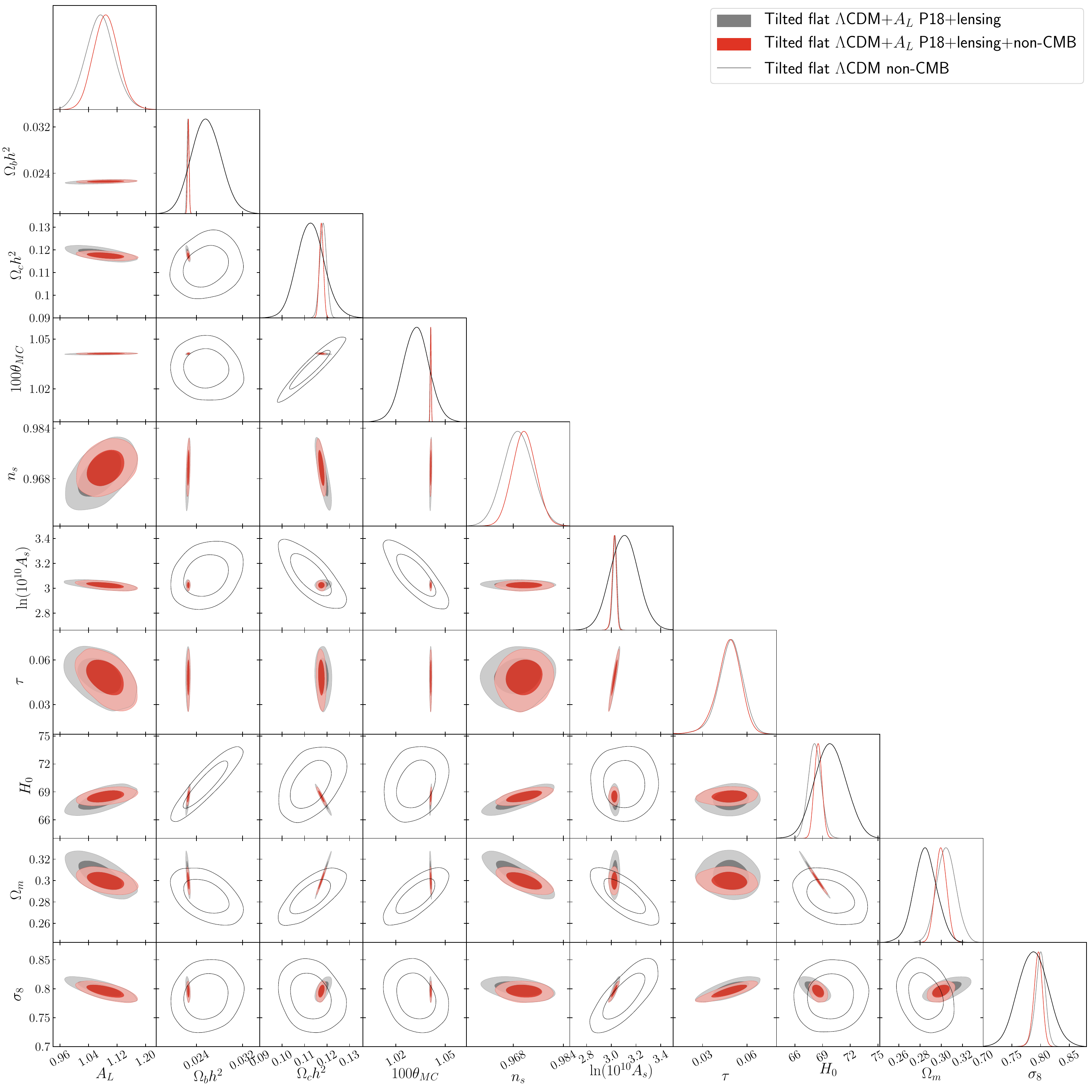}}
\caption{Likelihoods of the tilted flat $\Lambda$CDM$+A_L$ model parameters constrained by P18+lensing, non-CMB, and P18+lensing+non-CMB data sets. The likelihoods for the non-CMB data set, which do not depend on $A_L$, are the same as in Fig.\ \ref{fig:like_FL_P18_lensing_nonCMB}. 
}
\label{fig:like_FL_Alens_P18_lensing_nonCMB}
\end{figure*}

\begin{figure*}[htbp]
\centering
\mbox{\includegraphics[width=170mm]{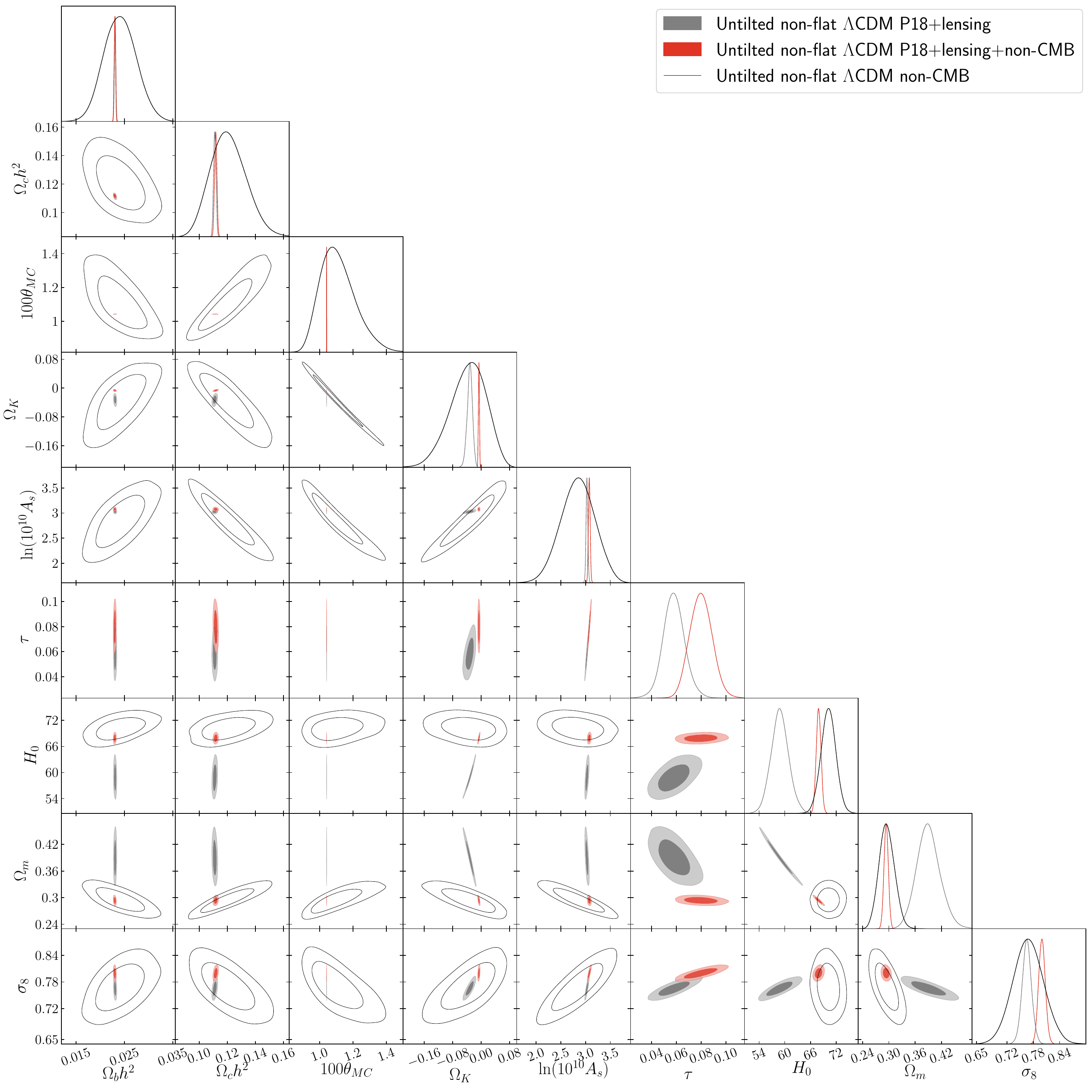}}
\caption{Likelihoods of the untilted non-flat $\Lambda$CDM model parameters constrained by P18+lensing, non-CMB, and P18+lensing+non-CMB data sets.
}
\label{fig:like_NL_P18_lensing_nonCMB}
\end{figure*}

\begin{figure*}[htbp]
\centering
\mbox{\includegraphics[width=170mm]{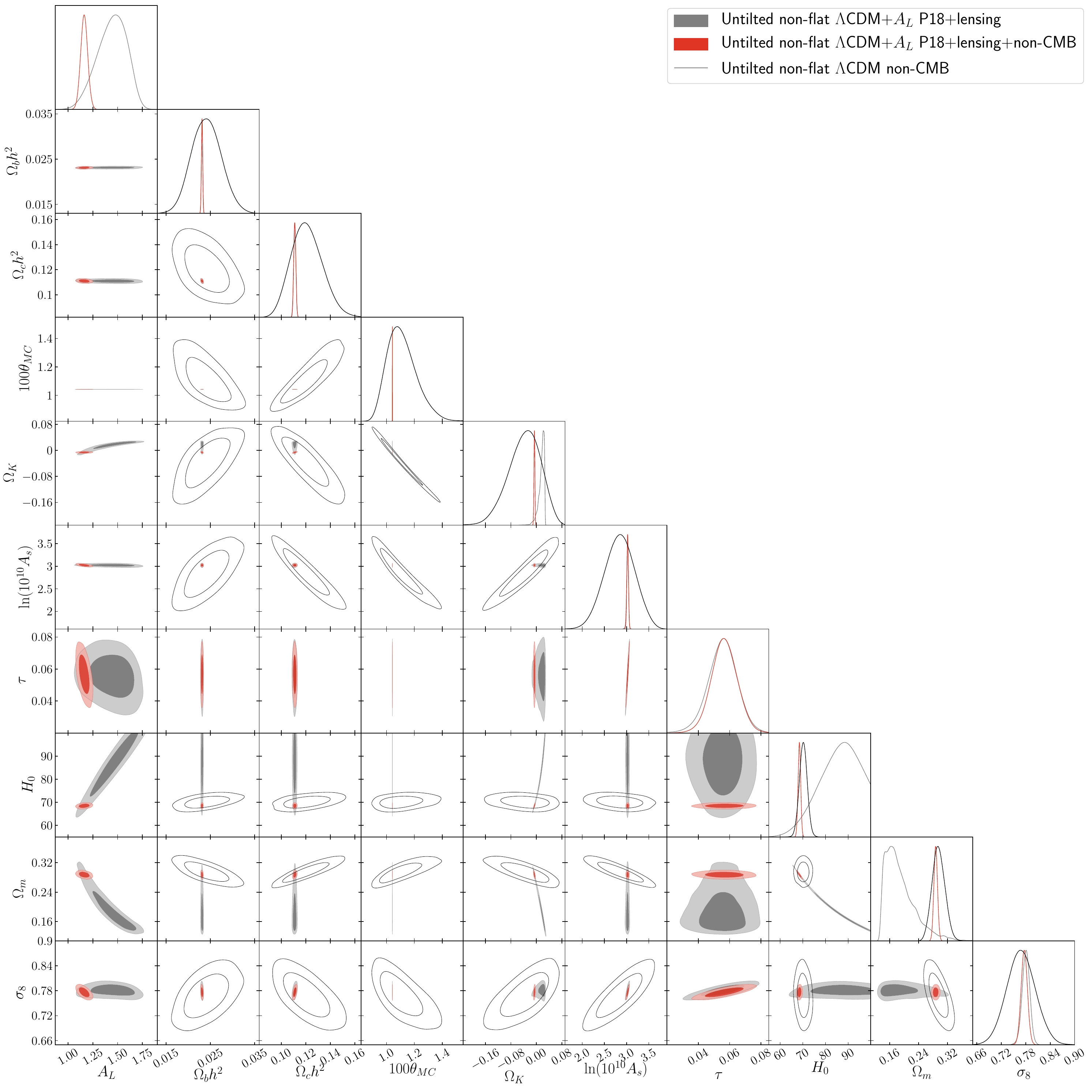}}
\caption{Likelihoods of the untilted non-flat $\Lambda$CDM$+A_L$ model parameters constrained by P18+lensing, non-CMB, and P18+lensing+non-CMB data sets. The likelihoods for the non-CMB data set, which do not depend on $A_L$, are the same as in Fig.\ \ref{fig:like_NL_P18_lensing_nonCMB}.
}
\label{fig:like_NL_Alens_P18_lensing_nonCMB}
\end{figure*}

\begin{figure*}[htbp]
\centering
\mbox{\includegraphics[width=170mm]{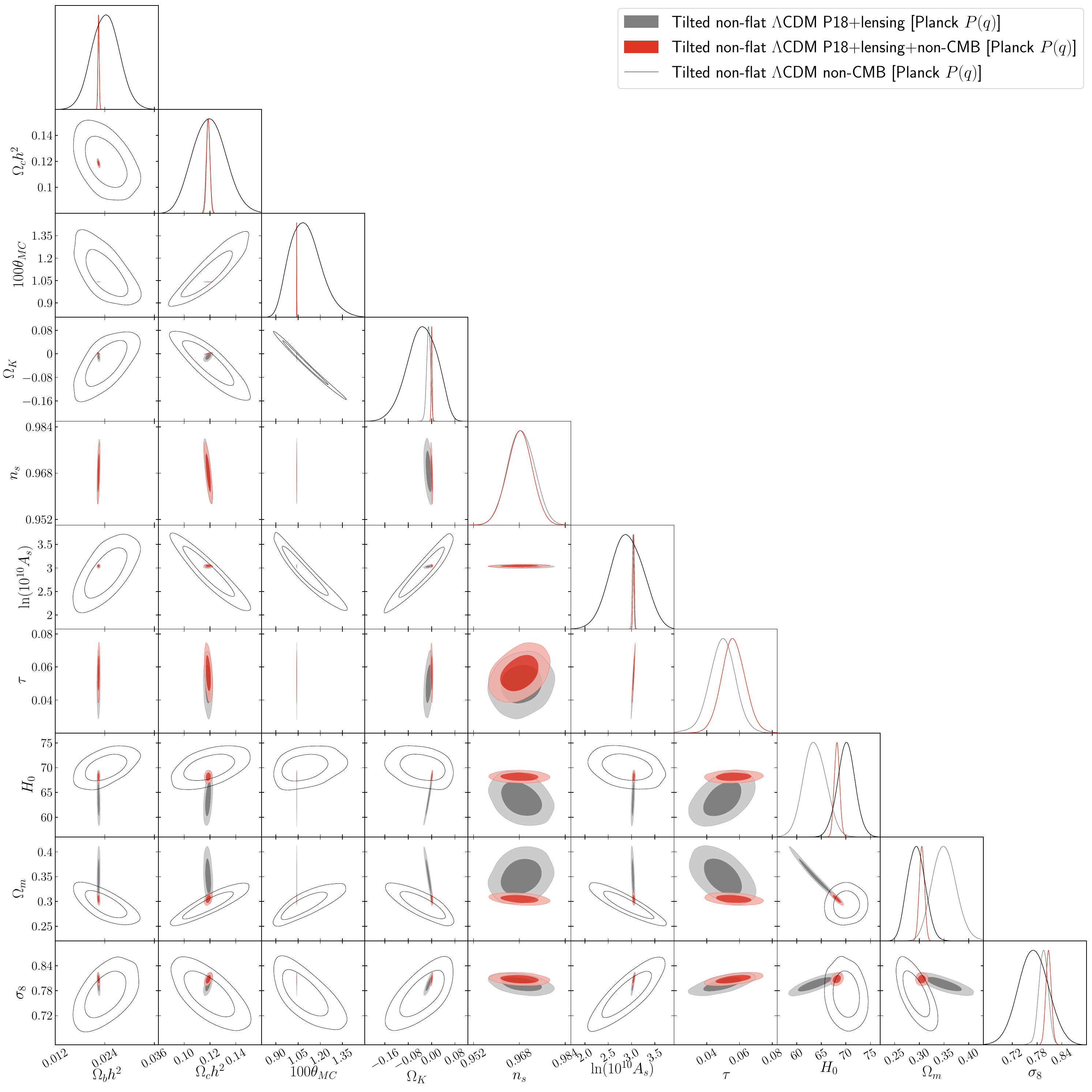}}
\caption{Likelihoods of the tilted non-flat $\Lambda$CDM model [with Planck $P(q)$] parameters constrained by P18+lensing, non-CMB, and P18+lensing+non-CMB data sets.
}
\label{fig:like_NL_ns_P18_lensing_nonCMB}
\end{figure*}

\begin{figure*}[htbp]
\centering
\mbox{\includegraphics[width=170mm]{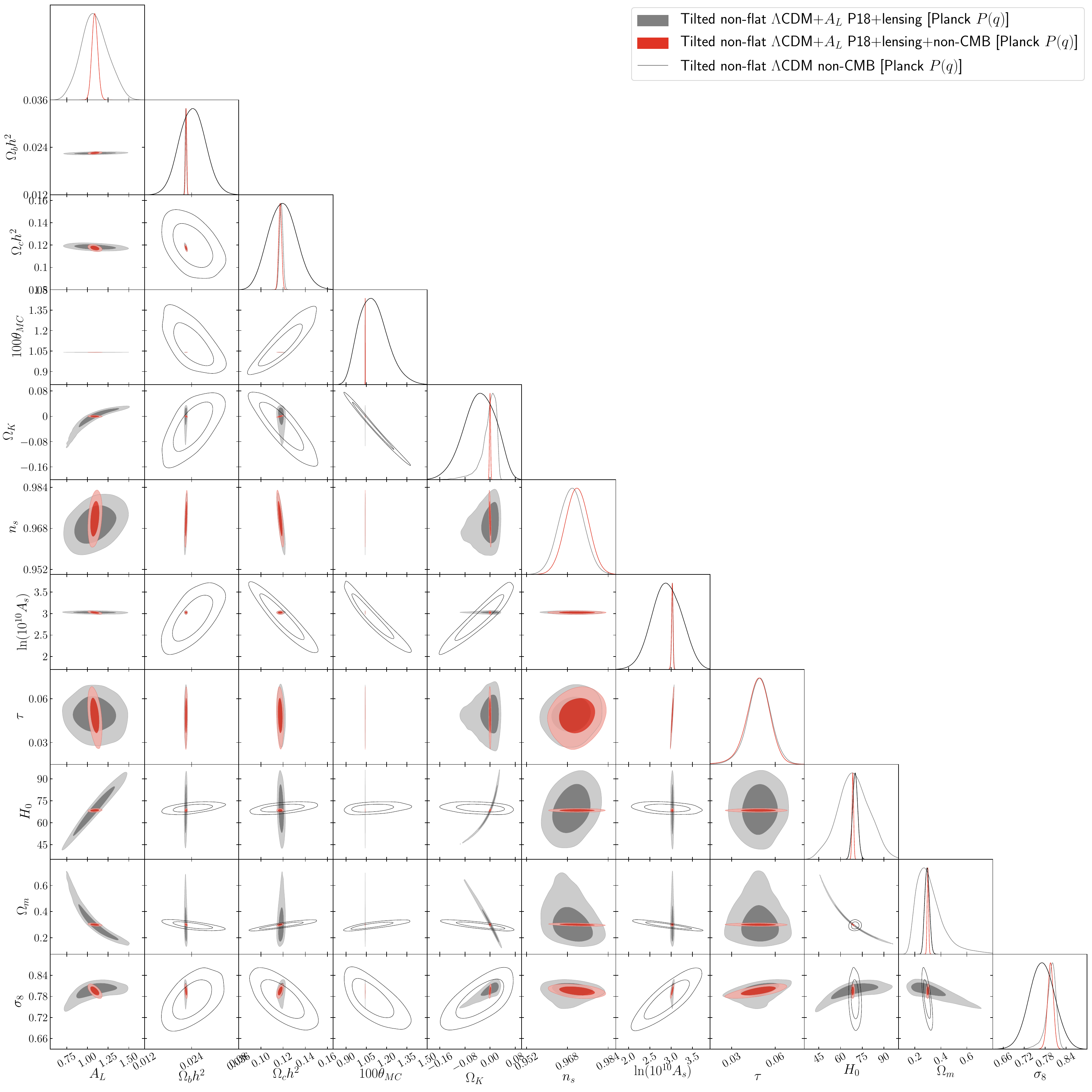}}
\caption{Likelihoods of the tilted non-flat $\Lambda$CDM$+A_L$ model [with Planck $P(q)$] parameters constrained by P18+lensing, non-CMB, and P18+lensing+non-CMB data sets. The likelihoods for the non-CMB data set, which do not depend on $A_L$, are the same as in Fig.\ \ref{fig:like_NL_ns_P18_lensing_nonCMB}.
}
\label{fig:like_NL_Alens_ns_P18_lensing_nonCMB}
\end{figure*}

\begin{figure*}[htbp]
\centering
\mbox{\includegraphics[width=170mm]{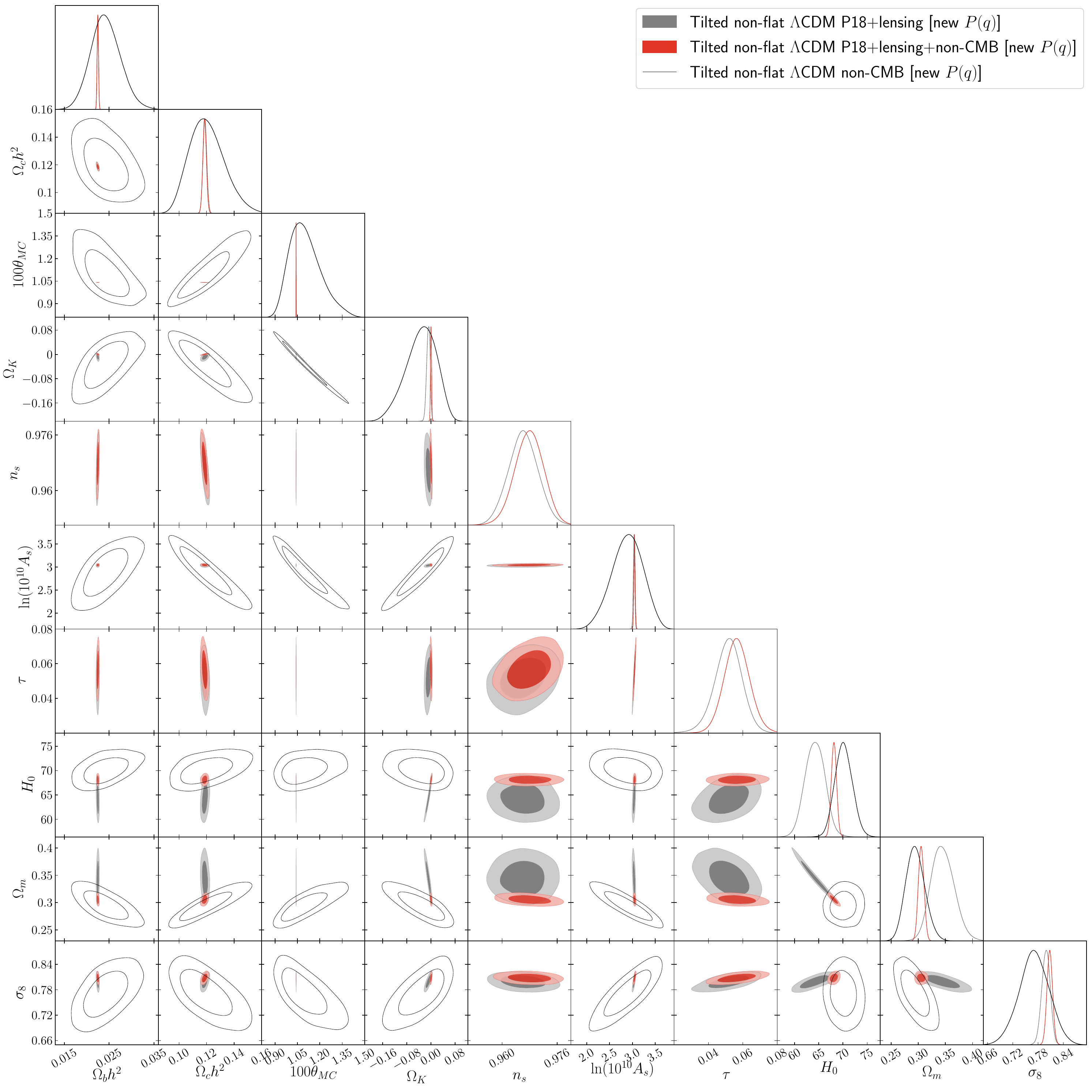}}
\caption{Likelihoods of the tilted non-flat $\Lambda$CDM model [with new $P(q)$] parameters constrained by P18+lensing, non-CMB, and P18+lensing+non-CMB data sets.
}
\label{fig:like_TNL_P18_lensing_nonCMB}
\end{figure*}

\begin{figure*}[htbp]
\centering
\mbox{\includegraphics[width=170mm]{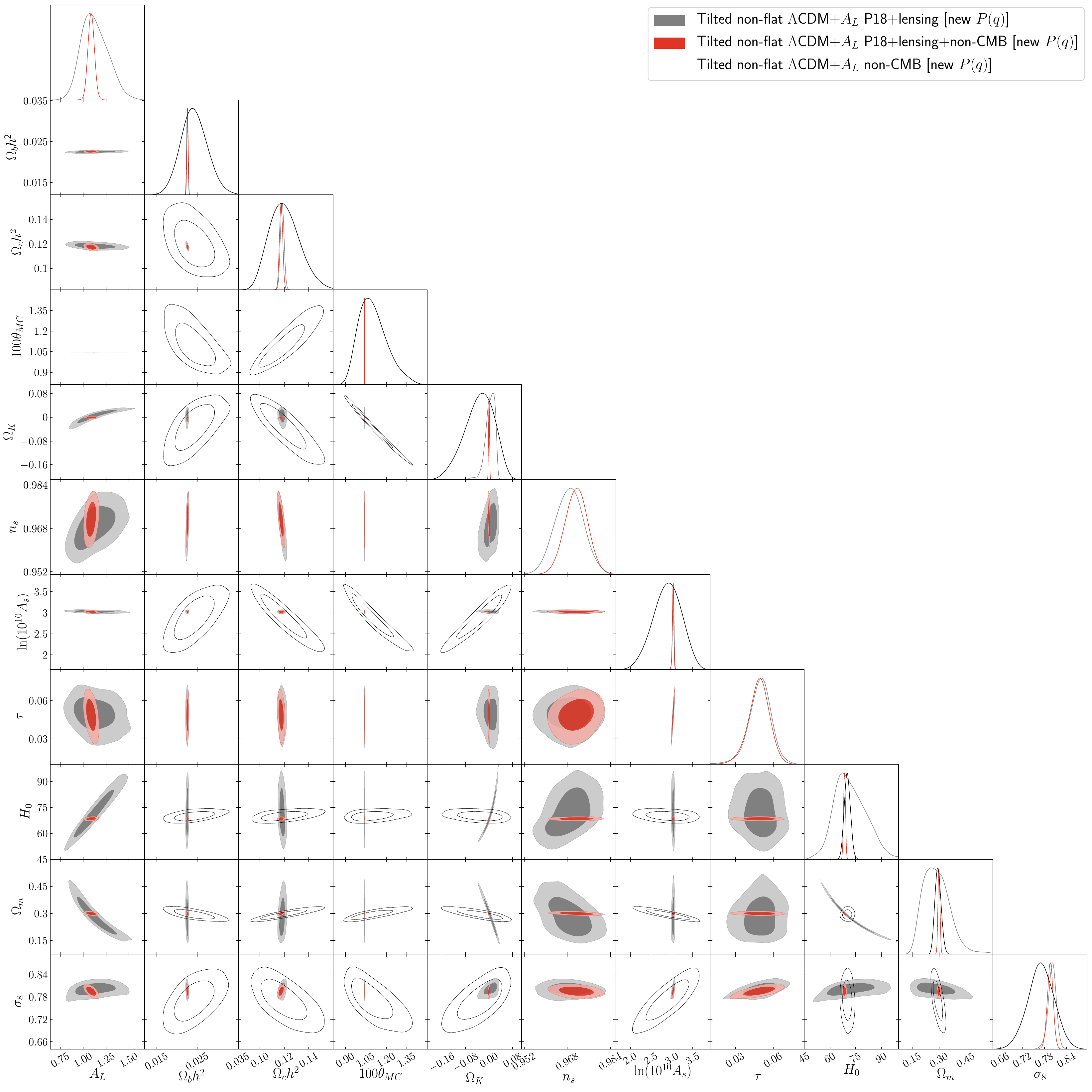}}
\caption{Likelihoods of the tilted non-flat $\Lambda$CDM$+A_L$ model [with new $P(q)$] parameters constrained by P18+lensing, non-CMB, and P18+lensing+non-CMB data sets. The likelihoods for the non-CMB data set, which do not depend on $A_L$, are the same as in Fig.\ \ref{fig:like_TNL_P18_lensing_nonCMB}.
}
\label{fig:like_TNL_Alens_P18_lensing_nonCMB}
\end{figure*}


%
\subsubsection{Comparing P18+lensing data and non-CMB data cosmological constraints}\label{sec:P18+lensing_vs_non-CMB}

In the previous subsubsection we compared non-CMB data cosmological constraints to those obtained from P18 data. We found significant tensions in the non-flat models with $A_L=1$ for the derived parameters $H_0$ and $\Omega_m$, and a 2.2$\sigma$ tension between the two $\Omega_m$ values in the flat $\Lambda$CDM model with $A_L = 1$. In view of these results additional tests are needed if one wants to know whether P18 and non-CMB data can be jointly analysed to determine cosmological constraints. We study this in Sec.\ \ref{subsec:data_set_tensions}. Interestingly, when the $A_L$ parameter is allowed to vary these tensions decrease significantly, with the largest tension being 2.3$\sigma$ between the two $H_0$ values in the tilted non-flat Planck $P(q)$ model, and the remaining tensions not exceeding 1.4$\sigma$, perhaps an indication that P18 and non-CMB data can be use jointly to constrain cosmological parameters when $A_L$ is allowed to vary.

In Secs.\ \ref{subsubsec:P18_data_constraints} and \ref{subsubsec:P18_lensing_data_constraints} we discussed the cosmological constrained obtained from P18 data and from P18+lensing data. We shall see, in Sec.\ \ref{subsec:data_set_tensions}, that, in the non-flat models, P18 data and lensing data are less mutually inconsistent than P18 data and non-CMB data are, however it is necessary to perform an additional test to determine whether or not P18, lensing, and non-CMB data can be jointly analyzed to derive cosmological constraints in the non-flat models with $A_L=1$. 

In this subsubsection we describe the results of this additional test that compares non-CMB data cosmological constraints to the ones obtained from P18+lensing data. P18+lensing+non-CMB data cannot be jointly used in the context of a given model unless the cosmological constraints obtained with P18+lensing data and with non-CMB data are consistent. While in the previous subsubsection we labelled the study of P18 vs.\ non-CMB as a study of high-redshift data cosmological constraints vs.\ low-redshift data cosmological constraints, we cannot do that in this subsubsection since most of the information in the lensing data is from low-redshift data. 

The cosmological parameter mean values and error bars favored by the P18+lensing, non-CMB, and P18+lensing+non-CMB data sets are summarized in Tables \ref{tab:para_FL_P18_lensing_nonCMB}-\ref{tab:para_TNL_P18_lensing_nonCMB} for the tilted flat $\Lambda$CDM (+$A_L$) model, the untilted non-flat $\Lambda$CDM (+$A_L$) model, the tilted non-flat $\Lambda$CDM (+$A_L$) Planck $P(q)$ model, and the tilted non-flat $\Lambda$CDM (+$A_L$) new $P(q)$ model. Likelihood distributions of cosmological parameters of the four models with $A_L=1$ and with varying $A_L$ are shown in Figs.\ \ref{fig:like_FL_P18_lensing_nonCMB}-\ref{fig:like_TNL_Alens_P18_lensing_nonCMB} for the P18+lensing, non-CMB and P18+lensing+non-CMB data sets. 

Since non-CMB data do not have the ability to constrain $\tau$ or $n_s$, in this subsubsection we set their values in the non-CMB data only analyses to those found in the corresponding P18+lensing data analyses. Note that in the previous subsubsection, where the case of P18 data versus non-CMB data was studied, the values of $\tau$ and $n_s$ in the non-CMB data only analyses were set to those found in the corresponding P18 data analyses, nevertheless, the cosmological parameter constraints from the two non-CMB data analyses are practically identical. This indicates that non-CMB data are mostly insensitive to changes in the values of $\tau$ and $n_s$, as we have assumed. Again, we do not include the $A_L$ parameter in the analyses when only non-CMB data are considered, since it does not play a role at low redshift. 

Similar to what happens in the previous subsubsection, when we compare cosmological constraints from P18 data and from non-CMB data, by looking at Tables \ref{tab:para_NL_P18_lensing_nonCMB}-\ref{tab:para_TNL_P18_lensing_nonCMB} we observe that for the six non-flat $\Lambda$CDM (+$A_L$) models the constraints imposed by non-CMB data on the $H_0$ and $\Omega_m$ parameters are tighter than the ones from P18+lensing data. P18+lensing data more restrictively constrain all other parameters in all eight cosmological models.

Comparing the six-parameter and the four-parameter tilted flat $\Lambda$CDM model primary cosmological parameter constraints for P18+lensing and non-CMB data, shown in the left part of Table \ref{tab:para_FL_P18_lensing_nonCMB}, we observe that the values of $\Omega_b h^2$, $\Omega_c{h^2}$, and $\theta_{\textrm{MC}}$ are in mild disagreement, at 1.3$\sigma$, 1.2$\sigma$, and 1.1$\sigma$, respectively. We also see tensions in the derived parameters. In particular, for the non-relativistic matter density parameter $\Omega_m$, the level of tension reaches 2.3$\sigma$, whereas the values of $H_0$ disagree by 1.4$\sigma$. 

From Table \ref{tab:para_FL_P18_lensing_nonCMB} we can compare the seven-parameter and the four-parameter tilted flat $\Lambda$CDM+$A_L$ model cosmological parameter constraints for P18+lensing data and for non-CMB data. Regarding the primary cosmological parameters we see that the values of $\Omega_b{h^2}$ and $\theta_{\textrm{MC}}$ disagree at 1.2$\sigma$ and 1.1$\sigma$ respectively. The inclusion of the $A_L$-varying parameter reduces significantly the tension found in the $A_L=1$ case for $\Omega_m$ and $H_0$, with them now disagreeing by only 1.4$\sigma$ and 0.96$\sigma$. 
We do not find any clear evidence that prevents us from jointly analyzing P18+lensing and non-CMB data, in the context of the tilted flat $\Lambda$CDM model, with and without a varying $A_L$ parameter. 

The results for the six-parameter and the four-parameter untilted non-flat $\Lambda$CDM model obtained from P18+lensing and non-CMB data are in Table \ref{tab:para_NL_P18_lensing_nonCMB}. While for the primary cosmological parameters we do not observe significant tensions, we do for the derived parameters. The primary spatial curvature parameter is $\Omega_k=-0.037\pm 0.050$ for non-CMB data, which is 0.74$\sigma$ away from flat hypersurfaces and in 0.095$\sigma$ tension with the P18+lensing analysis value $\Omega_k=-0.0322\pm 0.0075$, which is 4.3$\sigma$ away from flat. As for the derived parameters, we find that $H_0$, $\Omega_m$ and $\sigma_8$ values are in 4.2$\sigma$, 2.9$\sigma$, and 0.11$\sigma$ disagreement. The high levels of tensions reached for some of the parameters may indicate that P18+lensing and non-CMB data should not be jointly analyzed in the context of the untilted non-flat $\Lambda$CDM model. 

P18+lensing and non-CMB data results obtained for the seven-parameter and the four-parameter untilted non-flat $\Lambda$CDM+$A_L$ model are shown in Table \ref{tab:para_NL_P18_lensing_nonCMB}. Regarding the values of the primary cosmological parameters, except for $\Omega_k$ (discussed next), as was observed in the $A_L=1$ case, there are no significant tensions. The value of the curvature parameter is $\Omega_k=-0.037\pm 0.050$ (0.74$\sigma$ away from flat) for the non-CMB data and $\Omega_k=0.0161\pm 0.0094$ for the P18+lensing data, which indicates 1.7$\sigma$ evidence in favor of an open spatial geometry. The two $\Omega_k$ values disagree at 1.0$\sigma$. The disagreement in the values of the derived parameters $H_0$, $\Omega_m$ and $\sigma_8$ values are 1.8$\sigma$, 2.3$\sigma$, and 0.32$\sigma$, respectively, which clearly represents a reduction with respect to the $A_L=1$ case. This is due to the elongation of the error bars in the varying $A_L$ case compared to the $A_L=1$ case. Given these results, the P18+lensing and the non-CMB data should perhaps not be used together in the context of the untilted non-flat $\Lambda$CDM+$A_L$ model. It may be noticed however that when we do so, namely in the P18+lensing+non-CMB analysis, the obtained value for the curvature parameter is $\Omega_k=-0.0060\pm0.0014$, which is 4.3$\sigma$ away from flat. Nonetheless, according to the AIC and DIC this model is strongly disfavoured when it is compared with the tilted models, due to the lack of the degree of freedom contained in the $n_s$ parameter. 

The results that allow us to compare the seven-parameter and the five-parameter tilted non-flat $\Lambda$CDM Planck $P(q)$ model primary cosmological parameter constraints for P18+lensing data and non-CMB data can be seen in Table \ref{tab:para_NL_ns_P18_lensing_nonCMB}. There are no significant tensions in the values of the primary cosmological parameters. The non-CMB data value of the spatial curvature parameter, $\Omega_k=-0.033\pm 0.050$, is 0.66$\sigma$ away from flat and in 0.45$\sigma$ tension with the value found in the P18+lensing analysis, namely $\Omega_k=-0.0103\pm0.0066$ which represents a 1.6$\sigma$ deviation from flat hypersurfaces. As for the values of the derived parameters $H_0$, $\Omega_m$, and $\sigma_8$, the tensions are 2.2$\sigma$, 1.9$\sigma$, and 0.66$\sigma$, respectively. Given these results, further tests are probably necessary in order to decide whether P18+lensing and non-CMB data can be jointly analyzed in the context of the tilted non-flat $\Lambda$CDM Planck $P(q)$ model. 

P18+lensing and non-CMB data results obtained for the eight-parameter and the five-parameter tilted non-flat $\Lambda$CDM+$A_L$ Planck $P(q)$ model are shown in Table \ref{tab:para_NL_ns_P18_lensing_nonCMB}. Similar to the $A_L=1$ case we do not find significant disagreements in the values of the primary cosmological parameters. For the non-CMB data the value of the curvature parameter is $\Omega_k=-0.033\pm 0.050$ which is 0.66$\sigma$ away from flat and in 0.49$\sigma$ tension with the P18+lensing value, $\Omega_k=-0.005\pm 0.027$, which in turn is only 0.19$\sigma$ in favor of a closed geometry. An important reduction in the disagreements found in the derived parameters, with respect to the $A_L=1$ case, is observed. In particular, for $H_0$, $\Omega_m$, and $\sigma_8$ the disagreement found is 0.099$\sigma$, 0.23$\sigma$, and 0.63$\sigma$. We may say that in the context of the tilted non-flat $\Lambda$CDM+$A_L$ Planck $P(q)$ we are allowed to analyze together P18+lensing data and non-CMB data. By doing so, we get for P18+lensing+non-CMB data no evidence in favor of a non-flat geometry, $\Omega_k=-0.0002\pm 0.0017$, but still a clear 2.5$\sigma$ preference for $A_L\neq 1$ since $A_L=1.090\pm 0.036$. 

Comparing the seven-parameter and the five-parameter tilted non-flat $\Lambda$CDM new $P(q)$ model primary cosmological parameter constraints for P18+lensing data and non-CMB data, from the left part of Table \ref{tab:para_TNL_P18_lensing_nonCMB} we see no important differences in the values of the primary parameters. The value for the spatial curvature parameter is $\Omega_k=-0.033\pm 0.050$ for non-CMB data, which represents a 0.66$\sigma$ deviation from flat and it is in 0.48$\sigma$ tension with the value obtained in the P18+lensing analysis, $\Omega_k=-0.0086\pm 0.0057$, which is 1.5$\sigma$ away from flat hypersurfaces. Regarding the triad of derived parameters $H_0$, $\Omega_m$, and $\sigma_8$, the disagreement found for each of them is 2.2$\sigma$, 1.9$\sigma$, and 0.75$\sigma$ respectively. In light of these results we deem that more testing is required to decide whether the P18+lensing and non-CMB data can be jointly analyzed in the context of the tilted non-flat $\Lambda$CDM new $P(q)$ model. 

In Table \ref{tab:para_TNL_P18_lensing_nonCMB} we provide the results for the eight-parameter and the five-parameter tilted non-flat $\Lambda$CDM+$A_L$ new $P(q)$ model when P18+lensing data and non-CMB data are considered. The tensions found for the values of primary cosmological parameters are not significant, as in the $A_L=1$ case. When non-CMB data is considered we find $\Omega_k=-0.033\pm 0.050$, which shows a 0.66$\sigma$ evidence in favor of a closed geometry and is in 0.69$\sigma$ tension with the P18+lensing data value, $\Omega_k=0.003\pm 0.016$, which shows only a 0.19$\sigma$ preference for an open geometry. As for the derived parameters $H_0$, $\Omega_m$, and $\sigma_8$ the level of agreement is really good, with the corresponding values only in 0.20$\sigma$, 0.10$\sigma$, and 0.80$\sigma$ tension, respectively. These results seem to indicate that in the context of the tilted non-flat $\Lambda$CDM+$A_L$ new $P(q)$ model P18+lensing data and non-CMB data can be jointly analyzed. In the P18+lensing+non-CMB analysis we find $\Omega_k=-0.0002\pm 0.0017$, so no clear preference for an open or a closed geometry. On the other hand, we find $A_L=1.088\pm 0.035$ which is 2.5$\sigma$ away from the predicted value $A_L=1$. 

In Figs.\ \ref{fig:like_FL_P18_lensing_nonCMB}-\ref{fig:like_TNL_Alens_P18_lensing_nonCMB} we show the one-dimensional likelihoods and the two-dimensional contours for cosmological parameters obtained using P18+lensing, non-CMB, and P18+lensing+non-CMB data. The constraints coming from non-CMB data (shown with unfilled black lines) are less restrictive than P18+lensing constraints (shown in grey), except for the $H_0$ and $\Omega_m$ constraints in the six non-flat models. Except for the untilted non-flat model with $A_L=1$ we observe at least partial overlaps between the three sets of contours even when the $A_L$ parameter is not allowed to vary. 

The contour plots for the tilted flat $\Lambda$CDM (+$A_L$) models are in Figs.\ \ref{fig:like_FL_P18_lensing_nonCMB} and \ref{fig:like_FL_Alens_P18_lensing_nonCMB}. The aforementioned $\sim$1$\sigma$ disagreements (and the $\sim 2\sigma$ $\Omega_m$ disagreement in the $A_L = 1$ case) found when we compared the one-dimensional likelihood P18+lensing and non-CMB results can also be observed here. The largest tensions are seen in the panels containing one of the derived parameters and the inclusion in the analysis of the varying $A_L$ parameter clearly reduces them. 

Looking at the contour plots for the untilted non-flat $\Lambda$CDM (+$A_L$) models displayed in Figs.\ \ref{fig:like_NL_P18_lensing_nonCMB} and \ref{fig:like_NL_Alens_P18_lensing_nonCMB} we observe significantly non-overlapping contours either when the primary parameter $\Omega_k$ is involved or when the derived parameter $H_0$ or $\Omega_m$ is involved. This reinforces the idea that when $A_L$ is not allowed to vary the P18+lensing and non-CMB data sets cannot be analyzed together in the untilted non-flat $\Lambda$CDM model. Quite different results are found when we do allow $A_L$ to vary. The disagreements observed in the $A_L=1$ case largely disappear. Therefore we may say that in the context of this varying $A_L$ cosmological model we can jointly analyze P18+lensing and non-CMB data. 

Figures \ref{fig:like_NL_ns_P18_lensing_nonCMB} and \ref{fig:like_NL_Alens_ns_P18_lensing_nonCMB} show cosmological parameter constraints for the tilted non-flat $\Lambda$CDM (+$A_L$) models, while the ones for the tilted non-flat $\Lambda$CDM (+$A_L$) new $P(q)$ models are displayed in Figs. \ref{fig:like_TNL_P18_lensing_nonCMB} and \ref{fig:like_TNL_Alens_P18_lensing_nonCMB}. The contour plots for these tilted non-flat models are very similar, something that was not unexpected given the results discussed above in this subsubsection. In both cases, when $A_L$ is not allowed to vary and when it is allowed to vary, we observe overlaps between the primary parameter panels contours at 1$\sigma$. When $A_L=1$ we observe an improvement in the overlapping in the current P18+lensing data vs.\ non-CMB data case, compared to the P18 data vs.\ non-CMB data case of the previous subsubsection, where now for both the Planck $P(q)$ model and the new $P(q)$ model the contours do overlap at 2$\sigma$. On the other hand, in the varying $A_L$ case we observe overlaps, even in those panels that involve some of the derived parameters, at 1$\sigma$. 

As in the P18 data vs.\ non-CMB data cosmological constraints comparison discussed in the previous subsubsection, further tests are needed to determine whether or not P18+lensing data and non-CMB data can be jointly analyzed in the context of the non-flat models under study. We discuss this issue in detail in Sec. \ref{subsec:data_set_tensions}.

\begin{table*}
\caption{Individual and total $\chi^2$ values for the best-fit flat and non-flat $\Lambda\textrm{CDM}$ inflation models.
         The deviance information criterion (DIC) and the Akaike information criterion (AIC$_c$) are also listed.  }
{\scriptsize
\begin{ruledtabular}
\begin{tabular}{lcccccccccccccc}
    Data sets   & $\chi_{\textrm{plik}}^2$  & $\chi_{\textrm{lowl}}^2$  & $\chi_{\textrm{simall}}^2$  & $\chi_{\textrm{lensing}}^2$ &  $\chi_{\textrm{prior}}^2$  &  $\chi_{\textrm{SN}}^2$  & $\chi_{\textrm{BAO}}^2$  &  $\chi_{H(z)}^2$   &  $\chi_{f\sigma_8}^2$ &  $\chi^2_{\textrm{total}}$      & $\Delta\chi^2$  & DIC & $\Delta\textrm{DIC}$  & $\Delta\textrm{AIC}_c$ \\[+0mm]
 \hline \\[-2mm]
 \multicolumn{15}{c}{Tilted flat $\Lambda\textrm{CDM}$ model} \\
  \hline \\[-2mm]
   P18                       & 2344.71  & 23.39 & 396.05  &       & 1.66 &         &       &       &       & 2765.80  &         & 2817.93 &         & \\[+1mm]
   P18+lensing               & 2344.66  & 23.39 & 396.06  & 8.79  & 1.82 &         &       &       &       & 2774.71  &         & 2826.45 &         & \\[+1mm]
   P18+lensing+non-CMB       & 2346.61  & 22.64 & 396.34  & 8.94  & 1.84 & 1058.99 & 20.10 & 14.76 & 18.20 & 3888.41  &         & 3940.70 &         & \\[+1mm]
 \hline\\[-2mm]
    \multicolumn{15}{c}{Tilted flat $\Lambda$\textrm{CDM}$+A_L$ model}  \\
  \hline \\[-2mm]
   P18                       & 2337.23  & 21.92 & 395.66  &       & 1.31 &         &       &       &       & 2756.12  & $-9.68$ & 2812.41 & $-5.52$ & $-7.68$ \\[+1mm]
   P18+lensing               & 2341.62  & 22.29 & 395.68  & 9.94  & 1.71 &         &       &       &       & 2771.24  & $-3.47$ & 2825.53 & $-0.92$ & $-1.47$ \\[+1mm]
   P18+lensing+non-CMB       & 2342.43  & 21.99 & 395.68  & 9.74  & 2.06 & 1059.14 & 21.46 & 14.73 & 14.31 & 3881.55  & $-6.86$ & 3935.15 & $-5.55$ & $-4.86$ \\[+1mm]
   \hline \\[-2mm]
    \multicolumn{15}{c}{Untilted non-flat $\Lambda\textrm{CDM}$ model}  \\
  \hline \\[-2mm]
   P18                       & 2369.95  & 22.22 & 395.69  &       & 1.92 &         &       &       &       & 2789.77  & $23.97$  & 2847.14 & $29.21$ & $23.97$ \\[+1mm]
   P18+lensing               & 2383.06  & 20.88 & 396.13  & 10.63 & 2.43 &         &       &       &       & 2813.13  & $38.42$  & 2869.06 & $42.61$ & $38.42$ \\[+1mm]
   P18+lensing+non-CMB       & 2396.21  & 19.89 & 399.59  & 11.65 & 2.28 & 1059.51 & 20.65 & 15.68 & 12.77 & 3938.22  & $49.81$  & 3992.71 & $52.01$ & $49.81$ \\[+1mm]
 \hline\\[-2mm]
    \multicolumn{15}{c}{Untilted non-flat $\Lambda$\textrm{CDM}$+A_L$ model}  \\
  \hline \\[-2mm]
   P18                       & 2369.32  & 20.34 & 395.87  &       & 2.23 &         &       &       &       & 2787.76  & $21.96$  & 2846.45 & $28.52$ & $23.96$ \\[+1mm]
   P18+lensing               & 2378.87  & 20.09 & 395.65  & 11.25 & 2.05 &         &       &       &       & 2807.91  & $33.20$  & 2856.10 & $29.65$ & $35.20$ \\[+1mm]
   P18+lensing+non-CMB       & 2379.11  & 19.95 & 395.82  & 10.72 & 2.06 & 1060.16 & 22.50 & 15.47 &  9.26 & 3915.05  & $26.64$  & 3973.55 & $32.85$ & $28.64$ \\[+1mm]
   \hline \\[-2mm]
    \multicolumn{15}{c}{Tilted non-flat $\Lambda\textrm{CDM}$ model [Planck $P(q)$]}  \\
  \hline \\[-2mm]
   P18                       & 2336.45  & 21.29 & 395.60  &       & 1.38 &         &       &       &       & 2754.73  & $-11.07$ & 2810.59 & $-7.34$ & $-9.07$ \\[+1mm]
   P18+lensing               & 2342.29  & 21.86 & 395.66  & 10.09 & 1.63 &         &       &       &       & 2771.53  & $-3.18$  & 2826.17 & $-0.28$ & $-1.18$ \\[+1mm]
   P18+lensing+non-CMB       & 2345.82  & 22.90 & 396.53  & 8.92  & 1.88 & 1059.00 & 20.09 & 14.70 & 18.15 & 3887.99  & $-0.42$  & 3942.07 & $1.37$  & $1.58$ \\[+1mm]
 \hline\\[-2mm]
    \multicolumn{15}{c}{Tilted non-flat $\Lambda$\textrm{CDM}$+A_L$ model [Planck $P(q)$]}  \\
  \hline \\[-2mm]                
   P18                       & 2336.57  & 21.51 & 395.61  &       & 1.29 &         &       &       &       & 2754.99  & $-10.81$ & 2811.63 & $-6.30$ & $-6.81$  \\[+1mm]
   P18+lensing               & 2341.32  & 22.55 & 395.71  & 9.44  & 2.12 &         &       &       &       & 2771.14  & $-3.57$  & 2827.14 & $0.69$  & $0.43$   \\[+1mm]
   P18+lensing+non-CMB       & 2341.91  & 22.16 & 395.77  & 9.62  & 1.60 & 1059.06 & 20.61 & 14.74 & 15.90 & 3881.37  & $-7.04$  & 3936.85 & $-3.85$ & $-3.04$   \\[+1mm]
 \hline\\[-2mm]                  
    \multicolumn{15}{c}{Tilted non-flat $\Lambda\textrm{CDM}$ model [new $P(q)$]}  \\
  \hline \\[-2mm]                
   P18                       & 2338.26  & 21.42 & 396.28  &       & 1.42 &         &       &       &       & 2757.38  & $-8.42$  & 2811.54 & $-6.39$ & $-6.42$  \\[+1mm]
   P18+lensing               & 2342.99  & 21.18 & 395.90  & 9.92  & 1.76 &         &       &       &       & 2771.75  & $-2.96$  & 2825.74 & $-0.71$ & $-0.96$  \\[+1mm]
   P18+lensing+non-CMB       & 2346.63  & 22.53 & 396.30  & 8.91  & 1.53 & 1058.99 & 20.12 & 14.75 & 17.79 & 3887.55  & $-0.86$  & 3942.22 & $1.52$  & $1.14$   \\[+1mm]
 \hline\\[-2mm]                  
    \multicolumn{15}{c}{Tilted non-flat $\Lambda\textrm{CDM}$+$A_L$ model [new $P(q)$]}  \\
  \hline \\[-2mm]
   P18                       & 2337.56  & 21.31 & 395.93  &       & 1.52 &         &       &       &       & 2756.33  & $-9.47$  & 2814.83 & $-3.10$ & $-5.47$  \\[+1mm]
   P18+lensing               & 2341.21  & 22.62 & 395.75  & 9.49  & 1.37 &         &       &       &       & 2770.45  & $-4.26$  & 2827.29 & $0.84$  & $-0.26$  \\[+1mm]
   P18+lensing+non-CMB       & 2342.85  & 21.35 & 395.81  & 9.72  & 1.53 & 1059.13 & 21.27 & 14.77 & 14.27 & 3880.69  & $-7.72$  & 3937.52 & $-3.18$ & $-3.72$  \\[+1mm]
\end{tabular}
\\[+1mm]
Note: $\Delta\chi^2$, $\Delta\textrm{DIC}$, and $\Delta\textrm{AIC}_c$ indicate the values relative to those of the tilted flat $\Lambda\textrm{CDM}$ model for the same combination of data sets. For the tilted flat $\Lambda$CDM model AIC$_c=2819.8$ (P18), $2828.7$ (P18+lensing), and $3942.4$ (P18+lensing+non-CMB). All $\chi^2$ values are
computed at the corresponding model best-fit cosmological parameter values.
\end{ruledtabular}
}
\label{tab:chi2_lcdm}
\end{table*}

\subsection{Model selection}
\label{subsec:model_selection}

In Sec.\ \ref{subsec:cosmological_parameters} we determined and discussed the cosmological parameter mean values and error bars in eight cosmological models (with $A_L = 1$ and with varying $A_L$) from P18, P18+lensing, and P18+lensing+non-CMB data, as well as the differences in the values of the cosmological parameters obtained from P18 data and BAO/BAO$^\prime$ data, from P18 data and non-CMB data, and from P18+lensing data and non-CMB data. In this subsection we utilize the DIC, eq.\ \eqref{eq:DIC}, to determine which of these models best-fit some combinations of these data sets.

For the P18, P18+lensing, and P18+lensing+non-CMB data sets, the values of $\Delta \textrm{AIC}_c$, $\Delta \textrm{DIC}$, and the individual contributions to the $\chi^2_{\textrm{total}}$ for each model are in Table \ref{tab:chi2_lcdm}. Here the Planck CMB data $\chi^2$s are: $\chi^2_{\textrm{plik}}$ from the TT data power spectra $30\leq \ell\leq 2508$ multipoles, the TE data $30\leq \ell \leq 1996$ multipoles, and the EE data $30\leq \ell \leq 1996$ multipoles; $\chi^2_{\textrm{lowl}}$ from the TT data power spectra $2\leq \ell \leq 29$ multipoles; $\chi^2_{\textrm{simall}}$ from the EE data power spectra $2\leq \ell \leq 29$ multipoles; $\chi^2_{\textrm{lensing}}$ from the lensing potential data power spectrum; and $\chi^2_{\textrm{prior}}$ from the priors for the Planck calibration and dust foreground emission. The P18+BAO/BAO$^{\prime}$ data values of $\Delta \textrm{AIC}_c$ and $\Delta \textrm{DIC}$ are provided in Tables \ref{tab:para_FL_BAO}-\ref{tab:para_TNL_ns_BAO}, whereas the corresponding P18+non-CMB data results can be found in Tables \ref{tab:para_FL_P18_nonCMB}-\ref{tab:para_TNL_P18_nonCMB}. 

In this subsection we do not discuss the results obtained for the untilted non-flat $\Lambda$CDM models, without and with a varying $A_L$, since as seen in the results presented in Tables \ref{tab:chi2_lcdm}, \ref{tab:para_NL_BAO}, and \ref{tab:para_NL_P18_nonCMB} this model is not able to fit CMB data as well as the other (tilted) models do. According to the statistical criteria we use, the untilted non-flat $\Lambda$CDM model is very strongly disfavoured when it is compared with the rest of the models that allow for a tilt ($n_s$) degree of freedom. 

We also do not discuss results obtained when only BAO$^\prime$, BAO, (P18) lensing (but see Table \ref{tab:para_lensing} and the brief related discussion in the third paragraph in Sec.\ \ref{subsec:data_set_tensions}), or non-CMB data are considered, because these data sets do not much discriminate between models. From Tables \ref{tab:para_FL_BAO}-\ref{tab:para_TNL_ns_BAO} and \ref{tab:para_FL_P18_nonCMB}-\ref{tab:para_TNL_P18_nonCMB} one sees that for these three data sets the DIC values for all models, including the untilted non-flat $\Lambda$CDM model, are very similar. In order to find more significant differences among the models under study we must include CMB data.   

In what follows we summarize results we find in a number of different combinations of data sets for the three tilted models. For clarity we focus on DIC results, since this is a more reliable indicator, \cite{DIC, Liddle:2007fy}. The tables also list the AIC$_c$ values.

{\bf P18.} The results for these data are listed in Table \ref{tab:chi2_lcdm}. When $A_L = 1$, the non-flat Planck $P(q)$ and the non-flat new $P(q)$ models are strongly favored over the tilted flat model while the Planck $P(q)$ model is weakly favored over the new $P(q)$ model. When $A_L$ is allowed to vary, the non-flat Planck $P(q)$ model is weakly favored over the flat model, with both models being positively favored over the non-flat new $P(q)$ model. The flat$+A_L$ model is positively favored over the flat one, the Planck $P(q)$ model is weakly favored over the Planck $P(q)+A_L$ one, and the new $P(q)$ model is positively favored over the new $P(q)+A_L$ one. It is interesting that compared to the varying $A_L$ case, when $A_L=1$ both tilted non-flat models are strongly favored over the tilted flat $\Lambda$CDM model.

{\bf P18+lensing.} The results for these data are listed in Table \ref{tab:chi2_lcdm}. These data provide only weak discrimination between models. When $A_L = 1$, the non-flat new $P(q)$ model is weakly favored over the non-flat Planck $P(q)$ model and both are weakly favored over the flat model. When $A_L$ is allowed to vary, the tilted flat model is weakly favored over both non-flat models while the non-flat Planck $P(q)$ model is weakly favored over the non-flat new $P(q)$ model. The flat$+A_L$ model is weakly favored over the flat one, the Planck $P(q)$ model is weakly favored over the Planck $P(q)+A_L$ one, and the new $P(q)$ model is weakly favored over the new $P(q)+A_L$ one. 

{\bf P18+BAO/P18+BAO$^\prime$.} The results for these data are listed in Tables \ref{tab:para_FL_BAO}, \ref{tab:para_NL_ns_BAO}, and \ref{tab:para_TNL_ns_BAO}. We discuss the P18+BAO data and P18+BAO$^\prime$ data results together since the conclusions are very similar. When $A_L = 1$, the tilted flat model is weakly (positively) favored over the non-flat Planck and new $P(q)$ models with the non-flat new $P(q)$ model weakly (weakly) favored over the non-flat Planck $P(q)$ model for P18+BAO (P18+BAO$^\prime$) data. When $A_L$ is allowed to vary, the tilted flat model is positively (weakly) favored over the non-flat Planck (new) $P(q)$ model, and the non-flat new $P(q)$ model weakly is favored over the non-flat Planck $P(q)$ model, for P18+BAO data, while for P18+BAO$^\prime$ data the tilted flat model is weakly favored over both non-flat Planck and new $P(q)$ models, and the non-flat new $P(q)$ model is weakly favored over the non-flat Planck $P(q)$ model. The flat$+A_L$ model is strongly (positively) favored over the flat one, the Planck $P(q)+A_L$ model is positively (strongly) favored over the Planck $P(q)$ one, and the new $P(q)+A_L$ model is positively (strongly) favored over the new $P(q)$ one for P18+BAO (P18+BAO$^\prime$) data.

{\bf P18+non-CMB.} The results for these data are listed in Tables \ref{tab:para_FL_P18_nonCMB}, \ref{tab:para_NL_ns_P18_nonCMB}, and \ref{tab:para_TNL_P18_nonCMB}. Since the dominant component of non-CMB data is BAO/BAO$^\prime$ data, in the P18+non-CMB case here we find similar conclusions to the ones presented in the P18+BAO/P18+BAO$^\prime$ cases above. When $A_L = 1$, the tilted flat model is positively (weakly) favored over the non-flat Planck (new) $P(q)$ model with the non-flat new $P(q)$ model weakly favored over the non-flat Planck $P(q)$ model. When $A_L$ is allowed to vary, the tilted flat model is weakly favored over the non-flat Planck $P(q)$ and non-flat new $P(q)$ models, with the non-flat new $P(q)$ model weakly favored over the non-flat Planck $P(q)$ model. The flat$+A_L$ model is strongly favored over the flat one, the Planck $P(q)+A_L$ model is strongly favored over the Planck $P(q)$ one, and the new $P(q)+A_L$ model is strongly favored over the new $P(q)$ one.

{\bf P18+lensing+non-CMB.} The results for these data are listed in Table \ref{tab:chi2_lcdm}.  When $A_L = 1$, the tilted flat model is weakly favored over the non-flat Planck $P(q)$ and non-flat new $P(q)$ models with the non-flat Planck $P(q)$ model weakly favored over the non-flat new $P(q)$ model. When $A_L$ is allowed to vary, the tilted flat model is weakly (positively) favored over the non-flat Planck (new) $P(q)$ model, with the non-flat Planck $P(q)$ model weakly favored over the non-flat new $P(q)$ model. The flat$+A_L$ model is positively favored over the flat one, the Planck $P(q)+A_L$ model is positively favored over the Planck $P(q)$ one, and the new $P(q)+A_L$ model is positively favored over the new $P(q)$ one.

In summary: P18 data and P18+non-CMB data both strongly disfavor the tilted flat $\Lambda$CDM model with $A_L =1$ relative to some of the tilted $\Omega_k < 0$ or varying $A_L$ options; P18+lensing data are largely agnostic; and P18+lensing+non-CMB data, P18+BAO data, and P18+BAO$^\prime$ data all positively favor the varying $A_L$ options over the $A_L=1$ cases.

\begin{table*}
\caption{$\log_{10} \mathcal{I}$ and tension ($\sigma$ and $p$) parameters for P18 data versus lensing data, P18 data versus BAO (BAO$^\prime$) data, P18 data versus non-CMB data, and P18+lensing data versus non-CMB data in the six tilted flat and non-flat $\Lambda$CDM models. Table  \ref{tab:Priors} lists the Our, Handley, and Handley+$\Omega_k$ priors.
}
{\scriptsize
\begin{ruledtabular}
\begin{tabular}{lccccccc}
\\[-1mm]                   & \multicolumn{7}{c}{Tilted flat $\Lambda$CDM model}     \\[+1mm]
\cline{2-8}\\[-1mm]
	Data:                  & P18 vs.\ lensing  & P18 vs.\ lensing   & P18 vs.\ lensing      & P18 vs.\ BAO  & P18 vs.\ BAO$^\prime$ & P18 vs.\ non-CMB & P18+lensing vs.\ non-CMB    \\[+1mm]
	Prior:		           & Our               & Handley            & Handley+$\Omega_k$    & Our           & Our                   & Our              & Our              \\[+1mm]
 \hline \\[-1mm]
 $\log_{10} {\mathcal I}$  & $1.240$           & $1.166$            & $\ldots$              & $0.132$       & $0.707$               & $0.296$          & $0.029$ \\[+1mm]
 $\sigma$                  & $0.718$           & $0.390$            & $\ldots$              & $1.533$       & $0.426$               & $1.749$          & $1.747$ \\[+1mm]
 $p$ (\%)                  & $47.3$            & $69.7$             & $\ldots$              & $12.5$        & $67.0$                & $8.03$           & $8.06$ \\[+1mm]
 \hline \hline \\[-1mm]
\\[-1mm]                   & \multicolumn{7}{c}{Tilted flat $\Lambda$CDM$+A_L$ model}     \\[+1mm]
\cline{2-8}\\[-1mm]
        Data:              & P18 vs.\ lensing  & P18 vs.\ lensing   & P18 vs.\ lensing      & P18 vs.\ BAO  & P18 vs.\ BAO$^\prime$ & P18 vs.\ non-CMB & P18+lensing vs.\ non-CMB   \\[+1mm]
        Prior:             & Our  & Handley    & Handley+$\Omega_k$ & Our                   & Our           & Our                   & Our              \\[+1mm]
 \hline \\[-1mm]
 $\log_{10} {\mathcal I}$  & $\ldots$          & $\ldots$           &  $\ldots$             & $0.286$       & $0.810$               & $1.033$          & $1.033$  \\[+1mm]
 $\sigma$                  & $\ldots$          & $\ldots$           &  $\ldots$             & $1.402$       & $0.371$               & $0.835$          & $0.774$ \\[+1mm]
 $p$ (\%)                  & $\ldots$          & $\ldots$           &  $\ldots$             & $16.1$        & $71.0$                & $40.4$           & $43.9$ \\[+1mm]
 \hline \hline \\[-1mm]
\\[-1mm]                         & \multicolumn{7}{c}{Tilted non-flat $\Lambda$CDM model [Planck $P(q)$]}     \\[+1mm]
\cline{2-8}\\[-1mm]
	Data:                  & P18 vs.\ lensing  & P18 vs.\ lensing   & P18 vs.\ lensing      & P18 vs.\ BAO  & P18 vs.\ BAO$^\prime$ & P18 vs.\ non-CMB & P18+lensing vs.\ non-CMB   \\[+1mm]
    Prior:                 & Our  & Handley    & Handley+$\Omega_k$ & Our                   & Our           & Our                   & Our                        \\[+1mm]
 \hline \\[-1mm]
 $\log_{10} {\mathcal I}$  & $-0.486$          & $-0.316$           & $-0.360$              & $-1.236$      & $-0.891$              & $-1.263$         & $0.297$  \\[+1mm]
 $\sigma$                  & $2.479$           & $2.411$            & $2.403$               & $3.000$       & $2.478$               & $3.005$          & $1.837$\\[+1mm]
 $p$  (\%)                 & $1.32$            & $1.59$             & $1.63$                & $0.270$       & $1.32$                & $0.265$          & $6.62$  \\[+1mm]
 \hline \hline \\[-1mm]
\\[-1mm]                   & \multicolumn{7}{c}{Tilted non-flat $\Lambda$CDM$+A_L$ model [Planck $P(q)$]}     \\[+1mm]
\cline{2-8}\\[-1mm]
	Data:                  & P18 vs.\ lensing  & P18 vs.\ lensing   &  P18 vs.\ lensing     & P18 vs.\ BAO  & P18 vs.\ BAO$^\prime$ & P18 vs.\ non-CMB & P18+lensing vs.\ non-CMB  \\[+1mm]
    Prior:                 & Our               & Handley            &  Handley+$\Omega_k$   & Our           & Our                   & Our              & Our                   \\[+1mm]
 \hline \\[-1mm]
 $\log_{10} {\mathcal I}$  & $\ldots$          & $\ldots$           &  $\ldots$             & $0.182$       & $0.847$               & $0.972$          & $1.641$  \\[+1mm]
 $\sigma$                  & $\ldots$          & $\ldots$           &  $\ldots$             & $1.460$       & $0.465$               & $0.793$          & $0.516$ \\[+1mm]
 $p$  (\%)                 & $\ldots$          & $\ldots$           &  $\ldots$             & $14.4$        & $64.2$                & $42.8$           & $60.6$ \\[+1mm]
 \hline \hline \\[-1mm]
\\[-1mm]                   & \multicolumn{7}{c}{Tilted non-flat $\Lambda$CDM model [new $P(q)$]}     \\[+1mm]
\cline{2-8}\\[-1mm]
	Data:                  & P18 vs.\ lensing  & P18 vs.\ lensing   & P18 vs.\ lensing      & P18 vs.\ BAO  & P18 vs.\ BAO$^\prime$ & P18 vs.\ non-CMB & P18+lensing vs.\ non-CMB  \\[+1mm]
    Prior:                 & Our  & Handley    & Handley+$\Omega_k$ & Our                   & Our           & Our                   & Our                      \\[+1mm]
 \hline \\[-1mm]
 $\log_{10} {\mathcal I}$  & $-0.062$          & $-0.089$           & $-0.057$              & $-0.880$      & $-0.526$              & $-0.806$         & $0.143$  \\[+1mm]
 $\sigma$                  & $2.201$           & $1.887$            & $1.843$               & $2.604$       & $2.108$               & $2.577$          & $1.886$ \\[+1mm]
 $p$  (\%)                 & $2.77$            & $5.91$             & $6.54$                & $0.922$       & $3.50$                & $0.996$          & $5.93$ \\[+1mm]
 \hline \hline \\[-1mm]
\\[-1mm]                   & \multicolumn{7}{c}{Tilted non-flat $\Lambda$CDM$+A_L$ model [new $P(q)$]}     \\[+1mm]
\cline{2-8}\\[-1mm]
	Data:                  & P18 vs.\ lensing  & P18 vs.\ lensing   & P18 vs.\ lensing      & P18 vs.\ BAO  & P18 vs.\ BAO$^\prime$ & P18 vs.\ non-CMB & P18+lensing vs.\ non-CMB   \\[+1mm]
    Prior:                 & Our               & Handley            & Handley+$\Omega_k$    & Our           & Our                   & Our              & Our                \\[+1mm]
 \hline \\[-1mm]
 $\log_{10} {\mathcal I}$  & $\ldots$          & $\ldots$           & $\ldots$              & $1.066$       & $1.655$               & $1.798$          & $1.500$ \\[+1mm]
 $\sigma$                  & $\ldots$          & $\ldots$           & $\ldots$              & $1.052$       & $0.145$               & $0.402$          & $0.573$ \\[+1mm]
 $p$  (\%)                 & $\ldots$          & $\ldots$           & $\ldots$              & $29.3$        & $88.4$                & $68.7$           & $56.7$ \\[+1mm]
\end{tabular}
\\[+1mm]
\begin{flushleft}
Note: The statistical estimator values in the tilted flat $\Lambda$CDM model for the Handley+$\Omega_k$ priors are the same as for the Handley priors because $\Omega_k = 0$ in the flat model.
\end{flushleft}
\end{ruledtabular}
}
\label{tab:para_sigmap}
\end{table*}

\subsection{Data set tensions}
\label{subsec:data_set_tensions}

In this subsection we check whether there is concordance (discordance) between pairs of some of the data sets we study (in the context of a given cosmological model), as well as whether or not this concordance (discordance) is model independent. To do this, we use the two Sec.\ \ref{sec:method} statistical estimators, in eq.\ \eqref{eq:Tension_estimator_1} and in eqs.\ \eqref{eq:Tension_estimator_2} and  \eqref{eq:Tension_estimator_2_sigma}. The values of these statistical estimators for the six tilted flat and non-flat $\Lambda$CDM ($+A_L$) models are listed in Table \ref{tab:para_sigmap}; we do not compute these estimators in the untilted non-flat $\Lambda$CDM model which does not include the tilt ($n_s$) degree of freedom that is strongly favored by data. As in Sec.\ \ref{subsec:model_selection}, here we only study pairs of data sets in which one of the data sets is or includes the P18 data set. Conclusions based on either of the two statistical estimators qualitatively agree, for the five pairs of data sets we compare in this subsection, as discussed next.

\begin{table*}
\caption{Mean and 68.3\% confidence limits of tilted flat and non-flat $\Lambda\textrm{CDM}$ model parameters constrained by lensing data alone. Table  \ref{tab:Priors} lists the Our, Handley, and Handley+$\Omega_k$ priors. The Hubble constant $H_0$ has units of km s$^{-1}$ Mpc$^{-1}$.
}
{\tiny
\begin{ruledtabular}
\begin{tabular}{lccc}
\\[-1mm]                         & \multicolumn{3}{c}{Lensing data constraints with Our priors}     \\[+1mm] 
\cline{2-4}\\[-1mm]
	Parameter                    & Tilted flat $\Lambda$CDM   & Tilted non-flat $\Lambda$CDM [Planck $P(q)$] &  Tilted non-flat $\Lambda$CDM [new $P(q)$]    \\[+1mm]
 \hline \\[-1mm]
  $\Omega_b h^2$                 & $0.049 \pm 0.023$     & $0.052 \pm 0.027$      &  $0.048 \pm 0.026$  \\[+1mm]
  $\Omega_c h^2$                 & $0.125 \pm 0.032$     & $0.120 \pm 0.023$      &  $0.116 \pm 0.022$  \\[+1mm]
  $100\theta_\textrm{MC}$        & $1.016 \pm 0.022$     & $1.41 \pm 0.33$        &  $1.47 \pm 0.27$    \\[+1mm]
  $\tau$                         & $0.0542$              & $0.0483$               &  $0.0525$           \\[+1mm]
  $\Omega_k$                     & $\ldots$              & $-0.26 \pm 0.11$       &  $-0.279 \pm 0.095$ \\[+1mm]
  $n_s$                          & $0.9649$              & $0.9706$               &  $0.9654$           \\[+1mm]
  $\ln(10^{10} A_s)$             & $3.23 \pm 0.11$       & $3.10 \pm 0.19$        &  $3.13 \pm 0.16$    \\[+1mm]
 \hline \\[-1mm]
  $H_0$                          & $83 \pm 10$           & $65 \pm 17$            &  $66 \pm  16$        \\[+1mm]
  $\Omega_m$                     & $0.255 \pm 0.070$     & $0.54 \pm 0.48$        &  $0.48 \pm 0.36$  \\[+1mm]
  $\sigma_8$                     & $0.779 \pm 0.082$     & $0.85 \pm 0.16$        &  $0.88 \pm 0.15$  \\[+1mm]
 \hline \\[-1mm]
  $\chi_{\textrm{min}}^2$        & $3.67$                & $3.12$                 &  $3.38$             \\[+1mm]
  $\textrm{DIC}$ (lensing)       & $14.2$                & $13.3$                 &  $13.9$             \\[+1mm]
  $\textrm{AIC}$ (lensing)       & $13.7$                & $15.1$                 &  $15.4$             \\[+1mm]
 \hline \\[-1mm]
  $\textrm{DIC}$ (P18)           & $2817.9$              & $2810.6$               &  $2811.5$            \\[+1mm]
  $\textrm{DIC}$ (P18+lensing)   & $2826.5$              & $2826.2$               &  $2825.7$             \\[+1mm]
  $\log_{10} \mathcal{I}$        & $1.240$               & $-0.486$               &  $-0.062$            \\[+1mm]
 \hline \hline \\[-1mm]
                                 & \multicolumn{3}{c}{Lensing data constraints with Handley priors}        \\[+1mm]
\cline{2-4}\\[-1mm]
        Parameter                & Tilted flat $\Lambda$CDM   & Tilted non-flat $\Lambda$CDM [Planck $P(q)$]   &  Tilted non-flat $\Lambda$CDM [new $P(q)$]    \\[+1mm]
 \hline \\[-1mm]
  $\Omega_b h^2$                 & $0.0220 \pm 0.0018$   & $0.0221 \pm 0.0017$    &  $0.0220 \pm 0.0017$  \\[+1mm]
  $\Omega_c h^2$                 & $0.1121 \pm 0.0093$   & $0.1117 \pm 0.0099$    &  $0.1134 \pm 0.0097$  \\[+1mm]
  $100\theta_\textrm{MC}$        & $1.0397 \pm 0.0058$   & $1.0395 \pm 0.0058$    &  $1.0395 \pm 0.0059$    \\[+1mm]
  $\tau$                         & $0.21 \pm 0.11$       & $0.20 \pm 0.11$        &  $0.21 \pm 0.11$           \\[+1mm]
  $\Omega_k$                     & $\ldots$              & $-0.032 \pm 0.040$     &  $-0.029 \pm 0.040$ \\[+1mm]
  $n_s$                          & $0.957 \pm 0.043$     & $0.954 \pm 0.043$      &  $0.939 \pm 0.033$           \\[+1mm]
  $\ln(10^{10} A_s)$             & $3.26 \pm 0.15$       & $3.20 \pm 0.16$        &  $3.21 \pm 0.16$    \\[+1mm]
 \hline \\[-1mm]
  $H_0$                          & $69.7 \pm 3.9$        & $62 \pm 14$            &  $63 \pm  14$        \\[+1mm]
  $\Omega_m$                     & $0.281 \pm 0.050$     & $0.40 \pm 0.15$        &  $0.39 \pm 0.15$  \\[+1mm]
  $\sigma_8$                     & $0.869 \pm 0.064$     & $0.826 \pm 0.083$      &  $0.836 \pm 0.084$  \\[+1mm]
 \hline \\[-1mm]
  $\chi_{\textrm{min}}^2$        & $6.81$                & $6.89$                 &  $6.79$             \\[+1mm]
  $\textrm{DIC}$ (lensing)       & $13.9$                & $14.1$                 &  $13.8$             \\[+1mm]
  $\textrm{AIC}$ (lensing)       & $20.8$                & $22.9$                 &  $22.8$             \\[+1mm]
 \hline \\[-1mm]
  $\textrm{DIC}$ (P18)           & $2817.9$              & $2810.6$               &  $2811.5$            \\[+1mm]
  $\textrm{DIC}$ (P18+lensing)   & $2826.5$              & $2826.2$               &  $2825.7$             \\[+1mm]
  $\log_{10} \mathcal{I}$        & $1.166$               & $-0.316$               &  $-0.088$             \\[+1mm]
 \hline \hline \\[-1mm]
                                 & \multicolumn{3}{c}{Lensing data constraints with Handley$+\Omega_k$ priors }        \\[+1mm]
\cline{2-4}\\[-1mm]
        Parameter                & Tilted flat $\Lambda$CDM   & Tilted non-flat $\Lambda$CDM [Planck $P(q)$]   &  Tilted non-flat $\Lambda$CDM [new $P(q)$]    \\[+1mm]
 \hline \\[-1mm]
  $\Omega_b h^2$                 & $\cdots$             & $0.0221 \pm 0.0017$    &  $0.0221 \pm 0.0017$  \\[+1mm]
  $\Omega_c h^2$                 & $\cdots$             & $0.1088 \pm 0.0088$    &  $0.1104 \pm 0.0089$  \\[+1mm]
  $100\theta_\textrm{MC}$        & $\cdots$             & $1.0395 \pm 0.0058$    &  $1.0396 \pm 0.0059$    \\[+1mm]
  $\tau$                         & $\cdots$             & $0.20 \pm 0.11$        &  $0.20 \pm 0.11$           \\[+1mm]
  $\Omega_k$                     & $\cdots$             & $-0.123 \pm 0.095$     &  $-0.122 \pm 0.096$ \\[+1mm]
  $n_s$                          & $\cdots$             & $0.951 \pm 0.041$      &  $0.939 \pm 0.032$           \\[+1mm]
  $\ln(10^{10} A_s)$             & $\cdots$             & $3.11 \pm 0.16$        &  $3.11 \pm 0.16$    \\[+1mm]
 \hline \\[-1mm]
  $H_0$                          & $\cdots$             & $48 \pm 15$            &  $48 \pm  15$        \\[+1mm]
  $\Omega_m$                     & $\cdots$             & $0.70 \pm 0.33$        &  $0.71 \pm 0.33$  \\[+1mm]
  $\sigma_8$                     & $\cdots$             & $0.745 \pm 0.096$      &  $0.75 \pm 0.10$  \\[+1mm]
 \hline \\[-1mm]
  $\chi_{\textrm{min}}^2$        & $\cdots$             & $6.79$                 &  $6.77$             \\[+1mm]
  $\textrm{DIC}$ (lensing)       & $\cdots$             & $13.9$                 &  $13.9$             \\[+1mm]
  $\textrm{AIC}$ (lensing)       & $\cdots$             & $22.8$                 &  $22.8$             \\[+1mm]
 \hline \\[-1mm]
  $\textrm{DIC}$ (P18)           & $\cdots$             & $2810.6$               &  $2811.5$            \\[+1mm]
  $\textrm{DIC}$ (P18+lensing)   & $\cdots$             & $2826.2$               &  $2825.7$             \\[+1mm]
  $\log_{10} \mathcal{I}$        & $\cdots$             & $-0.360$               &  $-0.057$             \\[+1mm]
\end{tabular}
\\[+1mm]
\begin{flushleft}
Note: $\mathcal{I}=\exp(-\mathcal{F}/2)$ where $\mathcal{F}=\textrm{DIC(P18+lensing)}-\textrm{DIC(P18)}-\textrm{DIC(lensing)}$. The cosmological parameter values in the tilted flat $\Lambda$CDM model for the Handley+$\Omega_k$ priors are the same as for the Handley priors because $\Omega_k = 0$ in the flat model.
\end{flushleft}
\end{ruledtabular}
}
\label{tab:para_lensing}
\end{table*}

\begin{itemize}
\item 
{\bf P18 vs.\ lensing}. Since, as mentioned earlier, lensing data (see Sec.\ \ref{sec:data}) alone do not place significant constraints on cosmological parameters (even if we fix the values of some of them), the role played by the priors is more important in lensing data alone analyses than in other cases. Therefore, in this case, we use three different sets of priors (see Table \ref{tab:Priors}) in order to determine whether and how the lensing data alone cosmological parameter constraints and statistical estimator values depend on the priors used. In all three cases we report results obtained from converged chains. Due to the weak constraining power of lensing data alone, it is not possible to reach convergence when the $A_L$ parameter is allowed to vary. Consequently, we provide results only for the $A_L=1$ cases.

Here we first briefly comment on the lensing data alone cosmological parameter constraints, which do depend on the set of priors used, see Table \ref{tab:para_lensing}. For instance, if we look at the value of the curvature parameter $\Omega_k$ (which is most affected by the choice of prior) obtained by employing Our priors, for the titled non-flat $\Lambda$CDM Planck (new) $P(q)$ model, $\Omega_k= -0.26\pm 0.11$ ($\Omega_k=-0.279\pm 0.095$), we find a 1.9$\sigma$ (2.4$\sigma$) difference with the Handley priors analysis value $\Omega_k=-0.032\pm 0.040$ ($\Omega_k=-0.029\pm 0.040$) and a 0.94$\sigma$ (1.2$\sigma$) difference with the Handley+$\Omega_k$ priors analysis value $\Omega_k=-0.123\pm 0.095$ ($\Omega_k=-0.122\pm 0.096$). Reassuringly, we find that when we broaden the prior for $\Omega_k$, as we do when we move from Handley priors to Handley+$\Omega_k$ priors, the results get closer to those obtained with Our priors, the broadest priors we use. Additionally, our lensing data alone analysis (and cosmological parameter constraints) differ from those of the Planck team (Sec.\ 3.2.1 of Ref.\ \cite{Planck:2018lbu}) in that we fix $n_s$ and vary $\Omega_b h^2$ freely, whereas the Planck team use Gaussian priors for $n_s$ and $\Omega_b h^2$. Also, in our analysis $0.2 < h <1.0$ was chosen as the prior, while the Planck team used $0.4 < h < 1.0$. One notable difference is that when Our priors are used the value we find for $\Omega_b h^2$ is larger than the Gaussian prior value ($\Omega_b h^2 = 0.0222 \pm 0.0005$) adopted by the Planck team. In the tilted flat $\Lambda$CDM model we find $\Omega_b h^2=0.049 \pm 0.023$, and similar results are seen in the tilted non-flat models with the Planck and the new $P(q)$. However, when the Handley priors and the Handley+$\Omega_k$ priors are used, due to the very narrow range of $\Omega_b h^2$ (between $0.019$ and $0.025$) in these priors, such a deviation disappears, and $\Omega_b h^2$ is constrained with very consistent values in the tilted flat and the two tilted non-flat $\Lambda$CDM models and is also consistent with the Gaussian prior value adopted by the Planck team. Given the significant dependence on prior of the lensing data alone cosmological constraints, it is not possible to compare lensing data alone cosmological constraints to cosmological constraints we have derived from the other data sets. 

On the other hand, looking at Table \ref{tab:para_sigmap}, we do not see significant differences in the statistical estimator values from lensing only data analyses for the three different priors. This being the case, in the following, for the sake of consistency with our other discussions, we discuss only the lensing data alone results obtained using Our priors. 

For the tilted flat $\Lambda$CDM model we do not find discordance between P18 data and lensing data. We find $\textrm{log}_{10}\mathcal{I}=1.240$ which indicates a {\it strong} consistency between the two data sets. A similar conclusion is indicated by the other statistical estimator, $\sigma=0.718$ and $p=47.3\%$. We conclude that P18 and lensing data can be jointly analyzed in the context of the tilted flat $\Lambda$CDM model. 

Looking at the results for the tilted non-flat $\Lambda$CDM Planck $P(q)$ model, for the first statistical estimator $\textrm{log}_{10}\mathcal{I}=-0.486$ which is on the verge of indicating a {\it substantial} discordance while for the second one $\sigma=2.479$ and $p = 1.32\%$ which indicate a moderate tension. These results however may not be significant enough to conclude that P18 and lensing data cannot be used together in an analysis of the tilted non-flat $\Lambda$CDM Planck $P(q)$ model.  

In the tilted non-flat new $P(q)$ $\Lambda$CDM model, the two statistical estimators considered here point to somewhat different conclusions. While for the first one we get $\textrm{log}_{10}\mathcal{I}=-0.062$, which indicates neither consistency nor inconsistency between the two data sets, the second one, $\sigma=2.201$ and $p = 2.77\%$, indicates a moderate tension between the two data sets. Taken together these results indicate that there is at most moderate inconsistency between P18 and lensing data within the tilted flat new $P(q)$ $\Lambda$CDM model. 

\item 
{\bf P18 vs.\ BAO$^\prime$}. In the context of the tilted flat $\Lambda$CDM model there is no sign of discordance between these two data sets. We find $\textrm{log}_{10}\mathcal{I}=0.707$, which indicates a {\it substantial} consistency. The other statistical estimator points to a similar conclusion, with $\sigma=0.426$ and $p=67\%$. Very similar results are found for the tilted flat $\Lambda$CDM+$A_L$ model. The value $\textrm{log}_{10}\mathcal{I}=0.810$, once again, indicates a {\it substantial} consistency between P18 and BAO$^\prime$ data, whereas for the second estimator we find $\sigma=0.371$ and $p=71\%$. The P18 and BAO$^\prime$ data sets are mutually consistent and can be jointly analyzed in the tilted flat $\Lambda$CDM (+$A_L$) models. 

On the other hand, the opposite is true in the tilted non-flat $\Lambda$CDM models (with $A_L = 1$). The comparison of P18 and BAO$^\prime$ data in the tilted non-flat Planck $P(q)$ model results in $\textrm{log}_{10}\mathcal{I}=-0.891$ which indicates a {\it substantial} disagreement between these two data sets. Reassuringly the second statistical estimator points to the same conclusion, in particular, $\sigma =2.478$ and $p =1.32\%$. As expected  (see Sec.\ \ref{sec:P18_vs_BAO}) inclusion of the varying $A_L$ parameter reduces the tensions with respect to the $A_L=1$ case. For the Planck $P(q)$+$A_L$ model we find $\textrm{log}_{10}\mathcal{I}=0.847$, which indicates a {\it substantial} degree of consistency between the two data sets, and $\sigma=0.465$ and $p=64.2\%$, therefore, there is no tension between P18 data and BAO$^\prime$ data in this model. 

We noted in Sec.\ \ref{sec:P18_vs_BAO} that the tilted non-flat $\Lambda$CDM new $P(q)$ model better accommodates P18 and BAO$^\prime$ data than does the tilted non-flat $\Lambda$CDM Planck $P(q)$ model. In particular, in tilted non-flat $\Lambda$CDM new $P(q)$ model when $A_L=1$ we find $\textrm{log}_{10}\mathcal{I}=-0.526$, that is just in the range of {\it substantial} inconsistency. According to the values obtained for the other statistical estimator, $\sigma = 2.108$ and $p=3.50\%$, there is a moderate tension between the two data sets. The inclusion of a varying $A_L$ parameter in the analysis completely changes the conclusions with respect to the $A_L=1$ case. For the new $P(q)$+$A_L$ model we find $\textrm{log}_{10}\mathcal{I}=1.655$, indicating  {\it strong} agreement. The values $\sigma=0.145$ and $p=88.4\%$ support this conclusion. 

\item 
{\bf P18 vs.\ BAO}. We comment now on the results obtained when the tension between P18 data and BAO data is studied in the context of the different cosmological models. We note that the BAO data set includes some $f\sigma_8$ data points which, as we shall see, induces some changes in the results with respect to the P18 data and BAO$^\prime$ data case. 

Both statistical estimators do not indicate significant disagreement between P18 data and BAO data for the tilted flat $\Lambda$CDM model with $A_L=1$. For the first one we have $\textrm{log}_{10}\mathcal{I}= 0.132$, which neither indicates consistency nor inconsistency, and this is supported by the second one for which we obtain $\sigma=1.533$ and $p=12.5\%$. It is important to note that in this case the statistical estimators are closer to indicating a moderate tension than they are in the P18 data vs.\ BAO$^\prime$ data case. This is related to the previously mentioned $\sigma_8$ tension. We get similar results for the tilted flat $\Lambda$CDM+$A_L$ model, in which case we find $\textrm{log}_{10}\mathcal{I}=0.286$, which again neither indicates an agreement nor a disagreement, while for the second estimator $\sigma = 1.402$ and $p=16.1\%$, and again no tension is revealed. In view of these results we find no evidence that P18 and BAO data cannot be considered together in the analysis of the tilted flat $\Lambda$CDM (+$A_L$) models.

Given the P18 data vs.\ BAO$^\prime$ data comparison results in the tilted non-flat $\Lambda$CDM models, it should not come as a surprise that we find tensions when P18 data and BAO data are compared. In the tilted non-flat Planck $P(q)$ $\Lambda$CDM model with $A_L =1$ we find, for the first estimator $\textrm{log}_{10}\mathcal{I}=-1.236$, and $\sigma=3.000$ and $p=0.27\%$ for the second one. Both results indicate a {\it strong} inconsistency between the two data sets. This level of tension fades when the $A_L$ parameter is allowed to vary. For the Planck $P(q)$+$A_L$ model we obtain $\textrm{log}_{10}\mathcal{I}=0.182$, which does not indicate  consistency or inconsistency, and $\sigma=1.460$ and $p=14.4\%$. The P18 and BAO data can be jointly used in the Planck $P(q)$+$A_L$ model. As happens in the case of the P18 data vs.\ BAO$^\prime$ data comparison, the tilted non-flat new $P(q)$ $\Lambda$CDM model performs better than the Planck $P(q)$ when it comes to accommodating  the P18 and BAO data sets. For the $A_L=1$ case we find $\textrm{log}_{10}\mathcal{I}=-0.880$, revealing  {\it substantial} disagreement, while for the other estimator $\sigma=2.604$ and $p = 0.922\%$, which indicates a moderate tension. Once again, the tensions observed when $A_L=1$, in the context of non-flat models, disappear when this parameter is allowed to vary. For the tilted non-flat $\Lambda$CDM+$A_L$ new $P(q)$ model, we find $\textrm{log}_{10}\mathcal{I}=1.066$, which points out to a {\it strong} consistency between the two data sets, and for the other estimator we obtain $\sigma=1.052$ and $p=29.3\%$. The P18 and BAO data can be jointly used in the new $P(q)$+$A_L$ model.

In summary, in the tilted non-flat models, in the Planck $P(q)$ model P18 and BAO data should not be jointly analyzed unless the $A_L$ parameter is allowed to vary, while in the new $P(q)$ models these two data sets can be considered together to put constraints on the cosmological parameters even when $A_L =1$.

\item 
{\bf P18 vs.\ non-CMB}. We now discuss whether or not there is tension between P18 data and non-CMB data in the context of the different cosmological models. Similar results to the ones obtained in the P18 data and BAO$^\prime$/BAO data comparisons are expected, since BAO$^\prime$ data and BAO data are dominant components of non-CMB data. 

For the tilted flat $\Lambda$CDM model with $A_L = 1$ we find $\textrm{log}_{10}\mathcal{I}=0.296$, which neither indicates agreement nor disagreement, and $\sigma=1.749$ together with $p=8.03\%$, with neither of the two estimators pointing to tension between P18 and non-CMB data in this model. Including a varying $A_L$ in the model improves the agreement between the two data sets. For the tilted flat $\Lambda$CDM+$A_L$ model we find $\textrm{log}_{10}\mathcal{I}=1.033$ which points to {\it strong} consistency between the two data sets, and for the other estimator we get $\sigma=0.835$ and $p=40.4\%$, a result consistent with the first. There is no tension that prevents us from jointly analyzing P18 data and non-CMB data in the tilted flat $\Lambda$CDM (+$A_L$) models.

In the case of the tilted non-flat Planck $P(q)$ $\Lambda$CDM model with $A_L = 1$, the value $\textrm{log}_{10}\mathcal{I}=-1.263$ indicates a {\it strong} inconsistency between the P18 and non-CMB data sets. The second statistical estimator provides similar results, $\sigma = 3.005$ and $p=0.265\%$. In the light of these results, we conclude that P18 data and non-CMB data should not be jointly analyzed in the context of this tilted non-flat $A_L =1$ model. For the Planck $P(q)$+$A_L$ model, we get $\textrm{log}_{10}\mathcal{I}=0.972$, so {\it substantial} agreement is observed between P18 data and non-CMB data in this case. In agreement with the result obtained employing the first statistical estimator, for the second one we find $\sigma=0.793$ and $p=42.8\%$, which again does not indicate any tension. 

Once again the tilted non-flat $\Lambda$CDM new $P(q)$ model does better in jointly accommodating P18 and non-CMB data than does the tilted non-flat $\Lambda$CDM Planck $P(q)$ model. In the new $P(q)$ case with $A_L =1$,  the values obtained for both statistical estimators, $\textrm{log}_{10}\mathcal{I}=-0.806$ and $\sigma=2.577$ and $p=0.996\%$, indicate a {\it substantial} discordance between P18 data and non-CMB data in the context of this model. Allowing $A_L$ to vary reduces the tension found in the $A_L=1$ cases. For the new $P(q)$+$A_L$ model we get $\textrm{log}_{10}\mathcal{I}=1.798$, which points to a {\it strong} agreement between the two data sets, whereas for the second estimator we find $\sigma=0.402$ and $p=68.7\%$ and no tension. Therefore, we may say that in the context of the tilted non-flat $\Lambda$CDM (+$A_L$) new $P(q)$ models, P18 and non-CMB data can be jointly analyzed. 

\item 
{\bf P18+lensing vs.\ non-CMB}. In the previous cases we have detected some tensions in the context of the non-flat models. Here we study the possible disagreement between P18+lensing data and  non-CMB data. 

For the tilted flat $\Lambda$CDM model with $A_L =1$, both statistical estimators, with values $\textrm{log}_{10}\mathcal{I}=0.029$ and $\sigma=1.747$ and $p=8.06\%$, shed no light on a possible consistency or inconsistency between P18+lensing data and non-CMB data. For the tilted flat $\Lambda$CDM+$A_L$ model, we find $\textrm{log}_{10}\mathcal{I}=1.033$ which indicates a {\it strong} consistency between the two data sets. On the other hand, the second statistical estimator provides $\sigma=0.774$ and $p=43.9\%$, which do not indicate consistency or inconsistency. As we noted at the beginning of this subsection we do not always expect a perfect match in the conclusions from the two estimators. 

In the tilted non-flat $\Lambda$CDM Planck $P(q)$ model with $A_L=1$ we find $\textrm{log}_{10}\mathcal{I}=0.297$ which neither indicates consistency nor inconsistency between P18+lensing data and non-CMB data, whereas for the second estimator we find $\sigma=1.837$ and $p=6.62\%$, which does not reveal inconsistency. The consistency between P18+lensing and non-CMB data improves considerably in the context of the Planck $P(q)$+$A_L$ model. We get $\textrm{log}_{10}\mathcal{I}=1.641$ indicating a {\it strong} consistency between the two data sets, while the second one gives $\sigma=0.516$ and $p=60.6\%$, in agreement with the conclusion provided by the first estimator. 

Very similar conclusions are found for the new $P(q)$ (+$A_L$) and the Planck $P(q)$ (+$A_L$) models. When the $A_L$ parameter is not allowed to vary for the new $P(q)$ $A_L = 1$ model we find $\textrm{log}_{10}\mathcal{I}=0.143$ which does not reveal either an inconsistency or a consistency, with second estimator giving $\sigma =1.886$ and $p=5.927\%$ and again no tension is revealed. On the other hand, in the context of the new $P(q)$+$A_L$ model, we get $\textrm{log}_{10}\mathcal{I}=1.50$ indicating a {\it strong} consistency between the two data sets and reassuringly we find similar conclusions from the second statistical estimator, $\sigma=0.573$ and $p=56.7\%$. 

Unlike in the comparisons of P18 data and BAO$^\prime$/BAO data and the comparisons of P18 data and non-CMB data, we do not find tensions in the context of the non-flat models between P18+lensing data and non-CMB data, even when the $A_L$ parameter is not allowed to vary. This may be suggesting that if we want to jointly analyze P18 data and a low-redshift data set, such as BAO$^\prime$/BAO data or non-CMB data, we should either consider a varying $A_L$ parameter or include (P18) lensing data in the mix. 

\end{itemize}

We have studied the tensions between pairs of data sets, in the context of a given cosmological model, in three different ways based on Bayesian statistics. In Secs.\ \ref{sec:P18_vs_BAO}-\ref{sec:P18+lensing_vs_non-CMB} we quantified the level of tension by comparing the (one- and two-dimensional) cosmological parameter constraints favored by each of the pair of data sets. In the one-dimensional cases we estimated the tension by considering the quadrature sum of the two error bars for each parameter, while in the two-dimensional cases we looked at whether or not the two sets of contours shared a common parameter space area. In this subsection we study tensions between data set pairs by using the two more precise statistical estimators of Sec.\ \ref{sec:method}, see eq.\ \eqref{eq:Tension_estimator_1} and eqs.\ \eqref{eq:Tension_estimator_2} and  \eqref{eq:Tension_estimator_2_sigma}. Reassuringly, all three techniques employed result in similar conclusions in most cases.

Among all the data set comparisons we study there are two with significant enough discordances to be ruled out: we find in the tilted non-flat $\Lambda$CDM Planck $P(q)$ $A_L = 1$ model that P18 data and BAO data, as well P18 data and non-CMB data, are not mutually consistent. In the first case, when P18 and BAO data are compared, we observe a 2.7$\sigma$ tension between the derived cosmological parameter values of $\Omega_m$ and of $H_0$, obtained with P18 data and with BAO data. Additionally, in Fig.\ \ref{fig:like_NL_ns_BAO}, contour plot panels that contain one of these derived parameters show non-overlapping regions at more than 2$\sigma$. As for the P18 vs.\ non-CMB case, the tensions are even greater than for P18 vs.\ BAO. Comparing the derived cosmological parameter values of $\Omega_m$ and of $H_0$, obtained with P18 data and non-CMB data, we observe a disagreement at 2.9$\sigma$ and 3.9$\sigma$ respectively. Again the contour plot panels in Fig.\ \ref{fig:like_NL_ns_P18_nonCMB} containing $\Omega_m$ and/or $H_0$ show a non-overlapping region at more than 2$\sigma$. 

For the two statistical estimators of Sec.\ \ref{sec:method}, if we choose to say two data sets are mutually inconsistent (in a given model) when $\textrm{log}_{10}\mathcal{I}\leq -1$ or $\sigma\geq 3$, then this is true only in the two cases discussed in the previous paragraph. For the the tilted non-flat $\Lambda$CDM Planck $P(q)$ $A_L = 1$ model, in the P18 vs.\ BAO case we find $\textrm{log}_{10}\mathcal{I}=-1.236$ (meaning a {\it strong} disagreement between the two data sets) and $\sigma$ = 3.000 ($p=0.270\%$), while in the P18 vs.\ non-CMB analysis we find $\textrm{log}_{10}\mathcal{I}=-1.263$ (again a {\it strong} disagreement between the two data sets) and $\sigma$ = 3.005 ($p=0.265\%$). These results are qualitatively consistent with those of the previous paragraph. They mean that P18 data and BAO data, as well as P18 data and non-CMB data, cannot be jointly analyzed in this model, alternatively it means that the tilted non-flat $\Lambda$CDM Planck $P(q)$ $A_L = 1$ model is inconsistent with these data and ruled out at approximately 3$\sigma$, by them. We note that the level of tensions seen in the P18 vs.\ BAO and P18 vs.\ non-CMB comparisons are less severe in the context of the new $P(q)$ model, which does not strongly rule out the joint analyses of P18 data and BAO data, as well as P18 data and non-CMB data, in the tilted non-flat $\Lambda$CDM new $P(q)$ $A_L = 1$ model. What is more, none of the other combinations studied, namely, P18 data vs.\ lensing data, P18 data vs.\ BAO$^{\prime}$ data, and P18+lensing data vs.\ non-CMB data, are strongly mutually inconsistent in the tilted non-flat $\Lambda$CDM new $P(q)$ model, even when the $A_L$ parameter is not allowed to vary.

We now turn to a comparison between some of our results in Table \ref{tab:para_sigmap} and results presented in Refs.\ \cite{Handley:2019tkm} and \cite{DiValentino:2019qzk}. We emphasize that these are only semi-quantitative comparisons, since the data sets used are not identical and the priors used also differ.

Reference \cite{Handley:2019tkm} compares P18 data and lensing data, as well P18 data and BAO data (note that while we refer to both data sets as BAO there are some significant differences between the BAO data points used in Ref.\ \cite{Handley:2019tkm} and the updated BAO data we use here), in the tilted flat $\Lambda$CDM model and in the tilted non-flat $\Lambda$CDM Planck $P(q)$ model. As described in Sec.\ \ref{sec:method} we use the same ($p, \sigma$) statistical estimator as Ref.\ \cite{Handley:2019tkm} does and so these are the results we compare. For the tilted flat $\Lambda$CDM model, from the P18 vs.\ lensing analysis, Ref.\ \cite{Handley:2019tkm} Fig.\ 2 reports $\sigma \simeq 0.19$ and $p\simeq 85\%$, while we get $\sigma$=0.72 and $p=47\%$ (for Our priors) and $\sigma=0.39$ and $p = 70\%$ (for Handley priors). Some differences are expected due to the different set of data and priors used and this is reflected in these results. Reassuringly, when we employ the same priors for the lensing data (but not for P18 data) as used in Ref.\ \cite{Handley:2019tkm} the results get closer. From the P18 vs.\ BAO analysis in the tilted flat $\Lambda$CDM model Ref.\ \cite{Handley:2019tkm} finds $\sigma \simeq 0.95$ and $p\simeq 65\%$ while we get $\sigma = 1.5$ and $p = 13\%$, consequently the qualitative conclusions are the same, indicating that no tension is found. As for the tilted non-flat $\Lambda$CDM Planck $P(q)$ model, from the P18 vs.\ lensing analysis, Ref.\ \cite{Handley:2019tkm} reports $\sigma \simeq 2.5$ and $p\simeq 1.2\%$ and we find $\sigma=2.5$ and $p=1.3\%$ (for Our priors) and $\sigma=2.4$ and $p = 1.6\%$ (for Handley priors) so there is very good agreement between the results. Finally, in the tilted non-flat $\Lambda$CDM Planck $P(q)$ model, from a comparison of P18 data and BAO data, Ref.\ \cite{Handley:2019tkm} finds $\sigma \simeq 3.0$ and $p\simeq 0.3\%$ whereas we get $\sigma = 3.0$ and $p=0.3\%$. Considering all results, and the fact that somewhat different BAO data and priors are used in the two analyses, there is good agreement between the results and conclusions of Ref.\ \cite{Handley:2019tkm} and our results and conclusions.

Reference \cite{DiValentino:2019qzk} uses  $\textrm{log}_{10}\mathcal{I}$ to quantify tensions so here we compare our and their results for this statistical estimator. Reference \cite{DiValentino:2019qzk} compares P18 data and lensing data, as well as P18 and BAO$^{\prime}$ data (note that while we refer to both data sets as BAO$^{\prime}$ there are significant differences between the BAO$^{\prime}$ data used in Ref.\ \cite{DiValentino:2019qzk} and the updated BAO$^{\prime}$ data we use here), in the tilted flat $\Lambda$CDM model and in the tilted non-flat $\Lambda$CDM Planck $P(q)$ model. For the tilted flat $\Lambda$CDM model and the P18 data vs.\ lensing data analysis, Ref.\ \cite{DiValentino:2019qzk} find $\textrm{log}_{10}\mathcal{I}=0.6$ ({\it substantial} concordance) while we get $\textrm{log}_{10}\mathcal{I}= 1.24$ ({\it strong} concordance). For the P18 data vs.\ BAO$^{\prime}$ data analysis in the tilted flat $\Lambda$CDM model, Ref.\ \cite{DiValentino:2019qzk} report $\textrm{log}_{10}\mathcal{I}=0.2$ (neither a concordance nor a discordance) and we find $\textrm{log}_{10}\mathcal{I}=0.7$ ({\it substantial} concordance). On the other hand, in the tilted non-flat $\Lambda$CDM Planck $P(q)$ model, for the P18 vs.\ lensing data analysis, Ref.\ \cite{DiValentino:2019qzk} provide $\textrm{log}_{10}\mathcal{I}=-0.84$ ({\it substantial} discordance) while we obtain $\textrm{log}_{10}\mathcal{I}=-0.49$ which is on the verge of also indicating a {\it substantial} discordance between the two data sets. Finally, from the P18 vs.\ BAO$^{\prime}$ analysis
in the tilted non-flat $\Lambda$CDM Planck $P(q)$ model, Ref.\ \cite{DiValentino:2019qzk} report $\textrm{log}_{10}\mathcal{I}=-1.8$ ({\it strong} discordance) whereas we get $\textrm{log}_{10}\mathcal{I}=-0.89$ ({\it substantial} discordance). 

As can be appreciated from the preceding discussion, the agreement between our results and the results presented in Ref.\ \cite{DiValentino:2019qzk} is not as good as the one obtained from a comparison of our results and those of Ref.\ \cite{Handley:2019tkm}. It is important to note that the ($p, \sigma$) statistical estimator of eqs.\ \eqref{eq:Tension_estimator_2} and  \eqref{eq:Tension_estimator_2_sigma} is not as dependent on the priors as is the $\textrm{log}_{10}\mathcal{I}$ statistical estimator of eq.\ \eqref{eq:Tension_estimator_1}. This may explain the differences found in the comparisons of our results to those of Refs.\ \cite{Handley:2019tkm} and \cite{DiValentino:2019qzk}. All in all, we consider that there is reasonable, and so reassuring, agreement between our results and results available in the literature.

\section{Discussion}
\label{sec:discussion}

We have used P18 data, (P18) lensing data, BAO$^\prime$ data, BAO data, and non-CMB data to constrain cosmological parameters in eight cosmological models, the tilted flat $\Lambda$CDM (+$A_L$) model, the untilted non-flat $\Lambda$CDM (+$A_L$) model, the tilted non-flat $\Lambda$CDM (+$A_L$) Planck $P(q)$ model, and the tilted non-flat $\Lambda$CDM (+$A_L$) new $P(q)$ model, and to determine the goodness-of-fit of these models to the data sets. We have also used the models to examine whether or not pairs of data sets are mutually consistent, studying five cases: P18 data vs.\ lensing data, P18 data vs.\ BAO$^\prime$/BAO data, P18 data vs.\ non-CMB data, and P18+lensing data vs.\ non-CMB data. 

Assuming these data are correct and that there are no unaccounted systematic errors, three of the eight models we consider may be rejected because they are incompatible with some of these data at levels of significance discussed in Sec.\ \ref{sec:results} and summarized next. These rejected models are the two untilted non-flat $\Lambda$CDM (+$A_L$) models and the tilted non-flat $\Lambda$CDM Planck $P(q)$ model. 

When P18 data are included in the analyses the untilted non-flat $\Lambda$CDM (+$A_L$) models are, according to the DIC, {\it very strongly} disfavoured when compared with the tilted models. This is because the untilted models lack the degree of freedom encapsulated in the power spectrum tilt ($n_s$) parameter that is strongly favored by P18 data and so the untilted models are incompatible with P18 data. 

When we use the tilted non-flat $\Lambda$CDM Planck $P(q)$ model to compare cosmological parameter values from P18 data and BAO$^{\prime}$/BAO data, as well as from P18 data and non-CMB data, we find disagreements in the one-dimensional values of the $H_0$ and $\Omega_m$ derived parameters of 2.3$\sigma$ and 2.7$\sigma$ (BAO$^\prime$), 2.3$\sigma$ and 2.7$\sigma$ (BAO), and 2.9$\sigma$ and 2.9$\sigma$ (non-CMB). In Figs.\ \ref{fig:like_NL_ns_BAO} and \ref{fig:like_NL_ns_P18_nonCMB}, in those panels containing $H_0$ and $\Omega_m$, the two-dimensional contours do not overlap even at more than 2$\sigma$ significance. Additionally, in the P18 data vs.\ BAO data case we find $\textrm{log}_{10}\mathcal{I}=-1.236$ (meaning a {\it strong} disagreement between the two data sets) and $\sigma$ = 3.000 ($p=0.27\%$), while in the P18 data vs.\ non-CMB data analysis we get $\textrm{log}_{10}\mathcal{I}=-1.263$ (again a {\it strong} disagreement between the two data sets) and $\sigma$ = 3.005 ($p=0.265\%$). At their levels of significance, these results mean that the tilted non-flat $\Lambda$CDM Planck $P(q)$ model is unable to simultaneously accommodate P18 data and non-CMB data and so is ruled out at 3$\sigma$. Note that non-CMB data include BAO$^{\prime}$/BAO data and Refs.\ \citep{Handley:2019tkm, DiValentino:2019qzk} have previously noted the incompatibility of P18 data and older BAO$^{\prime}$/BAO data in the tilted non-flat $\Lambda$CDM Planck $P(q)$ model. We return to this point below.

The six-parameter tilted flat $\Lambda$CDM model is the simplest, (largely, see below) observationally consistent, general-relativistic cosmological model. It assumes the existence of cold dark matter, a non-evolving dark energy density $\Lambda$, flat spatial hypersurfaces ($\Omega_k = 0)$, and $A_L = 1$. This is the current standard cosmological model. We have found that this model passes all the consistency tests we use. The largest data set we have used is the P18+lensing+non-CMB data set. These data provide the most restrictive constraints on the parameters of this model, and if the tilted flat $\Lambda$CDM model is a reasonably good approximation of the Universe, the cosmological parameters values measured in this model from these data provide a reasonably good description of parameters of the Universe. From P18+lensing+non-CMB data we find, for the six primary cosmological parameters, $\Omega_b{h^2}=0.02250\pm 0.00013$, $\Omega_c{h^2}=0.11838\pm 0.00083$, $100\theta_{\textrm{MC}}= 1.04110\pm 0.00029$, $\tau=0.0569\pm 0.0071$, $n_s = 0.9688\pm 0.0036$, and $\ln(10^{10}A_s)= 3.046\pm 0.014$. We also provide the values of three derived parameters, $\Omega_m = 0.3053\pm 0.0050$, $H_0=68.09\pm 0.38$ km s$^{-1}$ Mpc$^{-1}$, and $\sigma_8 = 0.8072\pm 0.0058$. The least well-determined parameters are the reionization optical depth $\tau$ at 8.0$\sigma$ and the scalar spectral index $n_s$ which deviates from unity at 8.7$\sigma$. As we discuss below, the values of the cosmological parameters determined using any of the six tilted models we study are relatively independent of the cosmological model used, indicating that the values of the cosmological parameters listed above for the tilted flat $\Lambda$CDM model are relatively model independent.

It is interesting that the Hubble constant value measured using P18+lensing+non-CMB data in the tilted flat $\Lambda$CDM model, $H_0=68.09\pm 0.38$ km s$^{-1}$ Mpc$^{-1}$, is consistent with that from an early estimate from a median statistics analysis of a large compilation of Hubble constant measurements, $H_0=68\pm 2.8$ km s$^{-1}$ Mpc$^{-1}$, see Refs.\ \citep{ChenRatra2011, Gottetal2001, Calabreseetal2012}, as well as with some local measurements, e.g., $H_0=69.8\pm 1.7$ km s$^{-1}$ Mpc$^{-1}$ (quadrature sum of statistical and systematic uncertainties) from Ref.\ \citep{Freedman2021}, but not with some other local measurements, e.g, $H_0=73.04\pm 1.04$ km s$^{-1}$ Mpc$^{-1}$ from Ref.\ \citep{Riessetal2022}. 

As for the other derived parameter employed to quantify another tension affecting the tilted flat $\Lambda$CDM model, the $\sigma_8$ parameter, there are differences in its value depending on the data set considered. In the tilted flat $\Lambda$CDM model, using P18 data, we get $\sigma_8=0.8118\pm 0.0074$ whereas non-CMB data give $\sigma_8=0.787\pm 0.027$, with the two values differing by 0.89$\sigma$. In the P18+lensing+non-CMB data analysis case we obtain $\sigma_8=0.8072\pm 0.0058$ which is between the P18 value and the non-CMB value.

The shifts in the cosmological parameter values obtained by jointly analyzing non-CMB data with P18+lensing data, compared to the cosmological parameter values obtained from ``Planck'' P18+lensing data, for the tilted flat $\Lambda$CDM are: $-0.68\sigma$ ($\Omega_b{h^2}$), 1.1$\sigma$ ($\Omega_c{h^2}$), $-0.45\sigma$ (100$\theta_{\textrm{MC}}$), $-0.26\sigma$ ($\tau$), $-0.71\sigma$ ($n_s$),  $-0.10\sigma$ [$\ln(10^{10}A_s)$], $-1.1\sigma$  ($H_0$), 1.1$\sigma$  ($\Omega_m$), and 0.48$\sigma$ ($\sigma_8$), with the largest shifts being 1.1$\sigma$, suggesting again that in this model non-CMB data and P18+lensing data are not inconsistent. As for the reduction in the error bars obtained by jointly analyzing non-CMB data with P18+lensing data, compared to the error bars obtained from ``Planck'' P18+lensing data, we find 7.1$\%$ ($\Omega_b{h^2}$), 31$\%$ ($\Omega_c{h^2}$), 6.5$\%$ (100$\theta_{\textrm{MC}}$), 2.7$\%$ ($\tau$), 12$\%$ ($n_s$),  0$\%$ [$\ln(10^{10}A_s)$], 31$\%$  ($H_0$), 33$\%$  ($\Omega_m$), and 1.7$\%$ ($\sigma_8$), with the biggest reductions being the 33$\%$ $\Omega_m$ one and the 31$\%$ $\Omega_c{h^2}$ and $H_0$ ones; adding non-CMB data to the mix does quite significantly improve the constraints on some cosmological parameters.

We mentioned above that P18 data and non-CMB data are incompatible in the seven-parameter tilted non-flat $\Lambda$CDM Planck $P(q)$. When this model is used to analyze P18 data it favors a closed geometry at 2.5$\sigma$ with $\Omega_k = -0.043 \pm 0.017$, when it is used to analyze P18+lensing data it favors a closed geometry at 1.6$\sigma$ with $\Omega_k = -0.0103 \pm 0.0066$, and when it is used to analyze non-CMB data it favors a closed geometry at 0.63$\sigma$ with $\Omega_k = -0.032 \pm 0.051$. However, since P18 data and non-CMB data are incompatible in this model, the model is ruled out at the relevant levels of significance and so cannot be used to measure the geometry of spatial hypersurfaces from P18+lensing+non-CMB data.

On the other hand, the seven-parameter tilted non-flat $\Lambda$CDM new $P(q)$ model is not ruled out. According to the statistical estimators presented in Sec. \ref{sec:method} (see values in Table \ref{tab:para_sigmap}) for all the cases studied using the new $P(q)$ model, in none are our conditions to rule out a model, $\textrm{log}_{10}\mathcal{I}\leq -1$ or $\sigma\geq 3$, fulfilled. For the new $P(q)$ model, in the P18 data analysis, we find $\Omega_k=-0.033\pm 0.014$ which favors closed geometry at 2.4$\sigma$. When the new $P(q)$ model is used to analyze P18+lensing data the results indicate a 1.5$\sigma$ preference for closed geometry with $\Omega_k = -0.0086\pm 0.0057$, and when non-CMB data is analyzed alone we find $\Omega_k = -0.036\pm 0.051$ which is 0.71$\sigma$ in favor of  closed geometry. Contrary to what happens in the case of the Planck $P(q)$ model, in the new $P(q)$ model it is reasonable to jointly analyze P18 data, (P18) lensing data, and non-CMB data. And in the P18+lensing+non-CMB and P18+non-CMB analysis cases we obtain $\Omega_k = 0.0003\pm 0.0017$ favoring open geometry by only 0.18$\sigma$ in both cases. It may come as a surprise that even though each data set individually favors a closed geometry, some even with a somewhat significant level of evidence, the joint consideration of all three (or just two) of them reveals a result consistent with flat spatial hypersurfaces, and also more consistent with open than with closed geometry. This is because of the $H_0$-$\Omega_k$-$\Omega_m$ degeneracy and the fact that, in the non-flat models,  non-CMB data favor higher $H_0$ values and lower $\Omega_m$ values than do P18 data and P18+lensing data. 

We have found that with $A_L = 1$ the six-parameter untilted non-flat and the seven-parameter tilted non-flat $\Lambda$CDM Planck $P(q)$ models are incompatible with some data we consider. If these data are correct, these models are ruled out. On the other had, we find that the most restrictive data compilation we consider, the P18+lensing+non-CMB data set, indicates that the seven-parameter tilted non-flat $\Lambda$CDM new $P(q)$ model has flat (or very close to flat) spatial hypersurfaces. Yes, P18 data alone favor closed geometry at  2.4$\sigma$, and while it would be valuable to have a much better understanding of this result than is currently available, at this point we feel that the P18+lensing+non-CMB data support for flat geometry should be given more credence. Perhaps more and better future non-CMB might alter this conclusion, however current data are consistent with flat spatial hypersurfaces when $A_L = 1$. 

In the seven-parameter tilted flat $\Lambda$CDM$+A_L$ model $A_L$ is allowed to vary and is constrained by data. In this model P18 data favor $A_L = 1.181 \pm 0.067$, $A_L > 1$ at 2.7$\sigma$; P18+non-CMB data favor $A_L = 1.204 \pm 0.061$, $A_L > 1$ at 3.3$\sigma$; P18+lensing data favor $A_L = 1.073 \pm 0.041$, $A_L > 1$ at 1.8$\sigma$; and, P18+lensing+non-CMB data favor $A_L = 1.089 \pm 0.035$, $A_L > 1$ at 2.5$\sigma$. With P18+lensing+non-CMB data resulting in $\Delta{\rm DIC} = -5.55$ in favor of $A_L > 1$ over $A_L = 1$, just a little bit below the {\it strongly} favoring threshold of $-6$, the 2.5$\sigma$ $A_L > 1$ value indicates a more serious CMB weak lensing consistency issue than the preference for closed spatial geometry exhibited by some of the data sets. If these data are correct, these results are somewhat uncomfortable for the six-parameter tilted flat $\Lambda$CDM model --- the standard cosmological model. New, and better, data should help to clarify this issue. 

When $A_L$ is allowed to vary, the eight-parameter tilted non-flat $\Lambda$CDM+$A_L$ Planck $P(q)$ model is not ruled out by data sets incompatibilities, unlike what happens in the $A_L = 1$ seven-parameter tilted non-flat $\Lambda$CDM Planck $P(q)$ model. The eight-parameter tilted non-flat $\Lambda$CDM+$A_L$ new $P(q)$ model also does not suffer from data sets incompatibilities, similar to the $A_L = 1$ seven-parameter tilted non-flat $\Lambda$CDM new $P(q)$ model case. In the eight-parameter tilted non-flat $\Lambda$CDM$+A_L$ Planck (new) $P(q)$ model, P18 data favor $A_L = 0.88 \pm 0.15$ and $A_L < 1$ at 0.8$\sigma$ ($A_L = 0.94 \pm 0.20$ and $A_L < 1$ at 0.3$\sigma$) and $\Omega_k = -0.130 \pm 0.095$ and closed at 1.4$\sigma$ ($\Omega_k = -0.10 \pm 0.11$ and closed at 0.91$\sigma$); P18+non-CMB data favor $A_L = 1.203 \pm 0.062$ and $A_L > 1$ at 3.3$\sigma$ ($A_L = 1.204 \pm 0.061$ and $A_L > 1$ at 3.3$\sigma$) and $\Omega_k = -0.0006 \pm 0.0017$ and closed at 0.35$\sigma$ ($\Omega_k = -0.0006 \pm 0.0017$ and closed at 0.35$\sigma$); P18+lensing data favor $A_L = 1.089 \pm 0.16$ and $A_L > 1$ at 0.56$\sigma$ ($A_L = 1.13 \pm 0.15$ and $A_L > 1$ at 0.87$\sigma$) and $\Omega_k = -0.005 \pm 0.027$ and closed at 0.19$\sigma$ ($\Omega_k = 0.003 \pm 0.0016$ and open at 0.19$\sigma$); and, P18+lensing+non-CMB data favor $A_L = 1.090 \pm 0.036$ and $A_L > 1$ at 2.5$\sigma$ ($A_L = 1.088 \pm 0.035$ and $A_L > 1$ at 2.5$\sigma$) and $\Omega_k = -0.0002 \pm 0.0017$ and closed at 0.12$\sigma$ ($\Omega_k = -0.0002 \pm 0.0017$ and open at 0.12$\sigma$). With P18+lensing+non-CMB data in the eight-parameter tilted non-flat $\Lambda$CDM$+A_L$ Planck (new) $P(q)$ model resulting in  $\Delta{\rm DIC} = -5.22\ (-4.70)$, again (as in the seven-parameter tilted flat $\Lambda$CDM$+A_L$ model) {\it positively} favoring $A_L > 1$ over $A_L = 1$, there is a bit more evidence supporting the existence of a CMB weak lensing consistency issue, in all tilted, flat as well as non-flat, models, although the resulting $\Omega_k$ values in both non-flat cases are quite consistent with flat geometry. 

In the eight-parameter tilted non-flat $\Lambda$CDM$+A_L$ new $P(q)$ model, which unlike the Planck $P(q)$ model is not ruled out, allowing $A_L$ to vary reduces support for closed geometry. Compared to the seven-parameter new $P(q)$ model with $A_L = 1$, for P18 data, support for closed spatial hypersurfaces drops from 2.4$\sigma$ to 0.91$\sigma$, while for P18+lensing data the 1.5$\sigma$ support for closed geometry becomes 0.19$\sigma$ support for open geometry. We also note, from comparing P18 data results given in the two previous paragraphs for the seven-parameter tilted flat $\Lambda$CDM$+A_L$ model and for the eight-parameter tilted non-flat $\Lambda$CDM$+A_L$ Planck and new $P(q)$ models, as one goes from the first to either of the second models, $A_L$ values becomes consistent with unity while $\Omega_k$ values deviate from flat by only 1.4$\sigma$ and 0.91$\sigma$. So for P18 data both the tilted non-flat models cannot be ruled out while the seven-parameter tilted flat model with $A_L > 1$ at 2.7$\sigma$ and a lower DIC value indicates that the standard six-parameter tilted flat $\Lambda$CDM model with $A_L = 1$ is somewhat uncomfortably observationally squeezed. These and other results from our more comprehensive analyses and updated and more expansive data here support and extend the earlier results of Refs.\ \cite{Planck:2018vyg, Handley:2019tkm, DiValentino:2019qzk} that indicate that P18 data support either a closed geometry with $\Omega_k<0$ or $A_L>1$, both of which make the amount of CMB weak lensing higher than in the tilted flat $\Lambda$CDM model. We recall here the discussion in Sec.\ \ref{sec:intro} about the differences found in the values of the $A_L$ parameter from Planck data,  ACT CMB anisotropy data  \citep{ACT:2020gnv}, and from SPT CMB anisotropy data \cite{SPT:2017jdf}. Therefore, the possibility that CMB data employed in our paper are not completely correct remains open.

References \cite{Planck:2018vyg, Handley:2019tkm, DiValentino:2019qzk} have also noted that in the tilted non-flat Planck $P(q)$ model, when P18 data and BAO$^\prime$/BAO data are jointly analyzed, evidence for closed geometry dissipates, as we have found here for updated BAO$^\prime$/BAO data as well as for non-CMB data (even though, as we have found here, P18 data and to a lesser extent BAO$^\prime$/BAO data and non-CMB data are all by themselves not inconsistent with closed geometry). References \cite{Handley:2019tkm, DiValentino:2019qzk} have suggested that this might be because of a problem (possibly undetected systematic errors) with BAO$^\prime$/BAO data (and so also with non-CMB data) and so these results (from combinations of these data and P18 data) should not be taken to mean that spatial hypersurfaces are flat. Along these lines, we note that Ref.\  \cite{Glanville:2022xes} present results from a full-shape analysis (instead of the compressed BAO and $f\sigma_8$ data points analysis here) of the 6dFGS, BOSS, and eBOSS catalogs and find $\Omega_k = -0.0041^{+0.0026}_{-0.0021}$ (see their Table 6) when P18 data (not exactly the same P18 data used here) are jointly analyzed with the full-shape galaxy sample data, which is still in favor of a closed geometry, contrary to the conclusions we present here. New and better data and improved analysis techniques will help to shed some light on this issue. 

It is useful to determine which of the data sets we use are able to set model-independent constraints on the cosmological parameter values. Here we only consider the P18, P18+lensing, P18+non-CMB, and P18+lensing+non-CMB data sets, as the other data sets we study have less constraining power. In our analyses here we consider only the six tilted models, flat and non-flat, with $A_L = 1$ and varying $A_L$. In order to determine whether the constraints are model independent, we compute the shifts in the cosmological parameter value between pairs of models and say that the cosmological constraints are model-independent if almost all the shifts are $<1\sigma$. 

Neither P18 data nor P18+lensing data are able to place model-independent constraints on the cosmological parameter values. In the case of P18 data, when we compare the flat model with the flat+$A_L$ model, we observe disagreements in the values of the derived parameter $H_0$, $\Omega_m$, and $\sigma_8$ at $\sim 1\sigma$ confidence level. More significant are the discrepancies found when the flat model is compared with the tilted non-flat models. In particular for the Planck (new) $P(q)$ models, we get for $H_0$ a shift of $-3.5\sigma$ ($-2.8\sigma$), for $\Omega_m$ a shift of 2.6$\sigma$ (2.3$\sigma$), and for $\sigma_8$ a shift of $-2.2\sigma$ ($-1.7\sigma$). As expected, when the flat model is compared with the tilted non-flat models with varying $A_L$ the differences are smaller, though still significant. Comparing the flat model cosmological parameter values with the Planck (new) $P(q)$+$A_L$ cosmological parameter values we find for $H_0$ a shift of $-2.0\sigma$ ($-1.2\sigma$), for $\Omega_m$ a shift of 1.4$\sigma$ (0.89$\sigma$), and for $\sigma_8$ a shift of $-1.7\sigma$ ($-1.1\sigma$). Similar results are found when the flat+$A_L$ model is compared with the tilted non-flat models with and without a varying $A_L$ parameter. On the other hand, we do not find significant disagreements when we compare the cosmological parameter values of the four tilted non-flat models, the Planck $P(q)$ (+$A_L$) and the new $P(q)$ (+$A_L$) models, with each other, with the shifts always remaining below 1$\sigma$. The joint consideration of P18 data and (P18) lensing data reduces the disagreements discussed above though it is not possible to claim that P18+lensing data impose model-independent constraints. In this case when the cosmological parameter constraints for the flat and the flat+$A_L$ models are compared, the largest disagreement found is $-1.1\sigma$ for $\sigma_8$. When the flat model is compared with the Planck (new) $P(q)$ model, we get for $H_0$ a shift of  $-1.5\sigma$ ($-1.5\sigma$), for $\Omega_m$ a shift of 1.4$\sigma$ (1.3$\sigma$), and for $\sigma_8$ a shift of $-1.2\sigma$ ($-1.1\sigma$), while when the flat model is compared with the Planck (new) $P(q)$+$A_L$ model, all differences remain $<1\sigma$. When we compare the cosmological parameter values obtained for the flat+$A_L$ model with those obtained for the Planck (new) $P(q)$ model, we observe disagreements at $-1.9\sigma$ ($-1.9\sigma$) for $H_0$ and 1.8$\sigma$ (1.8$\sigma$) for $\Omega_m$. As happens in the P18 analysis, in the P18+lensing analysis no significant differences are observed when we compare the Planck $P(q)$ (+$A_L$) and new $P(q)$ (+$A_L$) models with each other. 

It is the inclusion of non-CMB data which results in model-independent constraints. When P18 data are jointly analyzed with non-CMB data we do not find discrepancies $>1\sigma$. The most important differences in this case, in absolute value, are 0.78$\sigma$-0.96$\sigma$ ($\Omega_b{h^2}$) and  0.87$\sigma$-0.98$\sigma$ ($\sigma_8$) that are found when the results for models with a varying $A_L$ parameter are compared with the results obtained when $A_L = 1$. In the P18+lensing+non-CMB data case almost no significant model-to-model discrepancies are found. The largest ones are found when the varying $A_L$ models are compared with those with $A_L=1$. In particular, the two largest shifts are in $\ln(10^{10}A_s)$ (the largest one being in absolute value 1$\sigma$) and in $\sigma_8$ (the largest one being in absolute value 1.3$\sigma$). We note that P18+non-CMB data cosmological parameter constraints are slightly more model-independent than those determined using P18+lensing+non-CMB data. This is partly because (P18) lensing data changes the $A_L$ parameter value, which in turn causes small shifts in some of the other parameter values. Consequently, when (P18) lensing data are included in the mix we observe larger differences between the cosmological parameter values of the varying $A_L$ models and those of the $A_L=1$ models. Also, P18+lensing+non-CMB cases error bars are smaller than the ones found in the P18+non-CMB analyses, and this contributes to increasing the significance of the differences in some of the cosmological parameter values in the P18+lensing+non-CMB cases. We may say that, as long as at least P18+non-CMB data are considered, if we start from the tilted flat $\Lambda$CDM and then vary $A_L$ and/or $\Omega_k$ (which implies the consideration of one of the non-flat $P(q)$s we have used in this work), we obtain model-independent constraints as a result, since the shifts in the cosmological parameter values remain within or just slightly above 1$\sigma$. In light of these results we can conclude that the P18+lensing+non-CMB data set is powerful enough to result in model-independent cosmological parameter constraints and, if these data are correct and include all systematic errors, this data set is able to accurately measure these parameters of the (reasonably accurate tilted flat $\Lambda$CDM approximation of the) real Universe.

\section{Conclusion}
\label{sec:conclusion}

In what follows we summarize our main conclusions.

If the data sets we use are correct and free from unknown systematics, three of the eight cosmological models are ruled out due to incompatibilities with some of the data sets employed in the analyses. The untilted non-flat $\Lambda$CDM (+$A_L$) models are unable to properly fit the P18 data while the tilted non-flat $\Lambda$CDM Planck $P(q)$ model is ruled out at 3$\sigma$ because it is not able to simultaneously accommodate P18 data and non-CMB (or some subset of these) data. 

Interestingly, the new $P(q)$ tilted non-flat inflation $\Lambda$CDM cosmological model, characterized by the primordial power spectrum in Eq.\ (\ref{eq:tilted_nonflat_new_PS}), does better than the Planck $P(q)$ model in being able to simultaneously accommodate P18 data and non-CMB data. In Sec.\ \ref{subsec:data_set_tensions} we study the mutual compatibility of pairs of data sets and in none of the cases studied is the level of tension high enough to rule out this model. The same holds true for the flat (+$A_L$) models and the Planck and new $P(q)$+$A_L$ models. 

P18 data do not break the geometrical $\Omega_m$-$H_0$-$\Omega_k$-$A_L$ degeneracy present in the Planck and the new $P(q)$ (+$A_L$) models. In the tilted non-flat $\Lambda$CDM new $P(q)$ model the P18 data analysis reveals a 2.4$\sigma$ evidence in favor of closed geometry with $\Omega_k=-0.033\pm 0.014$ and this model is {\it strongly} favored over the tilted flat $\Lambda$CDM model. In the tilted non-flat models when the $A_L$ parameter is allowed to vary the evidence in favor of closed geometry subsides yet they are either {\it strongly} favored (Planck $P(q)$+$A_L$) or {\it positively} favored (new $P(q)$+$A_L$) over the tilted flat model. The tilted flat $\Lambda$CDM+$A_L$ model better fits P18 data, compared to the tilted flat $\Lambda$CDM model fit, with an $A_L$ parameter value 2.7$\sigma$ larger than the theoretically expected value of $A_L=1$. These results update and strengthen those presented in Refs.\ \cite{Handley:2019tkm, DiValentino:2019qzk}; both options $\Omega_k<0$ and $A_L>1$ appear more indicative of a CMB weak lensing consistency issue.

The joint consideration of P18 data and (P18) lensing data does not result in significant changes in the values of most primary cosmological parameters with respect to those from the P18 data alone analysis, the exceptions being $\Omega_k$ and $A_L$. From P18+lensing data in the seven parameter tilted non-flat new $P(q)$ model we find 1.5$\sigma$ evidence in favor of closed geometry with $\Omega_k=-0.0086\pm 0.0057$, while in the seven-parameter tilted flat $\Lambda$CDM+$A_L$ model we find that $A_L>1$ is favored by 1.8$\sigma$  with $A_L=1.073\pm 0.041$. In these single parameter extensions of the tilted flat $\Lambda$CDM model, the addition of (P18) lensing data to P18 data does not favor $\Omega_k<0$ over $A_L>1$ or vice-versa. However, in the eight-parameter tilted non-flat Planck (new) $P(q)$ $\Lambda$CDM+$A_L$ models we find from P18+lensing data that $\Omega_k=-0.005\pm 0.027$ closed at 0.19$\sigma$ ($\Omega_k=0.003\pm 0.016$ open at 0.19$\sigma$), and $A_L=1.09\pm 0.16$ ($A_L=1.13\pm 0.15$) favoring $A_L>1$ at 0.56$\sigma$ (0.87$\sigma$), highlighting, if anything, the CMB weak lensing consistency issue. On the other hand, the values of the derived parameters $\Omega_m$ and $H_0$ are greatly affected by the inclusion of lensing data, and the geometrical degeneracy, when $A_L=1$, is partially broken. According to the DIC values, P18+lensing data do not strongly discriminate between models. The two statistical estimators ($\log_{10} {\mathcal I}$ and $\sigma$) tell us that there is only moderate tensions between P18 data and lensing data in the tilted non-flat models, and even less tension in the tilted flat model. 

Comparing the constraints from P18 data and non-CMB data allows for a robust test of the consistency of cosmological parameter values determined from high- and low-redshift data, respectively. For these data, the statistical estimators we consider do not show tensions between P18 data and non-CMB data, in the tilted flat model and in the varying $A_L$ models. Also, in the new $P(q)$ model with $A_L = 1$ we find $\textrm{log}_{10}\mathcal{I}=-0.806$ and $\sigma =2.577$ which indicates a non-negligible tension between P18 data results and non-CMB data results, but this is not high enough to rule out this model. No significant evidence is found in favor of non-flat hypersurfaces within the non-flat models. On the other hand, when the $A_L$ parameter is allowed to vary, the $A_L>1$ option is strongly preferred over the $A_L=1$ one. From P18+non-CMB data, for the flat+$A_L$ model we get $A_L=1.201\pm 0.061$ (3.3$\sigma$), for the Planck $P(q)$+$A_L$ model we find $A_L=1.203\pm 0.062$ (3.3$\sigma$), and for the new $P(q)$+$A_L$ model we obtain $A_L=1.204\pm 0.061$ (3.3$\sigma$). 

Amongst the data sets we consider in this paper, the P18+lensing+non-CMB data set provides the tightest constraints on cosmological parameters, and pins down the cosmological parameter values of the standard tilted flat $\Lambda$CDM model with impressive precision. (We emphasize that in most of the discussion in this paper we assume these data are accurate.) In fact, due to the great constraining power of this data set, almost all cosmological parameter values determined using this data set in the six tilted models considered are compatible at 1$\sigma$ (actually at slightly above 1$\sigma$ for the $\sigma_8$ parameter). Therefore we may say that the cosmological parameter values determined using P18+lensing+non-CMB data are very close to being model independent. From the P18+lensing+non-CMB analysis it is clear that the evidence in favor of $A_L>1$ remains while the evidence in favor of non-flat hypersurfaces subsides. We get $A_L=1.089\pm 0.035$ for the flat+$A_L$ model, $A_L=1.090\pm 0.036$ for the Planck $P(q)$+$A_L$ model, and $A_L=1.088\pm 0.035$ for the new $P(q)$+$A_L$ model, with a 2.5$\sigma$ deviation from $A_L=1$ in all cases.

It is interesting that the large (in absolute value) negative $\Omega_k$ values demanded by P18 data in order to deal with the lensing anomaly are not supported by non-CMB data (although the non-CMB data do mildly favor a closed geometry), and the larger $H_0$ and smaller $\Omega_m$ favored by non-CMB data (compared to those favored by P18 data) result in P18+lensing+non-CMB data favoring flat spatial hypersurfaces. This is at the heart of the tensions found, in the context of the tilted non-flat models, when comparing P18 data and BAO$^\prime$/BAO data cosmological parameter constraints and P18 data and non-CMB data constraints. It is interesting that the Hubble constant value measured using P18+lensing+non-CMB data in the tilted flat $\Lambda$CDM model, $H_0=68.09\pm 0.38$ km s$^{-1}$ Mpc$^{-1}$, is consistent with that from a median statistics analysis of a large compilation of Hubble constant measurements as well as with some local measurements. 

More and better cosmological data are needed in order to shed additional light on the issues studied in this paper. In the meantime the P18+lensing+non-CMB data set looks like the most reliable among all those considered and consequently we conclude that current observational data do not favor curved spatial geometry --- consistent with the standard tilted flat $\Lambda$CDM model --- but do favor $A_L>1$ and so somewhat uncomfortably squeeze the standard tilted flat $\Lambda$CDM model.

%
%
\acknowledgements
We thank H\'ector Gil-Mar\'in for useful discussions about BAO data. J.d.C.P.\ was supported by a FPI fellowship associated to the project FPA2016-76005-C2-1-P (MINECO). C.-G.P.\ was supported by National Research Foundation of Korea (NRF) grant funded by the Korea government (MSIT) (No.\ 2020R1F1A1069250). B.R.\ was supported by DOE grant DE-SC0011840.

\def\and{{and }}

\bibliography{references}


\end{document}